\documentclass[3p,11pt,sort&compress]{elsarticle}

\pdfoutput=1 

\journal{Physics Reports}

\usepackage{graphicx} %

\usepackage{amsmath,amsfonts,amssymb,slashed,float}
\usepackage{hyperref}
\usepackage[normalem]{ulem}
\usepackage{bbold,wasysym}
\usepackage{array,multirow}
\usepackage[utf8]{inputenc}
\usepackage{xspace}
\usepackage{placeins}
\usepackage{csquotes}
\usepackage{setspace}
\usepackage{tocloft}
\usepackage[noabbrev,capitalize]{cleveref}
\crefformat{figure}{Fig.~#2#1#3}
\usepackage{csvsimple} 

\hypersetup{colorlinks = true, linkcolor = blue, citecolor = blue,  urlcolor=blue}

\usepackage{caption}
\newcommand{\repeatcaption}[2]{%
  \renewcommand{\thetable}{\ref{#1}}%
  \captionsetup{list=no}%
  \caption{#2 (repeated from page \pageref{#1})}%
  \addtocounter{table}{-1}
}

\usepackage[usenames,dvipsnames]{xcolor} 
\definecolor{skobeloff}{rgb}{0.0, 0.48, 0.45}
\definecolor{tangerine}{rgb}{0.95, 0.52, 0.0}

\usepackage{soul}

\newcommand{\lcdm}{\ensuremath{\Lambda \text{CDM}}\xspace}
\newcommand{\eg}{e.g.~}
\newcommand{\ie}{i.e.~}
\newcommand{\lya}{\ensuremath{\mathrm{Lyman-}\alpha}}

\newcommand{\relpow}[3][]{\ensuremath{\left(\frac{#2}{#3}\right)^{#1}}}

\newcommand{\lcdmtable}{$\Lambda$CDM}
\newcommand{\Neffshort}{$\Delta N_{\rm ur}$}
\newcommand{\Neffmid}{free-streaming DR}
\newcommand{\Nefflong}{Free-streaming Dark Radiation}
\newcommand{\nefftable}{\Neffshort}
\newcommand{\idrshort}{SIDR}
\newcommand{\idrmid}{self-interacting DR}
\newcommand{\idrlong}{Self-Interacting Dark Radiation}
\newcommand{\idrtable}{\idrshort}
\newcommand{\Schmaltzshort}{DR-DM}
\newcommand{\Schmaltzmid}{self-interacting DR scattering on DM}
\newcommand{\Schmaltzlong}{Self-interacting Dark Radiation scattering on Dark Matter}
\newcommand{\schmaltztable}{\Schmaltzshort}
\newcommand{\Equishort}{mixed DR}

\newcommand{\Equilong}{Free-streaming plus self-interacting Dark Radiation}
\newcommand{\equitable}{\Equishort}
\newcommand{\SInushort}{SI$\nu$+DR}

\newcommand{\SInulong}{Self-interacting neutrinos plus free-streaming Dark Radiation}
\newcommand{\SINtable}{\SInushort}
\newcommand{\Majoronshort}{Majoron}
\newcommand{\Majoronmid}{Majoron}
\newcommand{\Majoronlong}{Majoron}
\newcommand{\majorontable}{\Majoronshort}
\newcommand{\Bshort}{primordial B}
\newcommand{\Bmid}{primordial magnetic field}
\newcommand{\Blong}{Recombination shifted by primordial magnetic field}
\newcommand{\primtable}{\Bshort}
\newcommand{\meshort}{varying $m_e$}
\newcommand{\memid}{varying effective electron mass}
\newcommand{\melong}{Recombination shifted by varying effective electron mass}
\newcommand{\varyingmetable}{\meshort}
\newcommand{\meOkshort}{varying $m_e$+$\Omega_k$}
\newcommand{\meOkmid}{varying effective electron mass in a curved universe}
\newcommand{\meOklong}{Recombination shifted by varying effective electron mass in a curved universe}
\newcommand{\varyingmeomegaktable}{\meOkshort}
\newcommand{\EDEshort}{EDE}
\newcommand{\EDEmid}{EDE}
\newcommand{\EDElong}{Early Dark Energy}
\newcommand{\edetable}{\EDEshort}
\newcommand{\NEDEshort}{NEDE}
\newcommand{\NEDEmid}{NEDE}
\newcommand{\NEDElong}{New Early Dark Energy}
\newcommand{\nedetable}{\NEDEshort}
\newcommand{\EMGshort}{EMG}
\newcommand{\EMGmid}{EMG}
\newcommand{\EMGlong}{Early Modified Gravity}
\newcommand{\emgtable}{\EMGshort}
\newcommand{\CPLshort}{CPL}
\newcommand{\CPLmid}{CPL DE}
\newcommand{\CPLlong}{Late Dark Energy with Chevallier-Linder-Polarski parametrization}
\newcommand{\cpltable}{\CPLshort}
\newcommand{\PEDEshort}{PEDE}
\newcommand{\PEDEmid}{Phenomenological Emergent DE}
\newcommand{\PEDElong}{Phenomenological Emergent Dark Energy}
\newcommand{\pedetable}{\PEDEshort}
\newcommand{\MEDEshort}{GPEDE}
\newcommand{\MEDEmid}{Generalized Phenomenological Emergent DE}
\newcommand{\MEDElong}{Generalized Phenomenological Emergent Dark Energy}
\newcommand{\medetable}{\MEDEshort}
\newcommand{\fracDMshort}{DM $\rightarrow$ DR}
\newcommand{\fracDMmid}{fraction of CDM decaying into DR}
\newcommand{\fracDMlong}{Fraction of Cold Dark Matter decaying into Dark Radiation}
\newcommand{\fracdmdecaytable}{\fracDMshort}
\newcommand{\DMDRWDMshort}{DM $\rightarrow$ DR+WDM}
\newcommand{\DMDRWDMmid}{CDM decaying into DR and WDM}
\newcommand{\DMDRWDMlong}{Cold Dark Matter decaying into Dark radiation and Warm Dark Matter}
\newcommand{\dmdecaytable}{\DMDRWDMshort}

\newcommand{\DA}{\ensuremath{\mathcal{D}_\mathrm{minimal}}\xspace}
\newcommand{\DB}{\ensuremath{\mathcal{D}_\mathrm{baseline}}\xspace}
\newcommand{\DCnoCMB}{\ensuremath{\mathcal{D}_\mathrm{no~CMB}}\xspace}
\newcommand{\DCnoP}{\ensuremath{\mathcal{D}_\mathrm{no~Planck}}\xspace}
\newcommand{\DD}{\ensuremath{\mathcal{D}_\mathrm{extended}}\xspace}
\usepackage[all]{nowidow}

\newcommand{\onlyprd}[1]{}

\begin{document}

\begin{frontmatter}

\title{The \texorpdfstring{$H_0$}{H0} Olympics: A fair ranking of proposed models}
\date{\today}

\author[1]{Nils Sch\"oneberg\corref{cor1}}
\cortext[cor1]{Corresponding author}
\ead{nils.science@gmail.com}

\author[2]{Guillermo Franco Abell\'an}
\ead{guillermo.franco-abellan@umontpellier.fr}

\author[1]{Andrea P\'erez S\'anchez}
\ead{andrea.perez@rwth-aachen.de}

\author[3]{Samuel J. Witte}
\ead{s.j.witte@uva.nl}

\author[2]{Vivian Poulin}
\ead{vivian.poulin@umontpellier.fr}

\author[1]{Julien Lesgourgues}
\ead{lesgourg@physik.rwth-aachen.de}

\address[1]{Institute for Theoretical Particle Physics and Cosmology (TTK), RWTH Aachen University, D-52056 Aachen, Germany.}
\address[2]{Laboratoire Univers \& Particules de Montpellier (LUPM), CNRS \& Universit\'e de Montpellier (UMR-5299),
Place Eug\`ene Bataillon, F-34095 Montpellier Cedex 05, France.}
\address[3]{GRAPPA Institute, Institute for Theoretical Physics Amsterdam and Delta Institute for Theoretical Physics, University of Amsterdam, Science Park 904, 1098 XH Amsterdam, The Netherlands}

\begin{abstract}
    Despite the remarkable success of the $\Lambda$ Cold Dark Matter ($\Lambda$CDM) cosmological model, a growing discrepancy has emerged (currently measured at the level of $\sim 4-6 \, \sigma$) between the value of the Hubble constant $H_0$ measured using the local distance ladder and the value inferred using the cosmic microwave background and galaxy surveys. While a vast array of $\Lambda$CDM extensions have been proposed to explain these discordant observations, understanding the (relative) success of these models in resolving the tension has proven difficult -- this is a direct consequence of the fact that each model has been subjected to differing, and typically incomplete, compilations of cosmological data. In this review, we attempt to make a systematic comparison of seventeen different models which have been proposed to resolve the $H_0$ tension (spanning both early- and late-Universe solutions), and quantify the relative success of each using a series of metrics and a vast array of data combinations. Owing to the timely appearance of this article, we refer to this contest as the \enquote{$H_0$ Olympics}; the goal being to identify which of the proposed solutions, and more broadly which underlying mechanisms, are most likely to be responsible for explaining the observed discrepancy (should unaccounted for systematics not be the culprit). This work also establishes a foundation of tests which will allow the success of novel proposals to be meaningfully \enquote{benchmarked}.
\end{abstract}


\end{frontmatter}


\setlength{\parskip}{0.5\baselineskip}
\newpage
\tableofcontents

\section{The Hubble tension circa July 2021}

The so-called \enquote{Hubble tension} refers to the inconsistency between local measurements of the current expansion rate of the Universe, \ie the Hubble constant $H_0$, and the value inferred from early-Universe data using the \lcdm model.
This tension is predominantly driven by the Planck collaboration's observation of the cosmic microwave background (CMB), which predicts a value in \lcdm of $H_0 = (67.27 \pm 0.60)$ km/s/Mpc \cite{Aghanim:2018eyx}, and the value measured by the SH0ES collaboration using the Cepheid-calibrated cosmic distance ladder, whose latest measurement yields $H_0 = (73.2\pm 1.3)$km/s/Mpc~\cite{Riess:2020fzl}. Taken at face value, these observations alone result in a $4.1\sigma$ tension.

The problem, however, is more severe than the na\"ive comparison between Planck and SH0ES may suggest. Today, there exist a variety of different techniques for calibrating \lcdm, and subsequently inferring the value of $H_0$, which do not involve Planck data -- for example, one can use alternative CMB data sets such as WMAP, ACT, or SPT, or one can remove observations of the CMB altogether and combine measurements of Big Bang nucleosynthesis (BBN) with data from baryonic acoustic oscillations (BAO) \cite{2019JCAP...10..029S,2013MNRAS.436.1674A,2015PhRvD..92l3516A,2018ApJ...853..119A,Blomqvist:2019rah,2019JCAP...10..044C} or with supernovae constraints \cite{2017MNRAS.467..731V,2021PhRvD.103j3533B,2015PhRvD..92l3516A}.\footnote{The robustness of such probes have been investigated for example in Refs.~\cite{2020MNRAS.494.2076C,2020PhRvD.102l3515B}.}
These probes, which invoke at least one measurement from high redshifts, are often dubbed \enquote{early-Universe calibrations}, and all result in $H_0$ values below $70$ km/s/Mpc (typically in strong agreement with the value inferred by Planck using \lcdm). 
Similarly, several alternative methods for directly measuring the local expansion rate have been proposed in the literature. A large number of these techniques offer alternative methods
for calibrating the cosmic distance ladder, removing any bias introduced from Cepheid observations. One example is the recent determination of $H_0$ obtained by the Chicago-Carnegie Hubble program (CCHP), which calibrates SNIa using the tip of the red giant branch (TRGB); this observation yielded a value of $H_0=(69.8 \pm 0.6~\mathrm{(stat)} \pm 1.6~\mathrm{(sys)})$ km/s/Mpc \cite{Freedman:2019jwv,Freedman:2021ahq}, in between the Planck CMB prediction and the SH0ES calibration measurement. However, alternative analyses using similar techniques have yielded values significantly closer to the value obtained by SH0ES, in particular the latest calibration of the TRGB using the parallax measurement of $\omega-$Centauri from GAIA DR3 leads to $H_0=(72.1 \pm2.0)$km/s/Mpc~\cite{Yuan:2019npk,Soltis:2020gpl}. Additional methods intended to calibrate SNIa at large distances include: surface brightness fluctuations of galaxies \cite{Khetan:2020hmh}, MIRAS \cite{Huang:2019yhh}, or the Baryonic Tully Fisher relation \cite{Schombert:2020pxm}. There also exists a variety of observations which do not rely on observations of SNIa -- these include e.g. time-delay of strongly lensed quasars \cite{Wong:2019kwg,Birrer:2020tax}, maser distances \cite{Pesce:2020xfe}, or gravitational waves as \enquote{standard sirens} \cite{Abbott:2019yzh}. While not all measurements are in tension with Planck, these direct probes tend to yield values of $H_0$ systematically larger than the value inferred by Planck\footnote{However, there is some debate about the robustness and independence of these additional measurements, see \eg Ref.~\cite{Freedman:2021ahq}.}. Depending on how one chooses to combine the various measurements, the tension may be elevated to as much as $6\sigma$ \cite{DiValentino:2021izs}. Intense experimental efforts are underway to establish whether this discrepancy can be caused by yet unknown systematic effects (appearing in either the early or late Universe measurements, or both). This includes (but is not limited to) issues in SNIa dust extinction modelling and intrinsic variations \cite{Mortsell:2021nzg,Mortsell:2021tcx}, the importance of Cepheid metallicity correction \cite{Efstathiou:2020wxn}, differences in TRGB calibration in the LMC \cite{Freedman:2019jwv,Freedman:2020dne,Yuan:2019npk} and in the Milky Way \cite{Cerny:2020inj,Soltis:2020gpl}, different population of SNIa at low-$z$ and high-$z$ \cite{Rigault:2014kaa,NearbySupernovaFactory:2018qkd,Jones:2018vbn,Brout:2020msh}, and the existence of a local void biasing $H_0$ estimates \cite{Lombriser:2019ahl,Kenworthy:2019qwq} (for a more complete review, see Ref.~\cite{DiValentino:2021izs}). Yet, the appearance of this discrepancy across a wide array of probes seems to suggest that a single systematic effect may not be sufficient to resolve this discrepancy. 

The alternative possibility is that the Hubble tension reflects a breakdown of the \lcdm model; properly accounting for new physics operating either in the early- or late-Universe could change the inference of $H_0$ from the \enquote{early-Universe} probes to be in agreement with the direct measurements. In order to understand how such a shift could occur within an extended cosmology, it is first useful to understand how $H_0$ is inferred from observations of the CMB. The key scales at play in the \lcdm prediction are the observed angular scale of sound horizon at recombination $\theta_s$\,, measured at {$\mathcal{O}(0.1\%)$} precision in CMB data, and the related angular scale of sound horizon at baryon drag $\theta_D$, measured at {$\mathcal{O}(1\%)$} precision within the latest BOSS data.\footnote{BAO also measure the redshift scale $H(z_s) r_s$ at a given survey redshift $z_s$\,, however the argument presented here holds since $H(z)$ and $1/D_A(z)$ are both directly proportional to $H_0$ if $H(z)/H_0$ is kept fixed.}
These angular scales are defined as $\theta_X\equiv r_X/D_A(z)$, where the numerator is either the sound horizon at recombination $r_s$ or sound horizon at baryon drag $r_D$, while the denominator is the angular diameter distance $D_A$ to recombination $z_*\simeq 1100$ or to the redshift of the survey $z_s \simeq 0.1-3$.  Given that the relationship between $r_s$ and $r_D$ is fixed within the \lcdm cosmology, it is common to say that one can calibrate BAO using the CMB sound horizon measurement, such that $r_s$ plays the role of a standard ruler that allows us to strongly constrain the angular diameter distance (and therefore the expansion rate) as a function of redshift.
The sound horizon is calculated as
\begin{equation}\label{eq:soundhorizon}
    r_s = \frac{1}{H_0}\int_{z_*}^\infty \frac{c_s dz}{H(z)/H_0}~,
\end{equation}
while the angular diameter distance is given by
\begin{equation}\label{eq:angdiamdistance}
    D_A(z_s) = \frac{1}{1+z_s} \cdot \frac{1}{H_0}\sin_K\left(\int_{0}^{z_s} \frac{dz}{H(z)/H_0}\right)~,
\end{equation}
with $\sin_K(x) = x$ in a flat universe, $\sin_K(x)=1/\sqrt{-\Omega_k} \cdot \sin(\sqrt{-\Omega_k} x)$ in a closed universe ($\Omega_k<0$), and finally $\sin_K(x)=1/\sqrt{\Omega_k} \cdot \sinh(\sqrt{\Omega_k} x)$ in an open universe ($\Omega_k>0$). 

There are now two ways to approach the question of how the angular scales measured by the CMB and BAO depend on the Hubble parameter. In a first approach, let's consider the physical densities $\rho_X$ contributing to the Hubble parameter $H(z)$ via the Friedmann law. One can immediately recognize that a change of the Hubble parameter today ($H_0$) exclusively impacts the density today, which is predominately dictated by dark energy; in other words, fixing all physical densities $\Omega_X h^2$, one observes that a change in $H_0$ is completely equivalent with a re-scaling of $\Omega_\mathrm{DE}$ (within the flat-\lcdm model). Thus, it would at first seem natural to attempt to increase the value of $H_0$ via a late-time modification of dark energy. However, since the sound horizon of \cref{eq:soundhorizon} is not impacted by $H_0$ in this view, one has to be careful to keep the angular diameter distance fixed in order not to change the given measured angular scales. This can be accomplished by requiring the energy density to be smaller in the past, in such a way so as to compensate for the higher energy density today. This is known as the \enquote{geometrical degeneracy} within CMB data. The problem for this class of solutions\footnote{A second problem, related to the \enquote{inverse-distance ladder} calibration of SNIa, will be discussed in more details later. We focus here on shortcomings of late-universe solutions that applies to all $H_0$ measurements.}, however, arises from the multitude of the low redshift measurements (such as supernovae, BAO, cosmic chronometers), which tightly constrain $H(z)/H_0$ in a redshift range of around $z \lesssim 2$. Keeping all of the angular diameter distances at redshifts from $\sim 1$ to $\sim 1000$ the same while varying $H_0$ is an exceptionally difficult task, especially since the integral involved in \cref{eq:angdiamdistance} is strongly dominated by the low redshift regime where $H(z) \sim H_0$ (since otherwise $H_0/H(z) \ll 1$). With this in mind, it is natural to instead consider solutions that modify the sound horizon, and to subsequently compensate any shift by varying $H_0$, which re-scales the angular diameter distance by a common factor for all probes. In other words, if the true $H_0$ were higher, all the observations would be physically closer (i.e. $D_A$ is smaller), but the intrinsic size of the acoustic pattern in the given model would appear smaller than the \lcdm prediction. This modification of the sound horizon can be accomplished by directly increasing the radiation density in the early Universe, by shifting the time of recombination, or by introducing a sharp increase of the expansion rate shortly before recombination. We categorize the models studied here depending on the mechanism via which they attempt to raise the inferred value of $H_0$; specifically, we consider models which alter the sound horizon via dark radiation (see \cref{sec:darkrad}), those which exploit an alternative early Universe mechanism that does not rely on dark radiation (see \cref{sec:early}), and the late-time solutions (see \cref{sec:late}).

There is an alternative approach to understanding how the CMB and BAO constrain the Hubble parameter. As we have just argued, most low redshift probes predominately constrain only the relative expansion law $H(z)/H_0$\,, not the absolute normalization of the expansion. From an observational perspective, this implies that these probes are sensitive to the fractional densities $\Omega_X=\rho_X/\rho_\mathrm{crit}$ rather than the physical densities $\rho_X$. When fixing $\Omega_X$, one finds that the explicit dependence on the parameter $H_0$ cancels out when computing the angular scales $\theta_x = r_s/D_A = (H_0 r_s)/(H_0 D_A)$; in this case, the dependence of the angular scales on $H_0$ is implicit, and arises subtly through the integrals. Due to our observation of the temperature of the CMB today ($T_0$), we can tightly constrain the physical density parameter $\omega_\gamma \propto T_0^4$. Instead, the fractional density $\Omega_\gamma = \omega_\gamma/h^2 \propto T_0^4/h^2$ is impacted by the Hubble parameter. From this perspective, it is clear that the impact of $h$ should, at least to some degree, be degenerate with the introduction of additional dark radiation (since they both directly alter $\Omega_r$). In addition, the impact of varying $H_0$ in this case appears primarily in the early time expansion integral ($H_0 r_s$), which is degenerate with any change of $r_s$ (which may appear \eg via shift in the recombination process). Moreover, from this perspective, in which the multitude of low redshift observations effectively fix the fractional density $\Omega_m$\,, the late time solutions are more difficult to justify. In other words, early solutions may directly correct for the change in $H_0 r_s$ so as keep the product constant, and thus do not require any modification to $H_0 D_A$, while late-time solutions must carefully tune $H(z)/H_0$ in order to ensure $H_0 D_A(z)$ is varied by exactly the same factor as $H_0 r_s$ at various different redshifts $z$.

\section{Condensed Summary}\label{sec:summary}
 
\noindent The primary goal of this work is to create a comprehensive and systematic comparison of the relative success of various proposed solutions to the $H_0$ tension -- the broader intent being to better understand the successes and drawbacks of each approach, and to generate a meaningful set of benchmarks for future proposals. The hasty reader may focus on this section for a brief overview of our approach and of the main findings. \\

\noindent {\bf Motivation.} The ever-growing tension between the value of $H_0$ (\ie the expansion rate of the Universe today) inferred from early Universe observations and the value measured locally  has been the focus of a great number of papers in the past decade. We refer the reader to Refs.~\cite{Verde:2019ivm,Riess:2020sih} for various reviews addressing different aspects of this tension, as well as Ref.~\cite{DiValentino:2021izs} for a summary of many of the proposed solutions. The aim of this paper is to ascertain the relative success of various cosmological models proposed to solve the $H_0$ tension; we do this by systematically confronting each of the considered models to data from the early and late Universe, assessing at each point the extent to which the tension between the Planck and local measurements remains. 

In particular, we will attempt to define \enquote{model success}, or equivalently tension reduction, by asking three distinct questions: Given a model, (i) to what extent does confrontation with data (without including a prior from local measurements) generate posteriors of $H_0$ (or intrinsic SNIa magnitude $M_b$) compatible with local measurements?; (ii) to what extent can one obtain a good fit to all combined data? and (iii) to what extent is that model favored over $\Lambda$CDM?
Answering these questions through a number of metrics (introduced later) leads to a fair comparison between the various models studied here, while providing a series of benchmarks for those wishing in the future to assess the quality of a given proposal.

\noindent {\bf Model-independent treatment of the local distance ladder measurement.} In this work, we focus on local measurements of $H_0$ using the cosmic distance ladder calibration of SNIa, such as the measurement by the SH0ES collaboration \cite{Riess:2016jrr,Riess:2018uxu,Riess:2019cxk,Riess:2020fzl}. These measurements do not directly determine the $H_0$ parameter. Instead, they calibrate the intrinsic magnitude of supernovae at redshifts of $0.02-0.15$ in order to infer the current expansion rate. 
Focusing on $H_0$ does not necessarily ensure that the predicted SNIa magnitudes are in agreement with what is observed. 
In particular, it was shown that for a wide range of smooth expansion histories at late-times -- assuming the \lcdm value of the sound horizon -- the \enquote{inverse distance ladder} calibration\footnote{The inverse distance ladder calibration of SNIa consists in making use of the determination of $D_A (z) \equiv D_L (z)/(1+z)^2$ as extracted from the BAO data that have been calibrated onto the CMB (or other determination of the sound horizon). The knowledge of the luminosity distance $D_L(0.1\lesssim z \lesssim 1)$, together with the (uncalibrated) SNIa magnitude from the Pantheon survey, allows for a determination of the intrinsic SNIa magnitude. This determination depends on the assumed cosmological model in the pre-recombination era.} of Pantheon SNIa data \cite{Scolnic:2017caz} cannot be made compatible with the direct calibration from SH0ES \cite{Benevento:2020fev,Camarena:2021jlr, Efstathiou:2021ocp}.
A more correct approach is to rather consider the calibration of the intrinsic magnitude $M_b$ from the local distance ladder, and to probe the consistency of the corresponding observed fluxes from the Pantheon catalogue with the underlying model -- this is the approach previously put forth in Refs.~\cite{Benevento:2020fev,Camarena:2021jlr, Efstathiou:2021ocp}. In this case, the value of $H_0$ is derived from the supernovae self-consistently within the given exotic expansion history. 
Consequently, the main conclusions of this work will be based on the inclusion of Pantheon supernovae data and we will assess the tension using $M_b$ rather than $H_0$, a point which is effectively irrelevant for early-Universe solutions but increasingly disfavors late-time solutions (consistent with the claims of Refs.~\cite{Benevento:2020fev, Camarena:2021jlr, Efstathiou:2021ocp}). Finally, while there is an abundant array of local $H_0$ measurements based on SNIa calibration \cite{Freedman:2019jwv,Freedman:2020dne,Yuan:2019npk,Freedman:2021ahq,Soltis:2020gpl,Huang:2019yhh}, we choose to focus exclusively on the latest result from the SH0ES collaboration, yielding $H_0 = 73.2 \pm 1.3$ km/s/Mpc~\cite{Riess:2020fzl}, as this is the data set most in tension with early Universe probes\footnote{When including the Pantheon dataset, we focus instead on $M_b=-19.2435\pm0.0373$ \cite{Riess:2020fzl}.}. Hence, a model significantly easing the tension with the SH0ES result is expected to also ease the tension with other local measurements.\\

\enlargethispage*{2\baselineskip}
\noindent {\bf Statistical approaches to quantify model success.} For each model $\mathcal{M}$ and compilation of data sets $\mathcal{D}$, we discuss three ways to quantify the tension between the cosmological inferred value and the SH0ES experiment, related to the questions (and definitions of model success) we introduced earlier. These tests will allow us to select a range of most successful models as a \enquote{finalist} sample, which we subsequently subject to further investigation. While there exist a wide array of distinct tests capable of addressing the aforementioned questions, we  choose these as they are (1) easy to understand, (2) easy to implement (thus allowing any reader interested to make comparisons using their favorite model), and (3) sufficiently accurate to allow for meaningful conclusions to be drawn. 
We outline the various approaches below:
\begin{enumerate}
    \item[\textbullet] Criterion 1: {\em When considering a data set $\mathcal{D}$ that does not include SH0ES, we ask, what is the residual level of tension between the posterior on $M_b$ or $H_0$ inferred using $\mathcal{D}$ and the SH0ES measurement? } 
    The tension on $x=H_0$ or $x=M_b$ can be quantified through the \enquote{rule of thumb difference in mean} \cite{Raveri:2018wln} or Gaussian Tension (GT), defined as
    \begin{equation}
    \frac{{\bar x}_{\mathcal{D}} - {\bar x}_{\rm SH0ES}}{({\sigma}_{\mathcal{D}}^2 + {\sigma}_{\rm SH0ES}^2)^{1/2}} \, ,
    \label{eq:crit_1}
    \end{equation}
    where $\overline{x}_i$ and $\sigma_i$ are the mean and standard deviation of observation $i$.
    When $\mathcal{D}$ includes (excludes) supernovae data, we quantify the tension on $M_b$ ($H_0$) using $M_b=-19.2435\pm 0.0373$  ($H_0=73.2\pm 1.3$ km/s/Mpc), with the uncertainties corresponding to the 68\%CL. 
The goal of this criterion is to answer question (i), namely, to quantify to what extent a model has a posterior compatible with a high $H_0$ given the data $\mathcal{D}$, independently of SH0ES, which represents perhaps the most optimistic definition of a successful model. 
Although this metric is only strictly valid if the parameter's posteriors are Gaussian, we adopt this \enquote{historical} approach since the reader might be most familiar with it.

We do note however a number of shortcomings: first, this may disfavor models with a probability distribution that deviates from Gaussian, e.g. due to the presence of long tails in the posterior\footnote{This is for instance the case of the \enquote{\EDElong{}} cosmologies.}. This can happen for instance if the data set $\mathcal{D}$ cannot disentangle between $\Lambda$CDM and a more complex model which has parameters that become irrelevant when others are close to their \lcdm limit.  As a result, the posterior is necessarily dominated by the Gaussian \lcdm limit, and the easing power of the model can only show up in the aforementioned tails of the probability distribution.
Secondly, and perhaps even more worryingly, this criterion does not quantify how good (or bad) the $\chi^2$ of the new model is. As a result, a model which does not contain the \lcdm best fit can appear arbitrarily good. 
One could, for instance, consider only \lcdm models with $\Omega_\mathrm{cdm} h^2$ fixed to $0.11$. In these models the posterior of $H_0$ is naturally centered around\footnote{We quote the numbers for the \DA data set (Planck 2018 TTTEEE + lensing + BAO).} $(71.84 \pm 0.16)\mathrm{km/s/Mpc}$ and the criterion would recognize this model as a good solution to the tension (1.0$\sigma$), which it is clearly not. Instead, the likelihood difference between this model and \lcdm is $\Delta \chi^2 \approx 106$ and clearly excludes the model. Less trivial examples of this can be found in the literature; for instance, in the case of interacting dark matter - dark radiation models, for which a theoretically motivated lower bound on the radiation density artificially induces higher values of $H_0$ \cite{Buen-Abad:2015ova,Lesgourgues:2015wza,Buen-Abad:2017gxg,Archidiacono:2019wdp}. 
Another example is that of \PEDElong{}~[\PEDEshort{}]~\cite{Li:2019yem,Pan:2019hac,Rezaei:2020mrj,Yang:2021egn}, which fixes the late-time expansion rate by hand so as to give the locally measured value of Hubble. 
In order to avoid such problems, we instead use the two additional tests listed below in order to identify successful models.

   \item[\textbullet] Criterion 2: {\em How does the addition of the SH0ES measurement to the data set $\mathcal{D}$ impact the fit within a particular model~$\mathcal{M}$?} 
    We compute the change in the effective best-fit chi-square $\chi^2=-2 \ln \mathcal{L}$ between the combined data set and the data set $\mathcal{D}$ as \begin{equation}
    \Delta \chi^2 = \chi^2_{\mathrm{min},\mathcal{D}+\mathrm{SH0ES}}-\chi^2_{\mathrm{min},\mathcal{D}}~,
    \label{eq:crit_2}
    \end{equation}
    where we use $\chi^2_{\mathrm{min},\mathrm{SH0ES}}=0$. 
    In the $\Lambda$CDM framework, the $\chi^2$ of the combined fit to $\mathcal{D}$+SH0ES is notably worse than the sum of the separate best-fitting $\chi^2$ to $\mathcal{D}$ and to SH0ES, reflecting the fact that the data sets are in tension.
    Since we are comparing the $\chi^2$ values within a given model, there is no change in the number of model parameters, and the tension can simply be expressed as Tension=$\sqrt{\Delta \chi^2}$ in units of $\sigma$. This tension metric is identical to the $Q_{\rm DMAP}$ (for \enquote{difference of the maximum a posteriori}) metric discussed in Ref.~\cite{Raveri:2018wln}. 

    This criterion attempts to answer the somewhat more modest question (ii) introduced above, namely, whether one can obtain a good fit to all data in a given model.
    Moreover, it naturally generalizes the commonly used criterion discussed in point 1 to the case of non-Gaussian posteriors. Indeed, for any Gaussian posterior, it is equivalent to that criterion.
    For \lcdm, for example, the contours are very Gaussian, and this generalization returns a tension of $4.5\sigma$ instead of $4.4\sigma$ in the case of criterion 1.
    However, the square root of \cref{eq:crit_2} is much better at capturing the long probability tails of the posteriors.\footnote{ 
    In practice,  in nearly all cases we find $\lesssim 10\%$ difference compared to criterion 1, with the exception of the \EDEmid{} and \NEDEmid{} models, which due to the highly non-Gaussian nature of their posterior perform better with this generalized criterion.} 
    Yet, this criterion has two potential problems: First, similarly to criterion 1, it does not quantify the intrinsic success of a model. 
    Second, it is sensitive to the effect of over-fitting (i.e. a model with arbitrarily large number of parameters could fit any features better than $\Lambda$CDM), which usually requires Bayesian methods to compute Occam's razor
    factors. For this reason, we also consider a third criterion, which attempts to quantify the intrinsic success of a model and to penalize overly complex models.
    

    \item[\textbullet] Criterion 3: {\em When the data set $\mathcal{D}$ includes the SH0ES likelihood (or, equivalently, a SH0ES driven prior on $M_b$ or $H_0$), does the fit within a particular model $\mathcal{M}$ significantly improve upon that of $\Lambda$CDM?}
    In order to assess the extent to which the fit is improved, we compute the Akaike Information Criterium (AIC) of the extended model $\mathcal{M}$ relative to that of \lcdm, defined as
    \begin{equation}
    \Delta {\rm AIC} = 
    \chi^2_{\mathrm{min},\mathcal{M}} - \chi^2_{\mathrm{min},\lcdm} + 2 (N_\mathcal{M}-N_{\lcdm})~, 
    \end{equation}
    where $N_\mathcal{M}$ stands for the number of free parameters of the model. Importantly, this metric attempts to penalize models which introduce new parameters that do not subsequently improve the fit.
    Thus, the ability of a model $\mathcal{M}$ to resolve the tension at a significant level despite having more parameters can be assessed through $\Delta$AIC, with more negative values indicating larger model success. While this metric does indicate whether a model is favored compared to \lcdm for the combined data set, it does not quantify whether this improvement stems from improving the Hubble tension or from simply fitting better other data sets such as Planck data. This criterion is thus especially useful if applied together with criterion 1 or 2 above.
    \enlargethispage*{3\baselineskip}
    The penalty of $2 N_\mathcal{M}$ to the $\Delta$AIC is certainly not perfect. 
   It can be circumvented by simply fixing model parameters to their best-fit in order to reduce the penalty. 
   Furthermore, it is not always clear that it correctly estimates and penalises the effect of over-fitting. In fact, several better Bayesian estimators have been proposed in the literature \cite{Raveri:2018wln,Raveri:2021wfz,schwarz1978estimating,2013PDU.....2..166V}, including the Bayes factor ratio \cite{Kass:1995loi}, or the related \enquote{suspiciousness} \cite{Handley:2019wlz}. 
   These approaches are usually computationally expensive (unless one uses the Savage-Dickey density ratio \cite{Trotta:2005ar}), and the result depends on the choice of priors on the parameters of the extended model. The $\Delta$AIC criterion offers the advantage of being numerically cheap and prior independent.
   In the spirit of setting easy benchmarks, the $\Delta$AIC applies a penalty that is particularly simple to apply to any model and that certainly provides an approximate metric to gauge the relevance of extra parameters. We leave an investigation of model comparison based on aforementioned Bayesian statistics to future work.
\end{enumerate}
For these criteria, we frequently use minimized $\chi^2$ values. Our minimization approach is further discussed in \ref{app:E}.

In the case of Test 1 (Gaussian Tension) and 2 ($Q_{\rm DMAP}$ tension), we require models to reduce the tension to the 3$\sigma$ level (instead of 4.41$\sigma$ in \lcdm). Thus we require eq. (\ref{eq:crit_1}) $\leq 3$ or $\sqrt{\mathrm{eq.~(}\ref{eq:crit_2}\mathrm{)}} \leq 3$.
This may not seem like a very stringent threshold for success since, at the end of the day, $3\sigma$ may still be considered as a significant tension. However, we will show that only a limited number of models are capable of reducing the tension to this rather meager level. 
In addition, it is important to bear in mind that some of the current data likelihoods might underestimate systematic errors. 
In the future, as long as systematic errors are revised slightly but not drastically, there is a good chance that the models that do not pass our 3$\sigma$ criteria will remain excluded, while the models not passing some possible 2$\sigma$ criterion could be rescued. 
For Test 3 ($\Delta$AIC), we demand that the preference for the extended cosmology $\mathcal{M}$ over \lcdm is larger than a \enquote{weak preference} on Jeffrey's scale \cite{Jeffreys61,Nesseris:2012cq}, that is, $p=10^{1.5}$. Using $\exp(-\Delta \mathrm{AIC}/2) = p$  in the AIC formalism, this leads to the criterion $\Delta$AIC $\leq - 6.91$.

Since the questions addressed by criteria 1 and 2 are very similar\footnote{The most notable difference is that one is based on average properties of the posterior distribution, as opposed to the $\chi^2$ of a single point, and the ability to capture the effect of long tails in non-Gaussian posterior distributions.}, and since criterion 2 does not assume Gaussian posteriors, we consider that criterion 2 supersedes criterion 1. On the other hand, criteria 2 and 3 address significantly different questions. Each of them has its pros and cons and they complement each other.
Thus, as long as the $\Delta \chi^2$ computed for the AIC test is negative (as long  as a model is not giving a worse combined fit than $\Lambda$CDM), we will conservatively consider that it is successful when one of criterion 2 or 3 is fulfilled.
To summarize the success of each suggested solution we attribute \enquote{medals} to models passing our tests: A model passing either criterion 2 (obtaining a good combined fit) or 3 (strongly improving the fit over $\Lambda$CDM) receives a bronze medal. A model passing both criteria receives a silver medal. We reserve the gold medal for models that additionally pass criterion 1, that is, whose posterior distributions allow for high values of $H_0$ (or $M_B$) independently of the inclusion of a local distance ladder prior.

\noindent {\bf Brief overview of the competitors.} 
First, our analysis includes six models featuring extra relativistic relics, with different assumptions concerning interactions in the sectors of dark radiation, neutrinos and dark matter: \Nefflong~[\Neffshort], \idrlong~[\idrshort], \Schmaltzlong~[\Schmaltzshort], \Equilong~[\Equishort], \SInulong~[\SInushort], and one model featuring a \Majoronlong~[\Majoronshort].
Second, we consider six models in which the sound horizon is shifted due to some other ingredient: a \Bmid~[\Bshort], a \memid~[\meshort], a \meOkmid~[\meOkshort], two models of \EDElong~[\EDEshort] (a phenomenological \EDEshort{} model and the theoretically better motivated \NEDElong{}~[\NEDEshort{}] model), and an \EMGlong{} model~[\EMGshort{}].
Finally, we compare four models with a modification of the late-time cosmological evolution, either due to Dark Energy (\CPLlong~[\CPLshort], \PEDElong~[\PEDEshort]) or late decaying Dark Matter (\fracDMmid~[\fracDMshort], \DMDRWDMmid~[\DMDRWDMshort]). 

We implement \PEDEshort{} in two ways: with a parametrisation of the DE density such that $\Lambda$CDM is recovered for a particular value of the additional parameter (\MEDElong~[\MEDEshort]), or with another fixed value of this parameter assumed in many previous works [\PEDEshort].

We emphasize that the models of \Neffmid{} and \CPLmid{} are primarily included as benchmarks to compare other models against, given that neither of the two is a viable model to ease the Hubble tension by itself. Any model in the given category should perform at least as well as these benchmark models.

We also stress that our choice of models is intended as a realistic illustration of the performance of the wide variety models presented in the literature, such as those listed in Ref.~\cite{DiValentino:2021izs}. While we naturally could  not include {\em all} of the suggested models in our study, we believe that our comprehensive MCMC analysis, the various statistics we introduce and the final ranking we present can provide useful benchmarks for future studies. We therefore encourage other authors to compare their models to the ones presented here.

\begin{figure}[ht]
    \centering
    \includegraphics[width=0.95\textwidth]{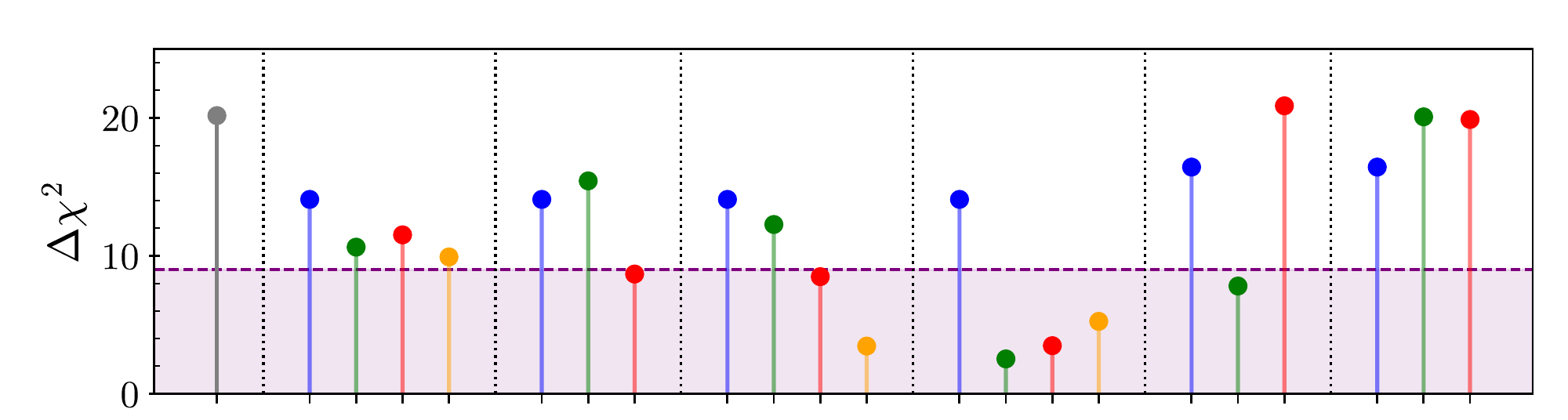}\\
    \includegraphics[width=0.96\textwidth]{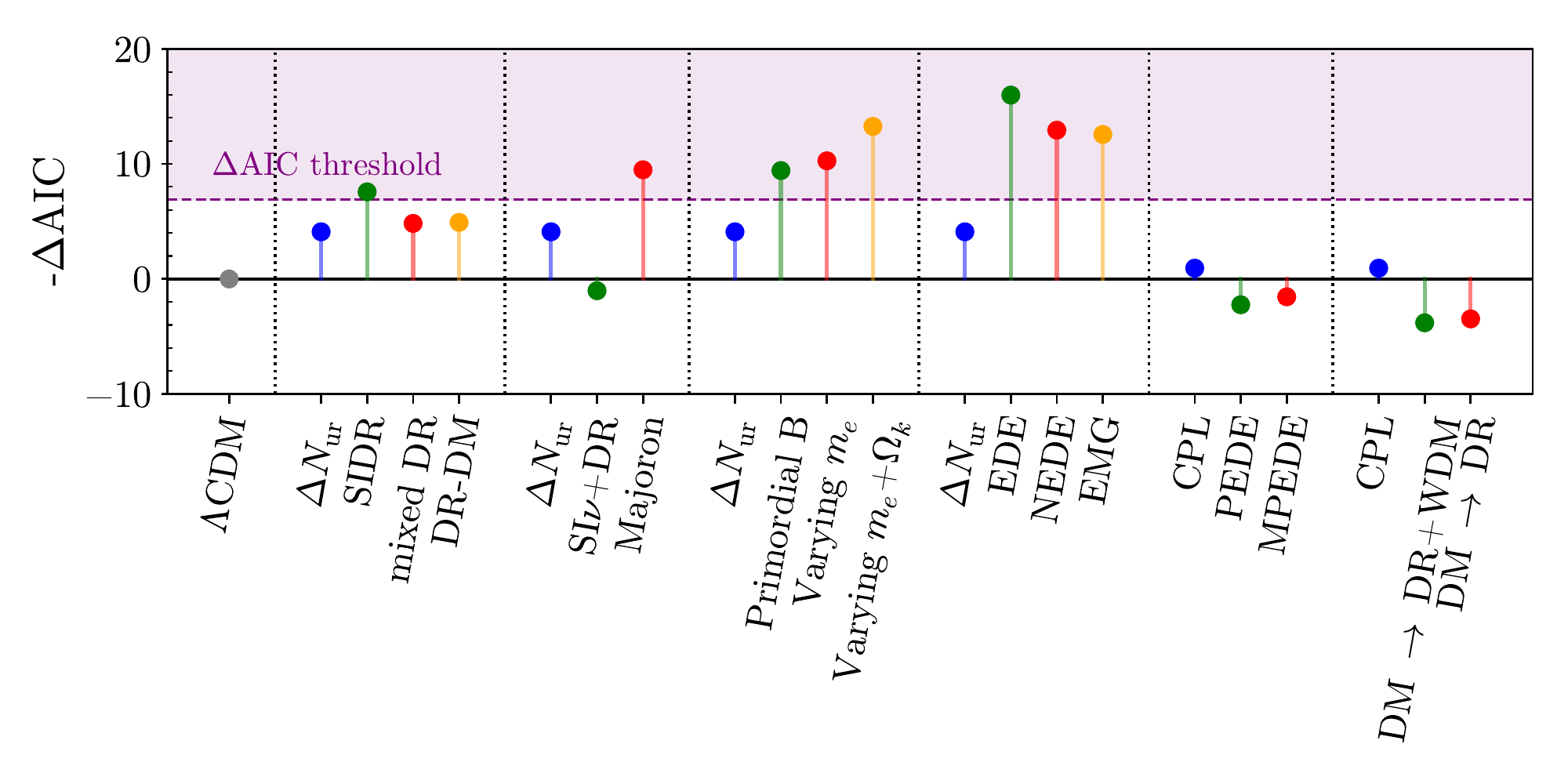}
    \caption{$\Delta \chi^2$ (Test 2) and $\Delta$AIC (Test 3) of the various models considered in this work, colored in the same way as in \cref{fig:moneyplots}. We additionally display the thresholds that have to be reached as purple dashed lines, and the regions of successful models as a purple region.
    \label{fig:deltaAIC}}
\end{figure}

\begin{figure}[hp]
    \centering
    \enlargethispage*{2\baselineskip}
    \textbf{\large Dark Radiation models}\\[0.5cm]
    \includegraphics[width=0.4\textwidth]{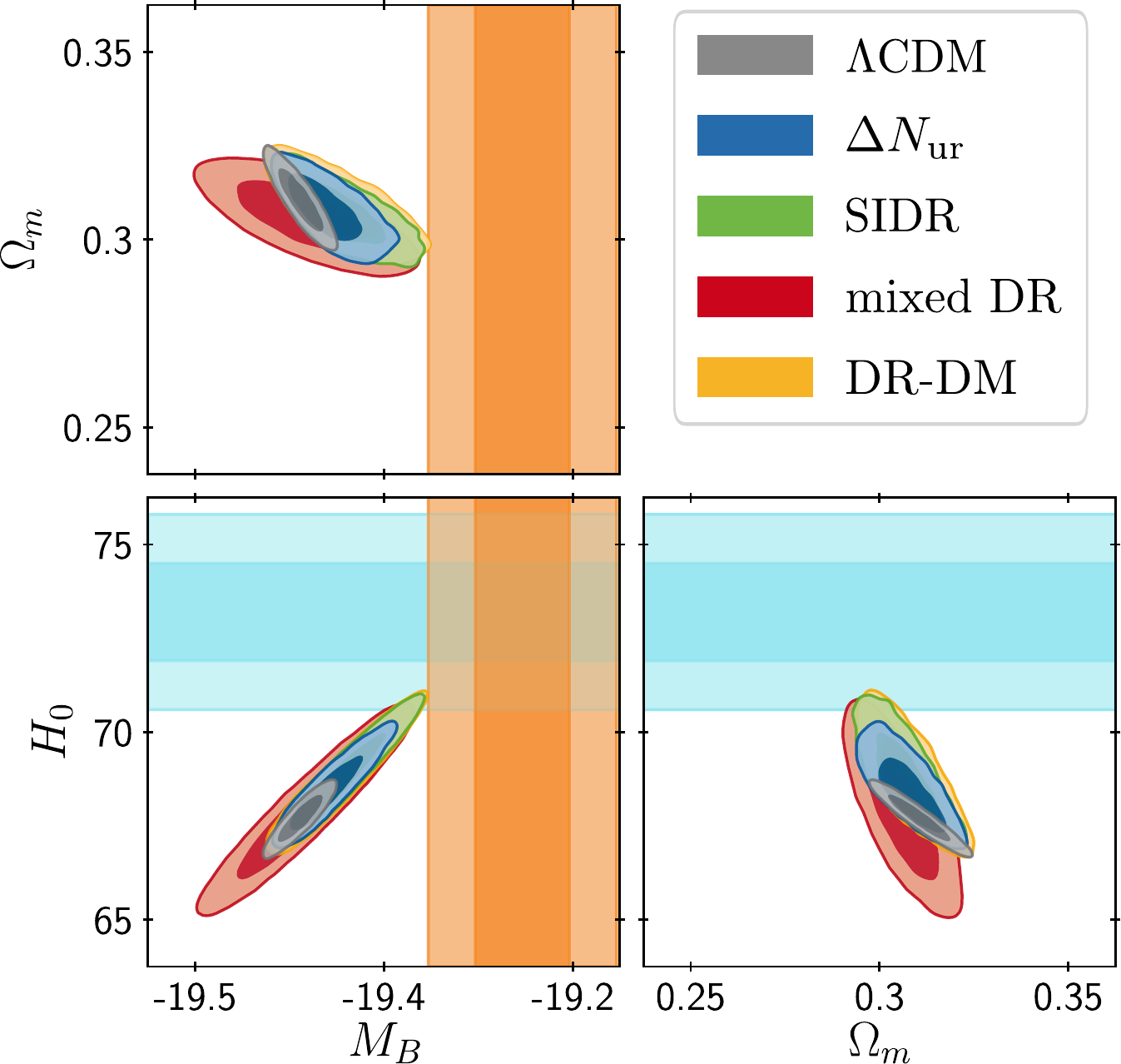}
    \includegraphics[width=0.4\textwidth]{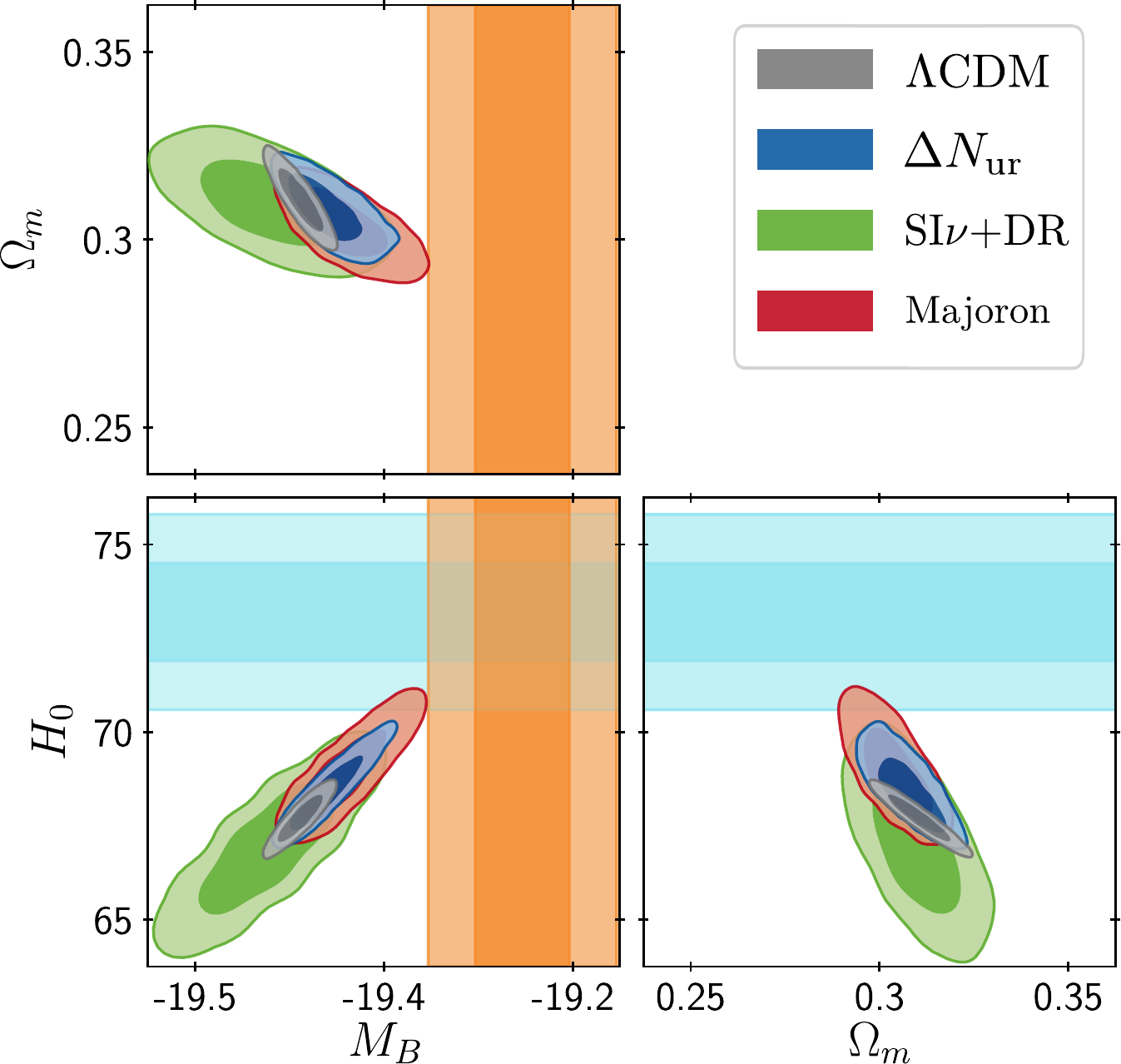} \\[0.25cm]
    \textbf{\large Other early universe models}\\[0.5cm]
    \includegraphics[width=0.4\textwidth]{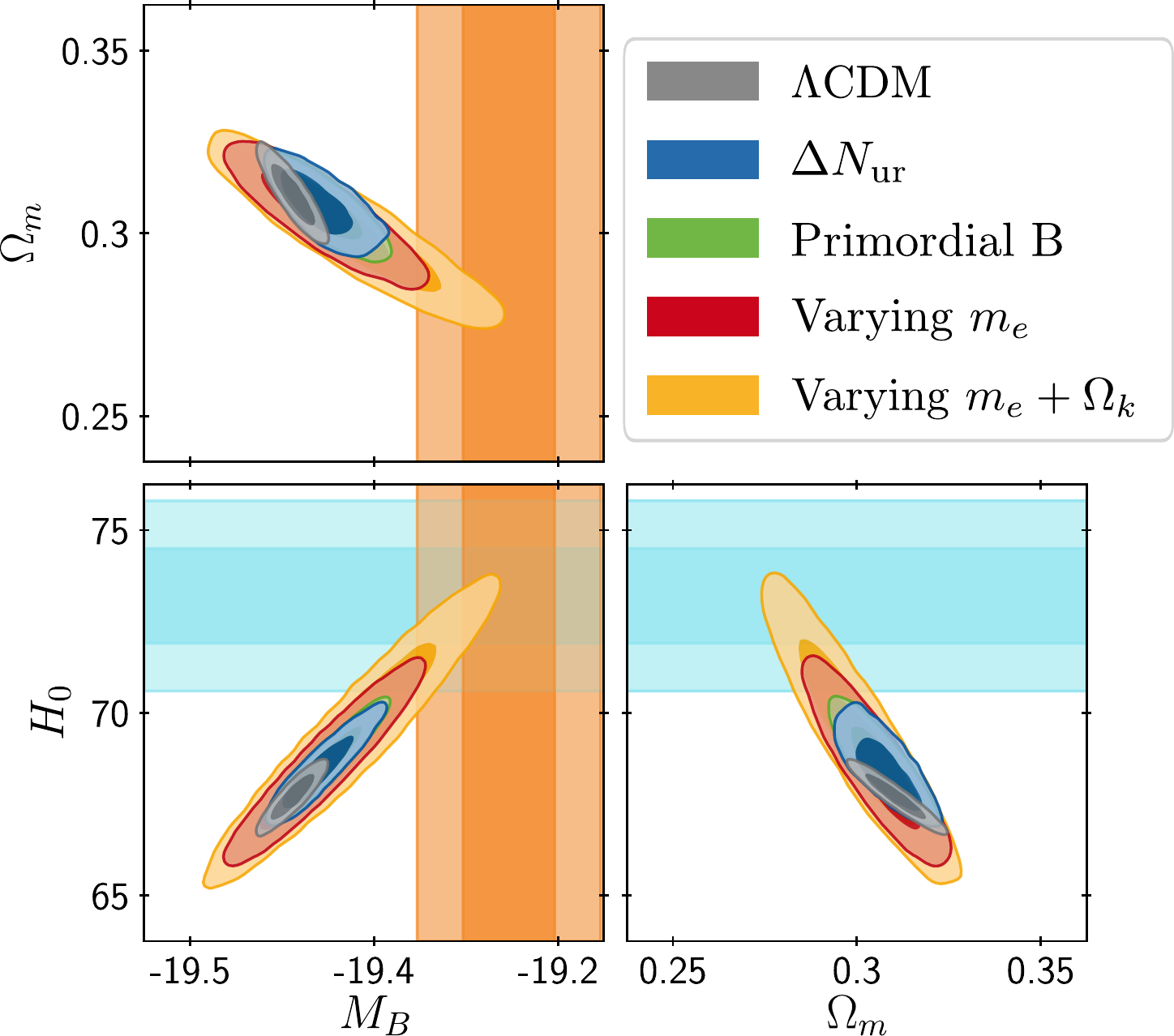}
    \includegraphics[width=0.4\textwidth]{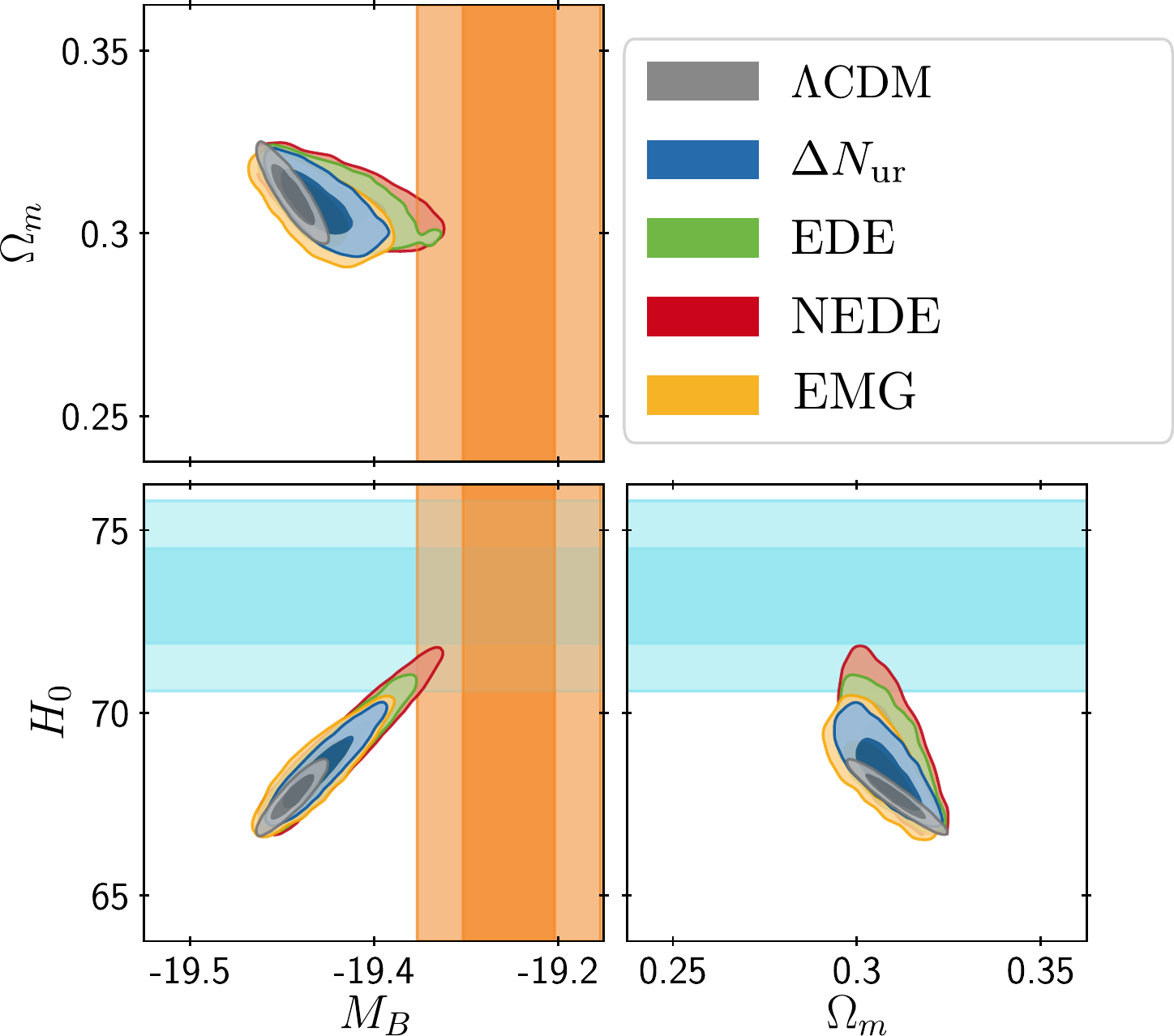}\\[0.25cm]
    \textbf{\large Late universe models}\\[0.5cm]
    \includegraphics[width=0.4\textwidth]{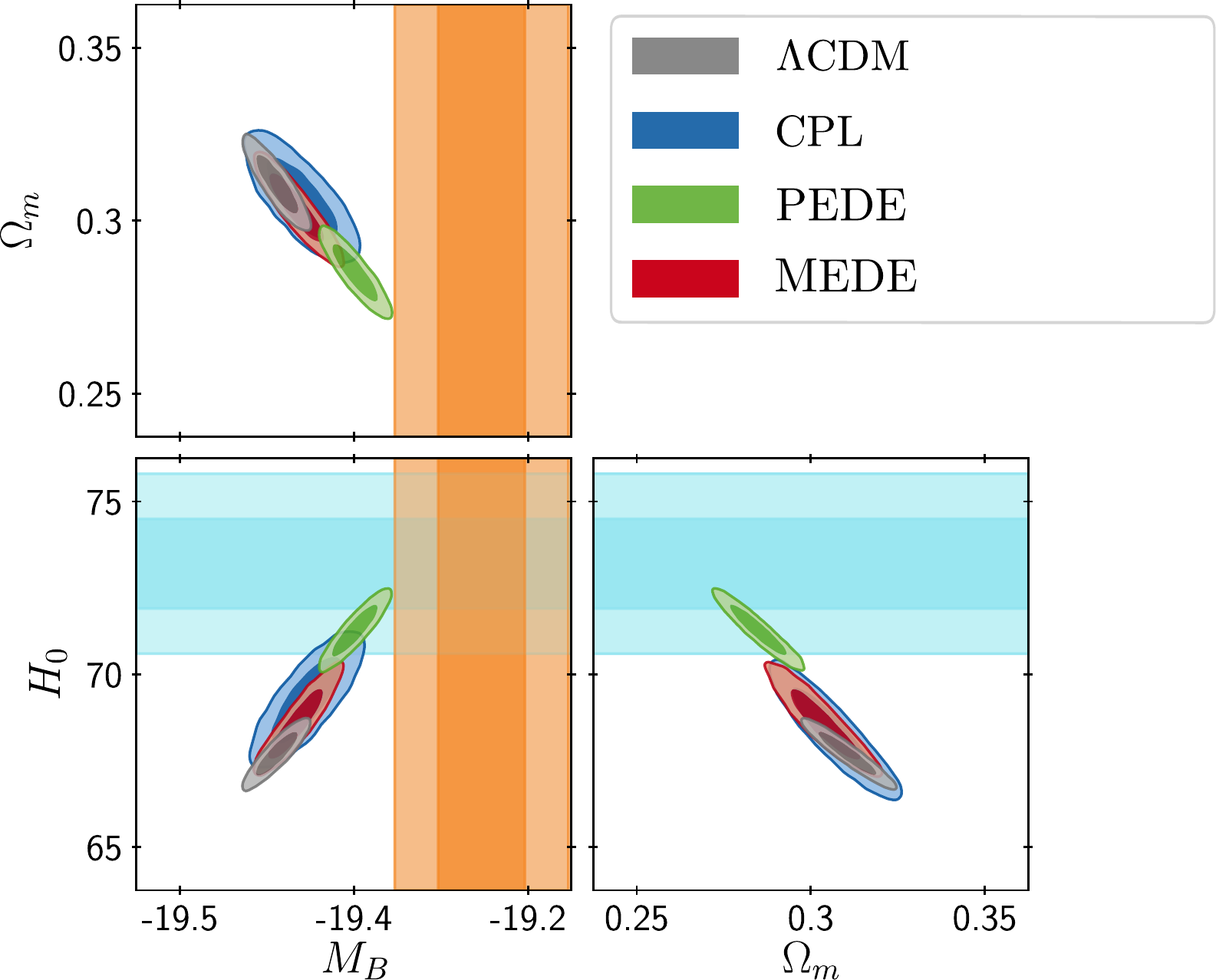}
    \includegraphics[width=0.4\textwidth]{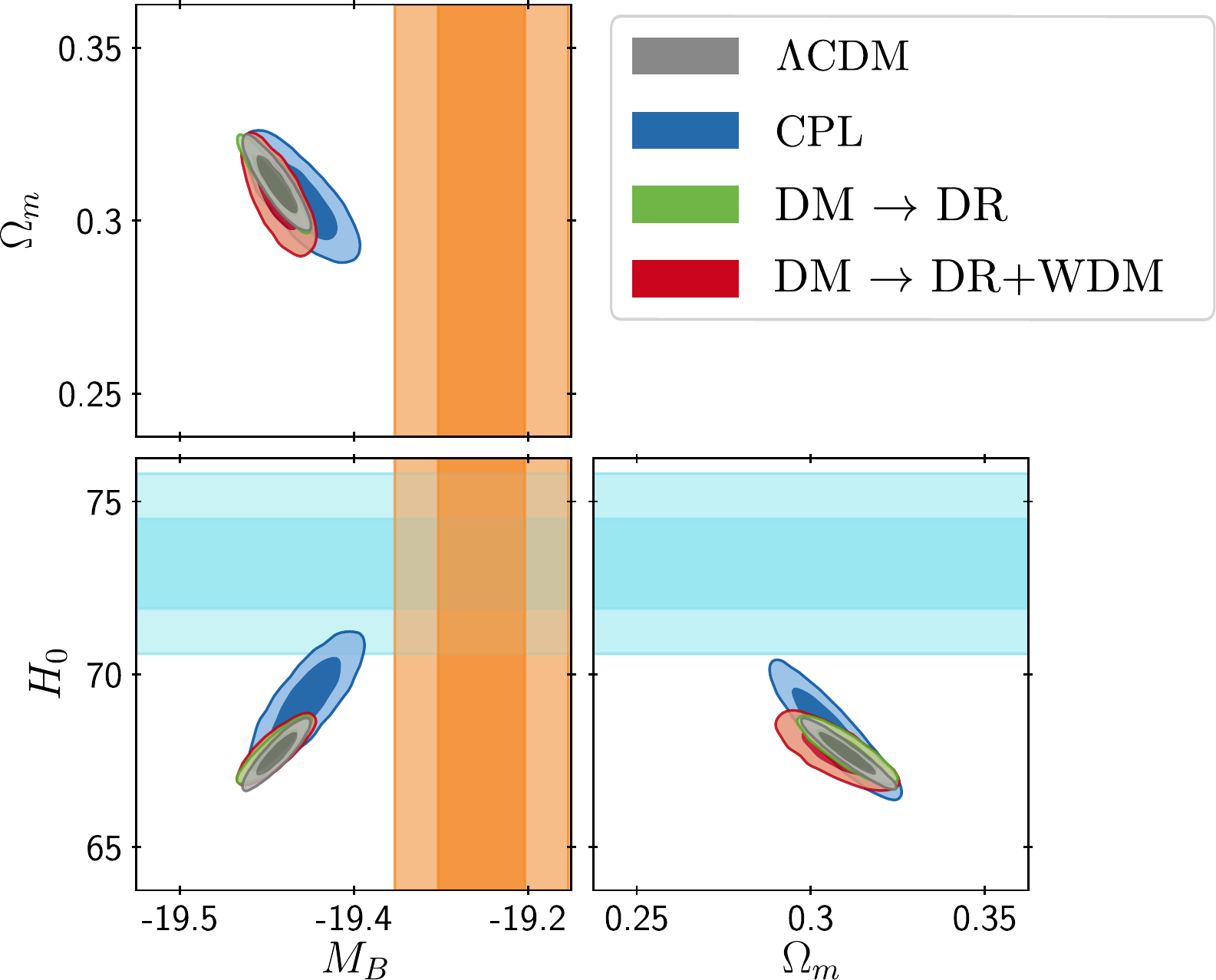}
    \caption{Contours ($68.3\%$ and $95.4\%$ C.L.) for the \DB dataset for the various considered models. For comparison, we show in each category either a \Neffmid{} model or a \CPLmid{} model. \label{fig:moneyplots}}
\end{figure}

\noindent {\bf Summary of \enquote{the qualifying round}.} For our baseline dataset \DB\,, where
\begin{center}
    \DB = Planck 2018 (including TTTEEE and lensing)~\cite{Planck:2019nip}\\
    + BAO (including BOSS DR12~\cite{BOSS:2016wmc} + MGS~\cite{Ross:2014qpa} + 6dFGS~\cite{Beutler:2011hx})
    + Pantheon~\cite{Scolnic:2017caz}~,
\end{center}
the result of our main tests ($Q_{\rm DMAP}$ for $M_b$ and Akaike Information Criterion) are summarized in \cref{tab:summary_MB} and represented graphically in \cref{fig:deltaAIC}. For the sake of completeness, we also present the results obtained using criterion 1 (Gaussian Tension on $M_b$), both in \cref{tab:summary_MB} and in the discussion. We show in \cref{fig:moneyplots} the 2D posterior distribution of $\{H_0, M_b, \Omega_m\}$ for all models.

First and foremost, no model is perfect -- in fact, none of the models studied here are capable of reducing the tension below the $\sim 1.6 \sigma$ level. A number of models, however, are capable of passing the criteria identified above (with varying levels of success). We enumerate the results for each of the test criteria below: 
 
\begin{itemize}
    \item Adopting the GT estimator, only four models can reduce the tension to the $3\sigma$ level, with the best model (\meOkshort{}) showing a residual 2.0$\sigma$ tension. From best to worse, they are: \meOkshort, \PEDEshort, \meshort{} in a flat universe, and the \Majoronshort{}.

    \item Making use instead of the more robust $Q_{\rm DMAP}$ criterion (reported in Fig.~\ref{fig:deltaAIC}), which compares $\chi^2$ of models with and without the inclusion of the SH0ES determination of $M_b$, we find that models with non-Gaussian tails perform significantly better.
    This most strongly impacts the two models of \EDEshort{} and the \EMGshort{} model, reducing their level of tension from roughly $3.1-3.7\sigma$ to $1.6-2.3\sigma$.
    From best to worse, models that pass criterion 2 are: \EDEshort{}, \meOkshort{}, \NEDEshort{}, \EMGshort{}, \PEDEshort{}, \meshort{}, and the \Majoronmid{}.
    
    \item Adopting the $\Delta$AIC criterion, which attempts at quantifying the role of enlarged model complexity in the improvement of the fit to \DB+SH0ES, we find that eight models are capable of significantly improving over \lcdm. They are, in decreasing level of success: \EDEshort{}, \meOkshort{}, \NEDEshort{}, \EMGshort{}, \meshort{}, the \Majoronmid{} model, a \Bmid{}, and \idrshort{}.
\end{itemize}

\begin{table}[h]
\renewcommand{\arraystretch}{1.18}
    \centering
    \hspace*{-0.5cm}
    \scalebox{0.85}{
    \begin{filecontents*}{sheets/MbTable_nohalofit2.csv}
Name,Npar,Mbstring,TensVal,Tension,Chi2,DChi2 wrt LCDM,AIK wrt LCDM,Passes DAIK Mb,String DAIK,Passes Sigma Mb,String SigmaH0,Passes both Mb,Chi2 wrt noSH0ES,Tension,Tensval,Passes Chi2,PassChi2,AIC or Chi2,PassBoth,Delta Crit 1 Crit 3,AIC AND Chi2,Passes all,Medal
\lcdmtable,0,-19.416 \pm 0.012,4.4092850522997,4.4 \sigma,3828.69,0.00,0.00,FALSE,\color{Red} X,FALSE,\color{Red} X,\color{Red} X ,20.16,4.48964488774017,4.5 \sigma,FALSE,\color{Red} X,FALSE,\color{Red} X,0.018225139560562,FALSE,FALSE,\nomedal
\nefftable,1,-19.395 \pm 0.019,3.61131137333291,3.6 \sigma,3822.59,-6.10,-4.10,FALSE,\color{Red} X,FALSE,\color{Red} X,\color{Red} X ,14.09,3.75315034408053,3.8 \sigma,FALSE,\color{Red} X,FALSE,\color{Red} X,0.039276306051868,FALSE,FALSE,\nomedal
\idrtable,1,-19.385 \pm 0.024,3.19023185828377,3.2 \sigma,3819.12,-9.57,-7.57,TRUE,\color{Green} \checkmark,FALSE,\color{Red} X,\color{Green} \checkmark,10.63,3.26069597211876,3.3 \sigma,FALSE,\color{Red} X,TRUE,\color{Green} \checkmark,0.022087458518734,FALSE,FALSE,\onzemedal
\equitable,2,-19.413 \pm 0.036,3.27411706314673,3.3 \sigma,3819.86,-8.83,-4.83,FALSE,\color{Red} X,FALSE,\color{Red} X,\color{Red} X ,11.52,3.39400305669518,3.4 \sigma,FALSE,\color{Red} X,FALSE,\color{Red} X,0.036616281958233,FALSE,FALSE,\nomedal
\schmaltztable,2,-19.388 \pm 0.026,3.19807688448604,3.2 \sigma,3819.76,-8.92,-4.92,FALSE,\color{Red} X,FALSE,\color{Red} X,\color{Red} X ,9.92,3.14971491910756,3.1 \sigma,FALSE,\color{Red} X,FALSE,\color{Red} X,-0.015122202225058,FALSE,FALSE,\nomedal
\SINtable,3,-19.440^{+0.037}_{-0.039},3.75518690429753,3.8 \sigma,3823.71,-4.98,1.02,FALSE,\color{Red} X,FALSE,\color{Red} X,\color{Red} X ,15.43,3.92810386828045,3.9 \sigma,FALSE,\color{Red} X,FALSE,\color{Red} X,0.046047498670445,FALSE,FALSE,\nomedal
\majorontable,3,-19.380^{+0.027}_{-0.021},2.95872601927562,3.0 \sigma,3813.20,-15.49,-9.49,TRUE,\color{Green} \checkmark,TRUE,\color{Green} \checkmark,\color{Green} \checkmark,8.68,2.9461839725313,2.9 \sigma,TRUE,\color{Green} \checkmark,TRUE,\color{Green} \checkmark,-0.004239002416114,TRUE,TRUE,\silvermedal
\primtable,1,-19.390^{+0.018}_{-0.024},3.53727518037797,3.5 \sigma,3817.27,-11.42,-9.42,TRUE,\color{Green} \checkmark,FALSE,\color{Red} X,\color{Green} \checkmark,12.27,3.5027846065666,3.5 \sigma,FALSE,\color{Red} X,TRUE,\color{Green} \checkmark,-0.009750605212369,FALSE,FALSE,\onzemedal
\varyingmetable,1,-19.391 \pm 0.034,2.92248859480687,2.9 \sigma,3816.42,-12.27,-10.27,TRUE,\color{Green} \checkmark,TRUE,\color{Green} \checkmark,\color{Green} \checkmark,8.49,2.9137400904898,2.9 \sigma,TRUE,\color{Green} \checkmark,TRUE,\color{Green} \checkmark,-0.002993511876355,TRUE,TRUE,\goldmedal
\varyingmeomegaktable,2,-19.368 \pm 0.048,2.04011204369655,2.0 \sigma,3811.43,-17.26,-13.26,TRUE,\color{Green} \checkmark,TRUE,\color{Green} \checkmark,\color{Green} \checkmark,3.46,1.85882092191266,1.9 \sigma,TRUE,\color{Green} \checkmark,TRUE,\color{Green} \checkmark,-0.088863316279141,TRUE,TRUE,\goldmedal
\edetable,3,-19.390^{+0.016}_{-0.035},3.60954610682971,3.6 \sigma,3806.71,-21.98,-15.98,TRUE,\color{Green} \checkmark,FALSE,\color{Red} X,\color{Green} \checkmark,2.53,1.59151832644684,1.6 \sigma,TRUE,\color{Green} \checkmark,TRUE,\color{Green} \checkmark,-0.559080759922836,TRUE,FALSE,\silvermedal
\nedetable,3,-19.380^{+0.023}_{-0.040},3.11493681206895,3.1 \sigma,3809.76,-18.93,-12.93,TRUE,\color{Green} \checkmark,FALSE,\color{Red} X,\color{Green} \checkmark,3.49,1.86761176234507,1.9 \sigma,TRUE,\color{Green} \checkmark,TRUE,\color{Green} \checkmark,-0.400433499932028,TRUE,FALSE,\silvermedal
\emgtable,3,-19.397^{+0.017}_{-0.023},3.74469334118738,3.7 \sigma,3810.13,-18.56,-12.56,TRUE,\color{Green} \checkmark,FALSE,\color{Red} X,\color{Green} \checkmark,5.25,2.29145425617874,2.3 \sigma,TRUE,\color{Green} \checkmark,TRUE,\color{Green} \checkmark,-0.388079597606739,TRUE,FALSE,\silvermedal
\cpltable,2,-19.400 \pm 0.020,3.697697597992,3.7 \sigma,3823.75,-4.94,-0.94,FALSE,\color{Red} X,FALSE,\color{Red} X,\color{Red} X ,16.43,4.05318442708935,4.1 \sigma,FALSE,\color{Red} X,FALSE,\color{Red} X,0.096137344841393,FALSE,FALSE,\nomedal
\pedetable,0,-19.349 \pm 0.013,2.67085183996509,2.7 \sigma,3830.93,2.24,2.24,FALSE,\color{Red} X,TRUE,\color{Green} \checkmark,\color{Green} \checkmark,7.81,2.79377410827846,2.8 \sigma,TRUE,\color{Green} \checkmark,FALSE,\color{Red} X,0.04602361930903,FALSE,FALSE,\nomedal
\medetable,1,-19.400 \pm 0.022,3.62104967146468,3.6 \sigma,3828.24,-0.45,1.55,FALSE,\color{Red} X,FALSE,\color{Red} X,\color{Red} X ,20.87,4.56823817368472,4.6 \sigma,FALSE,\color{Red} X,FALSE,\color{Red} X,0.261578433923253,FALSE,FALSE,\nomedal
\dmdecaytable,2,-19.420 \pm 0.012,4.50453016058739,4.5 \sigma,3828.50,-0.19,3.81,FALSE,\color{Red} X,FALSE,\color{Red} X,\color{Red} X ,20.07,4.48012629637705,4.5 \sigma,FALSE,\color{Red} X,FALSE,\color{Red} X,-0.005417626997787,FALSE,FALSE,\nomedal
\fracdmdecaytable,2,-19.410 \pm 0.011,4.28150738684098,4.3 \sigma,3828.16,-0.53,3.47,FALSE,\color{Red} X,FALSE,\color{Red} X,\color{Red} X ,19.88,4.45869936192166,4.5 \sigma,FALSE,\color{Red} X,FALSE,\color{Red} X,0.04138541851528,FALSE,FALSE,\nomedal
\end{filecontents*}
\csvreader[/csv/head=true,tabular={l c|  c c c c | r r c | c c},/csv/respect dollar=false,/csv/respect backslash=false,
    table head={ Model& $\Delta N_{\rm param}$ & $M_B$ &  \begin{tabular}{@{}c@{}}Gaussian \\ Tension\end{tabular} & \begin{tabular}{@{}c@{}}$Q_{\rm DMAP}$ \\ Tension\end{tabular} & & $\Delta \chi^2$  & $\Delta$AIC & & \begin{tabular}{@{}c@{}}Finalist\end{tabular} & \\ \hline}]{sheets/MbTable_nohalofit2.csv}{1=\name,2=\nparam,3=\mbsig,5=\mbtens,7=\Dchi,8=\DAIK,12=\Mbtest,10=\DAIKtest,13=\anytest,16=\newmbtens,18=\newMbtest,20=\newanytest,24=\mname}{ \name & \nparam & $\mbsig$ & $\mbtens$ & $\newmbtens$ & $\newMbtest$ & $\Dchi$  &  $\DAIK$ & $\DAIKtest $ & $\newanytest$ & \mname}}
    \caption{Test of the models based on dataset \DB (Planck 2018 + BAO + Pantheon), using the direct measurement of $M_b$ by SH0ES for the quantification of the tension (3rd column) or the computation of the AIC (5th column). Eight models pass at least one of these three tests at the 3$\sigma$ level. \label{tab:summary_MB}}
\renewcommand{\arraystretch}{1}
\end{table}

Before declaring the official {\em finalists}, let us briefly comment on models that do not make it to the final, starting with late-universe models.
The CPL parameterization, our \enquote{late-universe defending champion} only reduces the tension to $3.7\sigma$, inducing a minor improvement to the global fit. 
The \PEDEshort{} model noticeably degrades the $\chi^2$ of BAO and Pantheon data, leading to an overall worse fit than $\Lambda$CDM. Thus, according to the general rules defined at the end of the previous subsection, we must exclude \PEDEshort{} from the final. We further comment on this choice in \cref{sec:resultsB} and below. The \MEDEshort{} model, which generalises \PEDEshort{} to include \lcdm as a limiting case, does not pass any of the tests.
This shows the danger of using only criterion 1 or 2 for models that do not include $\Lambda$CDM as a limit. Ideally, one should always perform a test equivalent to the $\Delta$AIC or consider models in which $\Lambda$CDM is nested.
As emphasized above, for late-time modifications of $\Lambda$CDM, it is also important to treat the SH0ES observation as a model-independent measurement of $M_b$\,, rather than a model-dependent measurement of $H_0$\,. We checked explicitly that using a SH0ES likelihood on $H_0$ rather than $M_b$ incorrectly yields more favorable results for these late-time models, a result consistent with the claims of Refs.~\cite{Benevento:2020fev,Camarena:2021jlr,Efstathiou:2021ocp,Li:2019yem,Pan:2019hac,Yang:2021egn}. 
Finally, the models of decaying dark matter studied here are only capable of reducing the tension from $4.4\sigma$ to $4.2\sigma$, despite only introducing two new parameters. Consequently, the $\Delta$AIC criteria disfavors both DDM models. 
We thus conclude that the late-time DE and dark matter decay models considered in this work cannot resolve the Hubble tension.

Secondly, the class of models invoking extra relativistic degrees of freedom perform significantly better than late-universe models, but a majority are not successful enough to pass our pre-determined thresholds. \idrlong~[\idrshort], \Schmaltzlong~[\Schmaltzshort], and  \Equilong~[\Equishort], all improve upon the \enquote{early-universe defending champion}, that is, \Neffmid{} (for all three criteria). However, none of them reduces the tension below the $\sim 3.2\sigma$ level. 
Perhaps the most surprising case is that of \SInulong~[\SInushort], which has long been claimed as a promising solution to the Hubble tension, but performs 
worse on $\Delta$AIC and $Q_\mathrm{DMAP}$
than the benchmark of \Neffmid{}. It may also sound surprising that the \Schmaltzshort{} model does not perform significantly better than the \idrshort{} model (the latter model passing the $\Delta AIC$ criterion). We emphasize that in several previous papers, the success of this model was boosted by a lower prior on the amount of DR that excluded $\Lambda$CDM as a limit, a situation comparable to that of \PEDEshort.
The only model which successfully passes both criteria is that of the \Majoronlong, which reduces the tension to the level of $\sim 2.9\sigma$ and shows a significant improvement to the fit. 
It is perhaps interesting to point out that this is the only model in this categorization which invokes a non-trivial evolution of $H(z)$. It is thus in some ways more similar to \EDElong{} than to the other Dark Radiation models presented here. 

Thirdly, all the models that shift the sound horizon using an ingredient that is not dark radiation are successful in passing at least one of the tests.

\begin{figure}[ht]
    \centering
    \includegraphics[width=\textwidth]{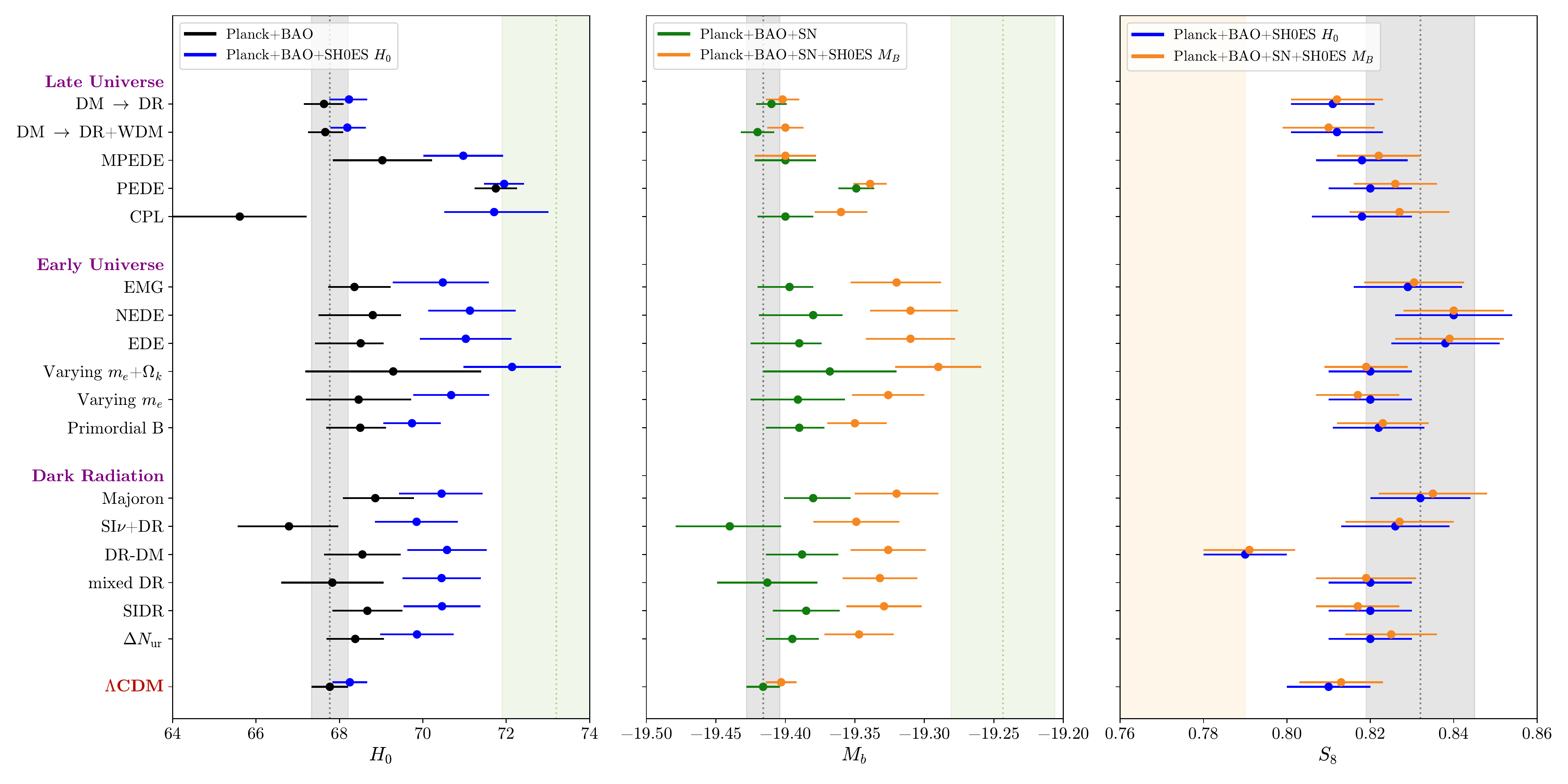}
    \caption{Level of tension on $H_0$, $M_b$ or $S_8$ (from left to right) for the sixteen cosmological models studied in this work. The grey vertical bands represent the respective predictions of $\Lambda$CDM without SH0ES. The green band represents the SH0ES calibration of $M_B$, while the orange band represents the $S_8$ measurement from \cite{Asgari:2020wuj}.
    \label{fig:whisker}}
\end{figure}

In summary, the models that pass at least one of criterion 2 or 3 without leading to a worse global fit, ranked from the best to worst $\Delta$AIC, are the following:
\begin{enumerate}
    \item \EDEmid{},
    \item \meOkmid{},
    \item \NEDEmid{},
    \item \EMGmid{},
    \item \Bmid{},
    \item \memid{},
    \item \Majoronmid{},
    \item \idrmid{}.
\end{enumerate}
These models constitute our \enquote{finalist} sample. The results obtained so far are summarized graphically in Figure~\ref{fig:whisker}. The left plot shows the tension on $H_0$ using the dataset $\mathcal{D}_{\rm minimal} \equiv \, $(Planck2018 + BAO, black, \cref{sec:resultsA}), and the middle plot the tension on $M_b$ obtained using dataset \DB\, (green, \cref{sec:resultsB}). For comparison, in each of these panels we also illustrate the impact of including the SH0ES likelihood in conjunction with $\mathcal{D}_{\rm minimal}$ and \DB. In addition, we illustrate in the right panel the extent to which each model is able to reduce the $S_8$ tension. The $S_8$ tension is not significantly impacted for most of our models, except for \Schmaltzlong. These two models return significantly lower values of $S_8$ than \lcdm.

\pagebreak
\noindent {\bf Summary of the \enquote{finale}.} Having identified a set of \enquote{finalist} models thanks to our baseline analysis, we proceed by investigating their performance against various data combinations.
We begin in \cref{sec:resultsC} by investigating the importance of Planck data in limiting the possibility of a resolution to the Hubble tension.
It is interesting to note that for most models, one could reach some qualitatively similar conclusions even with WMAP+ACT data used instead of Planck data. 
Dropping CMB data altogether, the combination of Big Bang Nucleosynthesis (BBN) + Pantheon + BAO +  Lya-$\alpha$-BAO + Cosmic Chronometers (CC) data is also limiting the possibility to accomodate a large $M_b$ or $H_0$. Since this dataset is less constraining, upper bounds on $M_b$ or $H_0$ are weaker than in presence of CMB data, especially for the \Bmid{} model. However, even in this case it is still not possible to reach $M_b\simeq -19.244$ (the value measured by SH0ES) at less than about 2$\sigma$.

The only notable exception occurs when replacing Planck by WMAP+ACT in the case of \EDEshort{} and \NEDEshort{}. In this case, larger values of $H_0$ and a non-zero fraction of EDE and NEDE are actually favored, even in the absence of information from SH0ES. This suggests that the higher precision of the Planck data in the range $1000<\ell<2000$ strongly impacts the \EDEshort{} contours. 
It is worth mentioning, however, that when \EDEshort{} is confronted with BBN + Lya-$\alpha$ BAO + CC + BAO + Pantheon, the results are more consistent with lower values of $H_0$, giving $M_b<-19.28$ (95\% CL). 

In the final series of tests (\cref{sec:resultsD}),  we recompute the posteriors for each finalist model using our baseline analysis (Planck+BAO+Pantheon) in combination with additional data from Redshift Space Distortions (RSD), CC, and Ly-$\alpha$-BAO. 
We show that in all cases the contours yield similar results to our baseline analysis, with the compatibility with SH0ES only undergoing minor modifications on a model-by-model basis.  We finally discuss in more details the status of the $S_8$-tension for the sample of finalists.

\FloatBarrier

\newpage

\section{Models entering the competition}\label{sec:models}
A large number of models have been proposed in order to explain the Hubble tension (for a recent overview of proposals see \cite{DiValentino:2021izs}). Our goal here is not to present an exhaustive study of all such models, but rather to compare a representative set in a systematic way so as to allow for an apples-to-apples comparison. In additional to allowing for a comparison of the relative success of various models, this approach allows us to make more generalized statements about the types of modified cosmologies most likely to explain the Hubble tension.

In this section we introduce the models selected for this analysis, and explain the justification as to why and how they may ameliorate the tension. For clarity, we split the model in three different categories: early Universe models that invoke extra relativistic dark relics (in addition to other ingredients) (\emph{Dark Radiation solutions}, see section~\ref{sec:darkrad}), alternative early Universe models that do not involve dark radiation (\emph{Other early Universe solutions}, see section~\cref{sec:early}), and finally, models modifying the cosmological expansion at late times (\ie well after recombination) (\emph{Late Universe solutions}, see section~\cref{sec:late}). Importantly, not all models fall nicely into one of the aforementioned categories; as such, it is important to bear in mind that conclusions drawn from this dichotomy should be treated with some level of flexibility.

It would be remiss not to point out that any selection of models is necessarily biased in two important ways. First, while we have attempted to select a broad variety of models, there are many not included -- the current selection is a consequence of \eg time/computational limitations, code availability, and incomplete descriptions at the level of the perturbations. In particular, we acknowledge that we have not been able to proportionally represent the large multitude of late-universe models; we do, however, believe that the late-Universe models considered reflect the broader features and shortcomings of this class of solutions\footnote{We also point out that our results are based on the standard assumptions of FLRW cosmology and on the standard model of particle physics, which could be incorrect or incomplete \cite{2021CQGra..38r4001K,2021MNRAS.500.5249A,2019PhRvD.100d3537D,2020PhRvD.102b3007D,2021PhRvD.103h3517A,2020MNRAS.499.2845H}.}. 
 Second, each model analyzed differs, sometimes arbitrarily, with regard to theoretical motivation, model complexity, and the robustness of implementation (to the extent that the cosmological analysis reflects the underlying microphysics). We attempt to remain largely neutral in this manner, implementing each model in the manner reflective of previously published work, and avoiding superfluous commentary on model motivation and/or complexity (although occasional commentary is unavoidable). Finally, it is necessary to emphasize that in some cases model parameters are fixed to particular values, rather than scanned, without any underlying motivation for doing so -- we caution the reader that this procedure may bias a particular models, in particular in the case of the $\Delta$AIC evaluation. 

\subsection{\texorpdfstring{$\Lambda$CDM}{LCDM}}

The success of the (minimal, flat) \lcdm model is well documented. This model describes most cosmological observations very well with a basis of 6 fundamental parameters, for instance:
$$\{\omega_{\rm cdm}\equiv\Omega_{\rm cdm} h^2, \,\, \omega_{\rm b} \equiv \Omega_{\rm b} h^2, \,\, \theta_s, \,\, \ln(10^{10} A_s), \,\, n_s, \,\, \tau_\mathrm{reio}\}.$$ 
In principle this model has a seventh fundamental parameter $T_0$ that we keep fixed to the FIRAS value, $T_0=2.7255$~K. In what follows, we will adopt flat priors\footnote{We do not need to specify our prior edges because they are always wide enough for the posteriors to drop exponentially before reaching them.}  on these parameters and keep the total neutrino mass\footnote{To match the Planck baseline model, we assume one neutrino species with mass 0.06eV and two massless neutrino species.} fixed to its minimal value of 0.06eV. Instead of reporting results on the sound horizon angle $\theta_s$\,, we will show posteriors and bounds for the expansion rate $H_0$\,, treated as a derived parameter.

\onlyprd{\pagebreak[40]}
\subsection{Dark radiation solutions}\label{sec:darkrad}

Massless dark relics (dark radiation) are a well known extension of the standard \lcdm model. For observations of the CMB anisotropies, there exists a strong degeneracy between increasing the Hubble parameter $h$ and enhancing the radiation density at early times. The latter effect can be achieved through the addition of massless free-streaming relics, captured by an increase in the effective neutrino number $N_\mathrm{eff}$\,. This degeneracy can be appreciated in light of the previous discussion, in which we have demonstrated that, for fixed fractional densities $\Omega_X$\,, the sound horizon scale $\theta_s$ is primarily impacted by $h$ through the fractional density of radiation, $\Omega_r \propto T_0^4 h^{-2}$. However, a more complete expression for the fractional density of radiation involves the total effective number of neutrino species $N_\mathrm{eff}$,
\begin{equation}\label{eq:omegar}
    \Omega_r = 4.18 \cdot 10^{-5} h^{-2} \relpow[4]{T_0}{2.7255\mathrm{K}} \relpow{1+\frac{7}{8} \relpow[4/3]{4}{11} N_\mathrm{eff}}{1+\frac{7}{8} \relpow[4/3]{4}{11} 3.044}~.
\end{equation}
The integral \cref{eq:soundhorizon} giving the sound horizon is strongly impacted by $\Omega_r$, and thus $h$ and $N_\mathrm{eff}$ inherit a large degeneracy. 
This degeneracy is broken primarily by the effect of increasing $h$ and $N_\mathrm{eff}$ on cosmological perturbations. In particular, at fixed $\theta_s$, increasing $h$ leads to an enhancement of Silk damping, while increasing $N_\mathrm{eff}$ produces a shift in the peak position and amplitude caused by the neutrino drag effect \cite{Bashinsky:2003tk,Lesgourgues:2018ncw,Baumann:2015rya,Follin:2015hya}.

\subsubsection{\Nefflong}

\textbf{Motivation:} We begin by considering one of the simplest extensions of \lcdm, in which we augment the standard cosmological model by additional free-streaming massless relics. The cosmological evolution of such relics is entirely equivalent to that of massless neutrinos, and thus can be fully characterized via a modification of the total effective number of neutrino species $N_\mathrm{eff}$\,.  In \lcdm, $N_\mathrm{eff} \approx 3.044$ \cite{Froustey:2020mcq,Bennett:2020zkv,2020JCAP...08..012A}, accounting for the three neutrino species (plus minor corrections arising from non-instantaneous neutrino decoupling and electroweak corrections). Consequently, we can parameterize an arbitrary number of massless free-streaming relics via their contribution to $N_\mathrm{eff}$\,. Let us emphasize again that it has been shown in many different studies (e.g. \cite{Bernal:2016gxb,Poulin:2018cxd, Aghanim:2018eyx,DiValentino:2019dzu}) that $\Delta N_\mathrm{eff}$~fails at resolving the Hubble tension. Nevertheless, as one of the simplest and best motivated extension to the \lcdm~model, we believe it is important to include it in our comparison study. Yet, rather than a real competitor, we will treat it as a useful benchmark model (the \enquote{early-universe defending champion}), in order to guide the reader in assessing the extent to which the additional complexity introduced by other models really helps for relieving the Hubble tension.

\textbf{Parameters:} We take $N_\mathrm{eff}$ as a free parameter, with $\Delta N_\mathrm{eff}=N_\mathrm{eff}-3.044$ defining the energy density in free streaming dark radiation. In most models where ordinary neutrinos are produced and decouple in the standard way, $\Delta N_\mathrm{eff}$ is strictly positive, and thus we adopt a flat prior on $N_\mathrm{eff}>3.044$. This model extends \lcdm by a single parameter.

\subsubsection{\idrlong}

\textbf{Motivation:} CMB constraints on free-streaming relics arise primarily because they increase the Silk damping and neutrino drag effects (leading to a small phase-shift and reduction in amplitude of the acoustic peaks \cite{Bashinsky:2003tk,Lesgourgues:2018ncw,Baumann:2015rya,Follin:2015hya}) . One of the simplest ways to counter-act these effects is to relax the assumption that these relics are free-streaming\footnote{Note that there have been other proposals, not studied here, which can counteract the effect of free-streaming radiation and help in alleviating the $H_0$ tension (see e.g. \cite{Ghosh:2021axu,Cyr-Racine:2021alc}).}. Since introducing interactions between dark radiation and the standard model plasma (baryons and photons) would unavoidably alter the energy density itself, we focus here on a scenario where dark radiation is strongly self-coupled and forms a relativistic perfect fluid. Its anisotropic stress vanishes and the Boltzmann hierarchy of dark radiation perturbations reduces to the continuity and Euler equations. Self-interacting dark radiation clusters more than free-streaming dark radiation on small scales, which counteracts the enhancement of Silk damping. Also, the dark radiation fluid has roughly the same sound speed $c_s\simeq 1/\sqrt{3}$ as the tightly-coupled baryon-photon fluid, which reduces the neutrino drag effect. The counteracting of the Silk damping and neutrino drag effects allows to reach larger values of $N_\mathrm{eff}$. In tables and figures, we will refer to this \idrlong{} model as \idrshort.

\textbf{Parameters:} To parametrize the density of \idrmid{}, we introduce its contribution $N_{\rm idr}$ to the effective neutrino number,  $N_\mathrm{eff} = 3.044 + N_\mathrm{idr}$ (here ``idr'' stands for ``interacting dark radiation''). This model features a single parameter extension of \lcdm, with a prior $N_\mathrm{idr}>0$.

\subsubsection{\Equilong}

\textbf{Motivation:} The authors of Refs.~\cite{Brust:2017nmv,Blinov:2020hmc} provide examples in which relativistic relics consist in an arbitrary mixture of free-streaming and self-interacting ultra-relativistic particles. One example is the \enquote{Dark Sector equilibration model} of Ref.~\cite{Blinov:2020hmc}. This class of models has more freedom than the one of the previous section, in which the contribution of free-streaming relics to $N_\mathrm{eff}$ was fixed to 3.044. The authors of Refs.~\cite{Brust:2017nmv,Blinov:2020hmc} argue that this additional freedom allows one to reach higher values of $N_\mathrm{eff}$, coming of course, at the expense of a second free parameter. Since the AIC criteria penalizes the introduction of novel parameters, it will be of interest to compare whether there exists any preference for this model over the far simpler \idrmid{} model. In tables and figures, we will refer to this model as \Equishort.

\textbf{Parameters:} This model is a two-parameter extension of \lcdm. Instead of varying the parameters $N_{\rm ur}$ (for free-streaming ultra-relativistic relics) and $N_{\rm idr}$ (for self-interacting ultra-relativistic relics), we choose to vary the total effective neutrino number
$N_{\rm tot} = N_{\rm eff} = N_{\rm ur}+N_{\rm idr}$ and the self-interacting dark radiation fraction $f_{\rm idr} = N_{\rm idr} / (N_{\rm ur}+N_{\rm idr})$, with flat priors on $N_{\rm tot}>0$ and $0<f_{\rm idr}<1$.

\subsubsection{\Schmaltzlong}

\textbf{Motivation:} Following Refs. \cite{Buen-Abad:2015ova,Lesgourgues:2015wza,Chacko:2016kgg,Buen-Abad:2017gxg}, we investigate how the \idrlong{}~solution changes in the presence of additional interactions with dark matter. The model is determined via (i) the amount of strongly self-coupled dark radiation $N_\mathrm{idr}$ and (ii) the redshift-dependent interaction rate with dark matter. We adopt a power law parameterization for the momentum transfer rate from dark matter to dark radiation, such that $\Gamma = \Gamma_{0,nadm} a^{-2}$, where $\Gamma_{0,nadm}$ is the interaction strength today. Indeed, this choice appears to be among the most phenomenologically promising in raising the inferred value of $H_0$ (see \eg \cite{Buen-Abad:2015ova,Lesgourgues:2015wza,Chacko:2016kgg, Ko:2016uft,Ko:2016fcd,Buen-Abad:2017gxg,Ko:2017uyb,Archidiacono:2019wdp}). The novel dark matter-dark radiation interactions introduced in this model enhance the growth of small-scale dark radiation perturbations and suppress the growth of small-scale dark matter perturbations. These features tend to counter-act the effects of a high $H_0$ and high $N_{\rm eff}$ on the small-scale CMB and matter power spectra.

\textbf{Parameters:} This model contains a two parameter extension of \lcdm, with the parameters characterizing the energy density in the strongly coupled fluid $N_\mathrm{idr}$ and the interaction strength today $\Gamma_{0,\mathrm{nadm}}$\,.

\subsubsection{\SInulong}

\textbf{Motivation:} It has been shown in Refs.~\cite{Kreisch:2019yzn,Park:2019ibn,2020PhRvD.102l3544G,2021JCAP...07..038D} that the presence of exotic neutrino self-interactions, additional radiation, and large neutrino masses can accommodate larger values of $H_0$ and smaller values of $\sigma_8$\,. The model presented in Refs.~\cite{Kreisch:2019yzn,Park:2019ibn,2021JCAP...07..038D} considered the inclusion of a four-point neutrino contact interaction mediated by an exotic massive particle ($m \gtrsim 1$ keV). In this scenario, neutrinos have self-interactions in the early Universe which proceed at a rate $\Gamma_{\nu} = G_{\rm eff}^2 T_\nu^5$; the presence of these interactions delays neutrino free-streaming, inducing an effect in the TT power spectrum that can largely offset the presence of exotic radiation (see \eg Refs.~\cite{Cyr-Racine:2013jua,Archidiacono:2013dua,Lancaster:2017ksf,Oldengott:2017fhy}). Intriguingly, fitting this model using the {\tt Planck2015 TT+lensing} and BAO data yields a bi-modal distribution~\cite{Lancaster:2017ksf,Oldengott:2017fhy,Kreisch:2019yzn}; following the notation of Ref.~ \cite{Kreisch:2019yzn}, we refer to these as the moderately interacting and strongly interacting modes. The former of these is roughly consistent with the $\Lambda$CDM limit, while the latter induces a significant upward shift, and broadening, of the $H_0$ posterior. In this latter mode of the posterior, neutrino free-streaming is suppressed and the neutrino drag effect is erased \cite{Bashinsky:2003tk,Baumann:2015rya,Follin:2015hya}. As a result the value of $\theta_s$ is larger than in the \lcdm model, which naturally leads to larger $H_0$\,. Consequently, we focus in this work purely on the strongly interacting mode.

Importantly, a number of caveats have been raised regarding the viability of this model. First, the solution prefers a value of $\Delta N_{\rm eff} \sim 1$, a value which is at least naively excluded by BBN~\cite{Aver:2020fon}. This is not necessarily devastating however, as (i) the constraint itself must be considered in the context of the novel neutrino interactions, which may allow for larger values of $\Delta N_{\rm eff}$~\cite{Huang:2021dba}, (ii) the contribution to $\Delta N_{\rm eff}$ can be generated after BBN~\cite{Berbig:2020wve}, and (iii) differing measurements of the helium abundance could point to systematic uncertainties which allow for larger values of $\Delta N_{\rm eff}$ (see \eg Refs.~\cite{Izotov:2014fga,Aver:2020fon}). The second concern is that the precise model under consideration, one in which all neutrinos interact equally, is strictly speaking excluded by various experiments~\cite{Blinov:2019gcj,Lyu:2020lps,Deppisch:2020sqh} (note that the preferred value of $G_{\rm eff}$ is roughly 10 orders of magnitude larger than the Fermi constant $G_F$). A small amount of parameter space may still exist should for example only $\tau$-type neutrinos interact, however this would likely require a highly tuned model. Finally, we emphasize that the strongly interacting mode is highly suppressed when CMB polarization data is included, and a recent analysis using the latest data from {\tt Planck 2018} suggests no clear preference for the strongly interacting mode~\cite{Brinckmann:2020bcn}. 

New investigations into this model have abandoned several assumptions of the interaction in order to re-establish preference even with {\tt Planck 2018} data. These include reducing the number of neutrinos participating in the self-interaction in Ref.~\cite{2021JCAP...07..038D} or investigating a temperature-independent interaction in Ref.~\cite{2020PhRvD.102l3544G}.

\textbf{Parameters:} This model is a three parameter extension of \lcdm. The parameters are the effective interaction strength $G_{\rm eff}$\, (in units of ${\rm MeV}^{-2}$), the sum of the neutrino masses $\sum m_\nu$\, (in units of eV), and the number ultra-relativistic species $N_{ur}$\,. We take a flat prior on  $\sum m_\nu \in [0, 1.5]$, a flat prior on  $N_{ur} \in [0, 5]$, and a flat prior on $\log_{10}(G_\mathrm{eff}/[\mathrm{MeV}^{-2}]) \in [-2.5 , -0.3]$. 
 
\subsubsection{The \Majoronlong}

\textbf{Motivation:} Motivated by the cosmological success (and the phenomenological and theoretical short-comings) of the strongly interacting neutrinos, Refs.~\cite{Escudero:2019gvw,Escudero:2020hkf,Escudero:2021rfi} proposed a solution driven by the presence of a $\sim$eV-scale majoron, a pseudo-Goldstone boson arising from the spontaneous symmetry breaking of a global $U(1)$ lepton number symmetry. The majoron arises as a byproduct of various neutrino mass models, and is expected to have weak couplings with active neutrinos (suppressed in the simplest models by the ratio of the active to sterile neutrino mass), and heavily suppressed (\ie phenomenologically irrelevant) couplings to other Standard Model fermions. In the weak coupling limit (which, as stated above, is the most motivated limit), the majoron cannot induce the 2-to-2 scatterings studied in the case of strongly self-interacting neutrinos; rather, the cosmology is expected to be entirely dominated by inverse neutrino decays and majoron decays, the interaction rate for which is given by 
\begin{equation}\label{eq:maj_int}
    \Gamma = \Gamma_\phi e^{\mu_\nu / T_\nu} \left(\frac{m_\phi}{T_\nu} \right)^3 \, K_2\left(\frac{m_\phi}{T_\nu}\right) \, ,
\end{equation}
where $\Gamma_\phi \sim \lambda_\phi^2 m_\phi / (16 \pi)$ is the majoron decay rate, $K_2(x)$ is a modified Bessel function of the second kind, $m_\phi$ is the majoron mass, $\lambda_\phi$ is the majoron-neutrino Yukawa coupling, and $\mu_\nu$ and $T_\nu$ are the neutrino chemical potential and temperature. The temperature dependence of the interaction rate shows that for sufficiently large couplings, inverse neutrino decay will thermalize the majoron population near $T \sim m_\phi$\,, and the majorons will then  subsequently decay back into the neutrinos at $T \sim m_\phi/3$. Should thermalization occur, neutrino free-streaming will be damped in a manner akin to that of the strongly self-interaction neutrino model -- the primary difference here being that the time-dependence of the interaction rate is contained to a narrow temperature range near the majoron mass.\footnote{It is worth pointing out that an eV scale majoron is not unmotivated from theoretical perspective. It has been shown that a perturbative breaking of the lepton number symmetry by quantum gravity at the Planck scale is roughly consistent with generating an eV scale majoron mass (see \eg~\cite{Akhmedov:1992hi,Rothstein:1992rh}).} In addition, the decays of the majoron will generate a subsequent shift in the energy density of neutrinos at the level of $\Delta N_{\rm eff} \sim 0.11$.\footnote{This value assumes no primordial majoron population is present. Should a sizeable primordial population exist, the late-time shift in $\Delta N_{\rm eff}$ due to the majoron decays will be enhanced~\cite{Escudero:2021rfi}.}

It was shown in Ref.~\cite{Escudero:2019gvw} that the presence of a $m_\phi \sim \mathcal{O}(1)$eV majoron arising from the breaking of lepton number at $\sim \mathcal{O}(1)$TeV, when combined with additional radiation $\Delta N_{ur}$, could largely broaden the $H_0$ posterior, increasing the compatibility with the measurement by SH0ES.\footnote{Note that the interaction strength in ~\cite{Escudero:2019gvw} was derived using the fractional change in neutrino number density. Here, we instead derive the rate in terms of the fractional change in energy density, consistent with the updated analysis of \cite{Escudero:2021rfi}.} A more recent follow-up performed in Ref.~\cite{Escudero:2021rfi} showed that the contribution to $\Delta N_{ur}$ can actually be sourced directly from a primordial majoron population produced from the decays of $\sim$GeV scale sterile neutrinos -- collectively, this entire model can be self-consistently embedded in low-scale models of leptogenesis, thus simultaneously addressing the origin of neutrino masses, the baryon asymmetry of the Universe, and the $H_0$ tension. Due to the increase in computational complexity, we do not include the presence of a primordial majoron population, but restrict our parameter space to $\Delta N_{ur} \geq 0$ in order to mimic this feature as closely as possible. 

\textbf{Parameters:} In summary, this model contains three new parameters: (1) the majoron mass $m_\phi$ (in units of eV), determining when the damping of neutrino free-streaming takes place, (2) an effective decay width $\Gamma_{\rm eff}$, defined as
\begin{equation}
    \Gamma_{\rm eff} = \left(\frac{\lambda_\nu}{4 \times 10^{-14}} \right)^2 \, \left(\frac{0.1 \, {\rm eV}}{m_\phi} \right) \, 
\end{equation}
where $\lambda_\nu$ is the majoron-neutrino Yukawa coupling, which determines the duration and strength of the damping, and (3) the amount of additional radiation $\Delta N_{ur}$\,. We take flat priors on $\log_{10} (m_\phi/\mathrm{eV})\in [-1, 1]$ and $\log_{10} (\Gamma_\mathrm{eff}) \in [-2,2]$, and a linear prior on $\Delta N_{ur} \in [0, 1]$.

\subsection{Other early universe solutions}\label{sec:early}

As mentioned in the introduction, the early Universe solutions primarily leave the late time expansion history (and the corresponding $H_0 D_A$) unchanged. Requiring a given angular diameter of the sound horizon from BAO or the CMB then necessitates fixing $H_0 r_s$\,. This implies that raising $h$ (via $\Omega_r$ through \cref{eq:omegar}) will increase the integral in \cref{eq:soundhorizon}, unless it is also balanced via a modification to \eg $z_*$ in the integration boundary (\ie a modification of the recombination history) or a change in the expansion rate $H(z)/H_0$ near recombination (\eg as realized in the early dark energy model of \cref{ssec:ede}).\footnote{In the context of \cite{2018arXiv181103624C,2015PhRvD..92l3516A,2021PhRvD.103h3533A} this kind of model has been shown to generically impact the Hubble tension.}


\subsubsection{\Blong}

\textbf{Motivation:} We consider the presence of small-scale, mildly non-linear inhomogeneities in the baryon density around recombination. One well-motivated mechanism to achieve 
such kind of inhomogeneities is via the introduction of primordial magnetic fields (PMFs) in the plasma before recombination. Simply put, the idea is that on scales much smaller than the photon mean free path the effective speed of sound is far lower than that of a relativistic plasma, and consequently PMFs of $\sim 0.1$ nano-Gauss (nG) strength can generate baryon inhomogeneities on $\sim \text{kpc}$ scales \cite{Jedamzik:2011cu, Jedamzik:2018itu}. Even though  inhomogeneities on such small scales do not directly source CMB temperature and polarization anisotropies, they can lead to strong signatures in the CMB anisotropy spectra by modifying the ionization history of the Universe. In particular, since the recombination rate is proportional to the electron density square $n_e^2$, the mean free electron fraction at a given epoch will be modified in the presence of an inhomogeneous plasma. This type of clumpy plasma recombines earlier, and thus reduces the sound horizon $r_{\ast}$ at recombination. The corresponding effect on the CMB spectra is a shift in the position of the peaks, which can be compensated by a larger value of $H_0$. 


It was recently shown in Ref.~\cite{Jedamzik:2020krr} that accounting for the aforementioned baryon inhomogeneities  can alleviate the $H_0$ tension, without significantly spoiling the fit to CMB data. Given the complete ignorance of the probability distribution functions (PDF) characterizing baryon over-densities in the presence of PMFs, the authors of Ref.~\cite{Jedamzik:2020krr}  choose to parameterize the PDF using a three-zone model. While this means that the baryon inhomogeneities can be studied agnostically with regard to their origin, alternative methods not invoking PMFs have not yet been proposed. It is also worth noting that the field strength required to solve the Hubble tension is of the right order to explain the existence of large-scale magnetic fields (without relying on dynamo amplification).  \\

\textbf{Parameters:} Following Ref.~\cite{Jedamzik:2020krr}, we parameterize the baryon density PDF using the three-zone model, in in which each zone is described by its baryon density $n_b^i = \langle n_b \rangle \Delta_i$ and its volume fraction $f_V^i$ ($i=1...3$). We introduce the clumping factor $b = (\langle n_b^2 \rangle-\langle n_b \rangle^2)/\langle n_b \rangle^2$, and emphasize that the parameters $\Delta_i$ and $f_V^i$ are subject to the following constraints
\begin{equation}
\sum_{i=1}^3 f_V^i = 1, \hspace{8mm} \sum_{i=1}^3 f_V^i \Delta_i = 1, \hspace{8mm} \sum_{i=1}^3 f_V^i \Delta_i^2 = 1+b \, .
\end{equation}
The above constraint equations imply that each baryon clumping model is specified by four free parameters, which can be taken to be $b, f_V^2, \Delta_1, \Delta_2$. In order to allow for a direct comparison with Ref.~\cite{Jedamzik:2020krr}, we consider their model M1, which is obtained by fixing $f_V^2 =1/3, \Delta_1 =0.1, \Delta_2 =1$, and leaving $b$ free to vary. We remark that this 
choice of values for the 1-parameter model M1 is very arbitrary, and is only motivated by the fact that it can significantly increase the extent of the high-$H_0$ tail. A more extensive analysis should consider 4-parameter models, which would likely be penalized under the $\Delta \rm{AIC}$ test\footnote{In \cite{Thiele:2021okz}, several extended 2-parameter models were considered, in which both $b$ and $\Delta_1$ were set as free parameters, and $f_V^2$ was sampled at three different points $f_V^2=1/3, 1/2, 2/3$. While the 2-parameter model $f_V^2=1/2$ was shown to allow for more clumping $b$ and higher values of $H_0$ than  the M1 model (when a SH0ES prior was included in the analysis), the fit to CMB likelihoods was still degraded with respect to $\Lambda$CDM, in a similar way to the 1-parameter models  M1 and M2.}.

In order to get the average ionization fraction $\langle \chi_e \rangle$, we compute the ionization fraction in each of the zones using the code \texttt{RECFAST}, and then take the average  $\langle \chi_e \rangle = \sum_i f_V^i \Delta_i \chi_e^i$. While  \texttt{RECFAST} has been calibrated on models close to $\Lambda$CDM, it has been recently shown \cite{Thiele:2021okz} that the more sophisticated \texttt{HyRec} code yields very similar results. Our modified version of CLASS is publicly available at \url{https://github.com/GuillermoFrancoAbellan/class_clumpy}.

\subsubsection{\melong}
\textbf{Motivation:} A variation of the fundamental properties of the hydrogen/helium atom, such as the electron mass or the fine structure constant, is one of the most effective ways to shift the time of recombination in the early Universe.\footnote{For the case of a fine structure variation together with $\Delta N_\mathrm{eff}$\,, it has been shown that there is also an impact on the Lithium problem \cite{2021arXiv210702243F}.} Essentially, by shifting the energy gap between successive excitation levels one directly changes the temperature at which photo-dissociation of hydrogen/helium becomes inefficient. This implies that there is a very strong degeneracy between variations of these fundamental parameters and the redshift of recombination. This degeneracy is only broken when one considers the secondary impact of varying these parameters, which can induce for example shifts in the radiative transfer at recombination, the two-photon decay rate, the photo-ionization cross section, the recombination coefficients, and Thomson scattering (see \eg \cite{Sekiguchi:2020teg}). 

In general, spacetime variations in fundamental parameters are expected to naturally arise from the interactions of exotic low mass particles appearing for example in theories of modified gravity or extra dimensions (see \eg\cite{Uzan:2002vq, Uzan:2010pm,Martins:2017yxk}). In order to phenomenologically parameterize the impact of light species in such models, it is necessary to specify the spacetime dependence of each parameter that is assumed to vary (in a more general theory, these quantities are set by the distribution and coupling of the background field). For simplicity, it is common to assume a spatially uniform and time-independent variation over the characteristic timescale at which the constant is being probed; under this assumption, cosmological probes such as from the \lya~forest and qausar absorption lines~\cite{King:2012id,Bagdonaite:2013sia,Kotus:2016xxb,Murphy:2017xaz,Wilczynska:2020rxx}, the CMB~\cite{Ade:2014zfo,Hart:2017ndk,Hart:2019gvj}, and 21cm observations~\cite{Lopez-Honorez:2020lno} would be largely complementary, as both the size and age of the Universe varies dramatically between these epochs. Here, we follow \cite{Hart:2019dxi} in allowing for a spatially uniform time-independent variation in the electron mass (the effect of simultaneously varying the electron mass and the fine structure constant has also been investigated in~\cite{Hart:2017ndk}). It is also worth pointing out that a variation of the electron mass also does not impact the Silk damping scale as the parameter dependencies exactly cancel out \cite{Sekiguchi:2020teg}. 

\textbf{Parameters:} This model contains one new parameter, the variation in the electron mass, defined by
\begin{equation}\label{eq:vary_me}
\delta m_e \equiv \frac{m_\mathrm{e,early}}{m_\mathrm{e,late}} - 1 \, ,    
\end{equation}
where $m^\mathrm{late}_{e}$ is the locally observed value, and we approximate the variation of $m_e^\mathrm{early}(z)$ as being redshift-independent over recombination. 

\subsubsection{\meOklong}

\textbf{Motivation:} It has recently been pointed out in Ref.~\cite{Sekiguchi:2020teg} that, in combination with a varying electron mass, the presence of non-zero curvature can further reduce the Hubble tension. This is because in addition to shifting the contribution to $r_s$\,, the original model is also simultaneously shifting the corresponding angular diameter distance $D_A$ for the CMB (as it impacts $z_*$) but not for the BAO and other late time probes. While either of these changes can be absorbed with the usual \lcdm parameters, it is impossible to absorb both changes simultaneously. However, here is where the additional freedom of a varying $\Omega_k$ comes into play. This is because the angular diameter distances to the BAO and CMB are impacted in distinct ways by an increase in $\Omega_k$\,, giving this model the additional freedom to preserve both angular scales of BAO and the CMB under a variation of the redshift of recombination. 

It is worth mentioning that we have explicitly checked whether the inclusion of non-zero curvature $\Omega_k$ would also impact other models, however it was found that this is not the case. This is likely due to the fact that these other models also impact other scales in the CMB, such as the diffusion damping scale.

\textbf{Parameters:} This model is a two parameter extension of \lcdm. The first parameter, characterizing the variation in the effective electron mass, is the same as described in the previous model (see \cref{eq:vary_me}). The second parameter is the curvature of the universe~$\Omega_k$\,. 

\subsubsection{\EDElong}\label{ssec:ede}

\textbf{Motivation:} Early dark energy refers to a model in which a scalar field is frozen-in at times prior to recombination, thus behaving during this epoch like a dark energy component. The idea that an anomalous era of expansion arising from EDE at such times might resolve the Hubble tension was first suggested in Ref.~\cite{Karwal:2016vyq}, where computation only at the level of the background was shown to partially alleviate the tension; it was later shown in Refs.~\cite{Poulin:2018cxd,Lin:2019qug,Smith:2019ihp} that {\it Planck} data is also strongly sensitive to the dynamics of the perturbations, favoring either a non-canonical kinetic term, whereby the equation of state $w$ is approximately equal to the effective sound speed $c_s^2$ \cite{Lin:2019qug}, or a potential that flattens close to the initial field value \cite{Smith:2019ihp}.

In this work, we study the modified axion potential introduced in Refs.~\cite{Kamionkowski:2014zda,Karwal:2016vyq,Poulin:2018dzj,Poulin:2018cxd,Smith:2019ihp,Murgia:2020ryi},
\begin{equation}\label{eq:potential}
    V_n(\Theta) = m^2 f^2[1-\cos (\Theta)]^n,
\end{equation} 
where $m$ represents the axion mass, $f$ the axion decay constant, and $\Theta \equiv \phi/f$ is a re-normalized field variable defined such that $-\pi \leq \Theta \leq \pi$. This potential is a phenomenological generalization of the well motivated axion-like potential (which can be recovered by setting $n=1$), arising generically in string theory \cite{Svrcek:2006yi,Douglas:2006es,Arvanitaki:2009fg,Marsh:2015xka}.  We assume that the field always starts in slow-roll (as enforced by the very high value of the Hubble rate at early times), and without loss of generality we restrict $0\leq \Theta_i \leq \pi$. The model at this point has four free parameters: $\{m,f,n,\Theta_i\}$. Given that the dynamics are relatively insensitive to changes of $2\lesssim n \lesssim 4.5$ \cite{Agrawal:2019lmo,Smith:2019ihp}, we further restrict the parameter space by following Refs. \cite{Poulin:2018cxd,Smith:2019ihp} in taking $n=3$. In order to make this model more physically accessible, instead of parameterizing the dynamics using the  mass $m$ and decay constant $f$, we use the fractional energy density $f_{\rm EDE}(z_c)$ (which is roughly the maximum energy contribution induced by EDE) and the critical redshift $z_c$ when the field becomes dynamical. The final degree of freedom is encoded in the dynamics of the linear perturbations, which is fully characterized via the effective sound speed $c_s^2$, and physically related to the curvature of the potential close to the initial field value -- note that after fixing all other phenomenological parameters, this is fully described by $\Theta_i$ \cite{Poulin:2018dzj,Smith:2019ihp}.

The evolution of the model can be summarized as follows (see Refs.~\cite{Poulin:2018dzj,Smith:2019ihp} for more details): at early times the scalar field is frozen-in due to Hubble friction and contributes as dark energy to the expansion; once the Hubble rate drops below its mass, Hubble friction is removed and the field begins to perform damped oscillations about the minimum; the rate of energy dilution is related to the period-averaged equation of state, roughly given by $w(n) = (n-1)/(n+1)$.

While this model is purely phenomenological, similar models set on a stronger theoretical ground have been proposed in the literature (see \eg \cite{Kaloper:2019lpl,Agrawal:2019lmo,Lin:2019qug,Sakstein:2019fmf,Berghaus:2019cls,Alexander:2019rsc,Braglia:2020bym,Ballesteros:2020sik,Gonzalez:2020fdy,Ballardini:2020iws,Niedermann:2019olb,Niedermann:2020dwg,Das:2020wfe,Allali:2021azp,Karwal:2021vpk}). Nevertheless, we stick to this phenomenological potential as it is an excellent proxy devised to capture the dynamics of the EDE properties required to resolve the Hubble tension, that has been shown to provide an excellent fit to both Planck and SH0ES data. 

To perform our analyses, we use the modified version of {\sc CLASS} presented in Ref.~\cite{Smith:2019ihp}. The code is publicly available at \url{https://github.com/PoulinV/AxiCLASS}.

\textbf{Parameters:} This model contains a three parameter extension of \lcdm. The considered parameters are: (1) the critical scale factor at which the field becomes dynamical $z_c$ (and related scale factor $a_c=1/(1+z_c)$), (2) the fractional energy density contributed by the field at the critical scale factor $f_{\rm EDE}(a_c)$, and (3) the initial re-normalized field variable $\theta_i$\,. These parameters control when the field begins to oscillate, the strength of the deviation from \lcdm, and the effective sound speed, respectively.  We take the priors $f_{\rm EDE}(a_c)\in[0,0.3]$, $\log_{10}(z_c)\in[3,4.5]$ and $\theta_i \in [0,3]$.  

\subsubsection{\NEDElong}\label{ssec:nede}

\textbf{Motivation:} We compare the phenomenological EDE model presented above to an alternative promising model dubbed \enquote{new early dark energy} [\NEDEshort{}]. In the original proposal discussed previously, EDE decays through a second-order phase transition when the Hubble rate drops below the effective axion-like mass. Here, NEDE is a component of vacuum energy at the electron volt scale, which decays through a first-order phase transition in a dark sector shortly before recombination due to the influence of a \enquote{trigger field} \cite{Niedermann:2019olb,Niedermann:2020dwg}. This model therefore complements our test of EDE cosmologies (we note that models involving non-minimal EDE couplings to gravity were recently proposed \cite{Sakstein:2019fmf,Allali:2021azp,Braglia:2020bym,Karwal:2021vpk}, but were shown to provide level of success very similar to the two EDE models studied here).  

The NEDE model contains two additional scalar field, the NEDE field $\psi$ of mass $M$ and the trigger (sub-dominant) field $\phi$ of mass $m_\mathrm{NEDE}$, whose potential is written as (with canonically normalized kinetic terms):
\begin{equation}
    V(\psi,\phi)=\frac{\lambda}{4}\psi^4+\frac{1}{2}\beta M^2\psi^2-\frac{1}{3}\alpha M\psi^3+\frac{1}{2}m_\mathrm{NEDE}^2\phi^2+\frac{1}{2}\gamma\phi^2\psi^2.
\end{equation}
where $\lambda$, $\beta$, $\alpha$, $\gamma$ are dimensionless couplings.
When $H\lesssim m$, $\phi$ rolls down the potential and if $\alpha^2>4\beta\lambda,$ $\beta>0$ a second minimum appears. As a result, the field configuration with $\psi=0$ becomes unstable once $\phi$ drops below a certain threshold, and a quantum tunneling to the lowest energy minimum eventually occurs. To put this into context, it was found in Ref.~\cite{Niedermann:2019olb} that a parameter choice $\lambda\sim0.1$, $\alpha\sim\beta\sim{\cal O}(1)$, $M\sim$ eV and $m\sim 10^{-27}$ eV leads to a transition at $z\sim 5000$ with a fractional density of NEDE $f_{\rm NEDE}\sim 0.1$. All necessary details about the model can be found in Ref.~\cite{Niedermann:2020dwg} and in practice, we make use of the modified class version presented therein and available at \url{https://github.com/flo1984/TriggerCLASS}.

\textbf{Parameters:} The NEDE field is described through a fluid formalism and specified via three parameters: the fraction of NEDE before the decay, $f_{\rm NEDE}\equiv\bar\rho_{\rm NEDE}(z_*)/\bar\rho_{\rm tot}(z_*)$ (where $z_*$ is given by the redshift at which $H=0.2 m$), the mass of the trigger field $m_\mathrm{NEDE}$ which controls the redshift of the decay $z_*$\,, and the equation of state after the decay $w_{\rm NEDE}$. Following  Ref.~\cite{Niedermann:2020dwg} we take the effective sound speed in the NEDE fluid $c_s^2$ to be equal to the equation of state after the decay, i.e. $c_s^2=w_{\rm NEDE}$\,. We take the flat priors on $f_{\rm NEDE}\in[0,0.3]$, ${\rm log}_{10}(m_\mathrm{NEDE}/{\rm eV})=[1.3,3.3]$, and $w_{\rm NEDE}\in[1/3,1]$. 

\subsubsection{\EMGlong}\label{ssec:emg}

\textbf{Motivation:} Many modified gravity (MG) models, typically changing General Relativity at early times, appear to have promising levels of success in explaining the $H_0$ tension \cite{Rossi:2019lgt,Braglia:2020iik,Zumalacarregui:2020cjh, Abadi:2020hbr,Ballardini:2020iws,Braglia:2020bym}. Here we consider one of such Early Modified Gravity (EMG) scenarios, which was introduced in Ref. \cite{Braglia:2020auw}. The model contains a scalar field $\sigma$ with quadratic non-minimal coupling (NMC) to gravity and a small effective mass induced by a quartic potential
\begin{equation}
S = \int d^4x \sqrt{-g} \left[ (M_{\rm pl}^2+\xi \sigma^2)\frac{R}{2} -\frac{g^{\mu \nu}}{2} \partial_\mu \sigma\partial_\nu \sigma -\Lambda -\frac{\lambda\sigma^4}{4} \right] +S_{\rm m}.
\end{equation}
Here $R$ is the Ricci scalar, $S_{\rm m}$ is the action for matter fields, and $\xi, \ \lambda$ denote dimensionless constants. For $\lambda = 0$, it reduces to the case of a non-minimally coupled massless scalar field considered in \cite{Braglia:2020iik}, while for $\xi = 0$ it reduces to Rock 'n' Roll model of Ref. \cite{Agrawal:2019lmo}, which is an example of an EDE model. 

In a similar way to EDE models, the scalar field is initially frozen deep in the radiation era, and when its effective mass becomes larger than the Hubble rate, it starts to perform damped oscillations about its minimum. However, due to the NMC parameter $\xi \geq 0$, the scalar field experiences a temporary growth before rolling down the potential, producing new features in the shape of the energy injection. In addition, the NMC predicts a weaker gravitational strength at early times, which leads to a suppression in the matter power spectrum at small scales. In Ref. \cite{Braglia:2020iik}, it was shown that this extra freedom introduced by $\xi$ allow the model to substantially relax the $H_0$ tension, even when Large Scale Structure (LSS) data is included in the fit in addition to CMB and supernovae data. Furthermore, thanks to the fast rolling of $\sigma$ towards the minimum, the tight constraints on the effective Newtonian constant from laboratory experiments and on post-Newtonian parameters from Solar System measurements are automatically satisfied. 

\textbf{Parameters:} This EMG model has three free parameters: (1) the non-minimal coupling to gravity $\xi$, (2) the initial field value in units of the Planck mass $\sigma_i [M_{\rm pl}]$, and (3) a constant $V_0$ measuring the amplitude of the quartic potential. This constant $V_0$ is related to $\lambda$ by
\begin{equation}
\lambda  =\frac{10^{2 V_0}}{3.156 \times 10^{109}},  
\end{equation}
where $3.156 \times 10^{109}$ is the numerical value of $M_{\rm pl}^4$ in units of $\rm{eV}^4$. To allow comparison with \cite{Braglia:2020auw}, we take flat priors $\xi \in [0,1]$, $\sigma_i [M_{\rm pl}] \in [0, 0.9]$ and $V_0 \in [0.6, 3.5]$. 

\pagebreak

\subsection{Late Universe solutions}\label{sec:late}
Since late time solutions cannot modify the sound horizon (by definition they only modify the post-recombination era), they
must attempt to balance an increase of $H_0$ and a corresponding increase of the energy density today by a decrease of the energy density at earlier times (such as through phantom dark energy $w<-1$).

\subsubsection{\CPLlong}

\textbf{Background:} The Chevallier- Polarski- Linder (CPL) parameterization of the Dark energy equation of state was first introduced in Refs~ \cite{Chevallier:2000qy,Linder:2002et}, and has become one of the most common choices for describing deviations from the cosmological constant with equation of state $w = -1$. Since the dark energy impacts primarily the later stages of expansion, the varying equation of state is expanded in the CPL model to first order, giving a linear function of the scale factor. We write
\begin{equation}
w(a) = w_0 + w_a \cdot (1-a)~,
\end{equation}
where it is assumed that today $a=1$. Similarly, to the case of $\Delta N_{\rm eff}$, it is already well known that this simple model does not resolve the Hubble tension \cite{Poulin:2018cxd,DiValentino:2019dzu}. Therefore, rather than a real competitor, we will consider the CPL parameterization of the Dark Energy equation of state as a historical benchmark model (the \enquote{late-universe defending champion}), useful for a reader to gauge the extent to which more complex late-universe models truly performs.

\textbf{Parameters:} This model introduces two new parameters, $w_0$ and $w_a$ describing the DE equation of state. We take the priors $w_0\in[-2,-1/3]$ and $w_a\in[-3,2]$.

\subsubsection{\PEDElong}

\textbf{Motivation:} The Phenomenologically Emergent Dark Energy (PEDE) Model~\cite{Li:2019yem,Pan:2019hac,Rezaei:2020mrj,Yang:2021egn} introduces a particular parameterization for dark energy of the form
\begin{equation}
    \Omega_{DE}(z) = \Omega_{DE, 0} \left[1 - \tanh(\log_{10}(1+z)) \right] \, ,
\end{equation}
which implies a time-evolving equation of state
\begin{equation}\label{eq:eqofstate}
    w_{DE}(z) = \frac{1}{3} \frac{d\ln\Omega_{DE}}{dz}(1+z) - 1\, .
\end{equation}
This model is motivated by the idea that dark energy could be an emergent phenomenon only arising at low redshift, though the particular form of the transition is chosen entirely by a phenomenological argument in that it has been observed to give relatively high $H_0$ values. In particular, $D_A$ is strongly decreased because of the phantom equation of state $w<-1$ that this model exhibits at later times. We stress here that this model does not reduce to \lcdm in some limit. This means that it has the potential to artificially decrease the tension while strongly degrading the fit to the CMB anisotropy data. To circumvent this issue, we also consider the \MEDEshort{} model below.

\textbf{Parameters:} This model introduces no new parameters, since the parametric form of dark energy is fully specified. 

\subsubsection{\MEDElong}

\textbf{Motivation:} The Generalized Phenomenologically Emergent Dark Energy (\MEDEshort{}) Model~\cite{2020ApJ...902...58L,2020MNRAS.497.1590H,2021arXiv210303815Y} generalizes the \PEDEshort{} model in order to also encompass \lcdm as a realization. For this purpose it introduces a parameter $\Delta$, which can be set to $1$ to recover the \PEDEshort{} model and to $0$ to recover the \lcdm model.\footnote{In their original work the authors also recommend the use of some characteristic transition redshift $z_t$\,. In order to keep the original \PEDEshort{} model as a limit of this generalization, we fix $z_t=0$ as was done in~\cite{Benaoum:2020qsi}.} The corresponding equation of state is described by 
\begin{equation}
    \Omega_{DE}(z) = \Omega_{DE, 0} \left[1 - \tanh(\Delta \log_{10}(1+z)) \right] \, ,
\end{equation}
and the equation of state is found from \cref{eq:eqofstate}. This model allows us to cross-check whether the apparent success of the \PEDEshort{} model at $\Delta=1$ is artificial (by degrading the Planck fit), or whether it remains even in \enquote{competition} with the \lcdm limit for $\Delta \to 0$. 

\textbf{Parameters:} This model introduces one new parameter $\Delta$, which switches between the \lcdm and \PEDEshort{} limits of this model. We only impose $\Delta\geq0$.

\enlargethispage*{1\baselineskip}
\subsubsection{\fracDMlong}

\textbf{Motivation:} We consider a scenario in which a fraction $f_{\rm dcdm}$ of dark matter is allowed to decay into invisible massless particles (\ie dark radiation (DR)). The cosmological impact of this class of model has been studied in Refs.~ \cite{Poulin:2016nat,Nygaard:2020sow,2016PhRvD..94b3528C,2018PhRvD..97h3508C}, extending upon previous work in which the entirety of dark matter was assumed to decay in a universal manner \cite{Audren:2014bca}. By allowing for a fraction of decaying dark matter smaller than unity, \ie $f_{\rm dcdm} < 1$, larger decay rates $\Gamma_\mathrm{dcdm}$ are allowed, leading to a richer phenomenology in the CMB and matter power spectra. This class of models has also invoked to both resolve the $\sigma_8$ \cite{Enqvist:2015ara} and the Hubble tension \cite{Berezhiani:2015yta,Bringmann:2018jpr,Pandey:2019plg}. We find little evidence for this behavior. Comparing \cite{Poulin:2016nat,Nygaard:2020sow} we observe that the constraints from Planck 2018 are simply stronger than those of Planck 2015. Additionally, compared to the earlier investigations \cite{Berezhiani:2015yta,Bringmann:2018jpr} we have included BAO and CMB lensing data.

\textbf{Parameters:} The two free parameters that characterize this scenario are the dark matter decay rate, $\Gamma_\mathrm{dcdm}$, and the fraction of dark matter that is allowed to decay,
\begin{equation}
f_{\rm dcdm} = \frac{ \Omega_{\rm dcdm}^{\rm ini} }{ \Omega_{\rm dcdm}^{\rm ini} + \Omega_{\rm sdm}}, 
\end{equation}
where $\Omega_{\rm sdm}$ is the present abundance of stable dark matter and $\Omega_{\rm dcdm}^{\rm ini}$ is the initial abundance of decaying dark matter, given by $\Omega_{\rm dcdm}^{\rm ini} = \lim_{a \rightarrow 0} a^3\rho_{\rm dcdm}(a)/\rho_c $. We take flat priors on $f_{\rm dcdm}\in[0,1]$ and $\Gamma_\mathrm{dcdm}\in [0,100]$ km/s/Mpc.

\subsubsection{\DMDRWDMlong}

\textbf{Motivation:} We study a 2-body Decaying Cold Dark Matter (DCDM) model where the entirety of dark matter is unstable due to the decay into a massless particle and a massive particle; the latter of these will then behave in a manner akin to warm dark matter.

The phenomenology of such a model has recently been reviewed in great details in Ref.~\cite{Abellan:2021bpx}. The model is characterized by two parameters, the DCDM lifetime, $\Gamma^{-1}$, and the fraction of DCDM rest mass energy converted into dark radiation, defined as \cite{Blackadder:2014wpa}:
\begin{equation}\label{eq:epsilon}
\varepsilon = \frac{1}{2} \left(1-\frac{m^2_{\rm wdm}}{m^2_{\rm dcdm}}\right),
\end{equation}
where $m^2_{\rm dcdm}$ and $m^2_{\rm wdm}$ refer to the mass of the parent particle and massive decay product respectively. Thus, $0 \leq \varepsilon \leq 1/2$, with the lower limit corresponding to the standard CDM case (so that $\Omega_{\rm{cdm}}=\Omega_{\rm{dcdm}}+\Omega_{\rm{wdm}}$) and $\varepsilon = 1/2$ corresponding to DM decaying solely into dark radiation. In general, small $\varepsilon$ values (\ie~heavy massive decay products) and small $\Gamma$ values (\ie~lifetimes much longer than the age of the universe) induce little departures from \lcdm. In the intermediate regime, the warm decay product imprints a characteristic suppression to the matter power spectrum, in a similar fashion as massive neutrinos or warm dark matter.

The authors of Ref.~\cite{Vattis:2019efj} suggested that such a model can provide a resolution of the Hubble tension. 
However, this conclusion was recently challenged in Refs.~\cite{Haridasu:2020xaa}, where it was shown that including CMB and SNIa data spoils the success of the model to resolve the Hubble tension. 
Still, in Ref.~\cite{Abellan:2020pmw}, it was shown thanks to a new fluid approximation to describe the perturbations of the massive decay products, that such a model can in fact resolve the $S_8$ tension if $\Gamma^{-1} \simeq 55~ (\varepsilon/0.007)^{1.4}$ Gyrs. 

To perform our analyses, we use the modified version {\sc CLASS} presented in Ref.~\cite{Abellan:2020pmw}. The code is publicly available at \url{https://github.com/PoulinV/class_decays}. 

\textbf{Parameters:}
We include two new parameters, the mass fraction fraction $\varepsilon$ guiding the thermal nature of the decay product, and the decay width $\Gamma$ guiding the time at which the decay occurs. We take logarithmic priors defined as ${\rm Log}_{10}(\Gamma / [\rm{km}/ \rm{s}/ \rm{Mpc}])\in[-1,4]$ and ${\rm Log}_{10}\varepsilon\in[-4,{\rm Log}_{10}(0.5)]$.

\newpage

\section{Results}\label{sec:results}

Having presented the individual models to be studied in this work, we proceed by systematically testing each model against a series of data sets, determining at each point the extent to which the model successfully passes each of the established criteria. We begin by analyzing the tension using the full Planck2018 likelihood in conjunction with data from BAO -- we call this \textit{Test A} and present the results from performing MCMCs, successively adding individual datasets one by one, in  \cref{sec:resultsA}. We then extend the previous test by including data from the Pantheon supernovae sample \cite{Scolnic:2017caz}, highlighting the tension in this case in terms of the calibration of the supernovae absolute magnitude (this test is referred to as \textit{Test B}, and is shown in \cref{sec:resultsB}). We use \textit{Test B} to then establish a set of models, the~\enquote{finalists}, which pass either the tension metric criterion (GT or $Q_{\rm DMAP}$) or the $\Delta$AIC criteria (\ie the model generates a $\Delta$AIC value relative to \lcdm of $\lesssim-6.91$ when including a SH0ES likelihood). For the sample of finalists, we proceed by investigating the importance of Planck data relative to other early Universe probes (shown in \textit{Test C} of \cref{sec:resultsC}); in particular, we consider replacing Planck by WMAP+ACT, removing information from CMB anisotropies altogether and using data from BBN. Finally, in \textit{Test~D} of \cref{sec:resultsD}, we analyze the impact of including information from redshift space distortions, cosmic chronometers, and the Lyman alpha forest.

\subsection{Test A -- the warm-up round: Planck and BAO data versus SH0ES} \label{sec:resultsA}

In our first test, we adopt the common approach of combining the CMB anisotropy data from Planck with a very robust probe of the late time cosmological evolution, the Baryonic Acoustic Oscillations (BAOs). Using these data sets, we investigate the extent to which each of our models eases the $H_0$ tension. In order to understand the importance of each data set, we successively apply each one by one starting from the Planck2018 TT + lowEE power spectrum; we then quantify the tension on $H_0$ with the SH0ES collaboration using a simple Gaussian likelihood, a common approach adopted in the literature (although, as stated above, not the correct approach for cosmologies with late-time modifications to the expansion rate). In \cref{sec:resultsB} we repeat this test, now properly modelling the SH0ES measurements as the magnitude calibration of the Pantheon SNIa sample.

For the dataset
\begin{center}
    \DA = Planck 2018 (including TTTEEE and lensing) + BAO (including BOSS DR12 + MGS + 6dFGS)\footnote{in the MontePython v3.1 package, this corresponds to the likelihoods {\tt bao\_boss\_dr12} + {\tt bao\_smallz\_2014}},
\end{center}
Our baseline analysis includes BAO measurements from 6dFGS at $z=0.106$~\cite{Beutler:2011hx}, SDSS DR7 at $z=0.15$~\cite{Ross:2014qpa} and BOSS DR12 at $z=0.38, 0.51, 0.61$~\cite{BOSS:2016wmc}.
Table~\ref{tab:summary_H0} summarizes the result of our two statistical tests, showing (1) the Gaussian na\"ive residual $H_0$ tension of each model using \DA and comparing with the SH0ES likelihood, (2) the $Q_\mathrm{DMAP}$ generalization of the former, and (3) the Akaike Information Criterion using both \DA and SH0ES for each model relative to $\Lambda$CDM. We identify whether each model has failed (red cross) or succeeded (green check mark) in reducing the tension as measured using our predefined thresholds. A discussion of the results for each model category can be found below.
\begin{table}[h]
    \hspace*{-0.5cm}
    \centering
    \begin{filecontents*}{sheets/H0Table_nohalofit.csv}
Name,,H0string,TensVal,Tension,Chi2,DChi2 wrt LCDM,AIK wrt LCDM,Passes DAIK H0,String DAIK,Passes Sigma H0,String SigmaH0,Passes both H0,Chi2 SH0ES,Chi2 Tension,Chi2 String,Chi2 passed,Chi2 string,Chi2 or DAIK,Chi2 vs H0
\lcdmtable,0,67.77 \pm 0.44,3.9564482421379,4.0 \sigma,2798.39,0.00,0.00,FALSE,\color{Red} X ,FALSE,\color{Red} X,\color{Red} X,16.7013530116783,4.08672888893774,4.1 \sigma,FALSE,\color{Red} X,\color{Red} X,-0.031878955110648
\nefftable,1,68.38 \pm 0.69,3.27497347651414,3.3 \sigma,2794.50,-3.89,-1.89,FALSE,\color{Red} X ,FALSE,\color{Red} X,\color{Red} X,11.824215070535,3.43863564085162,3.4 \sigma,FALSE,\color{Red} X,\color{Red} X,-0.047595087537959
\idrtable,1,68.67 \pm 0.84,2.92678654483803,2.9 \sigma,2791.46,-6.93,-4.93,FALSE,\color{Red} X ,TRUE,\color{Green} \checkmark,\color{Green} \checkmark,8.94766185792241,2.9912642574541,3.0 \sigma,TRUE,\color{Green} \checkmark,\color{Green} \checkmark,-0.021555338166929
\equitable,2,67.83 \pm 1.23,3.00056193806687,3.0 \sigma,2791.01,-7.38,-3.38,FALSE,\color{Red} X ,TRUE,\color{Green} \checkmark,\color{Green} \checkmark,8.16869176185674,2.85809232913437,2.9 \sigma,TRUE,\color{Green} \checkmark,\color{Green} \checkmark,0.049847797945578
\schmaltztable,2,68.55 \pm 0.92,2.91973933417524,2.9 \sigma,2791.27,-7.12,-3.12,FALSE,\color{Red} X ,TRUE,\color{Green} \checkmark,\color{Green} \checkmark,9.72945330420271,3.11920715955236,3.1 \sigma,FALSE,\color{Red} X,\color{Red} X,-0.06394824555537
\SINtable,3,66.79 \pm 1.18,3.65101427943275,3.7 \sigma,2794.46,-3.93,2.07,FALSE,\color{Red} X ,FALSE,\color{Red} X,\color{Red} X,13.98,3.73898381916799,3.7 \sigma,FALSE,\color{Red} X,\color{Red} X,-0.023527659917719
\majorontable,3,68.86 \pm 0.93,2.71421760417193,2.7 \sigma,2785.09,-13.30,-7.30,TRUE,\color{Green} \checkmark,TRUE,\color{Green} \checkmark,\color{Green} \checkmark,7.39000000000033,2.7184554438137,2.7 \sigma,TRUE,\color{Green} \checkmark,\color{Green} \checkmark,-0.001558914512065
\primtable,1,68.50 \pm 0.62,3.26325946281248,3.3 \sigma,2789.45,-8.94,-6.94,TRUE,\color{Green} \checkmark,FALSE,\color{Red} X,\color{Green} \checkmark,11.4399999999996,3.38230690505749,3.4 \sigma,FALSE,\color{Red} X,\color{Green} \checkmark,-0.035197114155137
\varyingmetable,1,68.46 \pm 1.26,2.61818521589857,2.6 \sigma,2790.89,-7.51,-5.51,FALSE,\color{Red} X ,TRUE,\color{Green} \checkmark,\color{Green} \checkmark,9.38807089463035,3.06399590316801,3.1 \sigma,FALSE,\color{Red} X,\color{Red} X,-0.145499766108856
\varyingmeomegaktable,2,69.29 \pm 2.11,1.5776778052337,1.6 \sigma,2784.29,-14.10,-10.10,TRUE,\color{Green} \checkmark,TRUE,\color{Green} \checkmark,\color{Green} \checkmark,2.29995297111873,1.51655958376805,1.5 \sigma,TRUE,\color{Green} \checkmark,\color{Green} \checkmark,0.040300573824997
\edetable,3,68.51 \pm 0.55,3.32256646731657,3.3 \sigma,2779.47,-18.92,-12.92,TRUE,\color{Green} \checkmark,FALSE,\color{Red} X,\color{Green} \checkmark,2.01368000000002,1.41904193031778,1.4 \sigma,TRUE,\color{Green} \checkmark,\color{Green} \checkmark,1.34141528613781
\nedetable,3,68.80 \pm 0.68,2.99910164335293,3.0 \sigma,2781.88,-16.52,-10.52,TRUE,\color{Green} \checkmark,TRUE,\color{Green} \checkmark,\color{Green} \checkmark,4.01567747400759,2.00391553564705,2.0 \sigma,TRUE,\color{Green} \checkmark,\color{Green} \checkmark,0.496620785658288
\emgtable,3,68.36 \pm 0.87,3.09412069913462,3.1 \sigma,2782.50,-15.89,-9.89,TRUE,\color{Green} \checkmark,FALSE,\color{Red} X,\color{Green} \checkmark,2.34387800000013,1.53097289329372,1.5 \sigma,TRUE,\color{Green} \checkmark,\color{Green} \checkmark,1.02101599100031
\cpltable,2,65.61 \pm 1.60,3.68169078805154,3.7 \sigma,2784.48,-13.92,-9.92,TRUE,\color{Green} \checkmark,FALSE,\color{Red} X,\color{Green} \checkmark,5.33817699999963,2.31044952336112,2.3 \sigma,TRUE,\color{Green} \checkmark,\color{Green} \checkmark,0.593495443560093
\pedetable,0,71.75 \pm 0.51,1.03833994359739,1.0 \sigma,2787.02,-11.37,-11.37,TRUE,\color{Green} \checkmark,TRUE,\color{Green} \checkmark,\color{Green} \checkmark,1.49187197998936,1.22142211376303,1.2 \sigma,TRUE,\color{Green} \checkmark,\color{Green} \checkmark,-0.149892627702305
\medetable,1,69.03 \pm 1.19,2.36772972764314,2.4 \sigma,2785.79,-12.60,-10.60,TRUE,\color{Green} \checkmark,TRUE,\color{Green} \checkmark,\color{Green} \checkmark,5.40676268755306,2.32524465111804,2.3 \sigma,TRUE,\color{Green} \checkmark,\color{Green} \checkmark,0.018271228580044
\dmdecaytable,2,67.66 \pm 0.43,4.04595185159281,4.0 \sigma,2798.35,-0.04,3.96,FALSE,\color{Red} X ,FALSE,\color{Red} X,\color{Red} X,16.6706100088518,4.08296583488667,4.1 \sigma,FALSE,\color{Red} X,\color{Red} X,-0.009065464858314
\fracdmdecaytable,2,67.63 \pm 0.47,4.02936133731452,4.0 \sigma,2797.91,-0.48,3.52,FALSE,\color{Red} X ,FALSE,\color{Red} X,\color{Red} X,15.3899999999999,3.92300904918659,3.9 \sigma,FALSE,\color{Red} X,\color{Red} X,0.027109875810759
\end{filecontents*}
\csvreader[/csv/head=true,tabular={l c|c c c c | r r c},
    table head={ Model & $\Delta N_{\rm param}$ & $H_0$ & \begin{tabular}{@{}c@{}}Gaussian \\ Tension\end{tabular} & \begin{tabular}{@{}c@{}}$Q_{\rm DMAP}$ \\ Tension\end{tabular}   & & $\Delta \chi^2$ & $\Delta$AIC &   \\ \hline}]{sheets/H0Table_nohalofit.csv}{1=\name,2=\nparam,3=\hsig,5=\htens,7=\Dchi,8=\DAIK,12=\Htest,10=\DAIKtest,16=\htensnew,18=\Htestnew}{ \name & \nparam & $\hsig$ & $\htens$ & ${\htensnew}$ & ${\Htestnew}$  & ${\Dchi}$  & ${\DAIK}$ & ${\DAIKtest}$ }
    \caption{Test of the models based on dataset \DA (Planck 2018 + BAO), using the indirect measurement of $H_0$ by SH0ES for the quantification of the tension (4th column) or the computation of the AIC (7th column). Eleven models pass at least one of the three tests at the 3$\sigma$ level.}
    \label{tab:summary_H0}
\end{table}

Table~\ref{tab:chi2_H0} provides further details on the AIC test: it shows a breakdown of each best-fit $\Delta \chi^2$ (relative to the \lcdm model) for the joint \DA+SH0ES dataset into contributions from the individual likelihoods.  This is useful to test the compatibility of data sets, with positive values reflecting that a shift toward the SH0ES measurement necessitates a degradation of the fit to another cosmological observation.
\begin{table}[h]
    \centering
    \begin{filecontents*}{sheets/dChi2_nohalofit.csv}
Name,Planck lowL,Planck highL,Planck lensing,BAO,Sh0es (H0),Total,Case 2: Up to Pantheon and using Mb,Planck lowL,Planck highL,Planck lensing,BAO,Pantheon,Sh0es (Mb),Total,Case 3: all data,Planck lowL,Planck highL,Planck lensing,BAO,Pantheon,RSD,CC,LBAO,Sh0es (Mb),Total
\lcdmtable,0.00,0.00,0.00,0.00,0.00,0.00,\lcdmtable,0.00,0.00,0.00,0.00,0.00,0.00,0.00,\lcdmtable,0.00,0.00,0.00,0.00,0.00,0.00,0.00,0.00,0.00,0.00
\nefftable,-2.19,4.10,0.40,-0.24,-5.96,-3.89,\nefftable,-1.06,5.61,-0.09,0.30,-0.05,-10.81,-6.10,\nefftable,-420.49,-2348.97,-8.93,-7.54,-1027.06,-0.52,-14.72,-4.71,-18.19,-3851.13
\idrtable,0.46,-2.72,0.02,-0.15,-4.55,-6.93,\idrtable,-0.27,3.25,-0.02,0.51,-0.05,-12.98,-9.57,\idrtable,-1.06,2.15,0.37,0.62,-0.18,-0.07,0.49,-0.23,-11.38,-9.30
\equitable,-0.87,-1.38,0.56,-0.07,-5.61,-7.38,\equitable,-1.23,2.53,0.12,0.14,0.24,-10.63,-8.83,\equitable,-420.49,-2348.97,-8.93,-7.54,-1027.06,-0.52,-14.72,-4.71,-18.19,-3851.13
\schmaltztable,-0.40,-0.10,0.22,-0.11,-6.74,-7.12,\schmaltztable,-0.34,3.61,-0.09,0.34,-0.05,-12.40,-8.92,\schmaltztable,-420.49,-2348.97,-8.93,-7.54,-1027.06,-0.52,-14.72,-4.71,-18.19,-3851.13
\SINtable,1.11,-1.83,0.03,1.86,-5.10,-3.93,\SINtable,1.34,1.02,-0.16,2.25,-0.03,-9.40,-4.98,\SINtable,-420.49,-2348.97,-8.93,-7.54,-1027.06,-0.52,-14.72,-4.71,-18.19,-3851.13
\majorontable,-1.04,-3.55,0.24,-0.10,-8.85,-13.30,\majorontable,-0.51,-0.84,-0.18,0.94,-0.03,-14.87,-15.49,\majorontable,-1.57,-1.33,0.30,2.80,-0.17,-0.25,0.92,-0.53,-13.39,-13.22
\primtable,0.60,-5.46,-0.21,-0.34,-3.53,-8.94,\primtable,1.19,-2.93,-0.65,0.67,0.06,-9.76,-11.42,\primtable,0.63,-3.72,-0.19,0.13,0.03,-0.07,-0.16,-0.02,-4.60,-7.97
\varyingmetable,-0.11,-2.92,0.42,-0.04,-4.85,-7.51,\varyingmetable,0.20,-1.23,-0.52,2.42,0.04,-13.19,-12.27,\varyingmetable,-1.08,0.97,0.15,2.63,-0.12,-0.14,0.40,-0.50,-11.52,-9.21
\varyingmeomegaktable,-1.69,-4.78,2.82,1.74,-12.19,-14.10,\varyingmeomegaktable,-1.47,-2.64,1.06,2.19,0.80,-17.20,-17.26,\varyingmeomegaktable,-1.68,-0.77,0.30,3.02,0.34,-0.19,1.03,-1.29,-14.57,-13.80
\edetable,-2.26,-5.49,0.34,-0.15,-11.35,-18.92,\edetable,-1.78,-3.80,-0.02,0.04,-0.22,-16.20,-21.98,\edetable,-2.97,-3.26,0.42,0.97,-0.38,-0.20,2.34,-0.19,-15.56,-18.84
\nedetable,-2.73,-2.80,0.39,-0.06,-11.32,-16.52,\nedetable,-2.67,-0.26,0.03,0.13,-0.24,-15.93,-18.93,\nedetable,-3.18,-2.09,0.33,0.75,-0.36,-0.19,1.31,-0.14,-13.67,-17.24
\emgtable,-0.95,-4.25,-0.09,-0.21,-7.28,-12.79,\emgtable,-0.55,-2.23,-0.53,0.11,-0.19,-15.17,-18.56,\emgtable,-1.35,-2.07,-0.01,0.79,-0.14,-0.17,1.50,-0.24,-14.38,-16.07
\cpltable,-0.96,-5.99,-0.01,4.70,-11.66,-13.92,\cpltable,0.18,-1.37,-0.79,4.72,0.74,-8.43,-4.94,\cpltable,-420.49,-2348.97,-8.93,-7.54,-1027.06,-0.52,-14.72,-4.71,-18.19,-3851.13
\pedetable,-0.81,-5.49,-0.18,6.63,-11.52,-11.37,\pedetable,0.05,-3.32,-0.25,5.61,11.14,-10.99,2.24,\pedetable,-0.88,-2.68,1.45,11.28,11.38,-0.39,0.42,-0.20,-11.57,8.80
\medetable,-0.81,-6.03,-0.37,4.51,-9.90,-12.60,\medetable,3.64,-3.32,-0.81,0.79,0.09,-0.83,-0.45,\medetable,-420.49,-2348.97,-8.93,-7.54,-1027.06,-0.52,-14.72,-4.71,-18.19,-3851.13
\dmdecaytable,-0.03,-2.29,0.04,-0.33,2.57,-0.04,\dmdecaytable,0.28,-0.54,-0.20,-0.03,-0.19,0.50,-0.19,\dmdecaytable,-420.49,-2348.97,-8.93,-7.54,-1027.06,-0.52,-14.72,-4.71,-18.19,-3851.13
\fracdmdecaytable,0.30,-2.19,-0.01,-0.31,1.73,-0.48,\fracdmdecaytable,0.27,-0.23,-0.28,-0.02,0.08,-0.35,-0.53,\fracdmdecaytable,-420.49,-2348.97,-8.93,-7.54,-1027.06,-0.52,-14.72,-4.71,-18.19,-3851.13
\end{filecontents*}
\csvreader[/csv/head=true,tabular={l|r r r r r | r},/csv/respect dollar=false,/csv/respect backslash=false,
    table head={ Model & low-$\ell$ & high-$\ell$ & lensing & BAO & SH0ES & Total \\ \hline}]{sheets/dChi2_nohalofit.csv}{1=\name,2=\lowl,3=\highl,4=\lens,5=\bao,6=\shoes,7=\total}{\name & $\lowl$ & $\highl$ & \lens & \bao & \shoes & \total}
    \caption{Contribution of each likelihood to the best-fit $\chi^2=-2\ln \mathcal{L}$ of each model to the dataset \DA+SH0ES, relative to the best-fit $\chi^2$ of $\Lambda$CDM to the same dataset. Here SH0ES is treated as a simple Gaussian measurement of $H_0$\,.
   \label{tab:chi2_H0}}
\end{table}

Finally, in \ref{app:AB}, the left panel of each figure shows the triangle plot for all our runs based on dataset \DA, compared to those without BAO, without Planck lensing and without Planck high-$\ell$ polarisation.

\noindent {\bf Dark Radiation models.} The first round of tests illustrates a mixed level of success among solutions invoking additional relativistic degrees of freedom. Only the \Majoronmid{} model passes the $\Delta$AIC test and three models are capable of reducing the GT to the $3\sigma$ level (the \idrmid{} model, the \Schmaltzmid{} model, and the \Majoronmid{} model), however the residual tension in the most promising of this category (the \Majoronmid{} model) is still at the $2.7\sigma$ level. Adopting the $Q_\mathrm{DMAP}$ criterion leads to qualitatively similar results, only shifting tensions by order of $\sim 0.1 \sigma$, as expected in the case of Gaussian posteriors. In previous works, most of these models had been presented as promising solutions to the Hubble tension. Many of these models, however, had not been confronted with the Planck 2018 polarisation data, which drastically reduces the extent to which this class of models can accommodate larger values of $\Delta N_{ur}$\,, and thus larger values of $H_0$. The triangle plots of \ref{app:AB} show that without the polarization data of Planck 2018, most models with Dark Radiation can indeed reach values of $H_0$ of the order of 74~km/s/Mpc. 

\enlargethispage*{1\baselineskip}
In general, models with dark radiation exploit the parameter degeneracy between $H_0$ and $N_{\rm eff}$ at the level of CMB data and BAO data. This degeneracy is mainly lifted by the measurement of the damping tail of the CMB spectra, which is suppressed when adopting higher values of $H_0$\,. To counteract the impact of a large $H_0$ on the damping tail, some of the previous models invoke additional ingredients, such as DR self-interactions, DM-DR scattering, or neutrino interactions. Our results suggest that the level of precision of Planck 2018 data (in particular in the polarisation spectrum) is sufficient to differentiate the subtle effects of these ingredients, limiting the preferred value of $H_0$ to be $\lesssim$ 70~km/s/Mpc at the 95\%CL.

\noindent {\bf Other early universe models.}
At this stage, early-Universe modifications that shift the sound horizon without relying on dark radiation tend to perform better (on average). All six models considered in this category pass at least one of the tests, with \EDEmid{}, \NEDEmid{}, \EMGshort{}, and \meOkshort{} passing all three. The residual GT in this class of models varies between 3.3$\sigma$ in the case of a \Bmid{}, to a remarkable 1.6$\sigma$ in the case of an \meOkmid{}. Instead, adopting the $Q_\mathrm{DMAP}$ criterion, four of the models reach a tension of less than $2\sigma$, with the \EDEshort{} model even reaching an agreement of $1.4\sigma$. The \EDEshort{} model shows an improvement in $\Delta \chi^2$ of around $-19$, easily passing the established threshold on $\Delta$AIC. It is worth mentioning that the relative success of \EDEshort{} in the $Q_{\rm DMAP}$ criterion, relative to that of the GT criterion, arises from a known sampling issue \cite{Smith:2019ihp,Murgia:2020ryi} we have already mentioned: given current precision, there exists a $\chi^2$ degeneracy within Planck and BAO data between $\Lambda$CDM and EDE. Since EDE has two parameters ($\Theta_i,z_c$) which once $f_{\rm EDE}\rightarrow0$ become irrelevant,  it leads to posteriors that are naturally weighted towards \lcdm\footnote{This was clearly demonstrated in Ref.~\cite{Smith:2019ihp} using mock Planck CMB data: even if a fake EDE signal (with $f_{\rm EDE} \!=\!0.12$ and $H_0=72$ km/s/Mpc) is injected in the data, Planck cannot break the degeneracy between EDE and $\Lambda$CDM and only obtains an upper limit on $f_{\rm EDE}$. 
A survey such as CMB-S4 would unambiguously break that degeneracy. Reducing the number of free parameters from three to $f_{\rm EDE}$-only make that degeneracy become evident \cite{Niedermann:2019olb,Murgia:2020ryi}.}. This preference for \lcdm is thus at least partially a result of the precise layout of the Bayesian prior volume, and not a result of a true preference (a similar caveat applies to \NEDEshort).
Therefore, this impacts the validity of the Gaussian approximation to quantify the level of tension with SH0ES: the complicated degeneracy in the multi-dimensional space leads to long tails in the 1D posteriors of $M_b$ (and $H_0$), and the distribution thus deviates from Gaussian (and becomes asymmetric). Concretely, in the case of P18+BAO+Pantheon (our canonical run to gauge of the tension level with SH0ES), we find that the error on $M_b$ (above the mean) at 68\% C.L. is $\sigma_{ 68\%} = 0.017$, at 95\% C.L. $\sigma_{ 95\%}=0.06$, and at  99.5\% C.L. $\sigma_{99.5\%} = 0.1$, which clearly shows that one cannot simply rescale the $\sigma_{ 68\%}$ to extract the limit at any given confidence level. 

Regarding the relevance of Planck polarization measurement, we note that it is of major importance in the case of a \Bmid{}. With Planck TT only, the \Bmid{} model would in fact provide a very promising resolution to {\em both} the $H_0$ and $S_8$ tension. However, the success of the solution is spoiled once other data sets are included in the analysis (we stress that more complex models might be worth investigating, although a recent study testing 5 different clumping models indicate that these are disfavored by the combination of Planck+ACT data \cite{Thiele:2021okz}). The situation is a bit different for the \memid{}, for which constraints are fairly similar, independently of the dataset considered (barring the inclusion of SH0ES). This 1-parameter model provides a remarkable tentative solution to both tensions, if purely based on the GT criterion (test 1). Yet, $\Delta$AIC (test 3) illustrates that the required value of $\delta m_e$ is not \enquote{strongly favored} (on the Jeffrey's scale) by Planck data.
However, once the extension to a non-flat universe is considered, constraints without late-time information dramatically change. 
In that case, the model provides a promising resolution to the $H_0$ tension, but does not perform as well regarding $S_8$. Once BAO data are included, the model shows similar performance as when $\Omega_k$ was set to 0 (indicating BAO data disfavors non-zero $\Omega_k$). However, allowing $\Omega_k$ to freely vary does lead to larger uncertainties on $H_0$, which indeed helps in reducing the tension. 

It was already noted that EDE models are more strongly constrained by polarization than temperature \cite{Lin:2019qug,Smith:2019ihp} (in particular regarding the dynamics of perturbations), a result that we confirm for both \EDEshort{} and \NEDEshort{}. We find here a similar behavior for the EMG model,  with polarization data providing most of the constraints on the model. We also note that BAO data do not significantly further affect the constraints to these models. This is of particular importance since it has been argued \cite{Jedamzik:2020zmd} that the degeneracy direction between $H_0-r_s(z_{\rm drag})$ and $H_0-r_s(z_{\rm CMB})$ is different, such that these data could potentially break the degeneracy that allows for high $H_0$. However, there is an interesting \enquote{coincidence} in that the impact of \EDEshort{} on the height of acoustic peaks is (partly) compensated by an increase in $\omega_{\rm cdm}$, which turns out to help \EDEshort{} models in providing an excellent fit to BAO data (see \cref{tab:chi2_H0}). However, that increase in $\omega_{\rm cdm}$ leads to a larger $S_8$, which can potentially be problematic regarding the $S_8$ tension \cite{Hill:2020osr,Murgia:2020ryi}. We will comment further about this in \cref{sec:resultsD}.

\noindent {\bf Late universe models.}
Of the late-Universe models, three involve modifications to DE and two rely on decaying dark matter. Two models invoking exotic DE each pass one of the two tests. 
Starting with the \enquote{benchmark} CPL extension to \lcdm, one can see that the apparent tension between Planck and SH0ES (and in fact even with $S_8$!) is resolved in this model, {\em in the absence of BAO and Pantheon data}. This is a perfect illustration of the importance of including late-time dynamics information when quantifying the tension. Indeed, when BAO data are included in the fit, the apparent resolution of both tensions disappear. Yet, this shows that any model whose only role is to relieve the \enquote{Planck+SH0ES} (no BAO) tension should not be considered successful (barring the presence of unknown systematic errors in BAO data analysis).

In the \PEDEmid{} model, the GT on $H_0$ disappears completely, reducing it to a spectacular level of just 1.0$\sigma$ (albeit the $Q_\mathrm{DMAP}$ residual tension is slightly larger). While the fit in the \PEDEmid{} model certainly improves upon \lcdm (although is not sufficient to pass the $\Delta$AIC cut), \Cref{tab:chi2_H0} shows that the fit comes at the expense of an increasing tension with BAO. At this stage (without uncalibrated SNIa data added), also the \MEDEmid{} model performs well, reducing the GT to around $2.4\sigma$ ($Q_\mathrm{DMAP}$ of 2.9$\sigma$) and reaching a similar $\Delta$AIC.  In the next section, we will see how the inclusion of supernovae data exacerbates this tension. 

Finally, neither of the two decaying dark matter models is successful in reducing the tension. Actually, \cref{fig:decay_std,fig:dcdm} show that for these models, Planck 2018 temperature and low-$\ell$ polarization data alone are already sufficient to constrain high values of $H_0$. 

\subsection{Test B -- the qualifying round: Adding Pantheon data}\label{sec:resultsB}

In our second test, we add the supernovae data from the Pantheon compilation to our test. 
In this case, it is important to properly treat the SH0ES measurement. 
In particular, as discussed in \cite{Camarena:2021jlr,Efstathiou:2021ocp} and in \cref{sec:summary} we abandon the simple Gaussian likelihood in $H_0$\,, and instead model the measurement as a Gaussian prior constraint on the absolute magnitude of the supernovae of type Ia ($M_B$).

For the dataset  
\begin{center}
\DB = \DA + Pantheon (marginalised over $M_B$),
\end{center}
the result of the three tests are summarized in \cref{tab:summary_MB}. More details on the contribution of each likelihood to the $\chi^2$ used for our AIC test (based on joint fits of \DB+SH0ES) are provided in \cref{tab:chi2_Mb}. 
The triangle plots of each of our MCMC runs are provided in \ref{app:AB}, in the right panel of each figure. Finally, the bounds on the extended parameters of each model are given in Table ~\ref{tab:additional_parameters}, except for cases for which parameter degeneracies do not allow to get bounds on individual parameters (e.g. there can be no limit on a decay rate when the decaying Dark Matter fraction goes to zero).

\begin{table}[h]
\renewcommand{\arraystretch}{1.2}
    \centering
    \hspace*{-0.5cm}
    \scalebox{0.85}{
    \csvreader[/csv/head=true,tabular={l c|  c c c c | r r c | c c},/csv/respect dollar=false,/csv/respect backslash=false,
    table head={ Model& $\Delta N_{\rm param}$ & $M_B$ &  \begin{tabular}{@{}c@{}}Gaussian \\ Tension\end{tabular} & \begin{tabular}{@{}c@{}}$Q_{\rm DMAP}$ \\ Tension\end{tabular} & & $\Delta \chi^2$  & $\Delta$AIC & & \begin{tabular}{@{}c@{}}One test \\ passed\end{tabular} & \\ \hline}]{sheets/MbTable_nohalofit2.csv}{1=\name,2=\nparam,3=\mbsig,5=\mbtens,7=\Dchi,8=\DAIK,12=\Mbtest,10=\DAIKtest,13=\anytest,16=\newmbtens,18=\newMbtest,20=\newanytest,24=\mname}{ \name & \nparam & $\mbsig$ & $\mbtens$ & $\newmbtens$ & $\newMbtest$ & $\Dchi$  &  $\DAIK$ & $\DAIKtest $ & $\newanytest$ & \mname}} 
    \repeatcaption{tab:summary_MB}{Test of the models based on dataset \DB (Planck 2018 + BAO + Pantheon), using the direct measurement of $M_B$ by SH0ES as the GT for the quantification of the tension (4th column), the $Q_\mathrm{DMAP}$ criterion (5th column), or the computation of the $\Delta$AIC (7th column). Eight models pass at least one of these three tests at the 3$\sigma$ level.}
\renewcommand{\arraystretch}{1}
\end{table}
\begin{table}[h]
    \centering
    \scalebox{0.95}{
    \csvreader[/csv/head=true,tabular={l|r r r r r r r r},/csv/respect dollar=false,/csv/respect backslash=false,
    table head={ Model & low-$\ell$ & high-$\ell$ & lensing & BAO & Pantheon & SH0ES & Total \\ \hline}]{sheets/dChi2_nohalofit.csv}{8=\name,9=\lowl,10=\highl,11=\lens,12=\bao,13=\pantheon,14=\shoes,15=\total}{\name & $\lowl$ & $\highl$ & \lens & \bao &\pantheon & \shoes & \total}}
    \caption{Contribution of each likelihood to the best-fit $\chi^2=-2\ln \mathcal{L}$ of each model to the dataset \DB+SH0ES, relative to the best-fit $\chi^2$ of $\Lambda$CDM to the same dataset. Here SH0ES is treated as a measurement of $M_B$. 
 \label{tab:chi2_Mb}}
\end{table}

\begin{table}[p]
    \centering
    \begin{tabular}{c|c|c|c}
    \hline
    Model & Parameter & ${\cal D}_\mathrm{baseline}$ & ${\cal D}_\mathrm{baseline}$ + $M_B$  \\
    \hline
    \nefftable &
    $N_\mathrm{eff}$ & 3.1557 $\pm$ 0.0677 & 3.3768 $\pm$ 0.1371\\
    \hline
    \idrtable &
    $N_\mathrm{idr}$ & 0.1463 $\pm$ 0.0926 & 0.3867 $\pm$ 0.1318 \\
    \hline
    \equitable &
    $N_\mathrm{tot}$ & 2.9604 $\pm$ 0.1736 & 3.3479 $\pm$ 0.1441 \\
    &
    $f_\mathrm{idr}$ & (8.9697 $\pm$ 5.6555)x10$^{-2}$ & (13.5157 $\pm$ 6.8720)x10$^{-2}$ \\
    \hline
    \schmaltztable &
    $N_\mathrm{idr}$ & 0.1501 $\pm$ 0.0965 & 0.4290 $\pm$ 0.1353 \\
    &
    $\Gamma_{0,\mathrm{nadm}}$ & (2.388 $\pm$ 1.447)x10$^{-8}$ & (2.371 $\pm$ 1.484)x10$^{-8}$ \\
    \hline
    \SINtable &
    $\log_{10}(G_\mathrm{eff}/[\mathrm{MeV}^{-2}])$ & $-1.77 \pm 0.15$ & $-1.66 \pm 0.16$ \\
    &
    $\Sigma m_\nu$ [eV] & $< 0.147$ & $< 0.154$ \\
    &
    $ N_{ur}$ & $1.74 \pm 0.19$ & $2.175 \pm 0.175$ \\
    \hline
    \majorontable &
    $\log_{10} (m_\phi/\mathrm{eV})$ & unconstrained & unconstrained \\
    &
    $\log_{10} (\Gamma_\mathrm{eff})$ & unconstrained & unconstrained \\
    &
    $\Delta N_{ur}$ & $< 0.363$ & $< 0.751$ \\
    \hline
    \primtable &
    $b$ & $<0.495$ & $0.48 \pm 0.22$ \\
    \hline
    \varyingmetable &
    ${m_\mathrm{e,early}}/{m_\mathrm{e,late}}$ & 1.0047 $\pm$ 0.0066 & 1.0169 $\pm$ 0.0053 \\
    \hline
    \varyingmeomegaktable &
    ${m_\mathrm{e,early}}/{m_\mathrm{e,late}}$ & 1.015 $\pm$ 0.018 & 1.0417 $\pm$ 0.0132 \\
    &
    $\Omega_k$ &  (-3.1870 $\pm$ 5.045)$\times10^{-3}$ & (-8.9719 $\pm$ 4.181) $\times 10^{-3}$ \\
    \hline
    \edetable &
    $\log_{10}(z_c)$ & unconstrained &  $3.614_{-0.13}^{+0.041}$  \\
    &
    $f_\mathrm{EDE}(a_c)$ & $< 0.084$ &  $0.104_{-0.028}^{+0.034}$  \\
    &
    $\theta_i$ & unconstrained & $2.602_{-0.0074}^{+0.32}$ \\
    \hline
    \nedetable &
    $f_\mathrm{NEDE}$ &$<0.117$ & $0.125_{-0.03}^{+0.038}$\\
    &
    $\log_{10}(m_\mathrm{NEDE})$ & $>1.839$ &  $2.5_{-0.1}^{+0.19}$ \\
    &
    $w_\mathrm{NEDE}$ & $>1.3$ & $2.106_{-0.21}^{+0.17}$  \\
    \hline  
    \emgtable &
    $\xi$ & $<0.75$ & $<0.31$ \\
    &
    $\sigma_i$ & $<0.38$ & $0.490_{-0.084}^{+0.14}$ \\
    &
    $V_0$ & unconstrained & $2.54_{-0.25}^{+0.11}$ \\
    \hline
    \cpltable &
    $w_0$ &  $-0.959_{-0.086}^{+0.08}$  & $-0.933_{-0.093}^{+0.088}$ \\
    &
    $w_a$ & $-0.279_{-0.28}^{+0.34}$ & $-0.5543_{-0.34}^{+0.41}$ \\
    \hline
    \medetable &
    $\Delta$ & unconstrained & $<0.2241$ \\
    \hline
    \fracdmdecaytable &
    $\Gamma_\mathrm{dcdm}$ [km/s/Mpc]& unconstrained & unconstrained \\
    &
    $f_\mathrm{dcdm}$ & unconstrained & unconstrained \\
    \hline
    \dmdecaytable &
    $\log_{10}(\Gamma/[\mathrm{km}/\mathrm{s}/\mathrm{Mpc}])$ & unconstrained & unconstrained \\
    &
    $\log_{10}(\epsilon)$ & unconstrained & unconstrained \\
    \hline
    \end{tabular}
    \caption{Confidence intervals on additional parameters in each model (beyond the six $\Lambda$CDM parameters) for test B, presented either as \enquote{mean $\pm$ 68\%CL} or \enquote{upper/lower 95\% bound}. For the unconstrained parameters, the contours can be seen in \ref{app:AB}.
 \label{tab:additional_parameters}}
\end{table}

\noindent {\bf Dark Radiation models.} In the previous section, three models invoking DR passed one of the tests, namely \idrmid{},  \Equishort, and \Majoronmid{}.  When including supernovae data, only two of these models still pass any test: namely \idrmid{} and the \Majoronmid{}.  The \Majoronmid{} model passes both the $Q_{\rm DMAP}$ criterion and the $\Delta$AIC, with a reduction of the $H_0$ tension to the 2.9$\sigma$ level and $\Delta$AIC=-9.49, while the \idrmid{} only pass the $\Delta$AIC criterion, with $\Delta$AIC=-7.56. Importantly, we note that for this class of models the results are qualitatively very similar regardless of whether one adopts a prior on $H_0$ or $M_b$.

 \noindent {\bf Other early universe models.} All models modifying the sound horizon through some other mechanism keep passing one of the three tests, although not as easily as in absence of supernovae data. In the case of \Bmid{}, the Hubble tension remains above the 3$\sigma$ level according to both GT and $Q_{\rm DMAP}$\,, but the AIC test is still passed ($\Delta$AIC=-9.42). The \meshort{} passes all three tests ($\Delta$AIC=-10.27), although both the GT and $Q_\mathrm{DMAP}$ tensions are still at the $2.9\sigma$ level. Finally, the \meOkshort{}\,, \EDEshort, \NEDEshort{}, and EMG models pass both the $Q_{\rm DMAP}$ and $\Delta$AIC criterion beyond $2.5\sigma$ (resulting in $1.9\sigma$, $1.6\sigma$, $1.9\sigma$ and $2.3\sigma$ tension, and $\Delta$AIC=-13.26, -15.98, -12.93 and -12.56  respectively). \EDEshort{}, \NEDEshort{} and EMG fail the GT criterion, however this is simply a consequence of the non-Gaussian posteriors.
We conclude that the \EDEshort{}, \NEDEshort{}, EMG and \meOkshort{} models perform the best in both the $Q_{\rm DMAP}$ tension and the $\Delta$AIC criterion. 
As in the DR case, we note that using a prior on $H_0$ or $M_b$ leads to similar results, except for EMG for which the inclusion of a prior on $M_b$ slightly restricts the success of the solution.
\pagebreak[20]

\noindent {\bf Late universe models.}
While the late-time DE models (\CPLshort{} and \PEDEshort{}) easily passed the $H_0$-based criterion of \cref{sec:resultsA}, the more consistent modeling of SH0ES via the calibration of the absolute supernova luminosity $M_B$ (rather than a measurement of $H_0$, as discussed in \cite{Camarena:2021jlr,Efstathiou:2021ocp}) puts these models under serious pressure. When including the Pantheon supernova data, these models still predict very high values of $H_0$, however they do not impact luminosity distances to the nearby supernovae, and thus do not ease the $M_B$ tension. This feature is nicely demonstrated already with our benchmark CPL scenario, which performs much worse when a prior on~$M_b$ is included, rather than on~$H_0$. As stressed in Refs.~\cite{Camarena:2021jlr,Efstathiou:2021ocp}, this is because the calibration of Pantheon SNIa intrinsic luminosity  based on the inverse-distance ladder method leads to values of $M_b$ incompatible with SH0ES\footnote{see also earlier discussions in Refs.~\cite{Bernal:2016gxb,Poulin:2018dzj,Knox:2019rjx} which showed that a reconstructed $H(z)$ history from either the Planck or SH0ES calibration of BAO and SNIa are already in tension.}.
Similarly, this is clearly seen for \PEDEshort{} in \cref{fig:PEDE_AB}, where the ($H_0-M_\mathrm{B}$) contour plot reveals that the \PEDEshort{} is in tension with the actual calibration of $M_B$ while agreeing with the na\"ive $H_0$ value derived from a \lcdm-based analysis. 
Nevertheless, the \PEDEshort{} model still passes the GT and $Q_{\rm DMAP}$ tests, albeit marginally, remaining at 2.7$\sigma$ GT and $2.8\sigma~Q_\mathrm{DMAP}$ tension in $M_B$\,.
However, the real issue is that the \PEDEshort{} model has a positive $\Delta$AIC (and $\Delta \chi^2$) with respect to \lcdm, and \PEDEshort{} provides in fact a worse fit to the combined data sets that $\Lambda$CDM. \Cref{tab:chi2_Mb} shows that this is caused by a strong degradation of the joint fit to BAO and Pantheon data (with $\Delta \chi^2_{\rm BAO}=5.61$ and $\Delta \chi^2_{\rm Panth.}=11.14$).
\enlargethispage*{2\baselineskip}
This is a consequence of the non-nested nature of the \PEDEshort{} model. Instead, the nested \MEDEshort{} model leads to a (slighlty) better fit than the \lcdm model ($\Delta \chi^2<0$) but is not reducing the tension according to criteria 1 or 2. This is similar to how a \lcdm model with $\Omega_\mathrm{cdm}h^2$ fixed to $0.11$ can reduce criteria 1 and 2 artificially (as described in \cref{sec:summary}), while giving a worse fit. The generalization to the full \lcdm model similarly to the generalized \MEDEshort{} model both do not reduce criteria 1 or 2.

In summary, the models that pass at least one of the three tests for data set \DB\ -- without degrading the fit to the combined dataset compared to $\Lambda$CDM -- and make it to the \enquote{finale} are (ranked from the best to the worst $\Delta$AIC): \EDElong, a \meOkmid{}, \NEDElong{}, EMG, the \Bmid{} model, a \memid{} in a flat universe, the \Majoronmid{} model and the SIDR model.

The \meshort{} (both with and without curvature) is the only model receiving a \enquote{gold medal} for passing all three criteria, and remarkably reducing the tension to the $\lesssim 2\sigma$ when non-zero curvature is considered. \EDEshort{}, \NEDEshort{}, EMG and \Majoronshort{}  all win \enquote{silver medals} for passing the $Q_{\rm DMAP}$ and $\Delta$AIC simultaneously, while the \Bmid{} and SIDR models win \enquote{bronze medals} for passing the $\Delta$AIC criterion.

\subsection{Test C -- first round of the finale: Role of Planck data}\label{sec:resultsC}

In our third test, we attempt to assess the extent to which our conclusions are robust under the exclusion of Planck data.
For this test, we decide to focus on the sample of finalists, which is well representative of the three categories of solutions studied earlier. Importantly, our goal is not to revive discarded models using data that are less constraining than Planck. Rather, we hope to understand the relative importance of Planck data in constraining the resolution to the $H_0$ tension - known to exist independently of Planck.
We thus repeat the analysis using alternatives to the dataset \DB, in which all of the Planck data is substituted by a combination of either non-Planck CMB data, or non-CMB data probing the sound horizon scale and the expansion history. 

Our non-Planck data combination \DCnoP includes:
\begin{itemize}
\item WMAP 9-year likelihood \cite{Hinshaw:2012aka}
\item ACT DR-4 likelihood \cite{Aiola:2020azj}
\item A Gaussian prior on $\tau_\mathrm{reio}=0.0543\pm 0.0073$
\end{itemize}
This is essentially the same combination of data sets as presented in the main analysis of Ref.~\cite{Aiola:2020azj}, except using a slightly more informative prior on $\tau_\mathrm{reio}$ inspired by SPT3G analysis \cite{Dutcher:2021vtw}. We do not perform the same analysis as in Ref.~\cite{Dutcher:2021vtw} as the official SPT3G likelihood is not yet released.

Our non-CMB data combination \DCnoCMB includes:
\begin{itemize}
\item the same BAO data as \DA and \DB
\item the same Pantheon supernova data as \DB
\item Big Bang Nucleosynthesis [BBN] data with Deuterium measured from \cite{Cooke:2017cwo} and with Helium as measured by \cite{Aver:2015iza}. We simply model each measurement as a one-dimensional Gaussian likelihood, and obtain the predictions for a given value of $\omega_b, N_\mathrm{eff}$ from a pre-tabulated table computed with \texttt{Parthenope 2.0}.
\item Additional Lyman-$\alpha$ (high-$z$) BAO data from \cite{Agathe:2019vsu,Blomqvist:2019rah}.
\item Cosmic Chronometers [CC] from table 1 of \cite{Vagnozzi:2020dfn}.
\end{itemize}
The combination of these datasets probe the expansion rate and thermal history of the Universe within each model, but is completely independent of the evolution of perturbations, providing an important cross-check at the level of the background expansion history. Thus, in this case, we remove $A_s$ and $n_s$ from our list of varied parameters (and $S_8$ from our list of derived parameters). 

\begin{table}[h]
    \centering
    \begin{tabular}{c|c|c|c}
    \hline
    Model & Parameter & \DCnoP & \DCnoCMB  \\
    \hline
    \idrtable &
    $N_\mathrm{idr}$ & (6.53 $\pm$ 3.91)x10$^{-2}$ & (1.73 $\pm$ 1.09)x10$^{-2}$ \\
    \hline
    \majorontable &
    $\log_{10} (m_\phi/\mathrm{eV})$ & unconstrained & unconstrained \\
    &
    $\log_{10} (\Gamma_\mathrm{eff})$ & unconstrained & unconstrained \\
    &
    $\Delta N_\mathrm{eff}$ & $<0.434$ & $< 0.411$ \\
    \hline
    \primtable &
    $b$ & $<0.463$ & $0.41 \pm 0.2$ \\
    \hline
    \varyingmetable &
    ${m_\mathrm{e,early}}/{m_\mathrm{e,late}}$ & 1.0052 $\pm$ 0.0057 & 1.005 $\pm$ 0.0108 \\
    \hline
    \varyingmeomegaktable &
    ${m_\mathrm{e,early}}/{m_\mathrm{e,late}}$ & unconstrained & 1.0088 $\pm$ 0.0388 \\
    &
    $\Omega_k$ & unconstrained & (-9.1932 $\pm$ 3.5651)x10$^{-3}$ \\
    \hline
    \edetable &
    $\log_{10}(z_c)$ & $3.326_{-0.093}^{+0.2}$ & unconstrained \\
    &
    $f_\mathrm{EDE}(a_c)$ & $0.1593_{-0.095}^{+0.051}$ &  $<0.24$ \\
    &
    $\theta_i$ & unconstrained &  unconstrained\\
    \hline
    \nedetable &
    $f_\mathrm{NEDE}$ &  $0.103_{-0.064}^{+0.021}$ & $<0.24$  \\
    &
    $\log_{10}(m_\mathrm{NEDE}/\mathrm{eV})$ & unconstrained &unconstrained  \\
    &
    $w_\mathrm{NEDE}$ &unconstrained  & unconstrained \\
    \hline  
    \emgtable & $\xi$& $-$ & unconstrained \\
    & $\sigma_i$ & $-$ & $<0.64$ \\
    & $V_0$ &  $-$ & unconstrained   \\
    \hline
    \end{tabular}
    \caption{Confidence intervals on additional parameters in finalist models (beyond the six $\Lambda$CDM parameters) for test C, presented either as \enquote{mean $\pm$ 68\%CL} or \enquote{upper/lower 95\% bound}.For the unconstrained parameters, the contours can be seen in \ref{app:C}. The EMG model was not confronted to WMAP+ACT, since the code available to us is not compatible with that likelihood.
 \label{tab:additional_parameters_noPlanck}}
\end{table}

The results of these runs are presented in the form of triangle plots in \ref{app:C}. For this case, our bounds on extended model parameters are given in Table~\ref{tab:additional_parameters_noPlanck}. The results show that, as far as the resolution of the $H_0$ tension is concerned, the role of CMB data in general, and of Planck data in particular, can vary from \enquote{essential} to \enquote{marginal} depending on the model. We summarize the results of each model here:
\begin{itemize}
\item Most impressively, for \EDEshort, the constraints significantly change when trading \DB for \DCnoP. In that case, the degeneracy between $f_{\rm EDE}(a_c)$-$M_B$ opens up, and we find $f_{\rm EDE}(a_c)=0.153^{+0.050}_{-0.085}$ (in fact, $f_{\rm EDE}(a_c)$ compatible with 0 only at $\sim 2.5 \sigma$ due to non-Gaussian posteriors) and $M_B = -19.252\pm 0.081$, which is in $0.1\sigma$ agreement with SH0ES. Barring potential residual systematic errors in ACT or Planck, this illustrates the importance of the high-accuracy measurement made by Planck in the $\ell$-range $1000-2000$ (that outperforms WMAP+ACT in TT) in constraining the EDE model. 
As a consequence of this potential mild inconsistency between Planck and ACT, when combining them, we find  only an upper limit on $f_{\rm EDE}(a_c) < 0.12$ with $M_b = -19.36_{-0.045}^{+0.023}$, which is significantly {\em  weaker} than the constraints from Planck only $f_{\rm EDE}(a_c) <0.092$, with  $M_b = -19.39_{-0.035}^{+0.016}$.
Further studies of the EDE model with ACT data are presented elsewhere \cite{Poulin:2021bjr}. 

The situation is different in the \DCnoCMB case; we find again a (relatively weak) upper limit on $f_{\rm EDE}(a_c)$ and $M_B$ is in tension with SH0ES at $2.2\sigma$, which is more in line with our \DB case. We conclude that besides ACT, none of the data we used favor a deviation from $\Lambda$CDM which brings $M_B$ in fully satisfying agreement with SH0ES. 

\item For \NEDEshort, we find a similar behavior as \EDEshort. When trading \DB for \DCnoP\,, a multi-modal distribution appears (corresponding to high or low trigger field mass, i.e., $z_{*}\sim10^4$ and $z_{*}\sim10^3$ respectively), both favoring non-zero $f_{\rm NEDE}$ at $\sim2\sigma$. Yet, only the low-mass part of the parameter space leads to $M_B$ in perfect agreement with SH0ES, with the high-mass part only compatible with SH0ES at the $2\sigma$ level. The high-mass mode, on the other hand, is closer to the parameter space favored by the analysis of \DB+SH0ES.
A deeper analysis of the \NEDEshort{} model in light of ACT data is performed in Ref.~\cite{Poulin:2021bjr}.
Nevertheless, when replacing CMB data by \DB, we find only a weak upper-limit on $f_{\rm NEDE}$, and $M_B$ is again in tension with SH0ES at $2\sigma$. 

\item For the \EMGshort{} model, the version of the code available is not compatible with the ACT likelihood, and further work is needed to confront this model against ACT data. Against the data \DCnoCMB, we find that the EMG model allows slightly higher $H_0$ and $M_b$ values, reducing the tension to around $2.4\sigma$, though the EMG free parameters are largely unconstrained.

\item For the \Majoronmid{} model, in addition to a general broadening of the posterior, there is also a slight upward shift in $H_0$ and $M_B$ when replacing Planck with ACT+WMAP. This marginally reduces the tension, but not by a significant amount. It is interesting that the WMAP+ACT result generates a slightly bi-modal distribution at high majoron mass, rather than the single low mass mode obtained with Planck data. This is perhaps not too surprising as the mass controls the time at which the majoron thermalizes with neutrinos (and damps neutrino free-streaming), and thus is correlated with the multipoles most influenced by the presence of the majoron; thus, changing CMB data is likely to induce a shift in this parameter. Using BBN rather than the CMB simply broadens the posteriors, but does not show a noticeable shift toward larger values of $H_0$.
\item For the \Bmid{} model, the Planck and ACT data are essentially interchangeable. Both constraints overlap very well. However, abandoning the CMB data altogether does allow for slightly higher $H_0$ and $M_B$ values, reducing the tension from 3.5$\sigma$ to 1.9$\sigma$. For the \DCnoCMB the \Bmid{} model has approximately the same constraints as other models shifting the sound horizon such as the \meshort{} or \meOkshort{} models, since in this case the additional effects caused by the changed recombination beyond a shift of the sound horizon are not probed by these data.
\item For the \memid{} model, the high-$\ell$ information stored in Planck data plays a moderate role in this case, since with \DCnoP the tension on $M_B$ decreases from 2.9$\sigma$ to 2.1$\sigma$. Note that, in this case, a small 2.1$\sigma$ tension remains without the use of any CMB data, so the CMB is not such a crucial probe of this model. Indeed, this models induces a shift of the sound horizon which can be probed almost as well by measuring the BAO scale as by measuring the CMB peak scale.

\item For the \meOkmid{} model, we reach similar conclusions. The tension on $M_B$ is nearly the same when using the data sets \DB, \DCnoP or \DCnoCMB (it is always around 2.0$\sigma$).

\item For the \idrmid{} model the role of Planck data also appears to be minor, with constraints from \DCnoP showing a strong tension ($\sim 4.0\sigma$), only degraded a slight bit when using \DCnoCMB ($\sim 2.7\sigma$).

\end{itemize}

We may conclude that as far as CMB measurements are concerned, the Planck data plays a crucial role only for the two models of EDE. The high-$\ell$ Planck temperature/polarization data around $\ell \in [1500,2500]$ appears to disfavor a large fraction of \EDEshort{} and \NEDEshort{} around recombination. Should new systematics be identified within the Planck data, the constraints on these models could radically change. The other models of our sample of finalists are not impacted significantly by the exchange of Planck with WMAP+ACT data, despite of the fact that the latter combination is less constraining and does not probe as accurately the details of the temperature and polarisation spectra at intermediate angular scales. Instead, these models are mostly constrained through many of the geometric probes contained in the Planck data, such as the angular scale of the acoustic oscillations. 
This means that most of our conclusions are robust against many possible Planck systematics.

\begin{figure}[h!]
    \centering
    \includegraphics[width=0.8\textwidth]{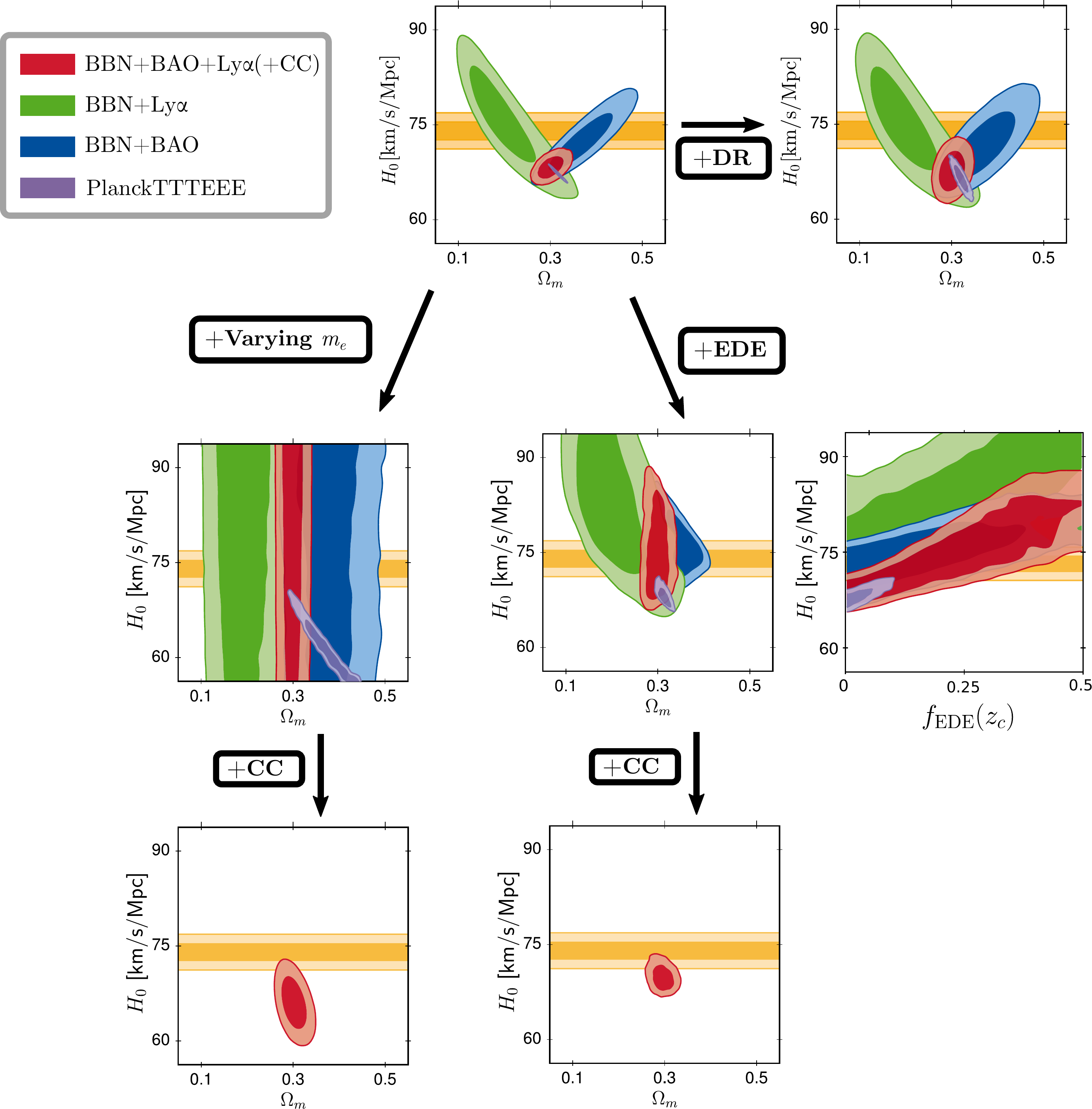}
    \caption{Constraints from the combination of BBN and BAO probes (galaxy BAO is denoted as \enquote{BAO}, while the BAO as derived from the Lyman-$\alpha$ forests from distant quasars are denoted as \enquote{Ly$\alpha$}) on various models. As a comparison we also show the Planck results in some of the panels. The orange bands represent the $H_0$ results from SH0ES. In the top row we show the \lcdm results on the left, and the \Neffshort{} on the right. In the middle row, we show the results for \meshort{} on the left and \EDEshort{} on the right. Finally, in the bottom row we additionally include Cosmic Chronometers (CC) for the respective models.}
    \label{fig:BAOBBNprinciple}
\end{figure}

A surprising amount of information can also be gleamed when abandoning the CMB anisotropy probes altogether, relying just on late universe observations and a calibration of the baryon density from BBN. In this case, most constraints are weakened a slight bit, but not by much.

In \cref{fig:BAOBBNprinciple} we demonstrate how the constraints from a combination of BAO and BBN data depend on the underlying cosmological model. In the \lcdm case the combined constraints are quite tight (top middle). The same is true when looking at solutions using for example dark radiation (top right), or when investigating late-time solutions. However, as soon as other early universe solutions are employed (middle row), the BAO+BBN constraints become very loose. In this case, the observational Hubble data in form of Cosmic Chronometers (CC) is crucial for establishing tight constraints (bottom row).

The BAO+BBN probe works on the basis of calibrating the sound horizon from BBN. This allows the BAO measurement of the sound horizon angle to be converted into one of the angular diameter distance at the given redshift. By combining probes from different redshifts (such as Lyman-$\alpha$ BAO and galaxy BAO), a joint measurement of $H_0$ and $\Omega_m$ can then be established. The crucial calibration of the sound horizon through BBN can only be established if the sound horizon is not impacted by the model (late solutions) or shifted in a way that can be constrained by BBN (Dark Radiation solutions). This is not the case for any of the other early universe solutions. In this case, the Cosmic Chronometers (CC) are crucial for calibrating the absolute distance scale of the BAO.

This is particularly visible in \cref{fig:BAOBBNprinciple} for the \memid{} model, for which there is virtually no constraints on $H_0$ in the absence of CC data. On the other hand, we note that the Hubble parameter appears to be partially constrained for the \EDEshort{} model. This is because, any additional early dark energy $f_\mathrm{EDE}(z_c)>0$ can only increase the Hubble parameter, thus leading to the same lower bound as \lcdm, while the upper constraint is purely artificial and arises from our imposed upper prior limit of $f_\mathrm{EDE}(z_c)<0.5$, most strongly impacting the galactic BAO constraints.

\subsection{Test D -- second round of the finale: Impact of additional data sets}\label{sec:resultsD}
\enlargethispage*{1\baselineskip}
While at this point we have already used a large fraction of the cosmological probes for our analysis, one might wonder if additional constraints can be derived from using complementary data sets.
For this purpose, in our final test, we go back to our (Planck + BAO + Pantheon) baseline data set \DB, and consider the impact of additional data on the models of our silver sample, including:
\begin{itemize}
    \item Redshift-Space-Distortions [RSD] (including the full-shape information of \cite{BOSS:2016wmc} and other $f \sigma_8$ measurements from the 6dFGS \cite{Beutler:2012px}, the MGS \cite{Howlett:2014opa}, and the eBOSS DR 14 quasars \cite{Zarrouk:2018vwy})
    \item Cosmic Chronometers [CC] (already introduced in the previous section),
    \item Lyman-$\alpha$ based high-$z$ BAO (already introduced in the previous section).
\end{itemize}
We successively add these three data sets to check if any of these additional late time probes of the expansion history add any information. We call this final dataset \DD, which includes all of the aforementioned datasets simultaneously.
For this case, our bounds on extended model parameters are given in Table~\ref{tab:additional_parameters_golden} while all figures are reported in~\ref{app:D}.
\begin{table}[h]
    \centering
    \begin{tabular}{c|c|c|c}
    \hline
    Model & Parameter & ${\cal D}_\mathrm{baseline}$ & ${\cal D}_\mathrm{baseline}$ + $M_B$  \\
    & & + RSD + CC + Ly$\alpha$ & + RSD + CC + Ly$\alpha$  \\
    \hline
    \idrtable &
    $N_\mathrm{idr}$ & (1.54 $\pm$ 0.97)x10$^{-2}$ & 0.36 $\pm$ 0.12 \\
    \hline
    \majorontable &
    $\log_{10} (m_\phi/\mathrm{eV})$ & unconstrained & unconstrained \\
    &
    $\log_{10} (\Gamma_\mathrm{eff})$ & unconstrained & unconstrained \\
    &
    $\Delta N_\mathrm{eff}$ & $< 0.310$ & $< 0.632$ \\
    \hline
    \primtable &
    $b$ & $<0.463$ & $0.41 \pm 0.2$ \\
    \hline
    \varyingmetable &
    ${m_\mathrm{e,early}}/{m_\mathrm{e,late}}$ & 1.0052 $\pm$ 0.0057 & 1.150 $\pm$ 0.0047 \\
    \hline
    \varyingmeomegaktable &
    ${m_\mathrm{e,early}}/{m_\mathrm{e,late}}$ & 1.0173 $\pm$ 0.0132 & 1.0392 $\pm$ 0.0110 \\
    &
    $\Omega_k$ & -4.1490x10$^{-3}$ & (-9.1932 $\pm$ 3.5651)x10$^{-3}$ \\
    \hline
    \edetable &
    $\log_{10}(z_c)$ & unconstrained & $3.628_{-0.2}^{+0.096}$ \\
    &
    $f_\mathrm{EDE}(a_c)$ & $< 0.0742$ &  $0.0824_{-0.026}^{+0.031}$ \\
    &
    $\theta_i$ & unconstrained &  $2.479_{0.058}^{+0.52}$\\
    \hline
    \nedetable &
    $f_\mathrm{NEDE}$ & $<0.098$ & $0.105_{-0.026}^{+0.031}$  \\
    &
    $\log_{10}(m_\mathrm{NEDE}/\mathrm{eV})$ & $>1.94$ &$2.52_{-0.13}^{+0.16}$  \\
    &
    $w_\mathrm{NEDE}$ &$>1.28$  & $2.13_{-0.26}^{+0.16}$ \\
    \hline  
       \emgtable & $\xi$& unconstrained & unconstrained   \\
    & $\sigma_i$ & $<0.31$  &$0.39_{-0.08}^{+0.17}$ \\
    & $V_0$ & unconstrained  & $2.64_{-0.35}^{+0.12}$  \\
    \hline
    \end{tabular}
    \caption{Confidence intervals on additional parameters in finalist models (beyond the six $\Lambda$CDM parameters) for test D, presented either as \enquote{mean $\pm$ 68\%CL} or \enquote{upper/lower 95\% bound}.For the unconstrained parameters, the contours can be seen in \ref{app:D}.
 \label{tab:additional_parameters_golden}}
\end{table}
\begin{table}[h]
    \centering
    \hspace*{-0.5cm}
    \scalebox{0.9}{
    \begin{filecontents*}{sheets/Alldata_nohalofit.csv}
Name,Npar,Mbstring,TensVal,Tension,Chi2,DChi2 wrt LCDM,AIK wrt LCDM,Passes DAIK Mb,String DAIK,Passes Sigma Mb,String SigmaH0,Passes both Mb,Chi2 wrt noSH0ES,Tension,Tensval,Passes Chi2,PassChi2,AIC or Chi2,PassBoth,Delta Crit 1 Crit 3,AIC AND Chi2,Passes all
\lcdmtable,0,-19.420 \pm 0.012,4.50453016058739,4.5 \sigma,3851.13,0.00,0.00,FALSE,\color{Red} X,FALSE,\color{Red} X,\color{Red} X ,18.95,4.3526439666024,4.4 \sigma,FALSE,\color{Red} X,FALSE,\color{Red} X,-0.033718543015634,FALSE,FALSE
\nefftable,1,.000 \pm 0.000,-515.911528150134,-515.9 \sigma,0.00,-3851.13,-3849.13,TRUE,\color{Green} \checkmark,TRUE,\color{Green} \checkmark,\color{Green} \checkmark,0.00,0,.0 \sigma,TRUE,\color{Green} \checkmark,TRUE,\color{Green} \checkmark,-1,TRUE,TRUE
\idrtable,1,-19.384 \pm 0.024,3.16768605009799,3.2 \sigma,3841.83,-9.30,-7.30,TRUE,\color{Green} \checkmark,FALSE,\color{Red} X,\color{Green} \checkmark,11.12,3.33426437774047,3.3 \sigma,FALSE,\color{Red} X,TRUE,\color{Green} \checkmark,0.052586754182071,FALSE,FALSE
\equitable,2,.000 \pm 0.000,-515.911528150134,-515.9 \sigma,0.00,-3851.13,-3847.13,TRUE,\color{Green} \checkmark,TRUE,\color{Green} \checkmark,\color{Green} \checkmark,0.00,0,.0 \sigma,TRUE,\color{Green} \checkmark,TRUE,\color{Green} \checkmark,-1,TRUE,TRUE
\schmaltztable,2,.000 \pm 0.000,-515.911528150134,-515.9 \sigma,0.00,-3851.13,-3847.13,TRUE,\color{Green} \checkmark,TRUE,\color{Green} \checkmark,\color{Green} \checkmark,0.00,0,.0 \sigma,TRUE,\color{Green} \checkmark,TRUE,\color{Green} \checkmark,-1,TRUE,TRUE
\SINtable,3,.000 \pm 0.000,-515.911528150134,-515.9 \sigma,0.00,-3851.13,-3845.13,TRUE,\color{Green} \checkmark,TRUE,\color{Green} \checkmark,\color{Green} \checkmark,0.00,0,.0 \sigma,TRUE,\color{Green} \checkmark,TRUE,\color{Green} \checkmark,-1,TRUE,TRUE
\majorontable,3,-19.380 \pm 0.021,3.19980275167169,3.2 \sigma,3837.91,-13.22,-7.22,TRUE,\color{Green} \checkmark,FALSE,\color{Red} X,\color{Green} \checkmark,10.79,3.28481354113143,3.3 \sigma,FALSE,\color{Red} X,TRUE,\color{Green} \checkmark,0.026567509330171,FALSE,FALSE
\primtable,1,-19.390^{+0.017}_{-0.021},3.57392556667074,3.6 \sigma,3843.16,-7.97,-5.97,FALSE,\color{Red} X,FALSE,\color{Red} X,\color{Red} X ,13.00,3.60527391469775,3.6 \sigma,FALSE,\color{Red} X,FALSE,\color{Red} X,0.008771404843838,FALSE,FALSE
\varyingmetable,1,-19.391 \pm 0.028,3.16252115005546,3.2 \sigma,3841.91,-9.21,-7.21,TRUE,\color{Green} \checkmark,FALSE,\color{Red} X,\color{Green} \checkmark,11.33,3.36673012245945,3.4 \sigma,FALSE,\color{Red} X,TRUE,\color{Green} \checkmark,0.064571575244774,FALSE,FALSE
\varyingmeomegaktable,2,-19.368 \pm 0.036,2.40166308866003,2.4 \sigma,3837.33,-13.80,-9.80,TRUE,\color{Green} \checkmark,FALSE,\color{Red} X,\color{Green} \checkmark,6.66,2.57990971842404,2.6 \sigma,TRUE,\color{Green} \checkmark,TRUE,\color{Green} \checkmark,0.074217999437822,TRUE,FALSE
\edetable,3,-19.390^{+0.018}_{-0.031},3.53727518037797,3.5 \sigma,3832.29,-18.84,-12.84,TRUE,\color{Green} \checkmark,FALSE,\color{Red} X,\color{Green} \checkmark,5.89,2.42682323212859,2.4 \sigma,TRUE,\color{Green} \checkmark,TRUE,\color{Green} \checkmark,-0.313928629135019,TRUE,FALSE
\nedetable,3,-19.380^{+0.019}_{-0.031},3.2608412791275,3.3 \sigma,3833.89,-17.24,-11.24,TRUE,\color{Green} \checkmark,FALSE,\color{Red} X,\color{Green} \checkmark,4.23,2.05679893531194,2.1 \sigma,TRUE,\color{Green} \checkmark,TRUE,\color{Green} \checkmark,-0.369242855063993,TRUE,FALSE
\emgtable,3,-19.400^{+0.014}_{-0.017},3.92813316228963,3.9 \sigma,3835.06,-16.07,-10.07,TRUE,\color{Green} \checkmark,FALSE,\color{Red} X,\color{Green} \checkmark,4.59,2.14242852856278,2.1 \sigma,TRUE,\color{Green} \checkmark,TRUE,\color{Green} \checkmark,-0.45459371155483,TRUE,FALSE
\cpltable,2,.000 \pm 0.000,-515.911528150134,-515.9 \sigma,0.00,-3851.13,-3847.13,TRUE,\color{Green} \checkmark,TRUE,\color{Green} \checkmark,\color{Green} \checkmark,0.00,0,.0 \sigma,TRUE,\color{Green} \checkmark,TRUE,\color{Green} \checkmark,-1,TRUE,TRUE
\pedetable,0,-19.353 \pm 0.012,2.79459519877802,2.8 \sigma,3859.93,8.80,8.80,FALSE,\color{Red} X,FALSE,\color{Red} X,\color{Red} X ,6.33,2.51554795317819,2.5 \sigma,TRUE,\color{Green} \checkmark,TRUE,\color{Green} \checkmark,-0.099852474419852,FALSE,FALSE
\medetable,1,.000 \pm 0.000,-515.911528150134,-515.9 \sigma,0.00,-3851.13,-3849.13,TRUE,\color{Green} \checkmark,TRUE,\color{Green} \checkmark,\color{Green} \checkmark,0.00,0,.0 \sigma,TRUE,\color{Green} \checkmark,TRUE,\color{Green} \checkmark,-1,TRUE,TRUE
\dmdecaytable,2,.000 \pm 0.000,-515.911528150134,-515.9 \sigma,0.00,-3851.13,-3847.13,TRUE,\color{Green} \checkmark,TRUE,\color{Green} \checkmark,\color{Green} \checkmark,0.00,0,.0 \sigma,TRUE,\color{Green} \checkmark,TRUE,\color{Green} \checkmark,-1,TRUE,TRUE
\fracdmdecaytable,2,.000 \pm 0.000,-515.911528150134,-515.9 \sigma,0.00,-3851.13,-3847.13,TRUE,\color{Green} \checkmark,TRUE,\color{Green} \checkmark,\color{Green} \checkmark,0.00,0,.0 \sigma,TRUE,\color{Green} \checkmark,TRUE,\color{Green} \checkmark,-1,TRUE,TRUE
\end{filecontents*}
\csvreader[/csv/head=true,tabular={l c|  c c c c | r r c | c},/csv/respect dollar=false,/csv/respect backslash=false,
    table head={ Model& $\Delta N_{\rm param}$ & $M_B$ &  \begin{tabular}{@{}c@{}}Gaussian \\ Tension\end{tabular} & \begin{tabular}{@{}c@{}}$Q_{\rm DMAP}$ \\ Tension\end{tabular} & & $\Delta \chi^2$  & $\Delta$AIC & & \begin{tabular}{@{}c@{}}One test \\ passed\end{tabular} \\ \hline},
    filter expr={
      test{\ifnumgreater{\thecsvinputline}{1}}
  and test{\ifnumless{\thecsvinputline}{3}} or
      test{\ifnumgreater{\thecsvinputline}{3}}
  and test{\ifnumless{\thecsvinputline}{5}} or
      test{\ifnumgreater{\thecsvinputline}{7}}
  and test{\ifnumless{\thecsvinputline}{15}} 
}]{sheets/Alldata_nohalofit.csv}{1=\name,2=\nparam,3=\mbsig,5=\mbtens,7=\Dchi,8=\DAIK,12=\Mbtest,10=\DAIKtest,13=\anytest,16=\newmbtens,18=\newMbtest,20=\newanytest}{ \name & \nparam & $\mbsig$ & $\mbtens$ & $\newmbtens$ & $\newMbtest$ & $\Dchi$  &  $\DAIK$ & $\DAIKtest $ & $\newanytest$}}
    \caption{Test of the models based on dataset \DD (Planck 2018 + BAO + Pantheon + RSD + LyaBAO + CC), using the direct measurement of $M_B$ by SH0ES for the quantification of the Gaussian tension (4th column), the $Q_\mathrm{DMAP}$ tension (5th column) or the computation of the $\Delta$AIC (5th column). Four models pass at least one of these two tests at the 3$\sigma$ level.\label{tab:summary_Alldata}}
\end{table}
A comparison of the tension metrics for all models in the finalists category using \DD is provided in Tab.~\ref{tab:summary_Alldata}. The additional data sets provide some interesting constraining power on the finalist models. Indeed, looking at the $\Delta\chi^2$ with respect to $\Lambda$CDM reported in Tab.~\ref{tab:chi2_alldata}, one can see that the fit to CC is somewhat degraded in all models, in particular for EDE, NEDE, EMG and \meOkshort{} ($\Delta\chi^2\sim 1-2$). Additionally, the BAO/RSD data further constrain the \Majoronshort{}, \meshort{} and \meOkshort{} models. As a consequence, none of the model now pass our criteria at better than $2.1\sigma$. In particular,  the \Bshort{} model now fails all criteria, while the other seven finalists solutions stay in the \enquote{winning} sample. 
The \Majoronshort{} and \meshort{} solution pass by virtue of the $\Delta$AIC (but fail both GT and $Q_{\rm DMAP}$ tension metric), while the other silver medalists (\idrmid{},\EDEshort{}, \NEDEshort{} and EMG) and gold medalist (\meOkshort{}) still pass at least two criteria.  
Our final test demonstrates that the additional information from RSD, CC, and Lyman-$\alpha$ BAO data is constraining enough to exclude (or include) solutions that are on the border of acceptance, while putting significant additional pressure on the gold and silver medalists.

\begin{table}[h]
    \centering
    \csvreader[/csv/head=true,tabular={l|r r r r r r r r r r},/csv/respect dollar=false,/csv/respect backslash=false,
    table head={ Model & low-$\ell$ & high-$\ell$ & lensing & BAO & Pantheon & RSD & CC & Ly$\alpha$ & SH0ES & Total \\ \hline},
    filter expr={
      test{\ifnumgreater{\thecsvinputline}{1}}
  and test{\ifnumless{\thecsvinputline}{3}} or
      test{\ifnumgreater{\thecsvinputline}{3}}
  and test{\ifnumless{\thecsvinputline}{5}} or
      test{\ifnumgreater{\thecsvinputline}{7}}
  and test{\ifnumless{\thecsvinputline}{15}} 
}]{sheets/dChi2_nohalofit.csv}{16=\name,17=\lowl,18=\highl,19=\lens,20=\bao,21=\pantheon,22=\rsd,23=\cc,24=\lbao,25=\shoes,26=\total}{\name & $\lowl$ & $\highl$ & \lens & \bao &\pantheon  & \rsd & \cc & \lbao & \shoes & \total}
    \caption{Contribution of each likelihood to the best-fit $\chi^2=-2\ln \mathcal{L}$ of each model to the dataset \DD, relative to the best-fit $\chi^2$ of $\Lambda$CDM to the same dataset. Here SH0ES is treated as a measurement of $M_B$. 
 \label{tab:chi2_alldata}}
\end{table}

\begin{figure}[h!]
    \centering
    \hspace*{-0.5cm}
    \includegraphics[width=0.45\textwidth]{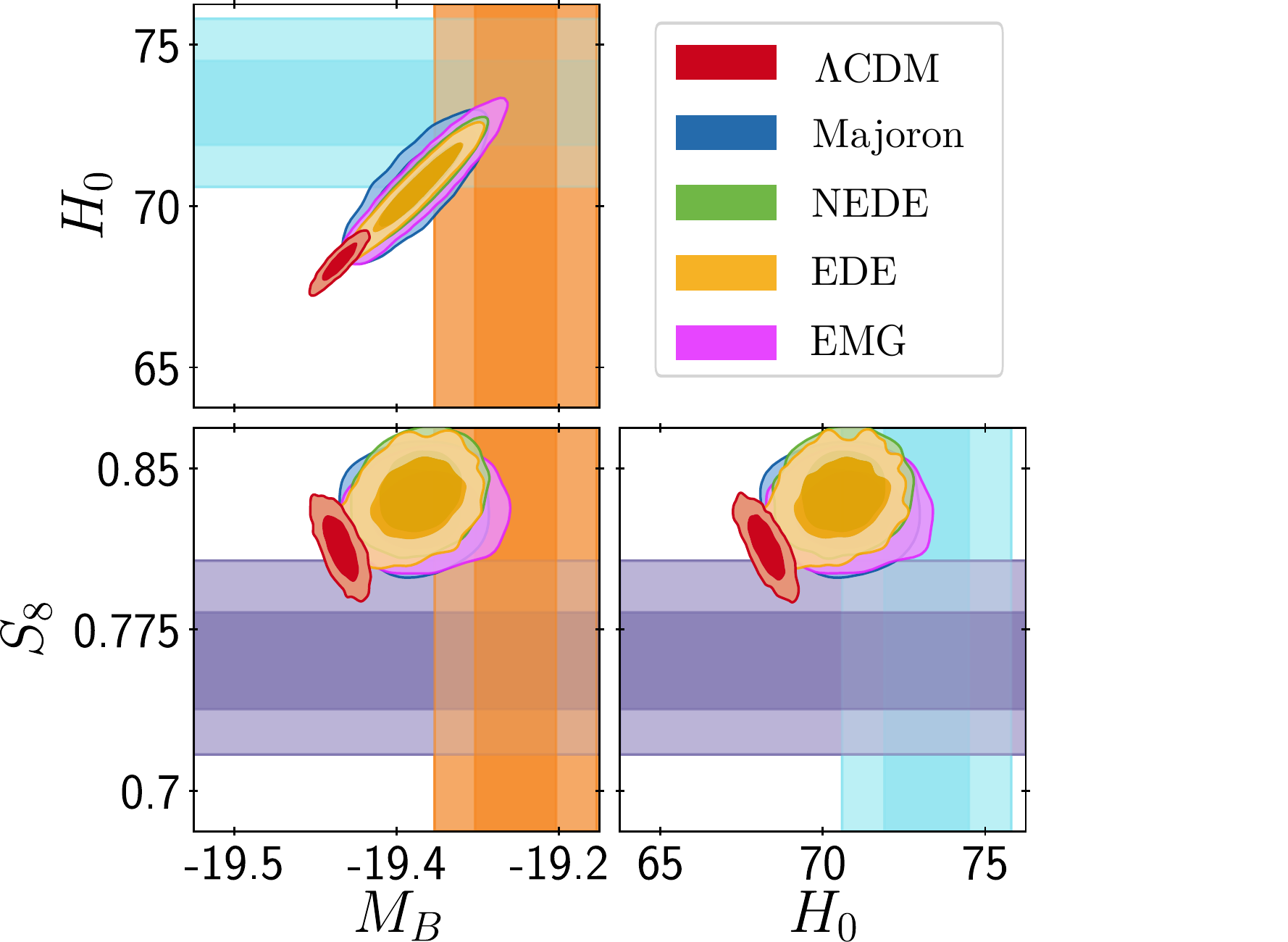}
    \includegraphics[width=0.45\textwidth]{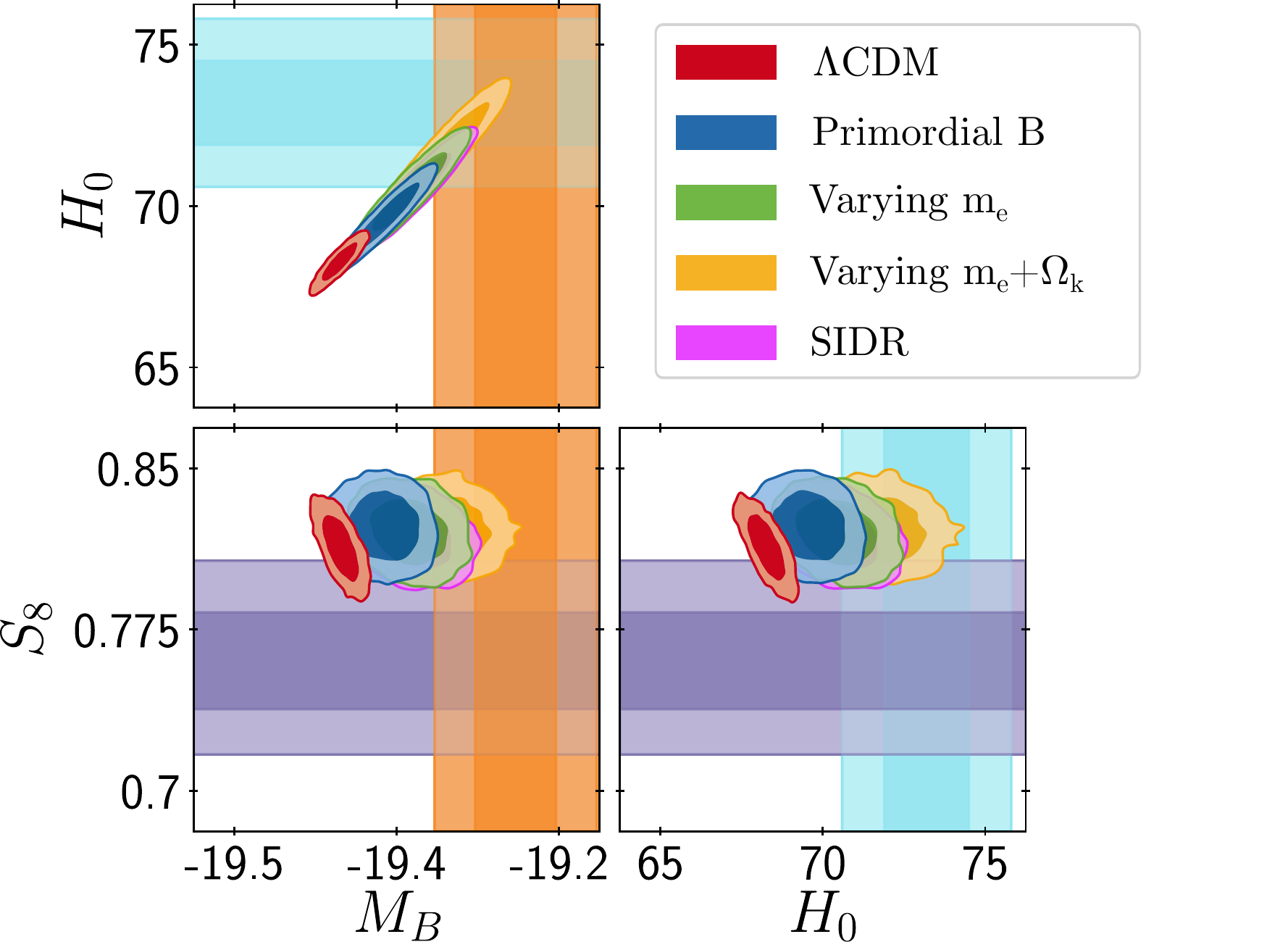}
    \caption{Contours ($68.3\%$ and $95.4\%$ C.L.) of $\{H_0,M_b,S_8\}$ obtained when considering the ${\cal D}_{\rm final}$ data set for the \enquote{finalists} models compared to $\Lambda$CDM.}
    \label{fig:S8_finalists}
\end{figure}
\FloatBarrier
\clearpage
Before concluding, we highlight the predicted distributions of $S_8$ using \DD for each of the finalist models. As illustrated in \cref{fig:S8_finalists}, none of the finalists are able to relieve the $S_8$ tension (the exotic recombination models being the most promising in mitigating both tensions), however neither do any of the models dramatically increase the tension. As discussed extensively in the literature \cite{Hill:2020osr,Ivanov:2020ril,DAmico:2020ods,Murgia:2020ryi,Niedermann:2020qbw,Smith:2020rxx}, the EDE and NEDE models increase the mean value of $S_8$, however error bars are increased as well, resulting in similar level of tension as \lcdm. This figure clearly illustrates that resolving the both $H_0$ and $S_8$ tension will likely require multiple extensions.

There has been attempts in the literature at resolving both tensions simultaneously. 
In particular, models of DM-DR interactions seemed particularly promising given that they predict both a higher $H_0$ and a smaller $S_8$ \cite{Buen-Abad:2015ova,Buen-Abad:2017gxg,Chacko:2016kgg,Heimersheim:2020aoc}. However, none passed the tests we considered in this work. Similarly, the primordial magnetic field model was advertised as promising to resolve both tensions \cite{Jedamzik:2020krr}. While it does pass our $\Delta$AIC criterion, it only reduces the $H_0$ tension at the $3.5\sigma$ level, and the $S_8$ tension is barely affected ($S_8$ is even slightly larger than in $\Lambda$CDM). 
Nevertheless, it is possible that the resolutions of the $S_8$ and $H_0$ tension lie in different sectors, or require more involve modifications than the ones discussed here. In fact, while we discuss the \enquote{$S_8$-tension}, it is becoming clear that the tension lies in the amplitude of fluctuation $\sigma_8$ rather than $\Omega_m$, in particular due to constraints on $\Omega_m$ from uncalibrated SNIa data \cite{Scolnic:2017caz}, and from the cross-correlation of weak lensing and galaxy surveys that breaks the $\Omega_m-\sigma_8$ degeneracy \cite{Heymans:2020gsg}. Therefore, one might expect that the $\sigma_8$ tension hints at new perturbation properties, while $H_0$, which is a measure of the total energy density, hints at a new background contribution. Many models discussed so far are mainly motivated by their impact at the background level, and it is perhaps not surprising that they fail at simultaneously reducing $\sigma_8$. In fact, the increase in $\sigma_8$ in the NEDE, EDE and Majoron case are related to \enquote{unexpected} effects within CMB power spectra, introduced by the increase in $H(z)$ at early-times  and compensated for by an increase in $\omega_{\rm cdm}$\,. This was already noted in Ref.~\cite{Poulin:2018cxd}, and further articulated in subsequent works \cite{Hill:2020osr,Murgia:2020ryi,Vagnozzi:2021gjh}. Additionally, it was noted that a similar increase in $\omega_{\rm cdm}$ is required to simultaneously preserve the BAO angular scale \cite{Jedamzik:2020zmd}. Therefore, this \enquote{accidental} degeneracy and increase in $\sigma_8$ could be due to those models having too little freedom at the level of perturbations. 
For instance, it was recently noted that extending the (N)EDE sector with an additional ultra-light axion of mass $m\sim10^{-26}$ eV representing $\sim 5\%$ of the DM would resolve both tensions \cite{Allali:2021azp}.
Alternatively, while less pleasing from the Occam's razor point of view, new properties of DM (e.g. decays \cite{Abellan:2020pmw,Abellan:2021bpx}, interactions with DR \cite{Buen-Abad:2015ova,Buen-Abad:2017gxg,Chacko:2016kgg,Heimersheim:2020aoc} or DE \cite{DiValentino:2019ffd,Lucca:2021dxo}) or neutrinos \cite{Poulin:2018dzj,Kreisch:2019yzn} could independently resolve the $\sigma_8$ tension, and leave unaffected the (relative) success  of the models studied here at resolving the $H_0$ tension. Along these lines, we note that the EMG model performs (slightly) better in letting $S_8$ unaffected than the EDE models. While the $S_8$ tension is not resolved in the EMG model we study here, it does give another interesting path to find a single solution to both tensions.
We leave the study of common resolutions to the $\sigma_8$ and $H_0$ tensions for future work.
\bibliographystyle{ieeetr}

\section{Discussion}

In this work, we have used a common analysis pipeline to compare and contrast the relative success of seventeen models proposed to ease the Hubble tension; this approach is thus intended as a fair comparison between proposed solutions, and provides a useful benchmark for those wishing to put forth novel ideals. We have broken down the various models into three generic categories: those  that modify the sound horizon by including a component of Dark Radiation (DR) impacting the early expansion history, solutions which modify the sound horizon through some other mechanism (such as a delay of recombination or some pre-recombination contribution to the expansion), and solutions that attempt to modify the late-time expansion history. While not all models fall perfectly into one of these categories, we find this to be a useful classification for the purpose of understanding how models respond to the inclusion of various datasets.

For each model and data-set, we quantify the residual tension using a series of metrics, each of which has both advantages and shortcomings, and attempting to answer slightly different questions, namely:  given a model, (i) to what extent does confrontation with data (other than SH0ES) generate posteriors compatible with high values of $H_0$, (ii) to what extent can one obtain a good combined fit to all data, and (iii) to what extent is that model favored over $\Lambda$CDM?

Specifically, our metrics involve the following: (1) when confronting a model to data other than SH0ES, we quantify the residual tension within a model assuming Gaussian posteriors; (2) we compute the $\chi^2_{\rm min}$ value with and without the SH0ES likelihood, and translate this into an estimate of the residual tension ($Q_{\rm DMAP}$ from Ref.~\cite{Raveri:2018wln}), which has a clear benefit over metric (1) in that it can more accurately account for non-Gaussian posteriors (in general, we assume the $Q_{\rm DMAP}$ supersedes the Gaussian posterior approximation, however present both for the sake of completeness); and (3) we quantify the model preference over that of \lcdm using the Akaike Information Criterion (AIC), which attempts to account both for the extent to which the new model improves the cosmological fit and the added model complexity (which can lead to over-fitting). The very short summary of our findings is that no model does exceptionally well in all of our tests -- all models are left with a residual tension, with the most promising reducing the tension to the $\sim 1.6\sigma$ level, and with very few models actually reducing the tension below $3\sigma$. Similarly, only a small subset of the models are capable of improving the fit sufficiently to the pass the AIC test. Six models, EDE, NEDE, EMG, \meshort{} (with and without curvature), and the Majoron, are able to simultaneously satisfy the $Q_{\rm DMAP}$ and $\Delta$AIC criteria. Furthermore, only the \meshort{} (with and without curvature) passes the Gaussian criterion and allows for high $H_0$ (or $M_B$) without a SH0ES prior, receiving the only \enquote{gold medal} of our tournament. We detail the relative success of the models in each category below.

{\bf Late-universe models:} For models affecting the late-time dynamics, we have concluded that the mechanism employed to ease the tension is disfavored by a combination of low redshift probes and measurements of the CMB anisotropies. In particular, following recent works \cite{Camarena:2021jlr,Efstathiou:2021ocp}, we have systematically  demonstrated that, even though late-time solutions employing modified dark energy can generate a high value of $H_0$\, -- and therefore reconcile the na\"ive \enquote{Planck vs SH0ES} tension -- they cannot reconcile the more fundamental problem of the difference in calibration of supernovae between the local distance ladder (such as from SH0ES using Cepheids) and the inverse distance ladder based on calibrating BAO onto \lcdm{} estimate of the sound horizon (irrespective of Planck).
Models invoking late-time decays of dark matter (either to dark radiation or a combination of warm dark matter and dark radiation), have virtually no effect on the $H_0$ tension, and are undoubtedly the least successful proposals. Yet, models of decays into warm dark matter are promising for resolving the $S_8$ tension \cite{Abellan:2020pmw,Abellan:2021bpx}.

{\bf Dark Radiation models:} Models that use dark radiation to shift the sound horizon at recombination had appeared relatively promising in the 2015 Planck data release; Planck 2018 data, however, seems to increasingly disfavor many of these models. With the exception of the Majoron solution (which marginally passes all tests) and the \idrshort{} model (which marginally passes only the AIC test) no other model in this category can reduce the tension below the $3\sigma$ level or significantly improve the $\Delta \chi^2$. Interestingly, the Majoron model is also the only model in this category, which -- in addition to dark radiation -- has a modified expansion history $H(z)$ before recombination, and thus shares similarities with the \EDElong{} in the final category of solutions (which appear to be one of the most successful class). The lackluster performance of dark radiation solutions appears to be related to the increase in the precision of the high-$\ell$ polarization data and its increased ability to constrain the various impacts of Dark Radiation such as an increase in Silk damping or other effects at the level of perturbations.

{\bf Other early-universe models:} In the final class of models, namely those which modify the sound horizon at recombination without dark radiation, all models considered are capable of passing at least one criteria. The model of a \memid{} appears to ease the tension to 3.2$\sigma$ or 1.7$\sigma$ depending on whether or not curvature is considered\footnote{It is perhaps rather surprising that the addition of curvature, having an entirely disjoint cosmological effect to that of a varying electron mass, produces such a strong effect in shifting $H_0$. The reason for this degeneracy is further explained in Ref.~\cite{Sekiguchi:2020teg}. We have verified explicitly that this is a peculiar degeneracy that uniquely appears in this model -- allowing curvature to float in other models seems to have no significant impact.}. On the other hand, the other model of exotic recombination, \Bmid{}, passes the $\Delta$AIC criterion, but fail both the GT and $Q_{\rm DMAP}$ criteria. Finally, \EDEshort{}, \NEDEshort{} and EMG pass the $\Delta$AIC criterion, and  significantly reduce the $Q_{\rm DMAP}$ tension (to $1.6\sigma$, $1.9\sigma$ and $2.3\sigma$ tension respectively), making them (together with \meOkshort{}) the most successful of the models studied in this work.

The most worrying aspect is perhaps that we find no preference for one of these models over $\Lambda$CDM without incorporating information from SH0ES. Yet, it is interesting that some models are able to give good simultaneous fits to all data considered in this work. This illustrates that the non-observation of a preference for a model beyond $\Lambda$CDM without SH0ES measurement of $M_B$ can still be the sign of a lack of precision within current surveys. It has been shown for several models in the literature that next-generation CMB experiments should unambiguously detect (or exclude) these models \cite{Smith:2019ihp,Park:2019ibn}.

In that context, we quantified the importance of Planck data in constraining solutions to the $H_0$ tension (in particular in the $\ell$-range $1000-2000$). For most models, constraints without Planck are similar to those with Planck. Yet, for EDE and NEDE we find that there exists a potentially large degeneracy in WMAP and ACT data that seems to favor EDE and NEDE over \lcdm, resulting in no residual $H_0$ tension, and a non-zero detection of $f_{\rm (N)EDE}$ (see also the recent Refs.~\cite{Poulin:2021bjr,Hill:2021yec}). Yet, these conclusions do not hold when CMB data are traded for BBN data, with the latter data being more in line with Planck results, albeit providing weaker constraints.

Additionally, we note that of the models of interest, none seem to either alleviate nor exacerbate the $S_8$ tension. The most promising models for $H_0$ -- \EDEshort{} and \NEDEshort{} -- do predict somewhat larger $S_8$, but the simultaneous increase in the size of the error bars result in no significant change in the tension level. 
Some models which had previously shown some success in reducing the $S_8$ tension, namely DM-DR and the strongly interaction neutrinos, are now disfavored by the data.
Finding a common resolutions to both tensions would certainly reinforce the degree of belief into the new concordance cosmology; however, one must bear in mind that the resolution of these tensions could arise from independent sectors -- either from new physics or systematics. 
We defer the study of simultaneous resolutions to both tensions, following the metrics we introduced here, for future work.

Another (potential) issue raised by cosmologies that resolve the Hubble tension by adjusting the sound horizon is that they predict a younger age of the universe $t_U$ than in $\Lambda$CDM \cite{Bernal:2021yli,Boylan-Kolchin:2021fvy,Vagnozzi:2021tjv}. While we do not report the reconstructed $t_U$ for all models, this is simple to understand since, in these models, $H_0$ is increased while $\Omega_m$ and $\Omega_\Lambda$ are left essentially unaffected. As a consequence, the age of the universe, which scales as $t_U\propto 1/H_0$, decreases roughly by the same amount as $H_0$ increases. These models can therefore be tested against measurements of the age of old objects such as globular clusters of stars, which was recently determined as $t_{\rm U}=13.5\pm0.027$ Gyrs \cite{Valcin:2020vav,Valcin:2021jcg,Bernal:2021yli}.
  This issue was discussed in great details in the context of the EDE cosmology, but current data are not accurate enough to play a decisive role in arbitrating the tension \cite{Bernal:2021yli,Boylan-Kolchin:2021fvy}. In this study, we did not include measurements of the age of the universe, however we have shown that the `cosmic chronometers' can provide significant  constraining power to the models (in particular when dropping Planck data).  Future measurements of the age of the universe, and the age-vs-redshift diagram, will certainly provide very important tests of cosmologies resolving the Hubble tension.

Collectively, this work illustrates the difficulty faced by theorists who are hoping to resolve the $H_0$ tension using modified cosmologies. Still, the fact that some models can provide a good combined fit to all data considered in this work provides a glimpse of hope for model builders, in the sense that it demonstrates that there at least exist potential solutions to this cosmic tension. However, it is fair to say that the current most \enquote{successful} models we identified likely raise more questions (\eg fine-tuning of EDE properties, the origin of non-zero curvature) than they really answer\footnote{An exception might be the Majoron model, which has been tied to UV-complete theories and can play a role in answering other fundamental questions, such as the origin of neutrino masses and the generation of the observed baryon asymmetry \cite{Escudero:2019gvw,Escudero:2020hkf,Escudero:2021rfi}.}, while being unable to simultaneously explain the growing $S_8$-tension. Further work must be done to establish whether these remaining theoretical and observational issues can be overcome in a new concordant cosmology, one that may either build upon the models studied here, or perhaps lie in a yet unexplored direction.




{\bf Acknowledgements: }
The authors would like to thank Christoph Weniger and Alex Cole for their participation in the early stages of this work. VP would like to warmly thank Tristan L. Smith for many useful discussions on EDE and NEDE models. SJW has received funding from the European Research Council (ERC) under the European Union’s Horizon 2020
research and innovation programme (Grant agreement No. 864035 - Un-Dark). NS and JL acknowledge support from the DFG grant LE 3742/4-1. Simulations were performed with computing resources granted by RWTH Aachen University under project jara0184.
 This work has been partly supported by the CNRS-IN2P3 grant Dark21. The authors acknowledge the use of computational resources from the the Dark Energy computing Center funded by the Excellence Initiative of Aix-Marseille University - A*MIDEX, a French "Investissements d'Avenir" programme (AMX-19-IET-008 - IPhU.

\newpage

\bibliography{references}

\providecommand{\noopsort}[1]{}\providecommand{\singleletter}[1]{#1}%
\begin{thebibliography}{100}

\bibitem{Aghanim:2018eyx}
N.~Aghanim {\em et~al.}, ``{Planck 2018 results. VI. Cosmological
  parameters},'' {\em Astron. Astrophys.}, vol.~641, p.~A6, 2020.

\bibitem{Riess:2020fzl}
A.~G. Riess, S.~Casertano, W.~Yuan, J.~B. Bowers, L.~Macri, J.~C. Zinn, and
  D.~Scolnic, ``{Cosmic Distances Calibrated to 1\% Precision with Gaia EDR3
  Parallaxes and Hubble Space Telescope Photometry of 75 Milky Way Cepheids
  Confirm Tension with $\Lambda$CDM},'' {\em Astrophys. J. Lett.}, vol.~908,
  no.~1, p.~L6, 2021.

\bibitem{2019JCAP...10..029S}
N.~{Sch{\"o}neberg}, J.~{Lesgourgues}, and D.~C. {Hooper}, ``{The BAO+BBN take
  on the Hubble tension},'' {\em JCAP}, vol.~2019, p.~029, Oct. 2019.

\bibitem{2013MNRAS.436.1674A}
G.~E. {Addison}, G.~{Hinshaw}, and M.~{Halpern}, ``{Cosmological constraints
  from baryon acoustic oscillations and clustering of large-scale structure},''
  {\em MNRAS}, vol.~436, pp.~1674--1683, Dec. 2013.

\bibitem{2015PhRvD..92l3516A}
{\'E}.~{Aubourg}, S.~{Bailey}, J.~E. {Bautista}, F.~{Beutler}, V.~{Bhardwaj},
  D.~{Bizyaev}, M.~{Blanton}, M.~{Blomqvist}, A.~S. {Bolton}, J.~{Bovy},
  H.~{Brewington}, J.~{Brinkmann}, J.~R. {Brownstein}, A.~{Burden}, N.~G.
  {Busca}, W.~{Carithers}, C.-H. {Chuang}, J.~{Comparat}, R.~A.~C. {Croft},
  A.~J. {Cuesta}, K.~S. {Dawson}, T.~{Delubac}, D.~J. {Eisenstein},
  A.~{Font-Ribera}, J.~{Ge}, J.~M. {Le Goff}, S.~G.~A. {Gontcho}, J.~R. {Gott},
  J.~E. {Gunn}, H.~{Guo}, J.~{Guy}, J.-C. {Hamilton}, S.~{Ho}, K.~{Honscheid},
  C.~{Howlett}, D.~{Kirkby}, F.~S. {Kitaura}, J.-P. {Kneib}, K.-G. {Lee},
  D.~{Long}, R.~H. {Lupton}, M.~V. {Maga{\~n}a}, V.~{Malanushenko},
  E.~{Malanushenko}, M.~{Manera}, C.~{Maraston}, D.~{Margala}, C.~K. {McBride},
  J.~{Miralda-Escud{\'e}}, A.~D. {Myers}, R.~C. {Nichol}, P.~{Noterdaeme},
  S.~E. {Nuza}, M.~D. {Olmstead}, D.~{Oravetz}, I.~{P{\^a}ris},
  N.~{Padmanabhan}, N.~{Palanque-Delabrouille}, K.~{Pan},
  M.~{Pellejero-Ibanez}, W.~J. {Percival}, P.~{Petitjean}, M.~M. {Pieri},
  F.~{Prada}, B.~{Reid}, J.~{Rich}, N.~A. {Roe}, A.~J. {Ross}, N.~P. {Ross},
  G.~{Rossi}, J.~A. {Rubi{\~n}o-Mart{\'\i}n}, A.~G. {S{\'a}nchez},
  L.~{Samushia}, R.~T. {G{\'e}nova-Santos}, C.~G. {Sc{\'o}ccola}, D.~J.
  {Schlegel}, D.~P. {Schneider}, H.-J. {Seo}, E.~{Sheldon}, A.~{Simmons}, R.~A.
  {Skibba}, A.~{Slosar}, M.~A. {Strauss}, D.~{Thomas}, J.~L. {Tinker},
  R.~{Tojeiro}, J.~A. {Vazquez}, M.~{Viel}, D.~A. {Wake}, B.~A. {Weaver}, D.~H.
  {Weinberg}, W.~M. {Wood-Vasey}, C.~{Y{\`e}che}, I.~{Zehavi}, G.-B. {Zhao},
  and {BOSS Collaboration}, ``{Cosmological implications of baryon acoustic
  oscillation measurements},'' {\em \prd}, vol.~92, p.~123516, Dec. 2015.

\bibitem{2018ApJ...853..119A}
G.~E. {Addison}, D.~J. {Watts}, C.~L. {Bennett}, M.~{Halpern}, G.~{Hinshaw},
  and J.~L. {Weiland}, ``{Elucidating {\ensuremath{\Lambda}}CDM: Impact of
  Baryon Acoustic Oscillation Measurements on the Hubble Constant
  Discrepancy},'' {\em \apj}, vol.~853, p.~119, Feb. 2018.

\bibitem{Blomqvist:2019rah}
M.~Blomqvist {\em et~al.}, ``{Baryon acoustic oscillations from the
  cross-correlation of Ly$\alpha$ absorption and quasars in eBOSS DR14},'' {\em
  Astron. Astrophys.}, vol.~629, p.~A86, 2019.

\bibitem{2019JCAP...10..044C}
A.~{Cuceu}, J.~{Farr}, P.~{Lemos}, and A.~{Font-Ribera}, ``{Baryon Acoustic
  Oscillations and the Hubble constant: past, present and future},'' {\em
  JCAP}, vol.~2019, p.~044, Oct. 2019.

\bibitem{2017MNRAS.467..731V}
L.~{Verde}, J.~L. {Bernal}, A.~F. {Heavens}, and R.~{Jimenez}, ``{The length of
  the low-redshift standard ruler},'' {\em MNRAS}, vol.~467, pp.~731--736, May
  2017.

\bibitem{2021PhRvD.103j3533B}
J.~L. {Bernal}, L.~{Verde}, R.~{Jimenez}, M.~{Kamionkowski}, D.~{Valcin}, and
  B.~D. {Wandelt}, ``{Trouble beyond H$_{0}$ and the new cosmic triangles},''
  {\em \prd}, vol.~103, p.~103533, May 2021.

\bibitem{2020MNRAS.494.2076C}
P.~{Carter}, F.~{Beutler}, W.~J. {Percival}, J.~{DeRose}, R.~H. {Wechsler}, and
  C.~{Zhao}, ``{The impact of the fiducial cosmology assumption on BAO distance
  scale measurements},'' {\em MNRAS}, vol.~494, pp.~2076--2089, May 2020.

\bibitem{2020PhRvD.102l3515B}
J.~L. {Bernal}, T.~L. {Smith}, K.~K. {Boddy}, and M.~{Kamionkowski},
  ``{Robustness of baryon acoustic oscillation constraints for early-Universe
  modifications of {\ensuremath{\Lambda}} CDM cosmology},'' {\em \prd},
  vol.~102, p.~123515, Dec. 2020.

\bibitem{Freedman:2019jwv}
W.~L. {Freedman}, B.~F. {Madore}, D.~{Hatt}, T.~J. {Hoyt}, I.~S. {Jang}, R.~L.
  {Beaton}, C.~R. {Burns}, M.~G. {Lee}, A.~J. {Monson}, J.~R. {Neeley}, M.~M.
  {Phillips}, J.~A. {Rich}, and M.~{Seibert}, ``{The Carnegie-Chicago Hubble
  Program. VIII. An Independent Determination of the Hubble Constant Based on
  the Tip of the Red Giant Branch},'' {\em \apj}, vol.~882, p.~34, 7 2019.

\bibitem{Freedman:2021ahq}
W.~L. {Freedman}, ``{Measurements of the Hubble Constant: Tensions in
  Perspective},'' {\em arXiv e-prints}, p.~arXiv:2106.15656, June 2021.

\bibitem{Yuan:2019npk}
W.~Yuan, A.~G. Riess, L.~M. Macri, S.~Casertano, and D.~Scolnic, ``{Consistent
  Calibration of the Tip of the Red Giant Branch in the Large Magellanic Cloud
  on the Hubble Space Telescope Photometric System and a Re-determination of
  the Hubble Constant},'' {\em Astrophys. J.}, vol.~886, p.~61, 2019.

\bibitem{Soltis:2020gpl}
J.~Soltis, S.~Casertano, and A.~G. Riess, ``{The Parallax of $\omega$ Centauri
  Measured from Gaia EDR3 and a Direct, Geometric Calibration of the Tip of the
  Red Giant Branch and the Hubble Constant},'' {\em Astrophys. J. Lett.},
  vol.~908, no.~1, p.~L5, 2021.

\bibitem{Khetan:2020hmh}
N.~Khetan {\em et~al.}, ``{A new measurement of the Hubble constant using Type
  Ia supernovae calibrated with surface brightness fluctuations},'' {\em
  Astron. Astrophys.}, vol.~647, p.~A72, 2021.

\bibitem{Huang:2019yhh}
C.~D. {Huang}, A.~G. {Riess}, W.~{Yuan}, L.~M. {Macri}, N.~L. {Zakamska},
  S.~{Casertano}, P.~A. {Whitelock}, S.~L. {Hoffmann}, A.~V. {Filippenko}, and
  D.~{Scolnic}, ``{Hubble Space Telescope Observations of Mira Variables in the
  SN Ia Host NGC 1559: An Alternative Candle to Measure the Hubble Constant},''
  {\em \apj}, vol.~889, p.~5, Jan. 2020.

\bibitem{Schombert:2020pxm}
J.~{Schombert}, S.~{McGaugh}, and F.~{Lelli}, ``{Using the Baryonic
  Tully-Fisher Relation to Measure H$_{o}$},'' {\em Astrophysical Journal},
  vol.~160, p.~71, Aug. 2020.

\bibitem{Wong:2019kwg}
K.~C. Wong {\em et~al.}, ``{H0LiCOW \textendash{} XIII. A 2.4 per cent
  measurement of H0 from lensed quasars: 5.3\ensuremath{\sigma} tension between
  early- and late-Universe probes},'' {\em Mon. Not. Roy. Astron. Soc.},
  vol.~498, no.~1, pp.~1420--1439, 2020.

\bibitem{Birrer:2020tax}
S.~Birrer {\em et~al.}, ``{TDCOSMO - IV. Hierarchical time-delay cosmography
  \textendash{} joint inference of the Hubble constant and galaxy density
  profiles},'' {\em Astron. Astrophys.}, vol.~643, p.~A165, 2020.

\bibitem{Pesce:2020xfe}
D.~W. Pesce {\em et~al.}, ``{The Megamaser Cosmology Project. XIII. Combined
  Hubble constant constraints},'' {\em Astrophys. J. Lett.}, vol.~891, no.~1,
  p.~L1, 2020.

\bibitem{Abbott:2019yzh}
B.~P. Abbott {\em et~al.}, ``{A Gravitational-wave Measurement of the Hubble
  Constant Following the Second Observing Run of Advanced LIGO and Virgo},''
  {\em Astrophys. J.}, vol.~909, no.~2, p.~218, 2021.

\bibitem{DiValentino:2021izs}
E.~{Di Valentino}, O.~{Mena}, S.~{Pan}, L.~{Visinelli}, W.~{Yang},
  A.~{Melchiorri}, D.~F. {Mota}, A.~G. {Riess}, and J.~{Silk}, ``{In the Realm
  of the Hubble tension $-$ a Review of Solutions},'' {\em arXiv e-prints},
  p.~arXiv:2103.01183, Mar. 2021.

\bibitem{Mortsell:2021nzg}
E.~{Mortsell}, A.~{Goobar}, J.~{Johansson}, and S.~{Dhawan}, ``{The Hubble
  Tension Bites the Dust: Sensitivity of the Hubble Constant Determination to
  Cepheid Color Calibration},'' {\em arXiv e-prints}, p.~arXiv:2105.11461, May
  2021.

\bibitem{Mortsell:2021tcx}
E.~{Mortsell}, A.~{Goobar}, J.~{Johansson}, and S.~{Dhawan}, ``{The Hubble
  Tension Revisited: Additional Local Distance Ladder Uncertainties},'' {\em
  arXiv e-prints}, p.~arXiv:2106.09400, June 2021.

\bibitem{Efstathiou:2020wxn}
G.~{Efstathiou}, ``{A Lockdown Perspective on the Hubble Tension (with comments
  from the SH0ES team)},'' {\em arXiv e-prints}, p.~arXiv:2007.10716, July
  2020.

\bibitem{Freedman:2020dne}
W.~L. {Freedman}, B.~F. {Madore}, T.~{Hoyt}, I.~S. {Jang}, R.~{Beaton}, M.~G.
  {Lee}, A.~{Monson}, J.~{Neeley}, and J.~{Rich}, ``{Calibration of the Tip of
  the Red Giant Branch},'' {\em \apj}, vol.~891, p.~57, Mar. 2020.

\bibitem{Cerny:2020inj}
W.~{Cerny}, W.~L. {Freedman}, B.~F. {Madore}, F.~{Ashmead}, T.~{Hoyt},
  E.~{Oakes}, N.~{Quang Hoang Tran}, and B.~{Moss}, ``{Multi-Wavelength,
  Optical (VI) and Near-Infrared (JHK) Calibration of the Tip of the Red Giant
  Branch Method based on Milky Way Globular Clusters},'' {\em arXiv e-prints},
  p.~arXiv:2012.09701, Dec. 2020.

\bibitem{Rigault:2014kaa}
M.~Rigault {\em et~al.}, ``{Confirmation of a Star Formation Bias in Type Ia
  Supernova Distances and its Effect on Measurement of the Hubble Constant},''
  {\em Astrophys. J.}, vol.~802, no.~1, p.~20, 2015.

\bibitem{NearbySupernovaFactory:2018qkd}
M.~Rigault {\em et~al.}, ``{Strong Dependence of Type Ia Supernova
  Standardization on the Local Specific Star Formation Rate},'' {\em Astron.
  Astrophys.}, vol.~644, p.~A176, 2020.

\bibitem{Jones:2018vbn}
D.~O. Jones {\em et~al.}, ``{Should Type Ia Supernova Distances be Corrected
  for their Local Environments?},'' {\em Astrophys. J.}, vol.~867, no.~2,
  p.~108, 2018.

\bibitem{Brout:2020msh}
D.~Brout and D.~Scolnic, ``{It\textquoteright{}s Dust: Solving the Mysteries of
  the Intrinsic Scatter and Host-galaxy Dependence of Standardized Type Ia
  Supernova Brightnesses},'' {\em Astrophys. J.}, vol.~909, no.~1, p.~26, 2021.

\bibitem{Lombriser:2019ahl}
L.~Lombriser, ``{Consistency of the local Hubble constant with the cosmic
  microwave background},'' {\em Phys. Lett. B}, vol.~803, p.~135303, 2020.

\bibitem{Kenworthy:2019qwq}
W.~D. Kenworthy, D.~Scolnic, and A.~Riess, ``{The Local Perspective on the
  Hubble Tension: Local Structure Does Not Impact Measurement of the Hubble
  Constant},'' {\em Astrophys. J.}, vol.~875, no.~2, p.~145, 2019.

\bibitem{Verde:2019ivm}
L.~Verde, T.~Treu, and A.~G. Riess, ``{Tensions between the Early and the Late
  Universe},'' {\em Nature Astron.}, vol.~3, p.~891, 7 2019.

\bibitem{Riess:2020sih}
A.~G. Riess, ``{The Expansion of the Universe is Faster than Expected},'' {\em
  Nature Rev. Phys.}, vol.~2, no.~1, pp.~10--12, 2019.

\bibitem{Riess:2016jrr}
A.~G. Riess {\em et~al.}, ``{A 2.4\% Determination of the Local Value of the
  Hubble Constant},'' {\em Astrophys. J.}, vol.~826, no.~1, p.~56, 2016.

\bibitem{Riess:2018uxu}
A.~G. Riess {\em et~al.}, ``{New Parallaxes of Galactic Cepheids from Spatially
  Scanning the Hubble Space Telescope: Implications for the Hubble Constant},''
  {\em Astrophys. J.}, vol.~855, no.~2, p.~136, 2018.

\bibitem{Riess:2019cxk}
A.~G. Riess, S.~Casertano, W.~Yuan, L.~M. Macri, and D.~Scolnic, ``{Large
  Magellanic Cloud Cepheid Standards Provide a 1\% Foundation for the
  Determination of the Hubble Constant and Stronger Evidence for Physics beyond
  $\Lambda$CDM},'' {\em Astrophys. J.}, vol.~876, no.~1, p.~85, 2019.

\bibitem{Scolnic:2017caz}
D.~M. Scolnic {\em et~al.}, ``{The Complete Light-curve Sample of
  Spectroscopically Confirmed SNe Ia from Pan-STARRS1 and Cosmological
  Constraints from the Combined Pantheon Sample},'' {\em Astrophys. J.},
  vol.~859, no.~2, p.~101, 2018.

\bibitem{Benevento:2020fev}
G.~Benevento, W.~Hu, and M.~Raveri, ``{Can Late Dark Energy Transitions Raise
  the Hubble constant?},'' {\em Phys. Rev. D}, vol.~101, no.~10, p.~103517,
  2020.

\bibitem{Camarena:2021jlr}
D.~{Camarena} and V.~{Marra}, ``{On the use of the local prior on the absolute
  magnitude of Type Ia supernovae in cosmological inference},'' {\em MNRAS},
  vol.~504, pp.~5164--5171, July 2021.

\bibitem{Efstathiou:2021ocp}
G.~{Efstathiou}, ``{To H0 or not to H0?},'' {\em arXiv e-prints},
  p.~arXiv:2103.08723, Mar. 2021.

\bibitem{Raveri:2018wln}
M.~Raveri and W.~Hu, ``{Concordance and Discordance in Cosmology},'' {\em Phys.
  Rev. D}, vol.~99, no.~4, p.~043506, 2019.

\bibitem{Buen-Abad:2015ova}
M.~A. Buen-Abad, G.~Marques-Tavares, and M.~Schmaltz, ``{Non-Abelian dark
  matter and dark radiation},'' {\em Phys. Rev. D}, vol.~92, no.~2, p.~023531,
  2015.

\bibitem{Lesgourgues:2015wza}
J.~Lesgourgues, G.~Marques-Tavares, and M.~Schmaltz, ``{Evidence for dark
  matter interactions in cosmological precision data?},'' {\em JCAP}, vol.~02,
  p.~037, 2016.

\bibitem{Buen-Abad:2017gxg}
M.~A. Buen-Abad, M.~Schmaltz, J.~Lesgourgues, and T.~Brinckmann, ``{Interacting
  Dark Sector and Precision Cosmology},'' {\em JCAP}, vol.~01, p.~008, 2018.

\bibitem{Archidiacono:2019wdp}
M.~Archidiacono, D.~C. Hooper, R.~Murgia, S.~Bohr, J.~Lesgourgues, and M.~Viel,
  ``{Constraining Dark Matter-Dark Radiation interactions with CMB, BAO, and
  Lyman-$\alpha$},'' {\em JCAP}, vol.~10, p.~055, 2019.

\bibitem{Li:2019yem}
X.~Li and A.~Shafieloo, ``{A Simple Phenomenological Emergent Dark Energy Model
  can Resolve the Hubble Tension},'' {\em Astrophys. J. Lett.}, vol.~883,
  no.~1, p.~L3, 2019.

\bibitem{Pan:2019hac}
S.~Pan, W.~Yang, E.~Di~Valentino, A.~Shafieloo, and S.~Chakraborty,
  ``{Reconciling $H_0$ tension in a six parameter space?},'' {\em JCAP},
  vol.~06, no.~06, p.~062, 2020.

\bibitem{Rezaei:2020mrj}
M.~Rezaei, T.~Naderi, M.~Malekjani, and A.~Mehrabi, ``{A Bayesian comparison
  between $\Lambda$CDM and phenomenologically emergent dark energy models},''
  {\em Eur. Phys. J. C}, vol.~80, no.~5, p.~374, 2020.

\bibitem{Yang:2021egn}
W.~Yang, E.~Di~Valentino, S.~Pan, and O.~Mena, ``{Emergent Dark Energy,
  neutrinos and cosmological tensions},'' {\em Phys. Dark Univ.}, vol.~31,
  p.~100762, 2021.

\bibitem{Raveri:2021wfz}
M.~{Raveri} and C.~{Doux}, ``{Non-Gaussian estimates of tensions in
  cosmological parameters},'' {\em arXiv e-prints}, p.~arXiv:2105.03324, May
  2021.

\bibitem{schwarz1978estimating}
G.~Schwarz, ``{Estimating the Dimension of a Model},'' {\em The Annals of
  Statistics}, vol.~6, no.~2, pp.~461 -- 464, 1978.

\bibitem{2013PDU.....2..166V}
L.~{Verde}, P.~{Protopapas}, and R.~{Jimenez}, ``{Planck and the local
  Universe: Quantifying the tension},'' {\em Physics of the Dark Universe},
  vol.~2, pp.~166--175, Sept. 2013.

\bibitem{Kass:1995loi}
R.~E. Kass and A.~E. Raftery, ``{Bayes Factors},'' {\em J. Am. Statist.
  Assoc.}, vol.~90, no.~430, pp.~773--795, 1995.

\bibitem{Handley:2019wlz}
W.~Handley and P.~Lemos, ``{Quantifying tensions in cosmological parameters:
  Interpreting the DES evidence ratio},'' {\em Phys. Rev. D}, vol.~100, no.~4,
  p.~043504, 2019.

\bibitem{Trotta:2005ar}
R.~Trotta, ``{Applications of Bayesian model selection to cosmological
  parameters},'' {\em Mon. Not. Roy. Astron. Soc.}, vol.~378, pp.~72--82, 2007.

\bibitem{Jeffreys61}
H.~Jeffreys, {\em Theory of Probability}.
\newblock Oxford, England: Oxford, third~ed., 1961.

\bibitem{Nesseris:2012cq}
S.~Nesseris and J.~Garcia-Bellido {\em JCAP}, vol.~1308, p.~036, 2013.

\bibitem{Planck:2019nip}
N.~Aghanim {\em et~al.}, ``{Planck 2018 results. V. CMB power spectra and
  likelihoods},'' {\em Astron. Astrophys.}, vol.~641, p.~A5, 2020.

\bibitem{BOSS:2016wmc}
S.~Alam {\em et~al.}, ``{The clustering of galaxies in the completed SDSS-III
  Baryon Oscillation Spectroscopic Survey: cosmological analysis of the DR12
  galaxy sample},'' {\em Mon. Not. Roy. Astron. Soc.}, vol.~470, no.~3,
  pp.~2617--2652, 2017.

\bibitem{Ross:2014qpa}
A.~J. Ross, L.~Samushia, C.~Howlett, W.~J. Percival, A.~Burden, and M.~Manera,
  ``{The clustering of the SDSS DR7 main Galaxy sample \textendash{} I. A 4 per
  cent distance measure at $z = 0.15$},'' {\em Mon. Not. Roy. Astron. Soc.},
  vol.~449, no.~1, pp.~835--847, 2015.

\bibitem{Beutler:2011hx}
F.~{Beutler}, C.~{Blake}, M.~{Colless}, D.~H. {Jones}, L.~{Staveley-Smith},
  L.~{Campbell}, Q.~{Parker}, W.~{Saunders}, and F.~{Watson}, ``{The 6dF Galaxy
  Survey: baryon acoustic oscillations and the local Hubble constant},'' {\em
  MNRAS}, vol.~416, pp.~3017--3032, Oct. 2011.

\bibitem{Asgari:2020wuj}
M.~Asgari {\em et~al.}, ``{KiDS-1000 Cosmology: Cosmic shear constraints and
  comparison between two point statistics},'' {\em Astron. Astrophys.},
  vol.~645, p.~A104, 2021.

\bibitem{2021CQGra..38r4001K}
C.~{Krishnan}, R.~{Mohayaee}, E.~{\'O}. {Colg{\'a}in}, M.~M. {Sheikh-Jabbari},
  and L.~{Yin}, ``{Does Hubble tension signal a breakdown in FLRW
  cosmology?},'' {\em Classical and Quantum Gravity}, vol.~38, p.~184001, Sept.
  2021.

\bibitem{2021MNRAS.500.5249A}
E.~{Asencio}, I.~{Banik}, and P.~{Kroupa}, ``{A massive blow for
  {\ensuremath{\Lambda}}CDM - the high redshift, mass, and collision velocity
  of the interacting galaxy cluster El Gordo contradicts concordance
  cosmology},'' {\em MNRAS}, vol.~500, pp.~5249--5267, Jan. 2021.

\bibitem{2019PhRvD.100d3537D}
H.~{Desmond}, B.~{Jain}, and J.~{Sakstein}, ``{Local resolution of the Hubble
  tension: The impact of screened fifth forces on the cosmic distance
  ladder},'' {\em \prd}, vol.~100, p.~043537, Aug. 2019.

\bibitem{2020PhRvD.102b3007D}
H.~{Desmond} and J.~{Sakstein}, ``{Screened fifth forces lower the
  TRGB-calibrated Hubble constant too},'' {\em \prd}, vol.~102, p.~023007, July
  2020.

\bibitem{2021PhRvD.103h3517A}
G.~{Alestas}, L.~{Kazantzidis}, and L.~{Perivolaropoulos}, ``{w -M phantom
  transition at z$_{t}$<0.1 as a resolution of the Hubble tension},'' {\em
  \prd}, vol.~103, p.~083517, Apr. 2021.

\bibitem{2020MNRAS.499.2845H}
M.~{Haslbauer}, I.~{Banik}, and P.~{Kroupa}, ``{The KBC void and Hubble tension
  contradict {\ensuremath{\Lambda}}CDM on a Gpc scale - Milgromian dynamics as
  a possible solution},'' {\em MNRAS}, vol.~499, pp.~2845--2883, Dec. 2020.

\bibitem{Bashinsky:2003tk}
S.~Bashinsky and U.~Seljak, ``{Neutrino perturbations in CMB anisotropy and
  matter clustering},'' {\em Phys. Rev. D}, vol.~69, p.~083002, 2004.

\bibitem{Lesgourgues:2018ncw}
J.~Lesgourgues, G.~Mangano, G.~Miele, and S.~Pastor, {\em {Neutrino
  Cosmology}}.
\newblock Cambridge University Press, 2 2013.

\bibitem{Baumann:2015rya}
D.~Baumann, D.~Green, J.~Meyers, and B.~Wallisch, ``{Phases of New Physics in
  the CMB},'' {\em JCAP}, vol.~01, p.~007, 2016.

\bibitem{Follin:2015hya}
B.~Follin, L.~Knox, M.~Millea, and Z.~Pan, ``{First Detection of the Acoustic
  Oscillation Phase Shift Expected from the Cosmic Neutrino Background},'' {\em
  Phys. Rev. Lett.}, vol.~115, no.~9, p.~091301, 2015.

\bibitem{Froustey:2020mcq}
J.~Froustey, C.~Pitrou, and M.~C. Volpe, ``{Neutrino decoupling including
  flavour oscillations and primordial nucleosynthesis},'' {\em JCAP}, vol.~12,
  p.~015, 2020.

\bibitem{Bennett:2020zkv}
J.~J. Bennett, G.~Buldgen, P.~F. De~Salas, M.~Drewes, S.~Gariazzo, S.~Pastor,
  and Y.~Y.~Y. Wong, ``{Towards a precision calculation of $N_{\rm eff}$ in the
  Standard Model II: Neutrino decoupling in the presence of flavour
  oscillations and finite-temperature QED},'' {\em JCAP}, vol.~04, p.~073,
  2021.

\bibitem{2020JCAP...08..012A}
K.~{Akita} and M.~{Yamaguchi}, ``{A precision calculation of relic neutrino
  decoupling},'' {\em JCAP}, vol.~2020, p.~012, Aug. 2020.

\bibitem{Bernal:2016gxb}
J.~L. Bernal, L.~Verde, and A.~G. Riess, ``{The trouble with $H_0$},'' {\em
  JCAP}, vol.~10, p.~019, 2016.

\bibitem{Poulin:2018cxd}
V.~Poulin, T.~L. Smith, T.~Karwal, and M.~Kamionkowski, ``{Early Dark Energy
  Can Resolve The Hubble Tension},'' {\em Phys. Rev. Lett.}, vol.~122, no.~22,
  p.~221301, 2019.

\bibitem{DiValentino:2019dzu}
E.~Di~Valentino, A.~Melchiorri, and J.~Silk, ``{Cosmological constraints in
  extended parameter space from the Planck 2018 Legacy release},'' {\em JCAP},
  vol.~01, p.~013, 2020.

\bibitem{Ghosh:2021axu}
S.~Ghosh, S.~Kumar, and Y.~Tsai, ``{Free-streaming and Coupled Dark Radiation
  Isocurvature Perturbations: Constraints and Application to the Hubble
  Tension},'' 7 2021.

\bibitem{Cyr-Racine:2021alc}
F.-Y. Cyr-Racine, F.~Ge, and L.~Knox, ``{A Symmetry of Cosmological
  Observables, and a High Hubble Constant as an Indicator of a Mirror World
  Dark Sector},'' 7 2021.

\bibitem{Brust:2017nmv}
C.~Brust, Y.~Cui, and K.~Sigurdson, ``{Cosmological Constraints on Interacting
  Light Particles},'' {\em JCAP}, vol.~08, p.~020, 2017.

\bibitem{Blinov:2020hmc}
N.~Blinov and G.~Marques-Tavares, ``{Interacting radiation after Planck and its
  implications for the Hubble Tension},'' {\em JCAP}, vol.~09, p.~029, 2020.

\bibitem{Chacko:2016kgg}
Z.~Chacko, Y.~Cui, S.~Hong, T.~Okui, and Y.~Tsai, ``{Partially Acoustic Dark
  Matter, Interacting Dark Radiation, and Large Scale Structure},'' {\em JHEP},
  vol.~12, p.~108, 2016.

\bibitem{Ko:2016uft}
P.~Ko and Y.~Tang, ``{Light dark photon and fermionic dark radiation for the
  Hubble constant and the structure formation},'' {\em Phys. Lett. B},
  vol.~762, pp.~462--466, 2016.

\bibitem{Ko:2016fcd}
P.~Ko and Y.~Tang, ``{Residual Non-Abelian Dark Matter and Dark Radiation},''
  {\em Phys. Lett. B}, vol.~768, pp.~12--17, 2017.

\bibitem{Ko:2017uyb}
P.~Ko, N.~Nagata, and Y.~Tang, ``{Hidden Charged Dark Matter and Chiral Dark
  Radiation},'' {\em Phys. Lett. B}, vol.~773, pp.~513--520, 2017.

\bibitem{Kreisch:2019yzn}
C.~D. Kreisch, F.-Y. Cyr-Racine, and O.~Dor\'e, ``{Neutrino puzzle: Anomalies,
  interactions, and cosmological tensions},'' {\em Phys. Rev. D}, vol.~101,
  no.~12, p.~123505, 2020.

\bibitem{Park:2019ibn}
M.~Park, C.~D. Kreisch, J.~Dunkley, B.~Hadzhiyska, and F.-Y. Cyr-Racine,
  ``{$\Lambda$CDM or self-interacting neutrinos: How CMB data can tell the two
  models apart},'' {\em Phys. Rev. D}, vol.~100, no.~6, p.~063524, 2019.

\bibitem{2020PhRvD.102l3544G}
S.~{Ghosh}, R.~{Khatri}, and T.~S. {Roy}, ``{Can dark neutrino interactions
  phase out the Hubble tension?},'' {\em \prd}, vol.~102, p.~123544, Dec. 2020.

\bibitem{2021JCAP...07..038D}
A.~{Das} and S.~{Ghosh}, ``{Flavor-specific interaction favors strong neutrino
  self-coupling in the early universe},'' {\em JCAP}, vol.~2021, p.~038, July
  2021.

\bibitem{Cyr-Racine:2013jua}
F.-Y. Cyr-Racine and K.~Sigurdson, ``{Limits on Neutrino-Neutrino Scattering in
  the Early Universe},'' {\em Phys. Rev. D}, vol.~90, no.~12, p.~123533, 2014.

\bibitem{Archidiacono:2013dua}
M.~Archidiacono and S.~Hannestad, ``{Updated constraints on non-standard
  neutrino interactions from Planck},'' {\em JCAP}, vol.~07, p.~046, 2014.

\bibitem{Lancaster:2017ksf}
L.~Lancaster, F.-Y. Cyr-Racine, L.~Knox, and Z.~Pan, ``{A tale of two modes:
  Neutrino free-streaming in the early universe},'' {\em JCAP}, vol.~07,
  p.~033, 2017.

\bibitem{Oldengott:2017fhy}
I.~M. Oldengott, T.~Tram, C.~Rampf, and Y.~Y.~Y. Wong, ``{Interacting neutrinos
  in cosmology: exact description and constraints},'' {\em JCAP}, vol.~11,
  p.~027, 2017.

\bibitem{Aver:2020fon}
E.~Aver, D.~A. Berg, K.~A. Olive, R.~W. Pogge, J.~J. Salzer, and E.~D.
  Skillman, ``{Improving helium abundance determinations with Leo P as a case
  study},'' {\em JCAP}, vol.~03, p.~027, 2021.

\bibitem{Huang:2021dba}
G.-y. {Huang} and W.~{Rodejohann}, ``{Solving the Hubble tension without
  spoiling big bang nucleosynthesis},'' {\em \prd}, vol.~103, p.~123007, June
  2021.

\bibitem{Berbig:2020wve}
M.~Berbig, S.~Jana, and A.~Trautner, ``{The Hubble tension and a renormalizable
  model of gauged neutrino self-interactions},'' {\em Phys. Rev. D}, vol.~102,
  no.~11, p.~115008, 2020.

\bibitem{Izotov:2014fga}
Y.~I. Izotov, T.~X. Thuan, and N.~G. Guseva, ``{A new determination of the
  primordial He abundance using the He i $\lambda$10830 \r{A} emission line:
  cosmological implications},'' {\em Mon. Not. Roy. Astron. Soc.}, vol.~445,
  no.~1, pp.~778--793, 2014.

\bibitem{Blinov:2019gcj}
N.~Blinov, K.~J. Kelly, G.~Z. Krnjaic, and S.~D. McDermott, ``{Constraining the
  Self-Interacting Neutrino Interpretation of the Hubble Tension},'' {\em Phys.
  Rev. Lett.}, vol.~123, no.~19, p.~191102, 2019.

\bibitem{Lyu:2020lps}
K.-F. Lyu, E.~Stamou, and L.-T. Wang, ``{Self-interacting neutrinos: Solution
  to Hubble tension versus experimental constraints},'' {\em Phys. Rev. D},
  vol.~103, no.~1, p.~015004, 2021.

\bibitem{Deppisch:2020sqh}
F.~F. Deppisch, L.~Graf, W.~Rodejohann, and X.-J. Xu, ``{Neutrino
  Self-Interactions and Double Beta Decay},'' {\em Phys. Rev. D}, vol.~102,
  no.~5, p.~051701, 2020.

\bibitem{Brinckmann:2020bcn}
T.~{Brinckmann}, J.~{Hyeok Chang}, and M.~{LoVerde}, ``{Self-interacting
  neutrinos, the Hubble parameter tension, and the Cosmic Microwave
  Background},'' {\em arXiv e-prints}, p.~arXiv:2012.11830, Dec. 2020.

\bibitem{Escudero:2019gvw}
M.~Escudero and S.~J. Witte, ``{A CMB search for the neutrino mass mechanism
  and its relation to the Hubble tension},'' {\em Eur. Phys. J. C}, vol.~80,
  no.~4, p.~294, 2020.

\bibitem{Escudero:2020hkf}
M.~{Escudero} and S.~J. {Witte}, ``{Could the Hubble Tension be Pointing
  Towards the Neutrino Mass Mechanism?},'' p.~arXiv:2004.01470, Apr. 2020.

\bibitem{Escudero:2021rfi}
M.~Escudero and S.~J. Witte, ``{The hubble tension as a hint of leptogenesis
  and neutrino mass generation},'' {\em Eur. Phys. J. C}, vol.~81, no.~6,
  p.~515, 2021.

\bibitem{Akhmedov:1992hi}
E.~K. Akhmedov, Z.~G. Berezhiani, R.~N. Mohapatra, and G.~Senjanovic, ``{Planck
  scale effects on the majoron},'' {\em Phys. Lett. B}, vol.~299, pp.~90--93,
  1993.

\bibitem{Rothstein:1992rh}
I.~Z. Rothstein, K.~S. Babu, and D.~Seckel, ``{Planck scale symmetry breaking
  and majoron physics},'' {\em Nucl. Phys. B}, vol.~403, pp.~725--748, 1993.

\bibitem{2018arXiv181103624C}
C.-T. {Chiang} and A.~{Slosar}, ``{Inferences of $H_0$ in presence of a
  non-standard recombination},'' {\em arXiv e-prints}, p.~arXiv:1811.03624,
  Nov. 2018.

\bibitem{2021PhRvD.103h3533A}
S.~{Alam}, M.~{Aubert}, S.~{Avila}, C.~{Balland}, J.~E. {Bautista}, M.~A.
  {Bershady}, D.~{Bizyaev}, M.~R. {Blanton}, A.~S. {Bolton}, J.~{Bovy},
  J.~{Brinkmann}, J.~R. {Brownstein}, E.~{Burtin}, S.~{Chabanier}, M.~J.
  {Chapman}, P.~D. {Choi}, C.-H. {Chuang}, J.~{Comparat}, M.-C. {Cousinou},
  A.~{Cuceu}, K.~S. {Dawson}, S.~{de la Torre}, A.~{de Mattia}, V.~d.~S.
  {Agathe}, H.~d.~M. {des Bourboux}, S.~{Escoffier}, T.~{Etourneau}, J.~{Farr},
  A.~{Font-Ribera}, P.~M. {Frinchaboy}, S.~{Fromenteau}, H.~{Gil-Mar{\'\i}n},
  J.-M. {Le Goff}, A.~X. {Gonzalez-Morales}, V.~{Gonzalez-Perez},
  K.~{Grabowski}, J.~{Guy}, A.~J. {Hawken}, J.~{Hou}, H.~{Kong}, J.~{Parker},
  M.~{Klaene}, J.-P. {Kneib}, S.~{Lin}, D.~{Long}, B.~W. {Lyke}, A.~{de la
  Macorra}, P.~{Martini}, K.~{Masters}, F.~G. {Mohammad}, J.~{Moon}, E.-M.
  {Mueller}, A.~{Mu{\~n}oz-Guti{\'e}rrez}, A.~D. {Myers}, S.~{Nadathur},
  R.~{Neveux}, J.~A. {Newman}, P.~{Noterdaeme}, A.~{Oravetz}, D.~{Oravetz},
  N.~{Palanque-Delabrouille}, K.~{Pan}, R.~{Paviot}, W.~J. {Percival},
  I.~{P{\'e}rez-R{\`a}fols}, P.~{Petitjean}, M.~M. {Pieri}, A.~{Prakash},
  A.~{Raichoor}, C.~{Ravoux}, M.~{Rezaie}, J.~{Rich}, A.~J. {Ross}, G.~{Rossi},
  R.~{Ruggeri}, V.~{Ruhlmann-Kleider}, A.~G. {S{\'a}nchez}, F.~J.
  {S{\'a}nchez}, J.~R. {S{\'a}nchez-Gallego}, C.~{Sayres}, D.~P. {Schneider},
  H.-J. {Seo}, A.~{Shafieloo}, A.~{Slosar}, A.~{Smith}, J.~{Stermer},
  A.~{Tamone}, J.~L. {Tinker}, R.~{Tojeiro}, M.~{Vargas-Maga{\~n}a},
  A.~{Variu}, Y.~{Wang}, B.~A. {Weaver}, A.-M. {Weijmans}, C.~{Y{\`e}che},
  P.~{Zarrouk}, C.~{Zhao}, G.-B. {Zhao}, and Z.~{Zheng}, ``{Completed SDSS-IV
  extended Baryon Oscillation Spectroscopic Survey: Cosmological implications
  from two decades of spectroscopic surveys at the Apache Point Observatory},''
  {\em \prd}, vol.~103, p.~083533, Apr. 2021.

\bibitem{Jedamzik:2011cu}
K.~{Jedamzik} and T.~{Abel}, ``{Weak Primordial Magnetic Fields and
  Anisotropies in the Cosmic Microwave Background Radiation},'' {\em arXiv
  e-prints}, p.~arXiv:1108.2517, Aug. 2011.

\bibitem{Jedamzik:2018itu}
K.~Jedamzik and A.~Saveliev, ``{Stringent Limit on Primordial Magnetic Fields
  from the Cosmic Microwave Background Radiation},'' {\em Phys. Rev. Lett.},
  vol.~123, no.~2, p.~021301, 2019.

\bibitem{Jedamzik:2020krr}
K.~Jedamzik and L.~Pogosian, ``{Relieving the Hubble tension with primordial
  magnetic fields},'' {\em Phys. Rev. Lett.}, vol.~125, no.~18, p.~181302,
  2020.

\bibitem{Thiele:2021okz}
L.~{Thiele}, Y.~{Guan}, J.~C. {Hill}, A.~{Kosowsky}, and D.~N. {Spergel},
  ``{Can small-scale baryon inhomogeneities resolve the Hubble tension? An
  investigation with ACT DR4},'' {\em arXiv e-prints}, p.~arXiv:2105.03003, May
  2021.

\bibitem{2021arXiv210702243F}
S.~A. {Franchino-Vi{\~n}as} and M.~E. {Mosquera}, ``{The cosmological lithium
  problem, varying constants and the $H_0$ tension},'' {\em arXiv e-prints},
  p.~arXiv:2107.02243, July 2021.

\bibitem{Sekiguchi:2020teg}
T.~Sekiguchi and T.~Takahashi, ``{Early recombination as a solution to the
  $H_0$ tension},'' {\em Phys. Rev. D}, vol.~103, no.~8, p.~083507, 2021.

\bibitem{Uzan:2002vq}
J.-P. Uzan, ``{The Fundamental Constants and Their Variation: Observational
  Status and Theoretical Motivations},'' {\em Rev. Mod. Phys.}, vol.~75,
  p.~403, 2003.

\bibitem{Uzan:2010pm}
J.-P. Uzan, ``{Varying Constants, Gravitation and Cosmology},'' {\em Living
  Rev. Rel.}, vol.~14, p.~2, 2011.

\bibitem{Martins:2017yxk}
C.~J.~A.~P. {Martins}, ``{The status of varying constants: a review of the
  physics, searches and implications},'' {\em Reports on Progress in Physics},
  vol.~80, p.~126902, Dec. 2017.

\bibitem{King:2012id}
J.~A. King, J.~K. Webb, M.~T. Murphy, V.~V. Flambaum, R.~F. Carswell, M.~B.
  Bainbridge, M.~R. Wilczynska, and F.~E. Koch, ``{Spatial variation in the
  fine-structure constant -- new results from VLT/UVES},'' {\em Mon. Not. Roy.
  Astron. Soc.}, vol.~422, pp.~3370--3413, 2012.

\bibitem{Bagdonaite:2013sia}
J.~Bagdonaite, M.~Dapr\`a, P.~Jansen, H.~L. Bethlem, W.~Ubachs, S.~Muller,
  C.~Henkel, and K.~M. Menten, ``{Robust Constraint on a Drifting
  Proton-to-Electron Mass Ratio at z=0.89 from Methanol Observation at Three
  Radio Telescopes},'' {\em Phys. Rev. Lett.}, vol.~111, p.~231101, 2013.

\bibitem{Kotus:2016xxb}
S.~M. Kotu\v{s}, M.~T. Murphy, and R.~F. Carswell, ``{High-precision limit on
  variation in the fine-structure constant from a single quasar absorption
  system},'' {\em Mon. Not. Roy. Astron. Soc.}, vol.~464, no.~3,
  pp.~3679--3703, 2017.

\bibitem{Murphy:2017xaz}
M.~T. Murphy and K.~L. Cooksey, ``{Subaru Telescope limits on cosmological
  variations in the fine-structure constant},'' {\em Mon. Not. Roy. Astron.
  Soc.}, vol.~471, no.~4, pp.~4930--4945, 2017.

\bibitem{Wilczynska:2020rxx}
M.~R. Wilczynska {\em et~al.}, ``{Four direct measurements of the
  fine-structure constant 13 billion years ago},'' {\em Sci. Adv.}, vol.~6,
  no.~17, p.~eaay9672, 2020.

\bibitem{Ade:2014zfo}
P.~A.~R. Ade {\em et~al.}, ``{Planck intermediate results - XXIV. Constraints
  on variations in fundamental constants},'' {\em Astron. Astrophys.},
  vol.~580, p.~A22, 2015.

\bibitem{Hart:2017ndk}
L.~Hart and J.~Chluba, ``{New constraints on time-dependent variations of
  fundamental constants using Planck data},'' {\em Mon. Not. Roy. Astron.
  Soc.}, vol.~474, no.~2, pp.~1850--1861, 2018.

\bibitem{Hart:2019gvj}
L.~{Hart} and J.~{Chluba}, ``{Improved model-independent constraints on the
  recombination era and development of a direct projection method},'' {\em
  MNRAS}, vol.~495, pp.~4210--4226, July 2020.

\bibitem{Lopez-Honorez:2020lno}
L.~Lopez-Honorez, O.~Mena, S.~Palomares-Ruiz, P.~Villanueva-Domingo, and S.~J.
  Witte, ``{Variations in fundamental constants at the cosmic dawn},'' {\em
  JCAP}, vol.~06, p.~026, 2020.

\bibitem{Hart:2019dxi}
L.~Hart and J.~Chluba, ``{Updated fundamental constant constraints from Planck
  2018 data and possible relations to the Hubble tension},'' {\em Mon. Not.
  Roy. Astron. Soc.}, vol.~493, no.~3, pp.~3255--3263, 2020.

\bibitem{Karwal:2016vyq}
T.~Karwal and M.~Kamionkowski, ``{Dark energy at early times, the Hubble
  parameter, and the string axiverse},'' {\em Phys. Rev. D}, vol.~94, no.~10,
  p.~103523, 2016.

\bibitem{Lin:2019qug}
M.-X. Lin, G.~Benevento, W.~Hu, and M.~Raveri, ``{Acoustic Dark Energy:
  Potential Conversion of the Hubble Tension},'' {\em Phys. Rev. D}, vol.~100,
  no.~6, p.~063542, 2019.

\bibitem{Smith:2019ihp}
T.~L. Smith, V.~Poulin, and M.~A. Amin, ``{Oscillating scalar fields and the
  Hubble tension: a resolution with novel signatures},'' {\em Phys. Rev. D},
  vol.~101, no.~6, p.~063523, 2020.

\bibitem{Kamionkowski:2014zda}
M.~Kamionkowski, J.~Pradler, and D.~G.~E. Walker, ``{Dark energy from the
  string axiverse},'' {\em Phys. Rev. Lett.}, vol.~113, no.~25, p.~251302,
  2014.

\bibitem{Poulin:2018dzj}
V.~Poulin, T.~L. Smith, D.~Grin, T.~Karwal, and M.~Kamionkowski,
  ``{Cosmological implications of ultralight axionlike fields},'' {\em Phys.
  Rev. D}, vol.~98, no.~8, p.~083525, 2018.

\bibitem{Murgia:2020ryi}
R.~Murgia, G.~F. Abell\'an, and V.~Poulin, ``{Early dark energy resolution to
  the Hubble tension in light of weak lensing surveys and lensing anomalies},''
  {\em Phys. Rev. D}, vol.~103, no.~6, p.~063502, 2021.

\bibitem{Svrcek:2006yi}
P.~Svrcek and E.~Witten, ``{Axions In String Theory},'' {\em JHEP}, vol.~06,
  p.~051, 2006.

\bibitem{Douglas:2006es}
M.~R. Douglas and S.~Kachru, ``{Flux compactification},'' {\em Rev. Mod.
  Phys.}, vol.~79, pp.~733--796, 2007.

\bibitem{Arvanitaki:2009fg}
A.~Arvanitaki, S.~Dimopoulos, S.~Dubovsky, N.~Kaloper, and J.~March-Russell,
  ``{String Axiverse},'' {\em Phys. Rev. D}, vol.~81, p.~123530, 2010.

\bibitem{Marsh:2015xka}
D.~J.~E. Marsh, ``{Axion Cosmology},'' {\em Phys. Rept.}, vol.~643, pp.~1--79,
  2016.

\bibitem{Agrawal:2019lmo}
P.~{Agrawal}, F.-Y. {Cyr-Racine}, D.~{Pinner}, and L.~{Randall}, ``{Rock 'n'
  Roll Solutions to the Hubble Tension},'' {\em arXiv e-prints},
  p.~arXiv:1904.01016, Apr. 2019.

\bibitem{Kaloper:2019lpl}
N.~Kaloper, ``{Dark energy, $H_0$ and weak gravity conjecture},'' {\em Int. J.
  Mod. Phys. D}, vol.~28, no.~14, p.~1944017, 2019.

\bibitem{Sakstein:2019fmf}
J.~Sakstein and M.~Trodden, ``{Early Dark Energy from Massive Neutrinos as a
  Natural Resolution of the Hubble Tension},'' {\em Phys. Rev. Lett.},
  vol.~124, no.~16, p.~161301, 2020.

\bibitem{Berghaus:2019cls}
K.~V. Berghaus and T.~Karwal, ``{Thermal Friction as a Solution to the Hubble
  Tension},'' {\em Phys. Rev. D}, vol.~101, no.~8, p.~083537, 2020.

\bibitem{Alexander:2019rsc}
S.~Alexander and E.~McDonough, ``{Axion-Dilaton Destabilization and the Hubble
  Tension},'' {\em Phys. Lett. B}, vol.~797, p.~134830, 2019.

\bibitem{Braglia:2020bym}
M.~Braglia, W.~T. Emond, F.~Finelli, A.~E. Gumrukcuoglu, and K.~Koyama,
  ``{Unified framework for early dark energy from $\alpha$-attractors},'' {\em
  Phys. Rev. D}, vol.~102, no.~8, p.~083513, 2020.

\bibitem{Ballesteros:2020sik}
G.~Ballesteros, A.~Notari, and F.~Rompineve, ``{The $H_0$ tension: $\Delta G_N$
  vs. $\Delta N_{\rm eff}$},'' {\em JCAP}, vol.~11, p.~024, 2020.

\bibitem{Gonzalez:2020fdy}
M.~Gonzalez, M.~P. Hertzberg, and F.~Rompineve, ``{Ultralight Scalar Decay and
  the Hubble Tension},'' {\em JCAP}, vol.~10, p.~028, 2020.

\bibitem{Ballardini:2020iws}
M.~Ballardini, M.~Braglia, F.~Finelli, D.~Paoletti, A.~A. Starobinsky, and
  C.~Umilt\`a, ``{Scalar-tensor theories of gravity, neutrino physics, and the
  $H_0$ tension},'' {\em JCAP}, vol.~10, p.~044, 2020.

\bibitem{Niedermann:2019olb}
F.~Niedermann and M.~S. Sloth, ``{New early dark energy},'' {\em Phys. Rev. D},
  vol.~103, no.~4, p.~L041303, 2021.

\bibitem{Niedermann:2020dwg}
F.~Niedermann and M.~S. Sloth, ``{Resolving the Hubble tension with new early
  dark energy},'' {\em Phys. Rev. D}, vol.~102, no.~6, p.~063527, 2020.

\bibitem{Das:2020wfe}
A.~{Gogoi}, R.~{Kumar Sharma}, P.~{Chanda}, and S.~{Das}, ``{Early Mass-varying
  Neutrino Dark Energy: Nugget Formation and Hubble Anomaly},'' {\em
  Astrophysical Journal}, vol.~915, p.~132, July 2021.

\bibitem{Allali:2021azp}
I.~J. {Allali}, M.~P. {Hertzberg}, and F.~{Rompineve}, ``{A Dark Sector to
  Restore Cosmological Concordance},'' {\em arXiv e-prints},
  p.~arXiv:2104.12798, Apr. 2021.

\bibitem{Karwal:2021vpk}
T.~{Karwal}, M.~{Raveri}, B.~{Jain}, J.~{Khoury}, and M.~{Trodden},
  ``{Chameleon Early Dark Energy and the Hubble Tension},'' {\em arXiv
  e-prints}, p.~arXiv:2106.13290, June 2021.

\bibitem{Rossi:2019lgt}
M.~Rossi, M.~Ballardini, M.~Braglia, F.~Finelli, D.~Paoletti, A.~A.
  Starobinsky, and C.~Umilt\`a, ``{Cosmological constraints on post-Newtonian
  parameters in effectively massless scalar-tensor theories of gravity},'' {\em
  Phys. Rev. D}, vol.~100, no.~10, p.~103524, 2019.

\bibitem{Braglia:2020iik}
M.~Braglia, M.~Ballardini, W.~T. Emond, F.~Finelli, A.~E. Gumrukcuoglu,
  K.~Koyama, and D.~Paoletti, ``{Larger value for $H_0$ by an evolving
  gravitational constant},'' {\em Phys. Rev. D}, vol.~102, no.~2, p.~023529,
  2020.

\bibitem{Zumalacarregui:2020cjh}
M.~Zumalacarregui, ``{Gravity in the Era of Equality: Towards solutions to the
  Hubble problem without fine-tuned initial conditions},'' {\em Phys. Rev. D},
  vol.~102, no.~2, p.~023523, 2020.

\bibitem{Abadi:2020hbr}
T.~Abadi and E.~D. Kovetz, ``{Can conformally coupled modified gravity solve
  the Hubble tension?},'' {\em Phys. Rev. D}, vol.~103, no.~2, p.~023530, 2021.

\bibitem{Braglia:2020auw}
M.~Braglia, M.~Ballardini, F.~Finelli, and K.~Koyama, ``{Early modified gravity
  in light of the $H_0$ tension and LSS data},'' {\em Phys. Rev. D}, vol.~103,
  no.~4, p.~043528, 2021.

\bibitem{Chevallier:2000qy}
M.~Chevallier and D.~Polarski, ``{Accelerating universes with scaling dark
  matter},'' {\em Int. J. Mod. Phys. D}, vol.~10, pp.~213--224, 2001.

\bibitem{Linder:2002et}
E.~V. Linder, ``{Exploring the expansion history of the universe},'' {\em Phys.
  Rev. Lett.}, vol.~90, p.~091301, 2003.

\bibitem{2020ApJ...902...58L}
X.~{Li} and A.~{Shafieloo}, ``{Evidence for Emergent Dark Energy},'' {\em
  \apj}, vol.~902, p.~58, Oct. 2020.

\bibitem{2020MNRAS.497.1590H}
A.~{Hern{\'a}ndez-Almada}, G.~{Leon}, J.~{Maga{\~n}a}, M.~A.
  {Garc{\'\i}a-Aspeitia}, and V.~{Motta}, ``{Generalized emergent dark energy:
  observational Hubble data constraints and stability analysis},'' {\em MNRAS},
  vol.~497, pp.~1590--1602, Sept. 2020.

\bibitem{2021arXiv210303815Y}
W.~{Yang}, E.~{Di Valentino}, S.~{Pan}, A.~{Shafieloo}, and X.~{Li},
  ``{Generalized Emergent Dark Energy Model and the Hubble Constant Tension},''
  {\em arXiv e-prints}, p.~arXiv:2103.03815, Mar. 2021.

\bibitem{Benaoum:2020qsi}
H.~B. {Benaoum}, W.~{Yang}, S.~{Pan}, and E.~{Di Valentino}, ``{Modified
  Emergent Dark Energy and its Astronomical Constraints},'' {\em arXiv
  e-prints}, p.~arXiv:2008.09098, Aug. 2020.

\bibitem{Poulin:2016nat}
V.~Poulin, P.~D. Serpico, and J.~Lesgourgues, ``{A fresh look at linear
  cosmological constraints on a decaying dark matter component},'' {\em JCAP},
  vol.~08, p.~036, 2016.

\bibitem{Nygaard:2020sow}
A.~Nygaard, T.~Tram, and S.~Hannestad, ``{Updated constraints on decaying cold
  dark matter},'' {\em JCAP}, vol.~05, p.~017, 2021.

\bibitem{2016PhRvD..94b3528C}
A.~{Chudaykin}, D.~{Gorbunov}, and I.~{Tkachev}, ``{Dark matter component
  decaying after recombination: Lensing constraints with Planck data},'' {\em
  \prd}, vol.~94, p.~023528, July 2016.

\bibitem{2018PhRvD..97h3508C}
A.~{Chudaykin}, D.~{Gorbunov}, and I.~{Tkachev}, ``{Dark matter component
  decaying after recombination: Sensitivity to baryon acoustic oscillation and
  redshift space distortion probes},'' {\em \prd}, vol.~97, p.~083508, Apr.
  2018.

\bibitem{Audren:2014bca}
B.~Audren, J.~Lesgourgues, G.~Mangano, P.~D. Serpico, and T.~Tram, ``{Strongest
  model-independent bound on the lifetime of Dark Matter},'' {\em JCAP},
  vol.~12, p.~028, 2014.

\bibitem{Enqvist:2015ara}
K.~Enqvist, S.~Nadathur, T.~Sekiguchi, and T.~Takahashi, ``{Decaying dark
  matter and the tension in $\sigma_8$},'' {\em JCAP}, vol.~09, p.~067, 2015.

\bibitem{Berezhiani:2015yta}
Z.~Berezhiani, A.~D. Dolgov, and I.~I. Tkachev, ``{Reconciling Planck results
  with low redshift astronomical measurements},'' {\em Phys. Rev. D}, vol.~92,
  no.~6, p.~061303, 2015.

\bibitem{Bringmann:2018jpr}
T.~Bringmann, F.~Kahlhoefer, K.~Schmidt-Hoberg, and P.~Walia, ``{Converting
  nonrelativistic dark matter to radiation},'' {\em Phys. Rev. D}, vol.~98,
  no.~2, p.~023543, 2018.

\bibitem{Pandey:2019plg}
K.~L. Pandey, T.~Karwal, and S.~Das, ``{Alleviating the $H_0$ and $\sigma_8$
  anomalies with a decaying dark matter model},'' {\em JCAP}, vol.~07, p.~026,
  2020.

\bibitem{Abellan:2021bpx}
G.~F. {Abell{\'a}n}, R.~{Murgia}, and V.~{Poulin}, ``{Linear cosmological
  constraints on 2-body decaying dark matter scenarios and robustness of the
  resolution to the $S_8$ tension},'' {\em arXiv e-prints},
  p.~arXiv:2102.12498, Feb. 2021.

\bibitem{Blackadder:2014wpa}
G.~Blackadder and S.~M. Koushiappas, ``{Dark matter with two- and many-body
  decays and supernovae type Ia},'' {\em Phys. Rev. D}, vol.~90, no.~10,
  p.~103527, 2014.

\bibitem{Vattis:2019efj}
K.~Vattis, S.~M. Koushiappas, and A.~Loeb, ``{Dark matter decaying in the late
  Universe can relieve the H0 tension},'' {\em Phys. Rev. D}, vol.~99, no.~12,
  p.~121302, 2019.

\bibitem{Haridasu:2020xaa}
B.~S. Haridasu and M.~Viel, ``{Late-time decaying dark matter: constraints and
  implications for the $H_0$-tension},'' {\em Mon. Not. Roy. Astron. Soc.},
  vol.~497, no.~2, pp.~1757--1764, 2020.

\bibitem{Abellan:2020pmw}
G.~F. {Abellan}, R.~{Murgia}, V.~{Poulin}, and J.~{Lavalle}, ``{Hints for
  decaying dark matter from $S_8$ measurements},'' {\em arXiv e-prints},
  p.~arXiv:2008.09615, Aug. 2020.

\bibitem{Jedamzik:2020zmd}
K.~Jedamzik, L.~Pogosian, and G.-B. Zhao, ``{Why reducing the cosmic sound
  horizon alone can not fully resolve the Hubble tension},'' {\em Commun. in
  Phys.}, vol.~4, p.~123, 2021.

\bibitem{Hill:2020osr}
J.~C. Hill, E.~McDonough, M.~W. Toomey, and S.~Alexander, ``{Early dark energy
  does not restore cosmological concordance},'' {\em Phys. Rev. D}, vol.~102,
  no.~4, p.~043507, 2020.

\bibitem{Knox:2019rjx}
L.~Knox and M.~Millea, ``{Hubble constant hunter\textquoteright{}s guide},''
  {\em Phys. Rev. D}, vol.~101, no.~4, p.~043533, 2020.

\bibitem{Hinshaw:2012aka}
G.~Hinshaw {\em et~al.}, ``{Nine-Year Wilkinson Microwave Anisotropy Probe
  (WMAP) Observations: Cosmological Parameter Results},'' {\em Astrophys. J.
  Suppl.}, vol.~208, p.~19, 2013.

\bibitem{Aiola:2020azj}
S.~Aiola {\em et~al.}, ``{The Atacama Cosmology Telescope: DR4 Maps and
  Cosmological Parameters},'' {\em JCAP}, vol.~12, p.~047, 2020.

\bibitem{Dutcher:2021vtw}
D.~{Dutcher}, L.~{Balkenhol}, P.~A.~R. {Ade}, Z.~{Ahmed}, E.~{Anderes}, A.~J.
  {Anderson}, M.~{Archipley}, J.~S. {Avva}, K.~{Aylor}, P.~S. {Barry}, R.~{Basu
  Thakur}, K.~{Benabed}, A.~N. {Bender}, B.~A. {Benson}, F.~{Bianchini}, L.~E.
  {Bleem}, F.~R. {Bouchet}, L.~{Bryant}, K.~{Byrum}, J.~E. {Carlstrom}, F.~W.
  {Carter}, T.~W. {Cecil}, C.~L. {Chang}, P.~{Chaubal}, G.~{Chen}, H.~M. {Cho},
  T.~L. {Chou}, J.~F. {Cliche}, T.~M. {Crawford}, A.~{Cukierman}, C.~{Daley},
  T.~{de Haan}, E.~V. {Denison}, K.~{Dibert}, J.~{Ding}, M.~A. {Dobbs},
  W.~{Everett}, C.~{Feng}, K.~R. {Ferguson}, A.~{Foster}, J.~{Fu}, S.~{Galli},
  A.~E. {Gambrel}, R.~W. {Gardner}, N.~{Goeckner-Wald}, R.~{Gualtieri},
  S.~{Guns}, N.~{Gupta}, R.~{Guyser}, N.~W. {Halverson}, A.~H.
  {Harke-Hosemann}, N.~L. {Harrington}, J.~W. {Henning}, G.~C. {Hilton},
  E.~{Hivon}, G.~P. {Holder}, W.~L. {Holzapfel}, J.~C. {Hood}, D.~{Howe},
  N.~{Huang}, K.~D. {Irwin}, O.~B. {Jeong}, M.~{Jonas}, A.~{Jones}, T.~S.
  {Khaire}, L.~{Knox}, A.~M. {Kofman}, M.~{Korman}, D.~L. {Kubik},
  S.~{Kuhlmann}, C.~L. {Kuo}, A.~T. {Lee}, E.~M. {Leitch}, A.~E. {Lowitz},
  C.~{Lu}, S.~S. {Meyer}, D.~{Michalik}, M.~{Millea}, J.~{Montgomery},
  A.~{Nadolski}, T.~{Natoli}, H.~{Nguyen}, G.~I. {Noble}, V.~{Novosad},
  Y.~{Omori}, S.~{Padin}, Z.~{Pan}, P.~{Paschos}, J.~{Pearson}, C.~M. {Posada},
  K.~{Prabhu}, W.~{Quan}, S.~{Raghunathan}, A.~{Rahlin}, C.~L. {Reichardt},
  D.~{Riebel}, B.~{Riedel}, M.~{Rouble}, J.~E. {Ruhl}, J.~T. {Sayre},
  E.~{Schiappucci}, E.~{Shirokoff}, G.~{Smecher}, J.~A. {Sobrin}, A.~A.
  {Stark}, J.~{Stephen}, K.~T. {Story}, A.~{Suzuki}, K.~L. {Thompson},
  B.~{Thorne}, C.~{Tucker}, C.~{Umilta}, L.~R. {Vale}, K.~{Vanderlinde}, J.~D.
  {Vieira}, G.~{Wang}, N.~{Whitehorn}, W.~L.~K. {Wu}, V.~{Yefremenko}, K.~W.
  {Yoon}, and M.~R. {Young}, ``{Measurements of the E-Mode Polarization and
  Temperature-E-Mode Correlation of the CMB from SPT-3G 2018 Data},'' {\em
  arXiv e-prints}, p.~arXiv:2101.01684, Jan. 2021.

\bibitem{Cooke:2017cwo}
R.~J. Cooke, M.~Pettini, and C.~C. Steidel, ``{One Percent Determination of the
  Primordial Deuterium Abundance},'' {\em Astrophys. J.}, vol.~855, no.~2,
  p.~102, 2018.

\bibitem{Aver:2015iza}
E.~Aver, K.~A. Olive, and E.~D. Skillman, ``{The effects of He I
  \ensuremath{\lambda}10830 on helium abundance determinations},'' {\em JCAP},
  vol.~07, p.~011, 2015.

\bibitem{Agathe:2019vsu}
V.~de~Sainte~Agathe {\em et~al.}, ``{Baryon acoustic oscillations at z = 2.34
  from the correlations of Ly$\alpha$ absorption in eBOSS DR14},'' {\em Astron.
  Astrophys.}, vol.~629, p.~A85, 2019.

\bibitem{Vagnozzi:2020dfn}
S.~Vagnozzi, A.~Loeb, and M.~Moresco, ``{Eppur \`e piatto? The Cosmic
  Chronometers Take on Spatial Curvature and Cosmic Concordance},'' {\em
  Astrophys. J.}, vol.~908, no.~1, p.~84, 2021.

\bibitem{Poulin:2021bjr}
V.~Poulin, T.~L. Smith, and A.~Bartlett, ``{Dark Energy at early times and ACT:
  a larger Hubble constant without late-time priors},'' 9 2021.

\bibitem{Beutler:2012px}
F.~Beutler, C.~Blake, M.~Colless, D.~H. Jones, L.~Staveley-Smith, G.~B. Poole,
  L.~Campbell, Q.~Parker, W.~Saunders, and F.~Watson, ``{The 6dF Galaxy Survey:
  $z \approx 0$ measurement of the growth rate and $\sigma_8$},'' {\em Mon.
  Not. Roy. Astron. Soc.}, vol.~423, pp.~3430--3444, 2012.

\bibitem{Howlett:2014opa}
C.~Howlett, A.~Ross, L.~Samushia, W.~Percival, and M.~Manera, ``{The clustering
  of the SDSS main galaxy sample \textendash{} II. Mock galaxy catalogues and a
  measurement of the growth of structure from redshift space distortions at $z
  = 0.15$},'' {\em Mon. Not. Roy. Astron. Soc.}, vol.~449, no.~1, pp.~848--866,
  2015.

\bibitem{Zarrouk:2018vwy}
P.~Zarrouk {\em et~al.}, ``{The clustering of the SDSS-IV extended Baryon
  Oscillation Spectroscopic Survey DR14 quasar sample: measurement of the
  growth rate of structure from the anisotropic correlation function between
  redshift 0.8 and 2.2},'' {\em Mon. Not. Roy. Astron. Soc.}, vol.~477, no.~2,
  pp.~1639--1663, 2018.

\bibitem{Ivanov:2020ril}
M.~M. Ivanov, E.~McDonough, J.~C. Hill, M.~Simonovi\'c, M.~W. Toomey,
  S.~Alexander, and M.~Zaldarriaga, ``{Constraining Early Dark Energy with
  Large-Scale Structure},'' {\em Phys. Rev. D}, vol.~102, no.~10, p.~103502,
  2020.

\bibitem{DAmico:2020ods}
G.~D'Amico, L.~Senatore, P.~Zhang, and H.~Zheng, ``{The Hubble Tension in Light
  of the Full-Shape Analysis of Large-Scale Structure Data},'' {\em JCAP},
  vol.~05, p.~072, 2021.

\bibitem{Niedermann:2020qbw}
F.~Niedermann and M.~S. Sloth, ``{New Early Dark Energy is compatible with
  current LSS data},'' {\em Phys. Rev. D}, vol.~103, no.~10, p.~103537, 2021.

\bibitem{Smith:2020rxx}
T.~L. Smith, V.~Poulin, J.~L. Bernal, K.~K. Boddy, M.~Kamionkowski, and
  R.~Murgia, ``{Early dark energy is not excluded by current large-scale
  structure data},'' {\em Phys. Rev. D}, vol.~103, no.~12, p.~123542, 2021.

\bibitem{Heimersheim:2020aoc}
S.~Heimersheim, N.~Sch\"oneberg, D.~C. Hooper, and J.~Lesgourgues,
  ``{Cannibalism hinders growth: Cannibal Dark Matter and the $S_8$ tension},''
  {\em JCAP}, vol.~12, p.~016, 2020.

\bibitem{Heymans:2020gsg}
C.~Heymans {\em et~al.}, ``{KiDS-1000 Cosmology: Multi-probe weak gravitational
  lensing and spectroscopic galaxy clustering constraints},'' {\em Astron.
  Astrophys.}, vol.~646, p.~A140, 2021.

\bibitem{Vagnozzi:2021gjh}
S.~{Vagnozzi}, ``{Consistency tests of $\Lambda$CDM from the early ISW effect:
  implications for early-time new physics and the Hubble tension},'' {\em arXiv
  e-prints}, p.~arXiv:2105.10425, May 2021.

\bibitem{DiValentino:2019ffd}
E.~Di~Valentino, A.~Melchiorri, O.~Mena, and S.~Vagnozzi, ``{Interacting dark
  energy in the early 2020s: A promising solution to the $H_0$ and cosmic shear
  tensions},'' {\em Phys. Dark Univ.}, vol.~30, p.~100666, 2020.

\bibitem{Lucca:2021dxo}
M.~{Lucca}, ``{Dark energy-dark matter interactions as a solution to the $S_8$
  tension},'' {\em arXiv e-prints}, p.~arXiv:2105.09249, May 2021.

\bibitem{Hill:2021yec}
J.~C. Hill {\em et~al.}, ``{The Atacama Cosmology Telescope: Constraints on
  Pre-Recombination Early Dark Energy},'' 9 2021.

\bibitem{Bernal:2021yli}
J.~L. Bernal, L.~Verde, R.~Jimenez, M.~Kamionkowski, D.~Valcin, and B.~D.
  Wandelt, ``{The trouble beyond $H_0$ and the new cosmic triangles},'' {\em
  Phys. Rev. D}, vol.~103, no.~10, p.~103533, 2021.

\bibitem{Boylan-Kolchin:2021fvy}
M.~Boylan-Kolchin and D.~R. Weisz, ``{Uncertain times: the
  redshift\textendash{}time relation from cosmology and stars},'' {\em Mon.
  Not. Roy. Astron. Soc.}, vol.~505, no.~2, pp.~2764--2783, 2021.

\bibitem{Vagnozzi:2021tjv}
S.~Vagnozzi, F.~Pacucci, and A.~Loeb, ``{Implications for the Hubble tension
  from the ages of the oldest astrophysical objects},'' 5 2021.

\bibitem{Valcin:2020vav}
D.~Valcin, J.~L. Bernal, R.~Jimenez, L.~Verde, and B.~D. Wandelt, ``{Inferring
  the Age of the Universe with Globular Clusters},'' {\em JCAP}, vol.~12,
  p.~002, 2020.

\bibitem{Valcin:2021jcg}
D.~Valcin, R.~Jimenez, L.~Verde, J.~L. Bernal, and B.~D. Wandelt, ``{The age of
  the Universe with globular clusters: reducing systematic uncertainties},''
  {\em JCAP}, vol.~08, p.~017, 2021.

\bibitem{Takahashi:2012em}
R.~Takahashi, M.~Sato, T.~Nishimichi, A.~Taruya, and M.~Oguri, ``{Revising the
  Halofit Model for the Nonlinear Matter Power Spectrum},'' {\em Astrophys.
  J.}, vol.~761, p.~152, 2012.

\bibitem{Ali-Haimoud:2012fzp}
Y.~Ali-Haimoud and S.~Bird, ``{An efficient implementation of massive neutrinos
  in non-linear structure formation simulations},'' {\em Mon. Not. Roy. Astron.
  Soc.}, vol.~428, pp.~3375--3389, 2012.

\bibitem{James:1975dr}
F.~James and M.~Roos, ``{Minuit: A System for Function Minimization and
  Analysis of the Parameter Errors and Correlations},'' {\em Comput. Phys.
  Commun.}, vol.~10, pp.~343--367, 1975.

\bibitem{iminuit}
H.~Dembinski, P.~Ongmongkolkul, C.~Deil, D.~M. Hurtado, H.~Schreiner,
  M.~Feickert, Andrew, C.~Burr, J.~Watson, F.~Rost, A.~Pearce, L.~Geiger, B.~M.
  Wiedemann, Gonzalo, J.~Drotleff, J.~Eschle, L.~Neste, M.~E. Gorelli, M.~Baak,
  and O.~Zapata, ``scikit-hep/iminuit:,'' July 2021.

\bibitem{Lesgourgues:2011re}
J.~Lesgourgues, ``{The Cosmic Linear Anisotropy Solving System (CLASS) I:
  Overview},'' 4 2011.

\bibitem{Blas:2011rf}
D.~Blas, J.~Lesgourgues, and T.~Tram, ``{The Cosmic Linear Anisotropy Solving
  System (CLASS) II: Approximation schemes},'' {\em JCAP}, vol.~07, p.~034,
  2011.

\bibitem{Audren:2012wb}
B.~Audren, J.~Lesgourgues, K.~Benabed, and S.~Prunet, ``{Conservative
  Constraints on Early Cosmology: an illustration of the Monte Python
  cosmological parameter inference code},'' {\em JCAP}, vol.~02, p.~001, 2013.

\bibitem{Brinckmann:2018cvx}
T.~Brinckmann and J.~Lesgourgues, ``{MontePython 3: boosted MCMC sampler and
  other features},'' {\em Phys. Dark Univ.}, vol.~24, p.~100260, 2019.

\end{thebibliography}

\begin{appendix}
\newpage
In the following we present the triangle plots for each of the tests described in the main text. Specifically, these include subjecting all models to the various data sets discussed in \cref{sec:resultsA,sec:resultsB,sec:resultsC,sec:resultsD}. In particular in \ref{app:AB} we present on the left panels the model contours from the \DA{} data set and on the right panels the contours from the \DB{} data set. Furthermore, in \ref{app:C} we consider the \DCnoP{} and \DCnoCMB{} data sets, while in \ref{app:D} we consider the \DD{} data set.


\section{Tests A and B: triangle plots\label{app:AB}}

\begin{figure}[H]
    \centering
    \includegraphics[height=2cm]{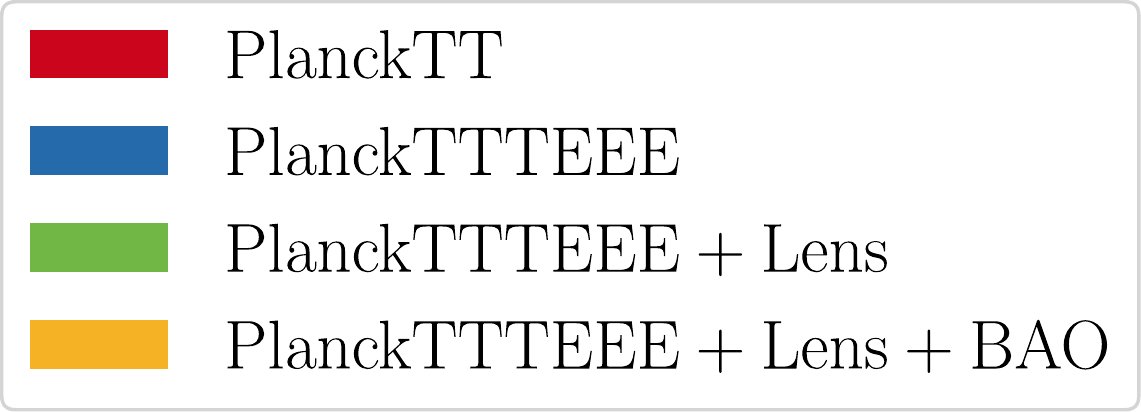}
    \hfil
    \includegraphics[height=1.5cm]{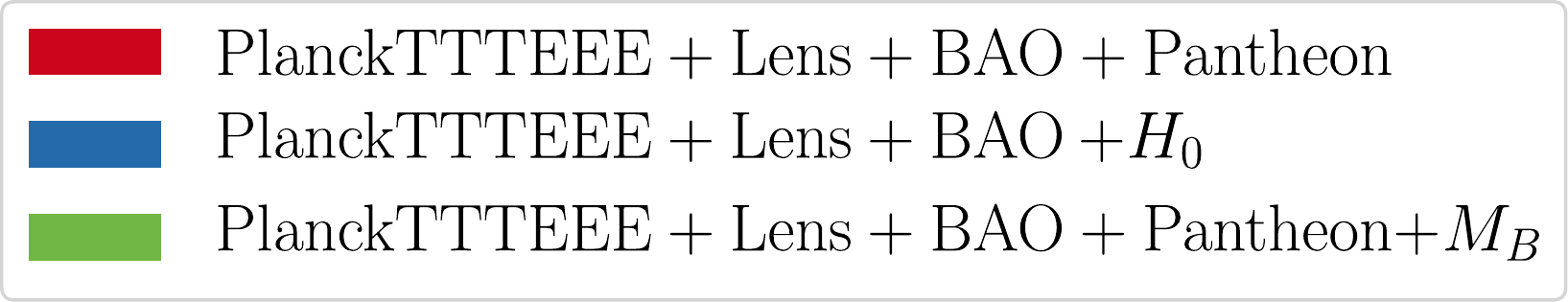}
\end{figure}
\subsection{\texorpdfstring{$\Lambda$CDM}{LCDM}}

\begin{figure}[H]
    \centering
    \includegraphics[height=3.6cm]{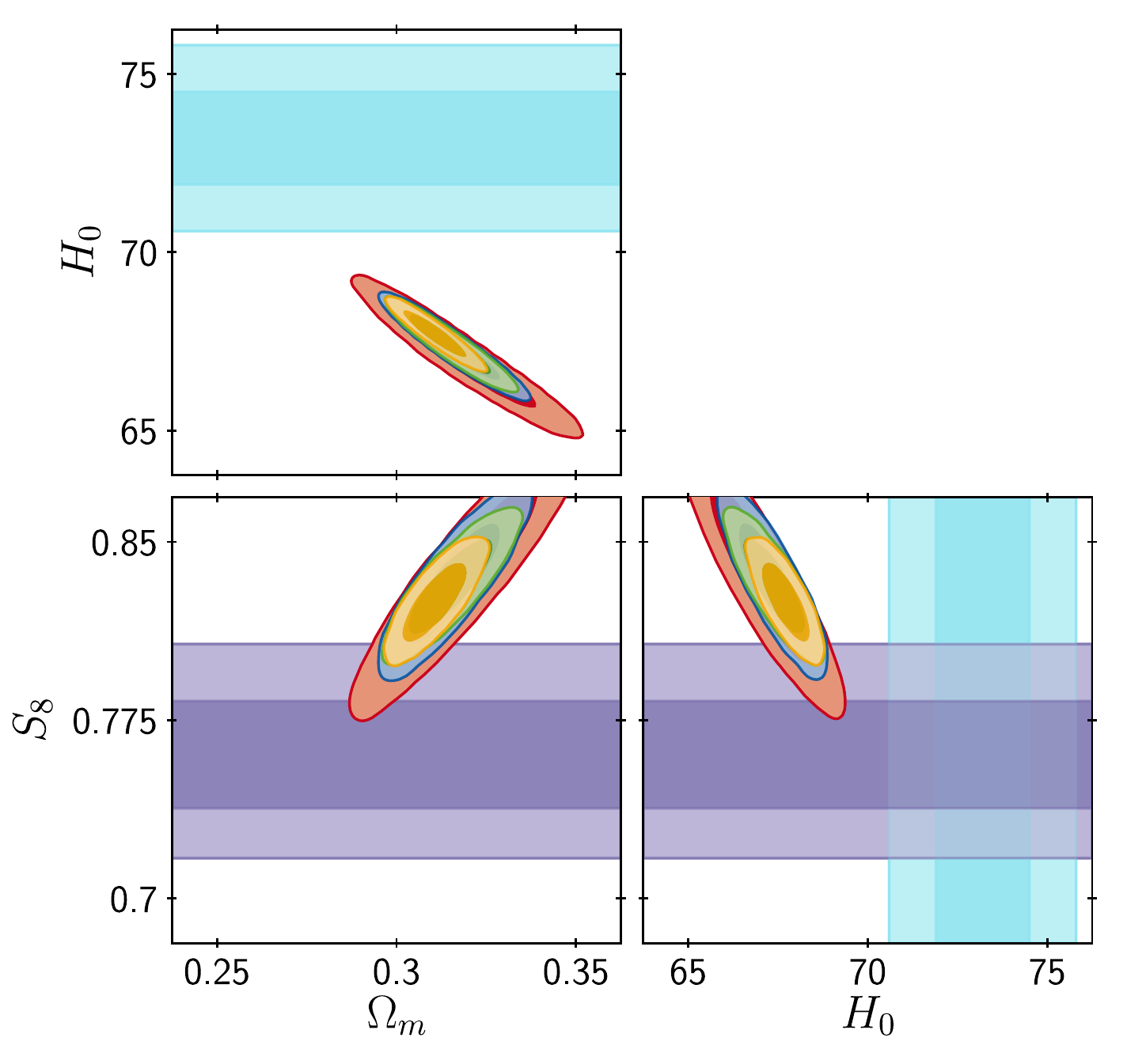}
    \hspace{3cm}
    \includegraphics[height=3.6cm]{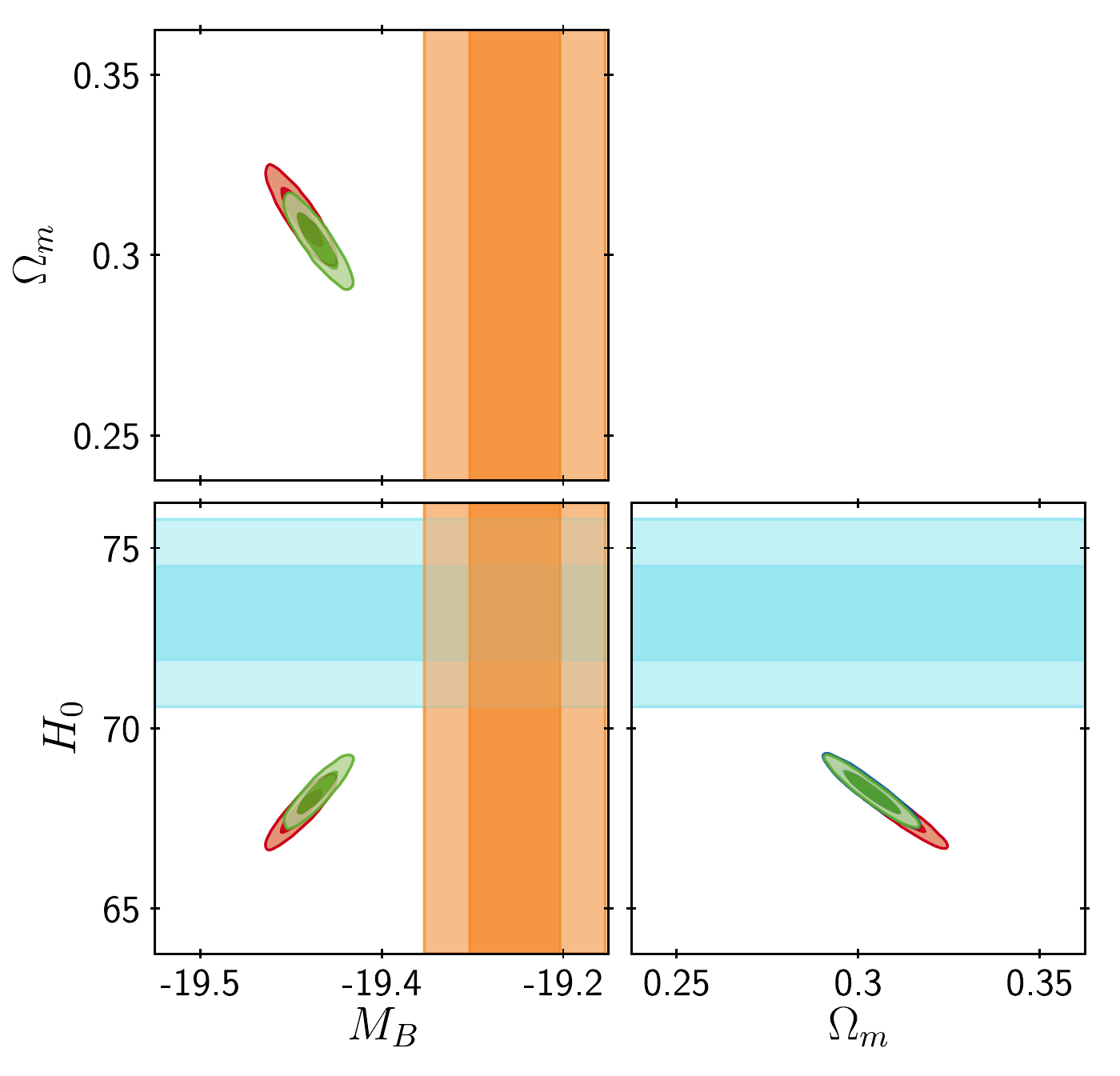}
    \caption{$\Lambda$CDM model: {\it (Left)} 2D contours on ($\Omega_{\rm m}$, $H_0$, $S_8$), for Planck + BAO data.
    {\it (Right)} 2D contours on ($\Omega_{\rm m}$, $M_{\rm B}$, $H_0$), for Planck +BAO + Pantheon and/or SH0ES (treated as a measurement of either $H_0$ or $M_B$).}
    \label{fig:lcdm_lens_bao_sh0es}
\end{figure}


\subsection{\Nefflong}

\begin{figure}[h]
    \centering
    \includegraphics[height=4.5cm]{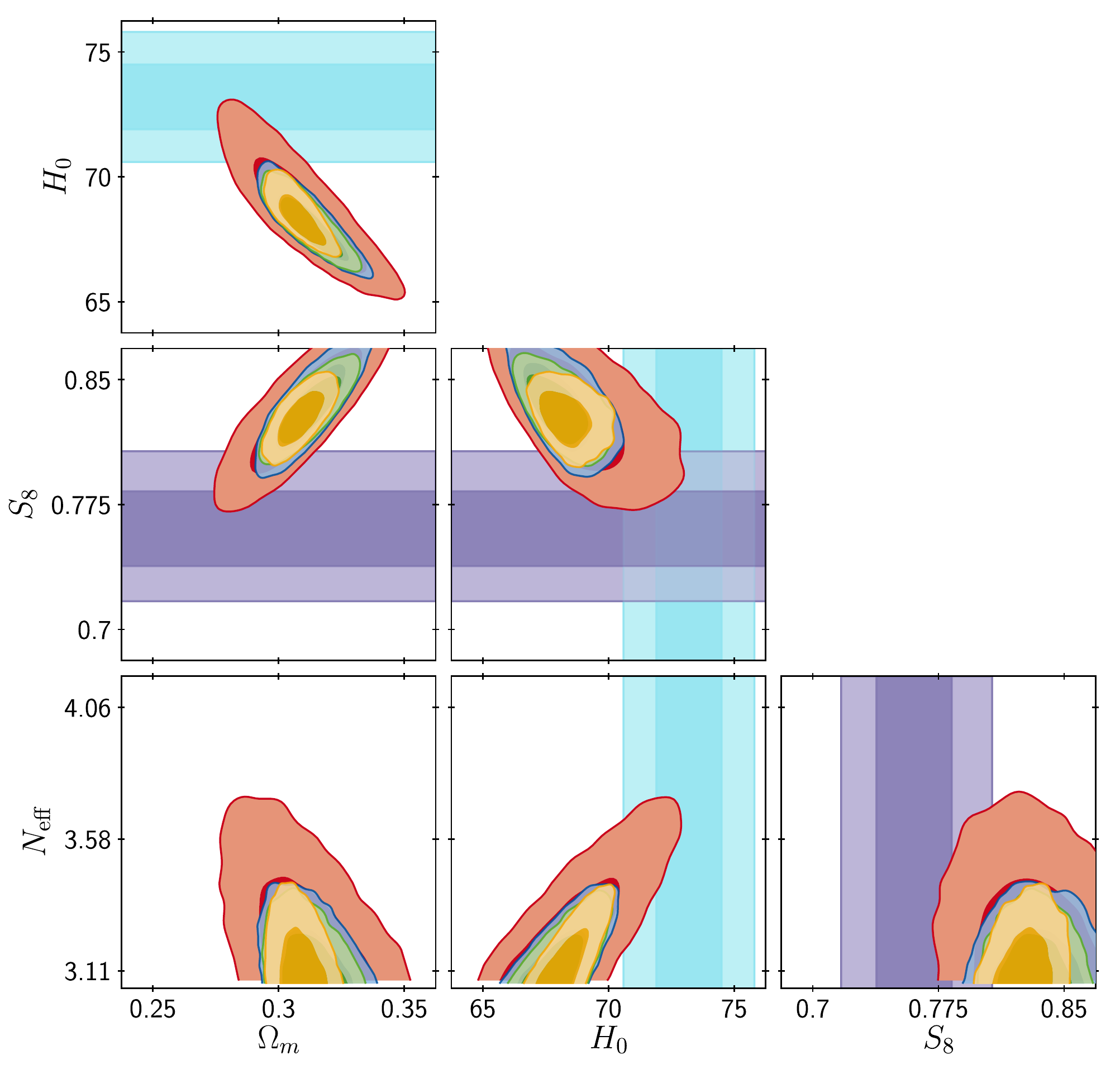}
    \hspace{1cm}
    \includegraphics[height=4.5cm]{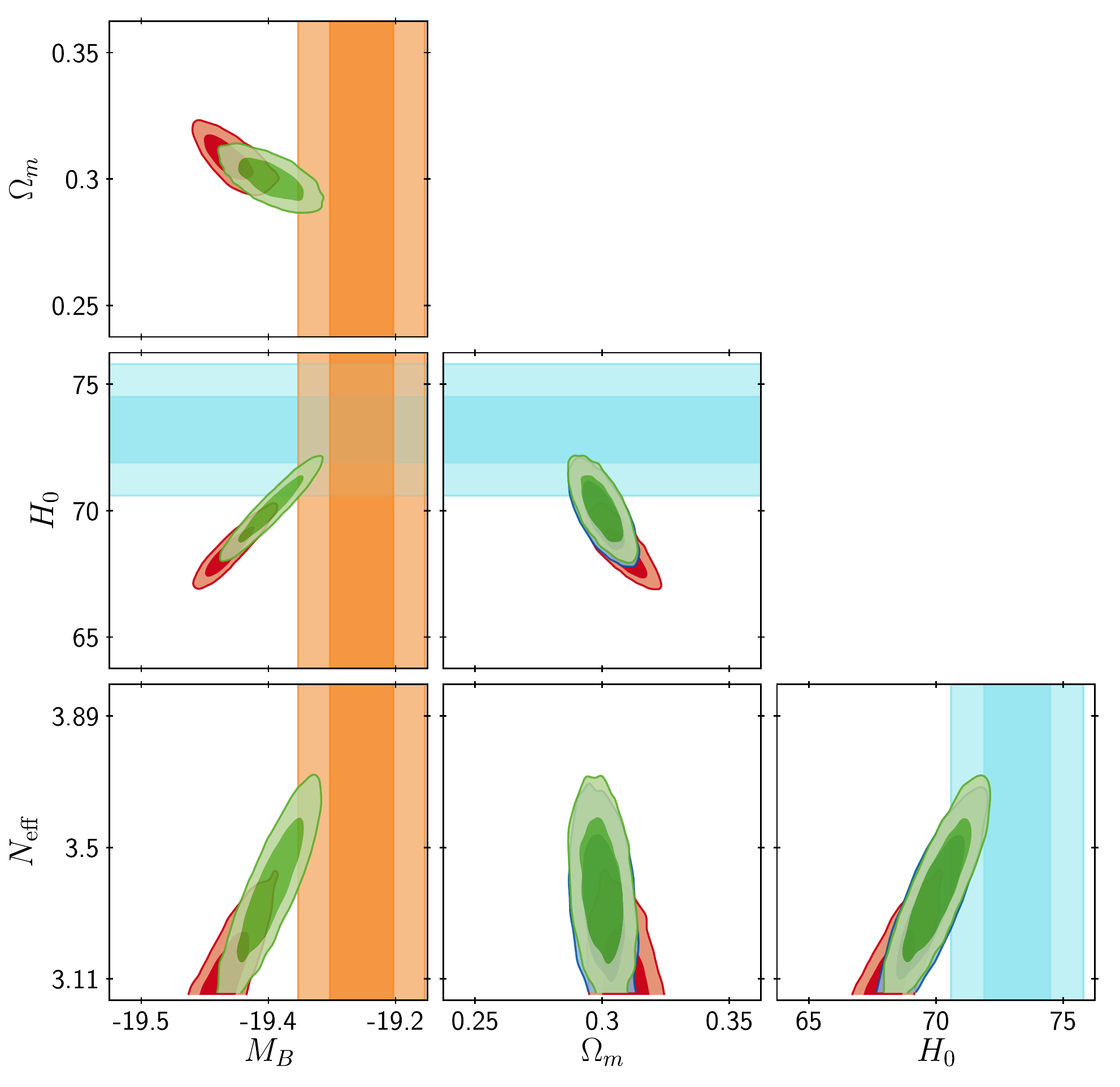}
    \caption{$\Lambda$CDM + $N_{\rm eff}$ model: {\it (Left)} 2D contours on ($\Omega_{\rm m}$, $H_0$, $S_8$, $N_{\rm eff}$), for Planck +BAO data. {\it (Right)} 2D contours on ($\Omega_{\rm m}$, $M_{\rm B}$, $H_0$, $N_{\rm eff}$), for Planck +BAO + Pantheon and/or SH0ES (treated as a measurement of either $H_0$ or $M_B$).}
    \label{fig:neff_lens_bao_sh0es}
\end{figure}

\FloatBarrier

\newpage 
\begin{figure}[H]
    \centering
    \includegraphics[height=2cm]{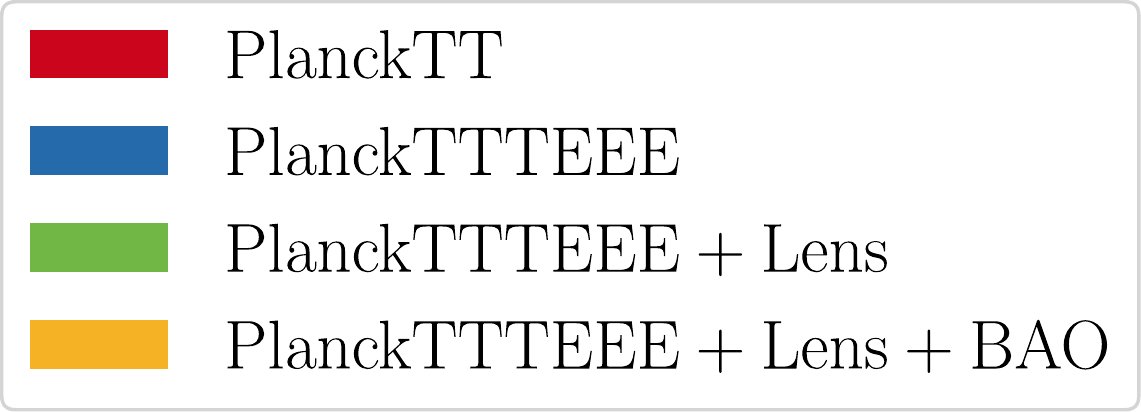}
    \hfil
    \includegraphics[height=1.5cm]{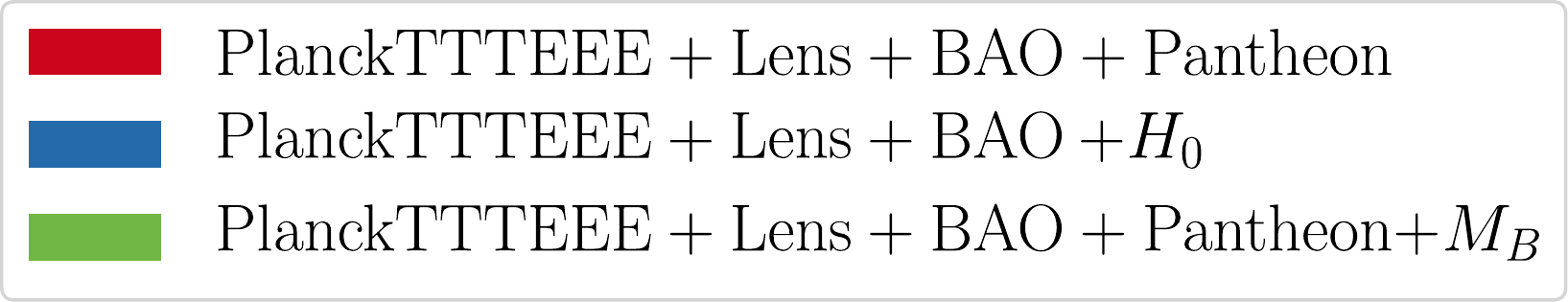}
\end{figure}

\subsection{\idrlong}

\begin{figure}[h]
    \centering
    \includegraphics[width=5.3cm]{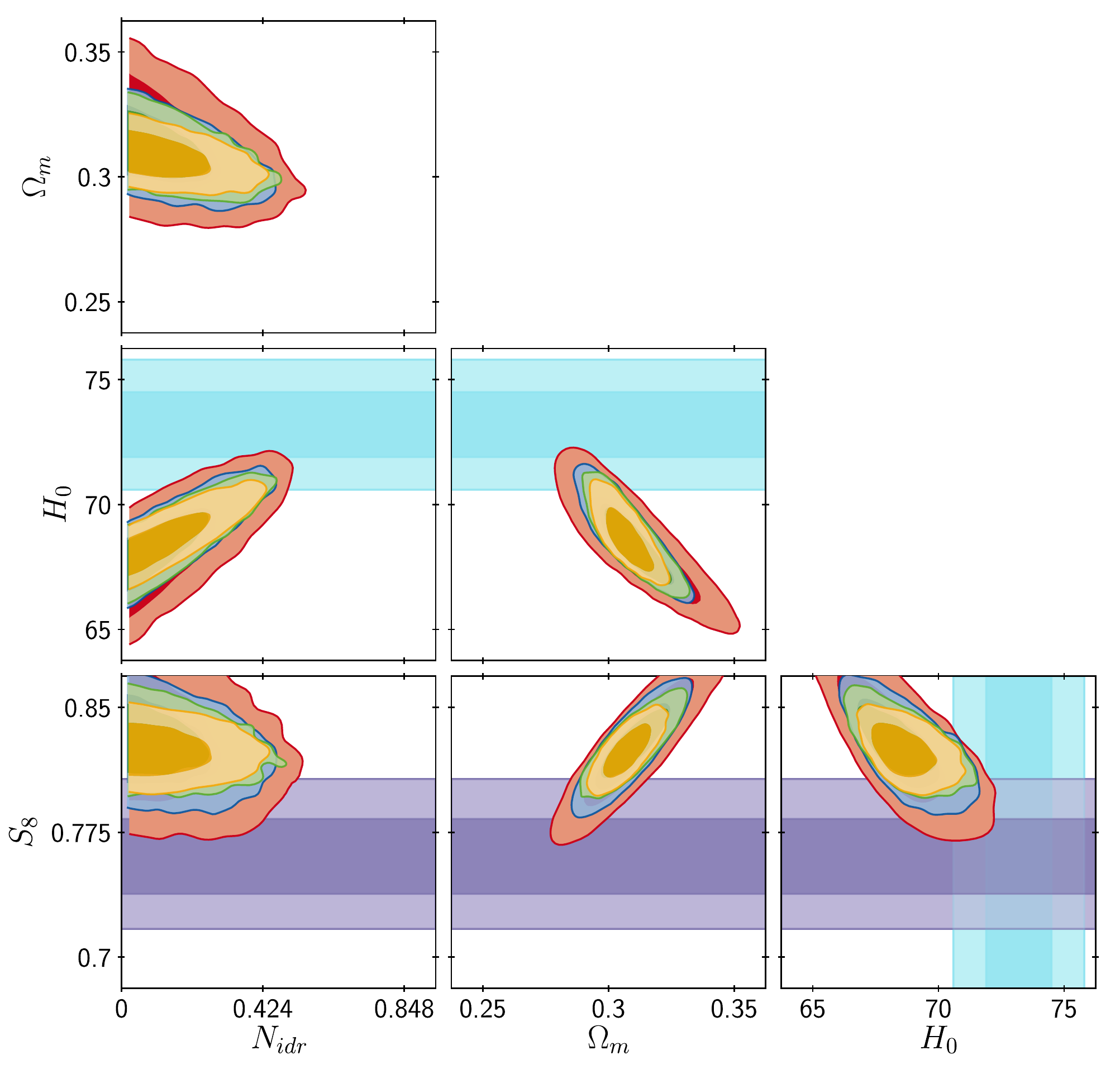}
    \hspace{1cm}
    \includegraphics[height=5.3cm]{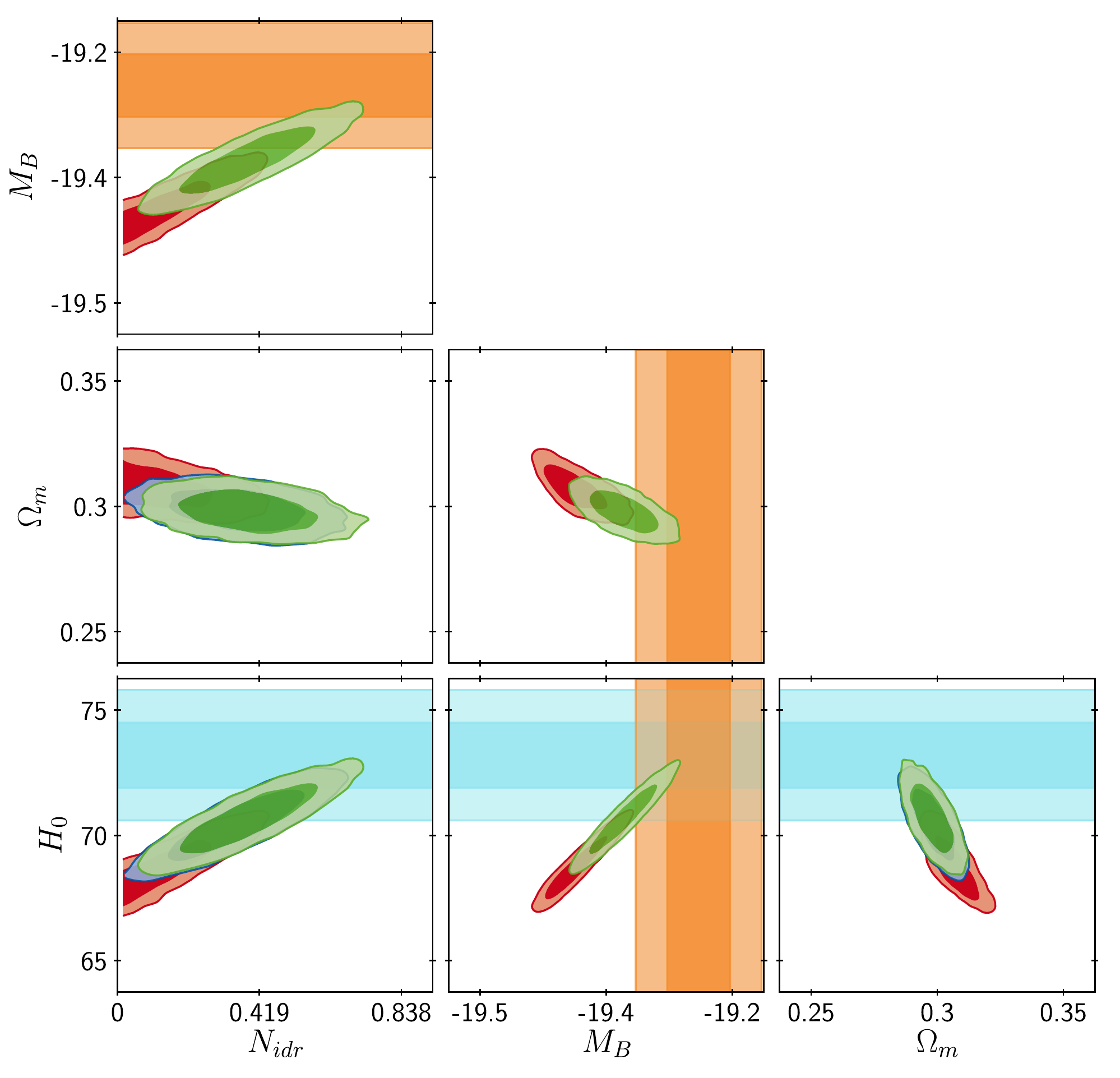}
    \caption{$\Lambda$CDM + $N_{\rm fluid}$ model: {\it (Left)} 2D contours on ($\Omega_{\rm m}$, $H_0$, $S_8$, $N_{\rm idr}$), for Planck +BAO data. {\it (Right)} 2D contours on ($\Omega_{\rm m}$, $M_{\rm B}$, $H_0$, $N_{\rm idr}$), for Planck +BAO + Pantheon and/or SH0ES (treated as a measurement of either $H_0$ or $M_B$).}
    \label{fig:sidr_lens_bao_sh0es}
\end{figure}

\FloatBarrier

\subsection{\Equilong}

\begin{figure}[h]
    \centering
    \includegraphics[height=6.5cm]{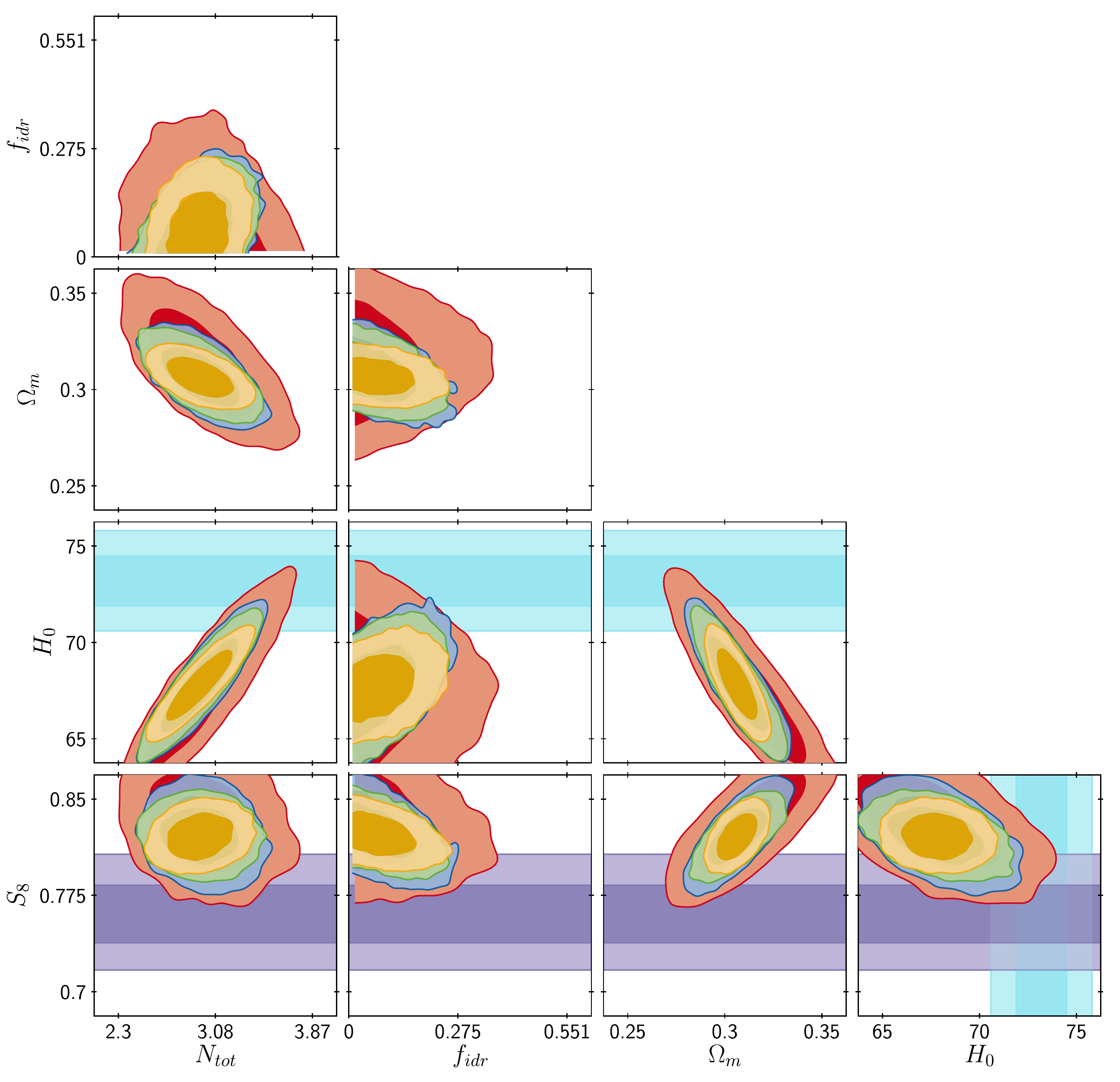}
    \includegraphics[height=6.5cm]{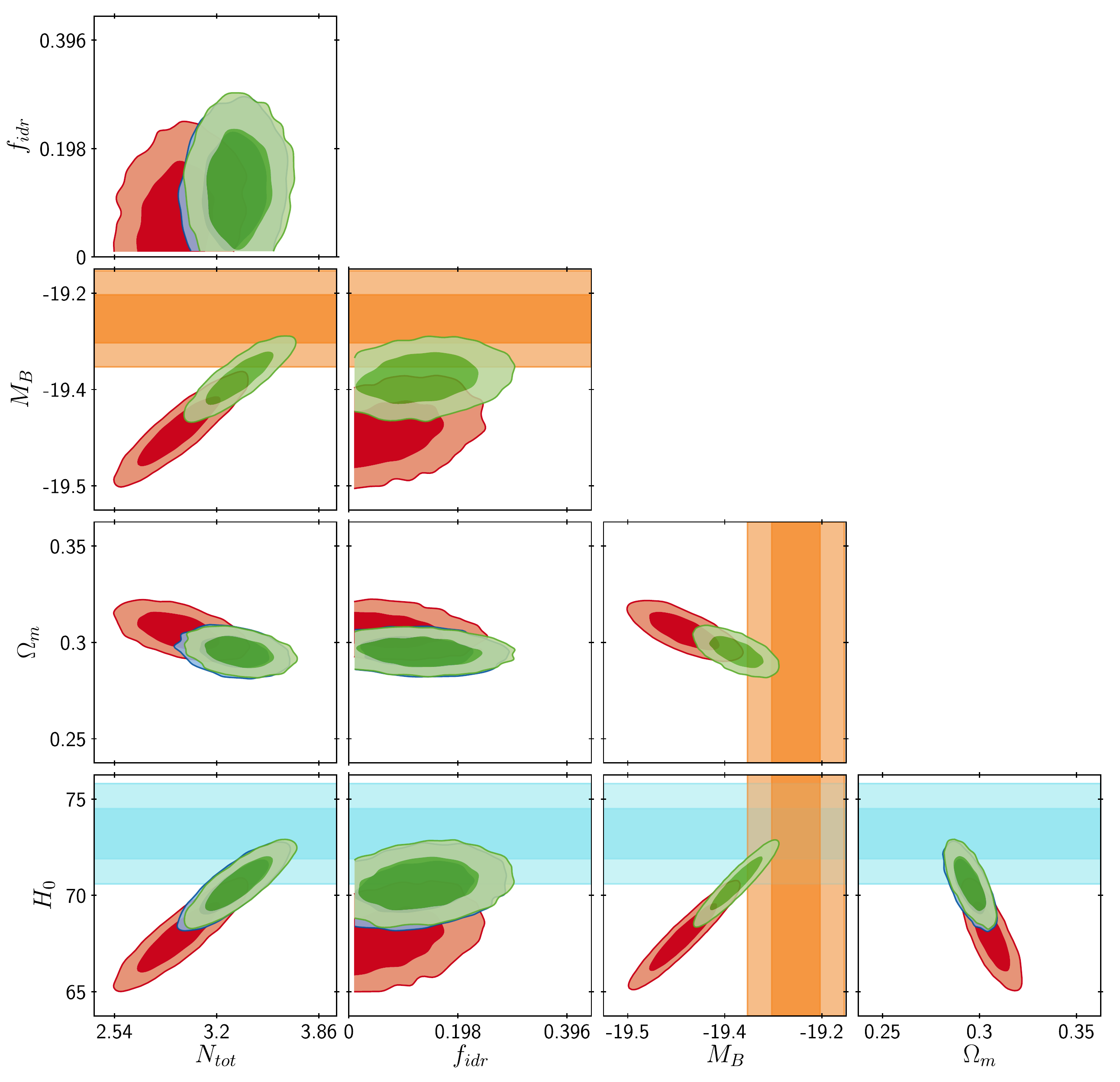}
    \caption{Equilibration model: {\it (Left)} 2D contours on ($\Omega_{\rm m}$, $H_0$, $S_8$, $N_{\rm tot}$, $f_{\rm idr}$), for Planck + BAO data. {\it (Right)} 2D contours on ($\Omega_{\rm m}$, $M_{\rm B}$, $H_0$, $N_{\rm tot}$, $f_{\rm idr}$), for Planck +BAO + Pantheon and/or SH0ES (treated as a measurement of either $H_0$ or $M_B$).}
    \label{fig:equilibriation_lens_bao_sh0es}
\end{figure}

\FloatBarrier

\newpage

\enlargethispage*{2cm}
\begin{figure}[H]
    \centering
    \includegraphics[height=1.5cm]{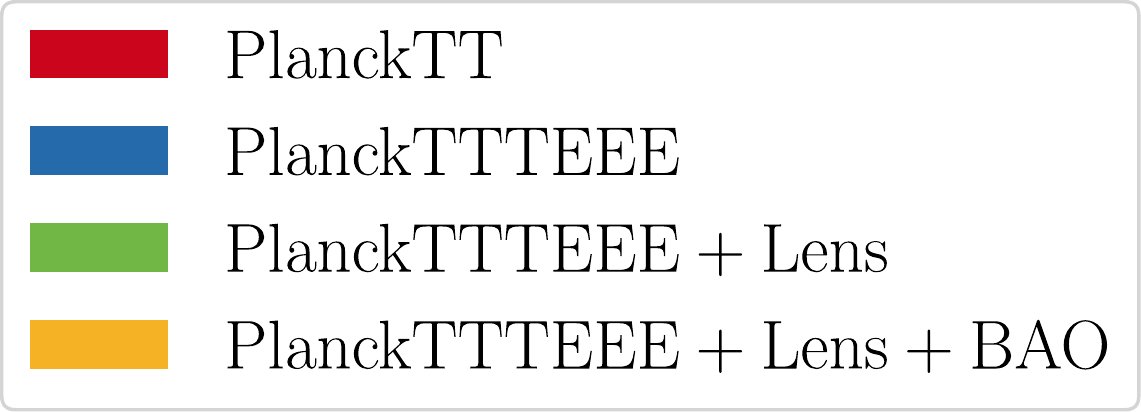}
    \hfil
    \includegraphics[height=1.125cm]{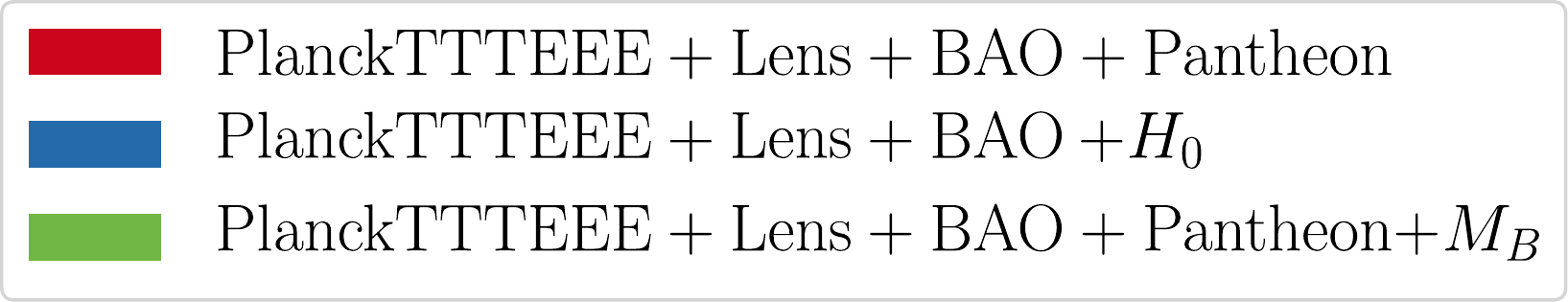}
\end{figure}

\subsection{\Schmaltzlong}

\begin{figure}[h]
    \centering
    \includegraphics[height=6.5cm]{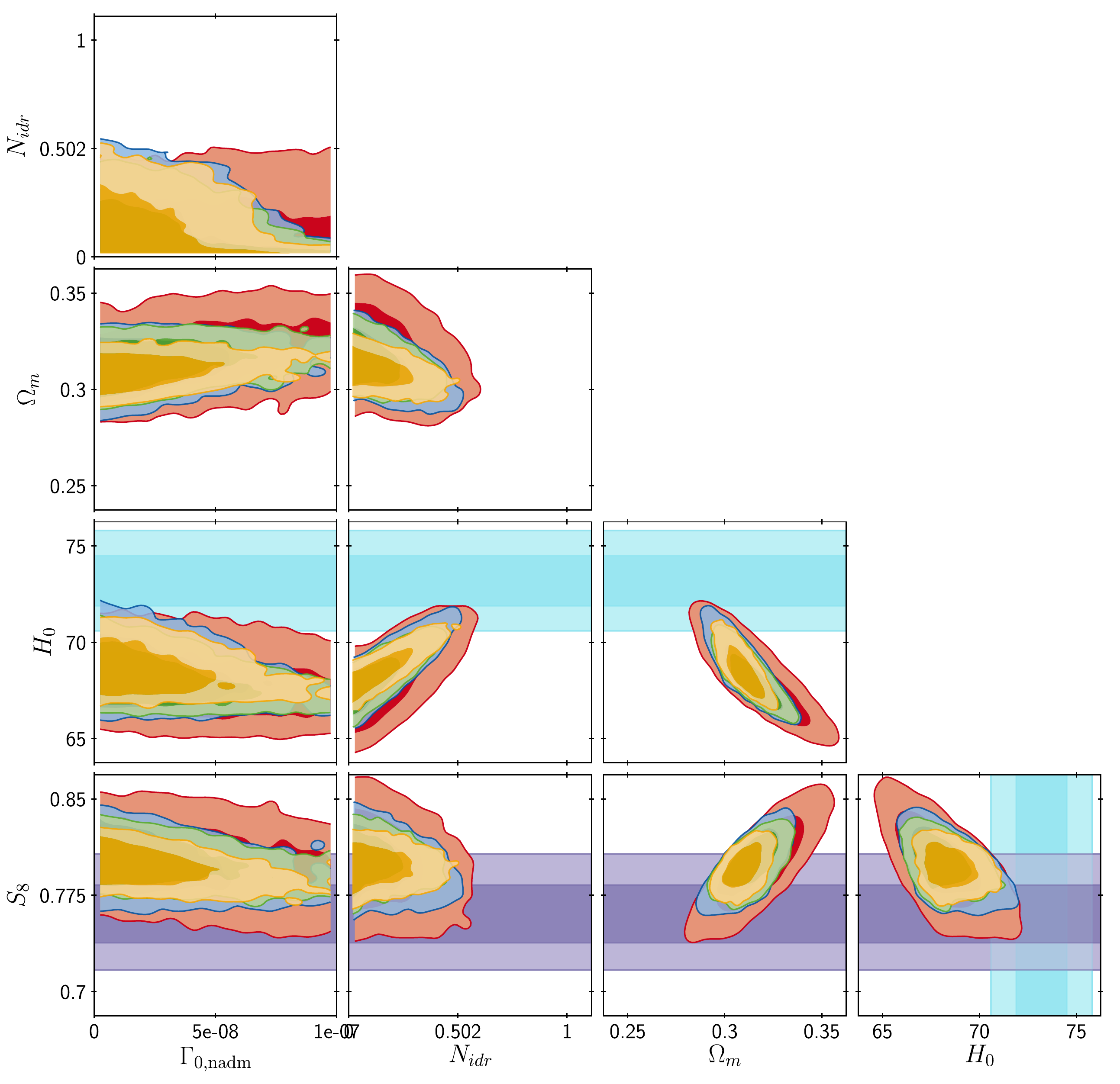}
    \includegraphics[height=6.5cm]{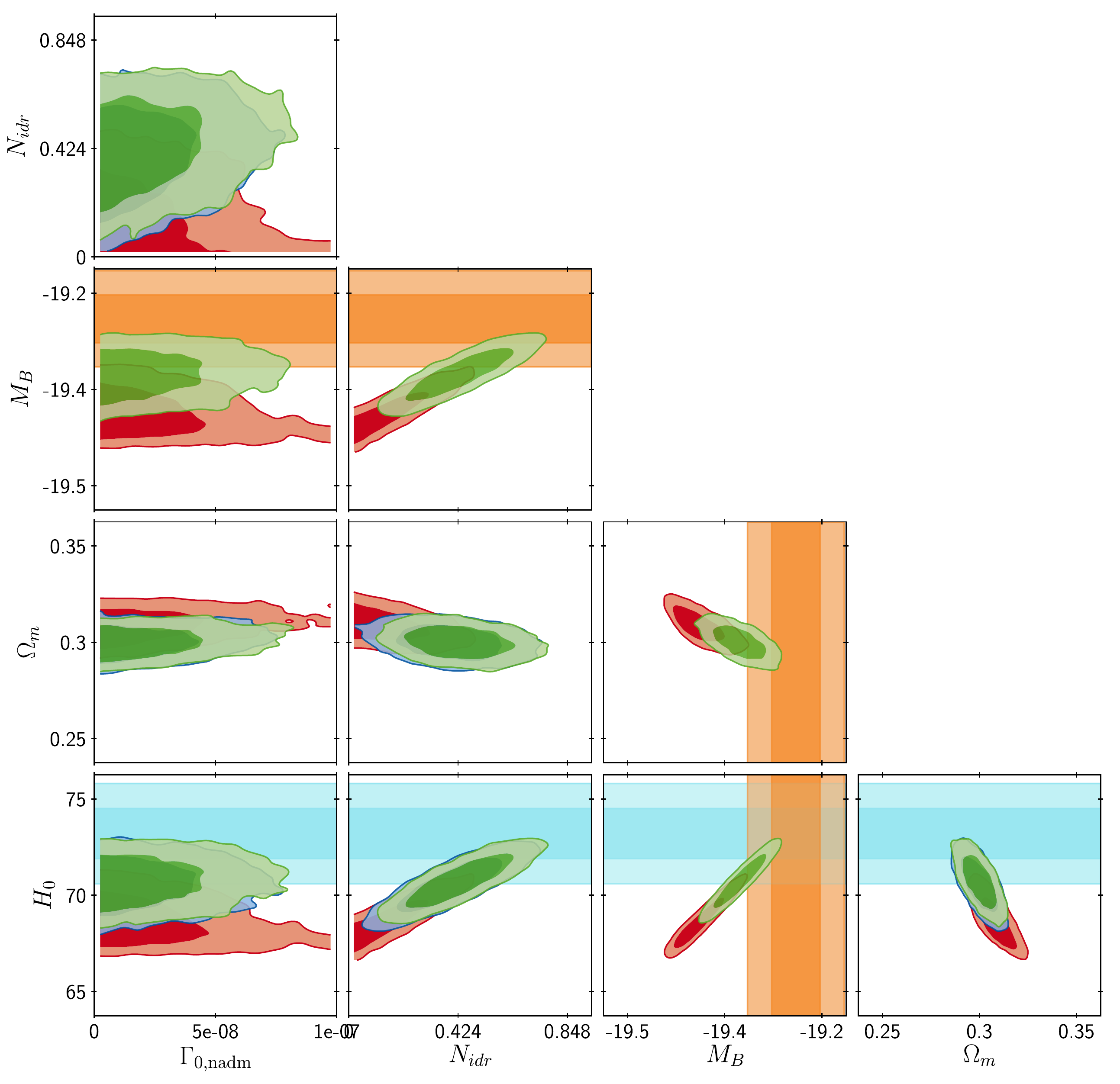}
    \caption{Interacting DM-DR: {\it (Left)} 2D contours on ($\Omega_{\rm m}$, $H_0$, $S_8$, $N_{\rm idr}$, $\Gamma_{0,{\rm nadm}}$), for Planck + BAO data. {\it (Right)} 2D contours on ($\Omega_{\rm m}$, $M_{\rm B}$, $H_0$, $N_{\rm idr}$, $\Gamma_{0,{\rm nadm}}$), for Planck +BAO + Pantheon and/or SH0ES (treated as a measurement of either $H_0$ or $M_B$).}
    \label{fig:schmaltz_lens_bao_sh0es}
\end{figure}


\subsection{\SInulong}
 
  \begin{figure}[h]
    \centering
    \includegraphics[height=7cm]{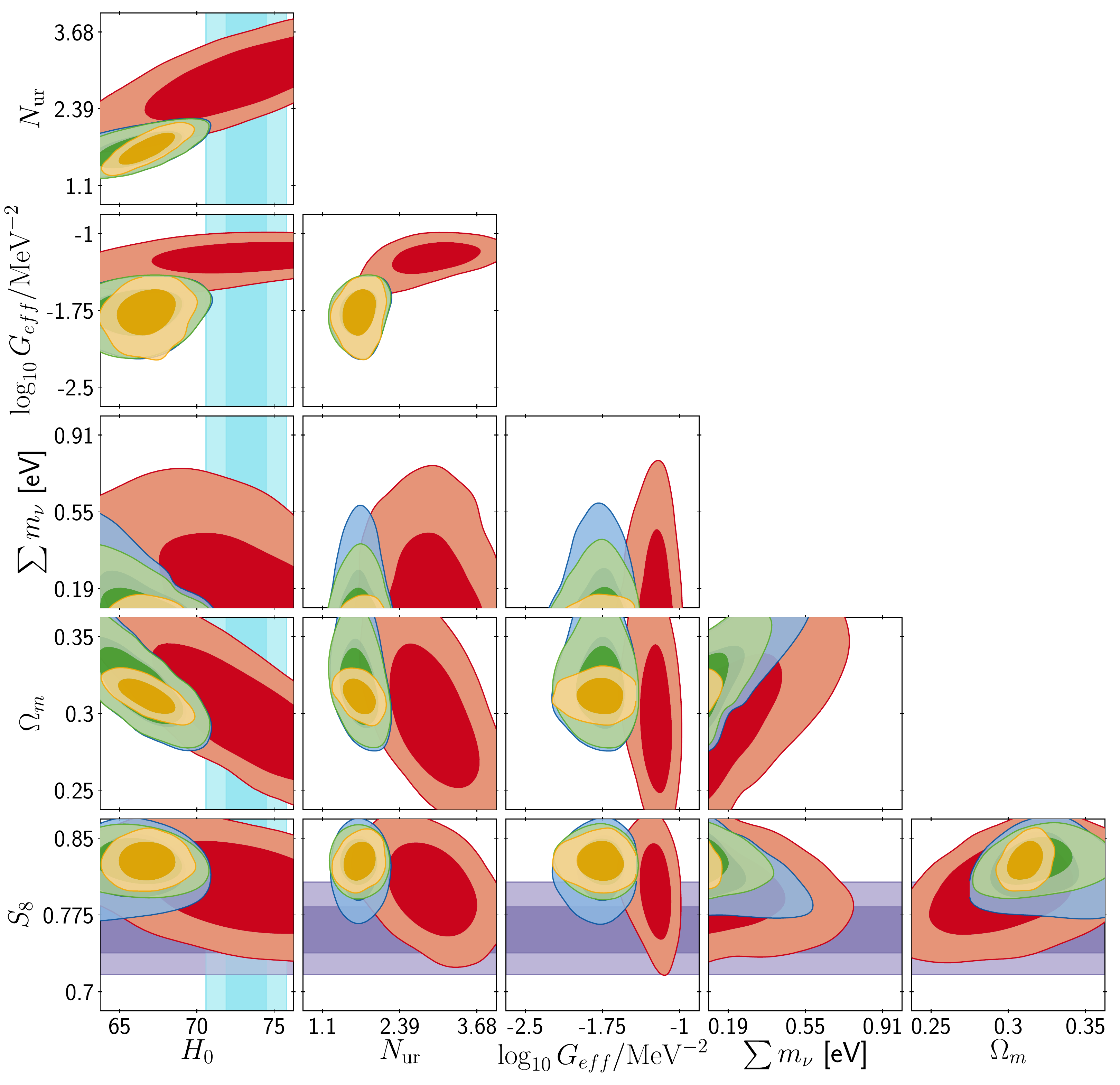}
    \includegraphics[height=7cm]{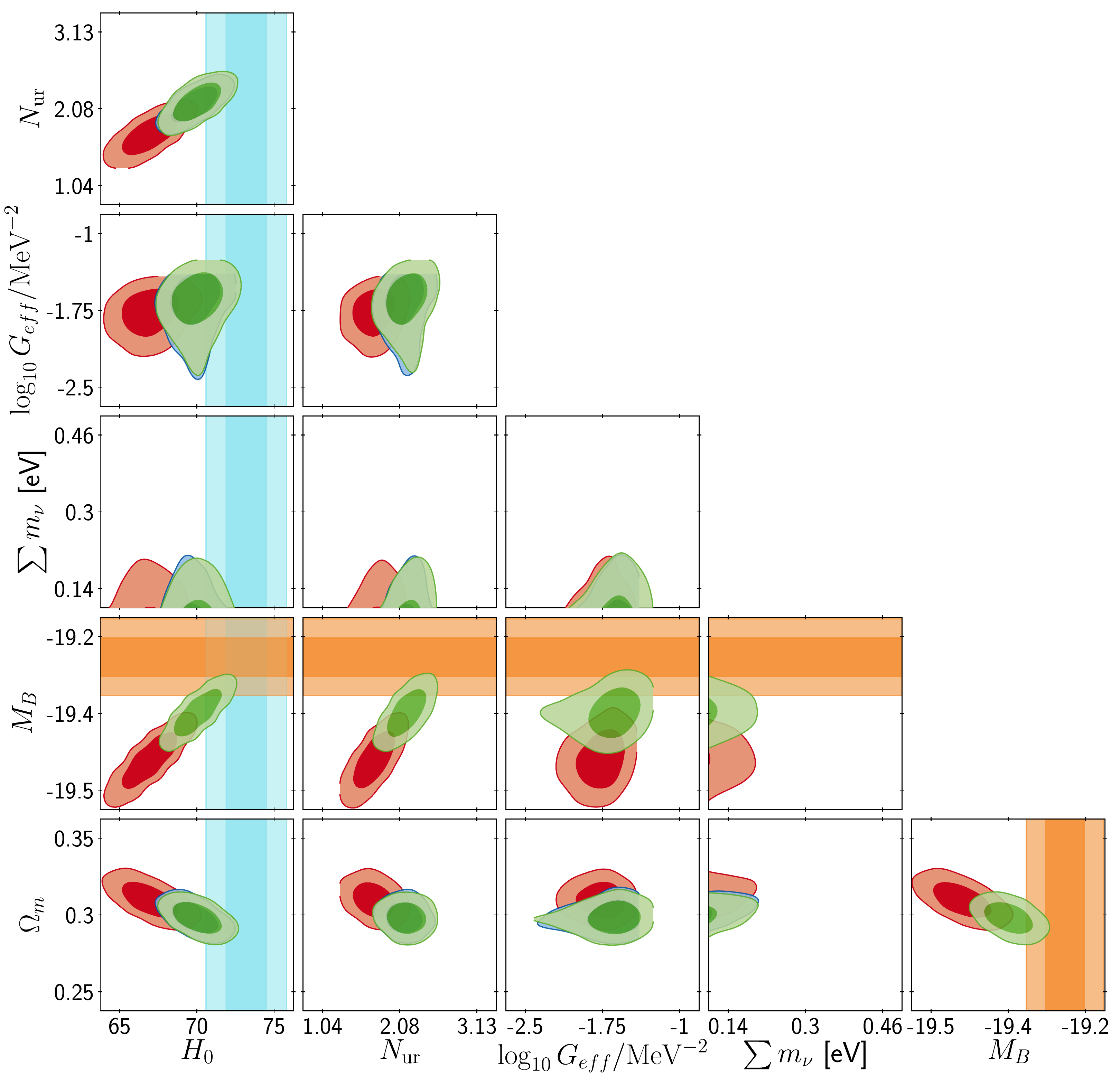}
    \caption{Self-interacting $\nu$s + $\Delta N_{\rm ur}$: {\it (Left)} 2D contours on ($\Omega_{\rm m}$, $H_0$, $S_8$, $N_{\rm ur}$, $\log_{10} G_{\rm eff}$, $m_\nu$), for Planck + BAO data. {\it (Right)} 2D contours on ($\Omega_{\rm m}$, $M_{\rm B}$, $H_0$, $N_{\rm ur}$, $\log_{10} G_{\rm eff}$, $m_\nu$), for Planck +BAO + Pantheon and/or SH0ES (treated as a measurement of either $H_0$ or $M_B$).
\vspace*{-1cm}}
    \label{fig:SI_sn}
\end{figure}

\FloatBarrier

\newpage

\enlargethispage*{1cm}
\begin{figure}[H]
    \centering
    \includegraphics[height=1.5cm]{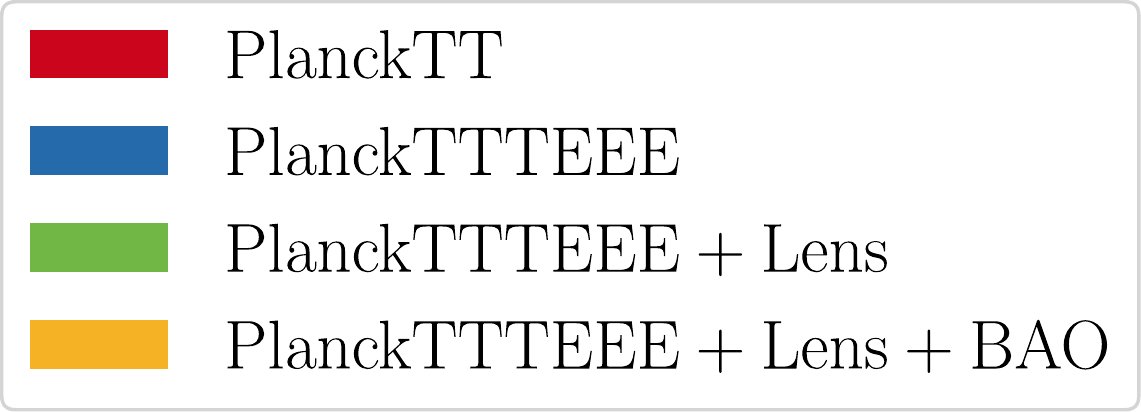}
    \hfil
    \includegraphics[height=1.125cm]{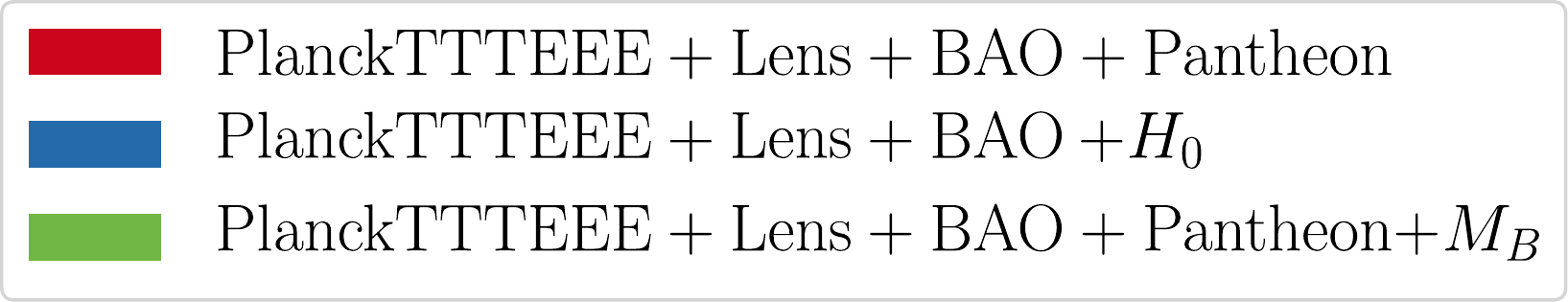}
\end{figure}

\subsection{\Majoronlong}
 
   \begin{figure}[h]
    \centering
    \includegraphics[height=7cm]{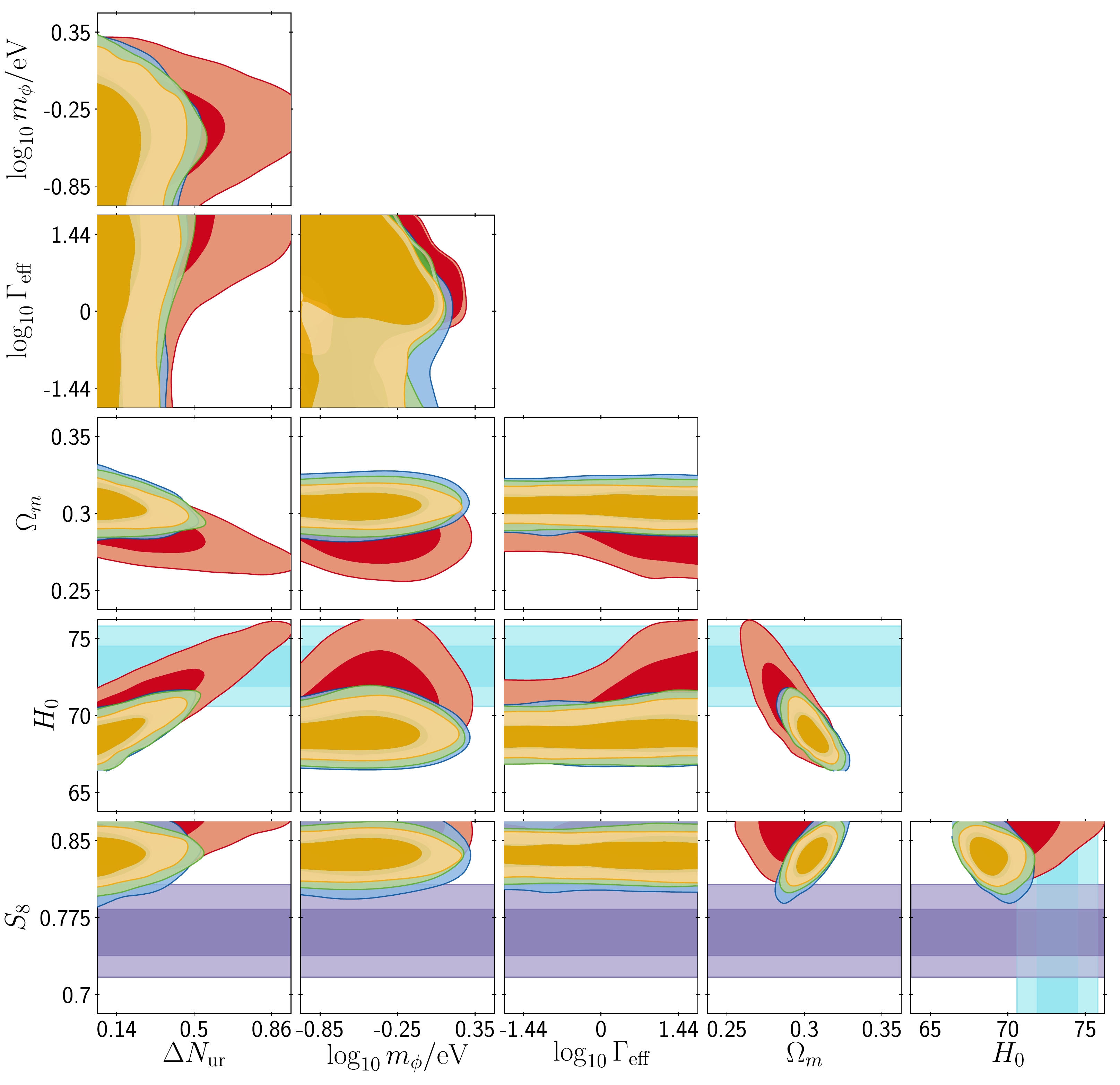}
    \includegraphics[height=7cm]{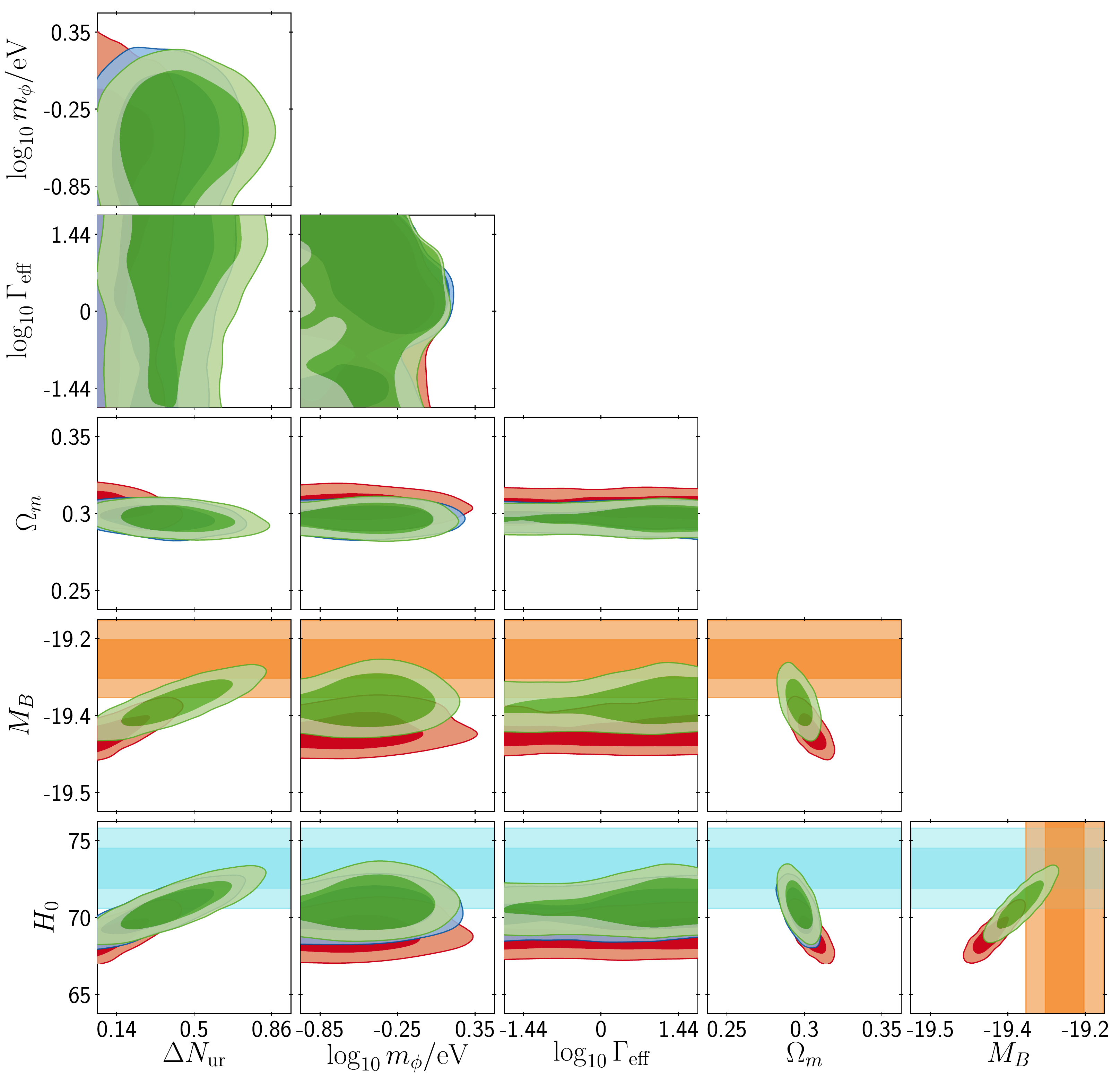}
    \caption{The majoron: {\it (Left)} 2D contours on ($\Omega_{\rm m}$, $H_0$, $S_8$, $\Delta N_{\rm ur}$, $\log_{10} m_\phi$, $\log_{10} g_{\rm eff}$), for Planck + BAO data. {\it (Right)} 2D contours on ($\Omega_{\rm m}$, $M_{\rm B}$, $H_0$, $\Delta N_{\rm ur}$, $\log_{10} m_\phi$, $\log_{10} \Gamma_{\rm eff}$), for Planck +BAO + Pantheon and/or SH0ES (treated as a measurement of either $H_0$ or $M_B$).}
    \label{fig:Maj_std}
\end{figure} 

\FloatBarrier

\subsection{\Blong}

\begin{figure*}[h]
	\centering
	\includegraphics[height=5.3cm]{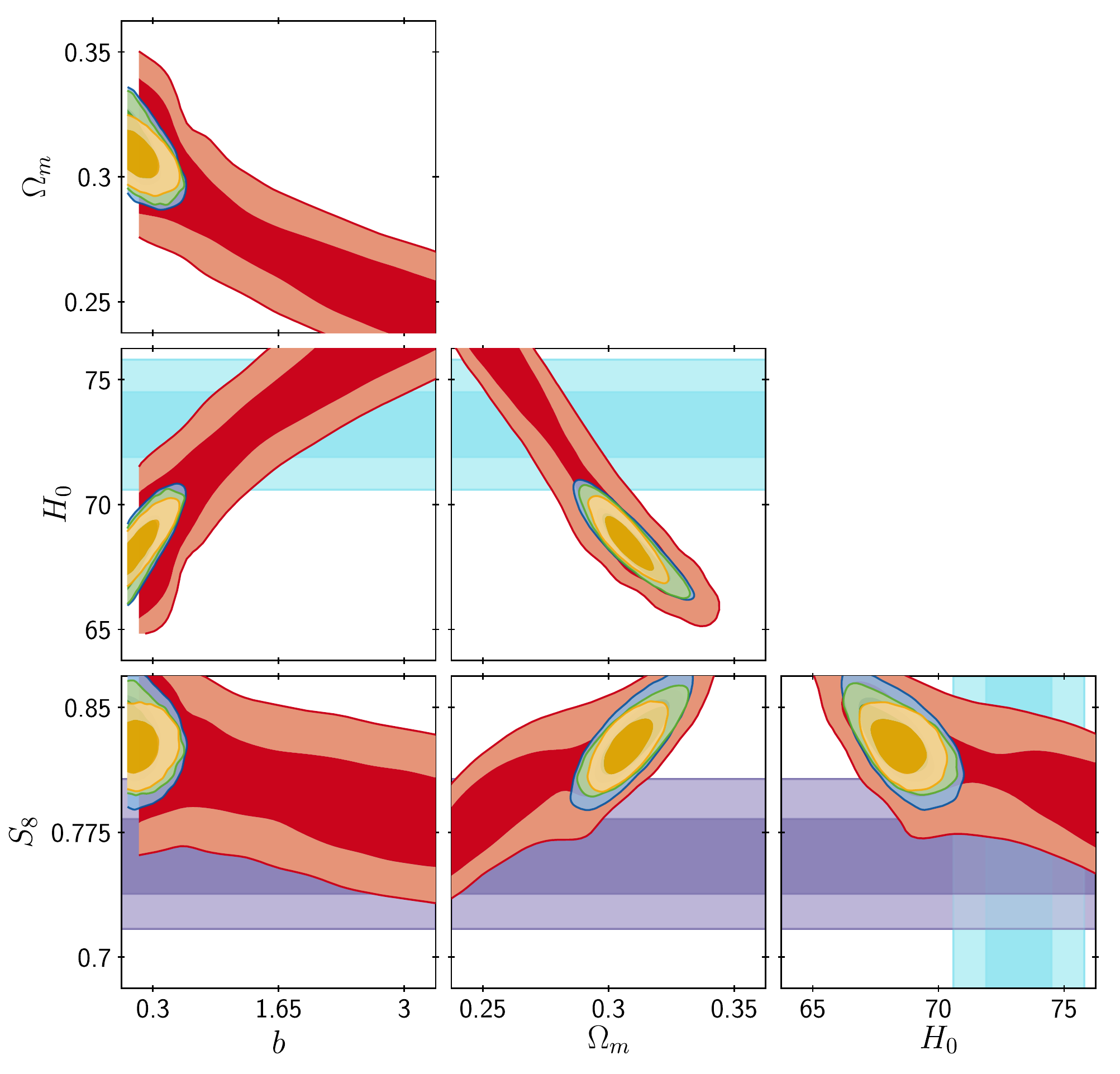}
	\includegraphics[height=5.3cm]{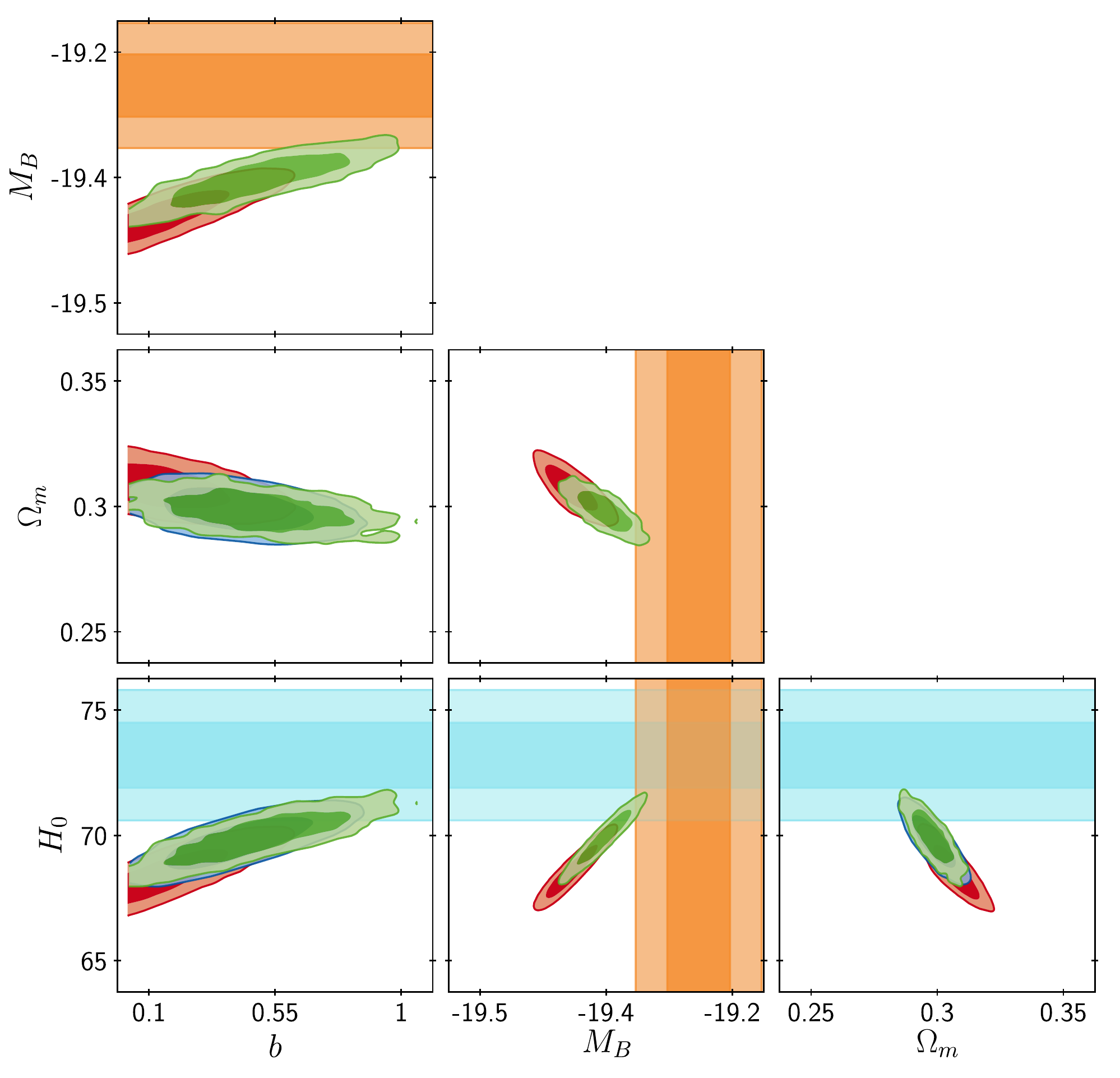}
	\caption{\Bshort: {\it (Left)} 2D contours on ($\Omega_{\rm m}$, $H_0$, $S_8$, $b$), for Planck + BAO data. {\it (Right)} 2D contours on ($\Omega_{\rm m}$, $M_{\rm B}$, $H_0$, $b$), for Planck +BAO + Pantheon and/or SH0ES (treated as a measurement of either $H_0$ or $M_B$).}
	\label{3-zones_1}
\end{figure*}

\FloatBarrier

\newpage

\begin{figure}[H]
    \centering
    \includegraphics[height=2cm]{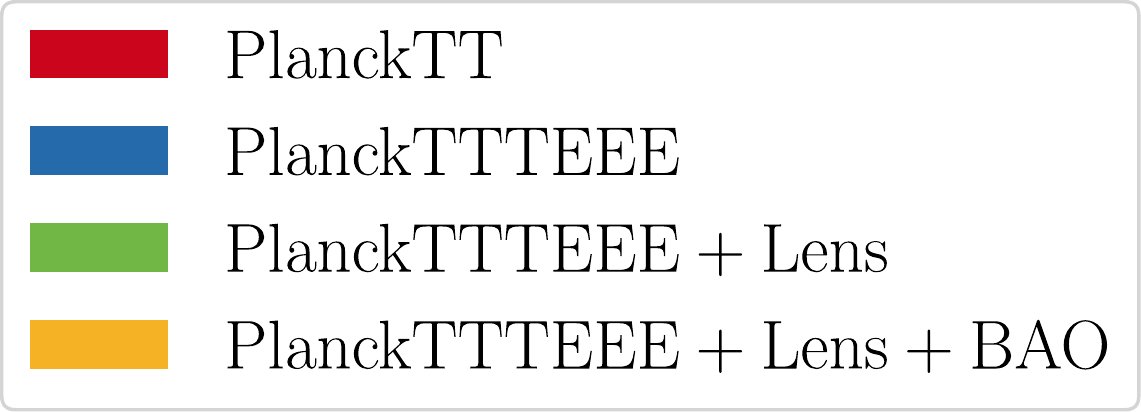}
    \hfil
    \includegraphics[height=1.5cm]{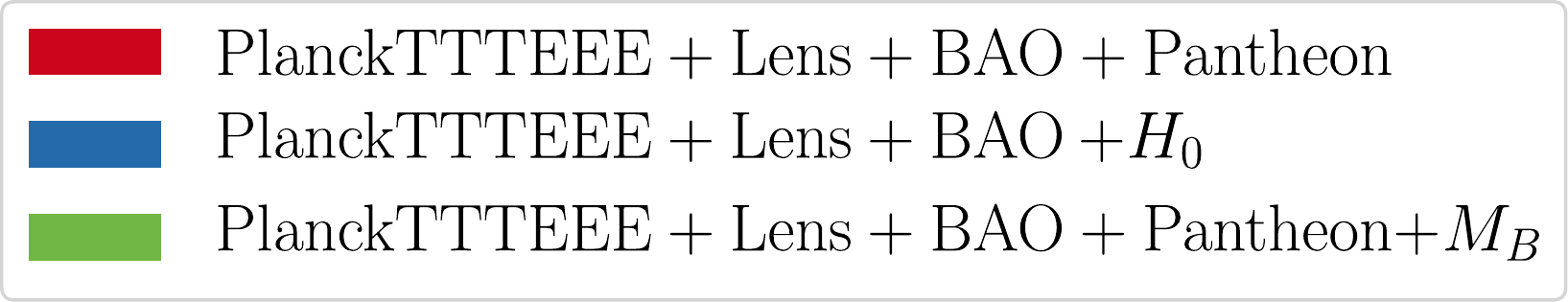}
\end{figure}

\subsection{\melong}

\begin{figure}[h]
    \centering
    \includegraphics[height=5.3cm]{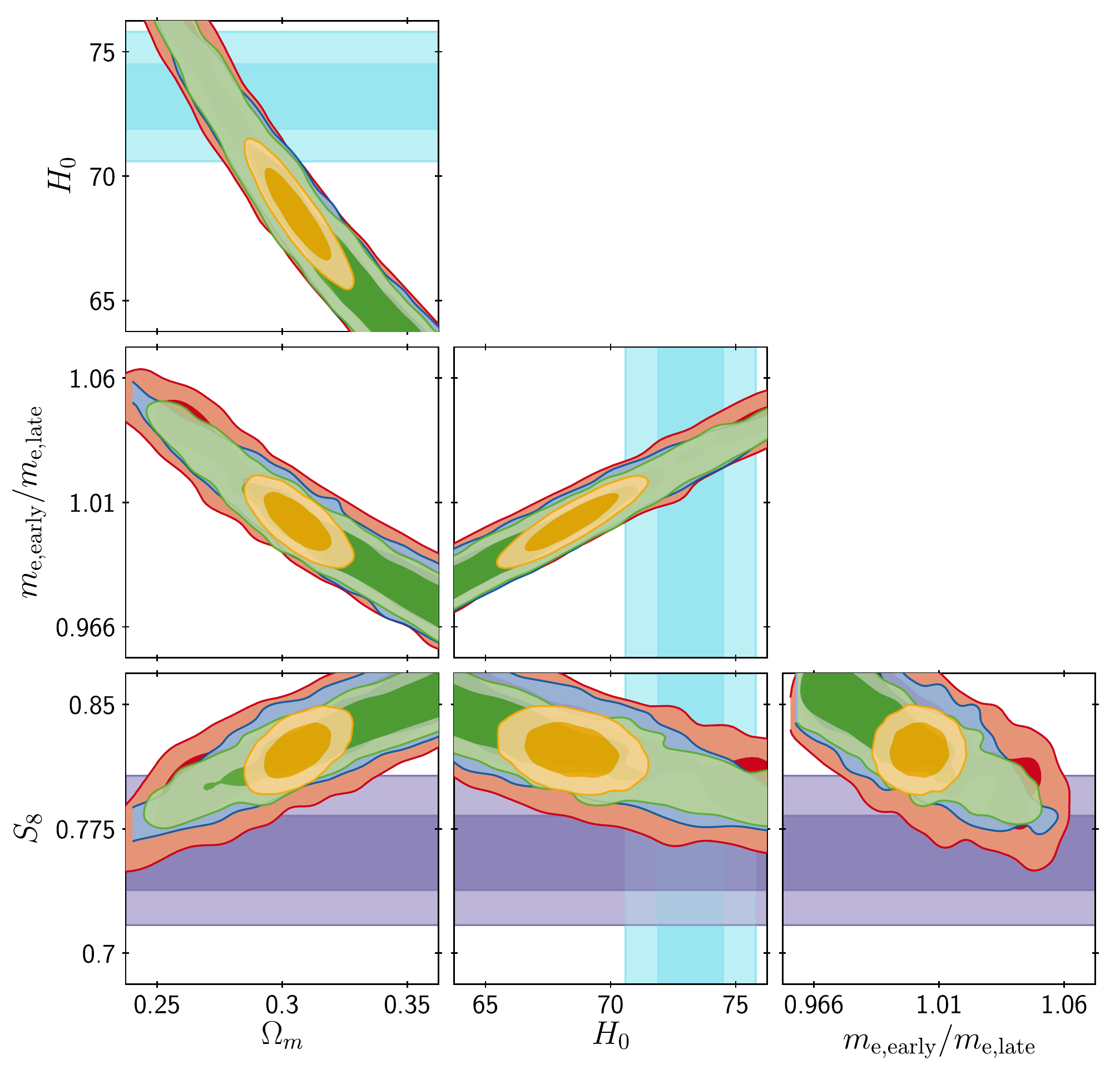}
    \includegraphics[height=5.3cm]{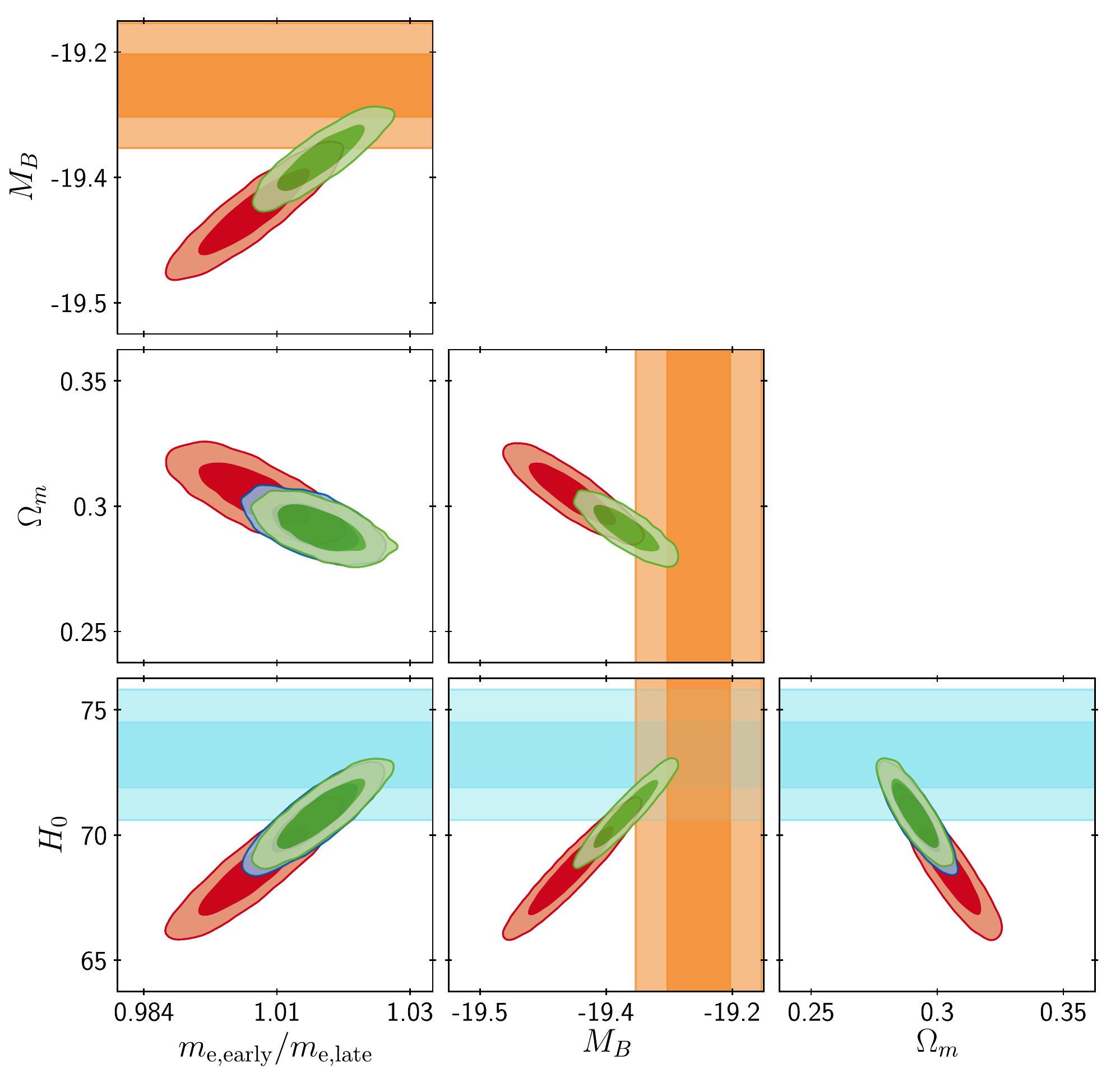}
    \caption{Varying $m_e$: {\it (Left)} 2D contours on ($\Omega_{\rm m}$, $H_0$, $S_8$, $m_{\rm e, early}/m_{\rm e, late}$), for Planck + BAO data. {\it (Right)} 2D contours on ($\Omega_{\rm m}$, $M_{\rm B}$, $H_0$, $m_e$), for Planck +BAO + Pantheon and/or SH0ES (treated as a measurement of either $H_0$ or $M_B$).}
    \label{fig:varyingme_lens_bao_sh0es}
\end{figure}

\FloatBarrier

\subsection{\meOklong}

\begin{figure}[h]
    \centering
    \includegraphics[height=6.5cm]{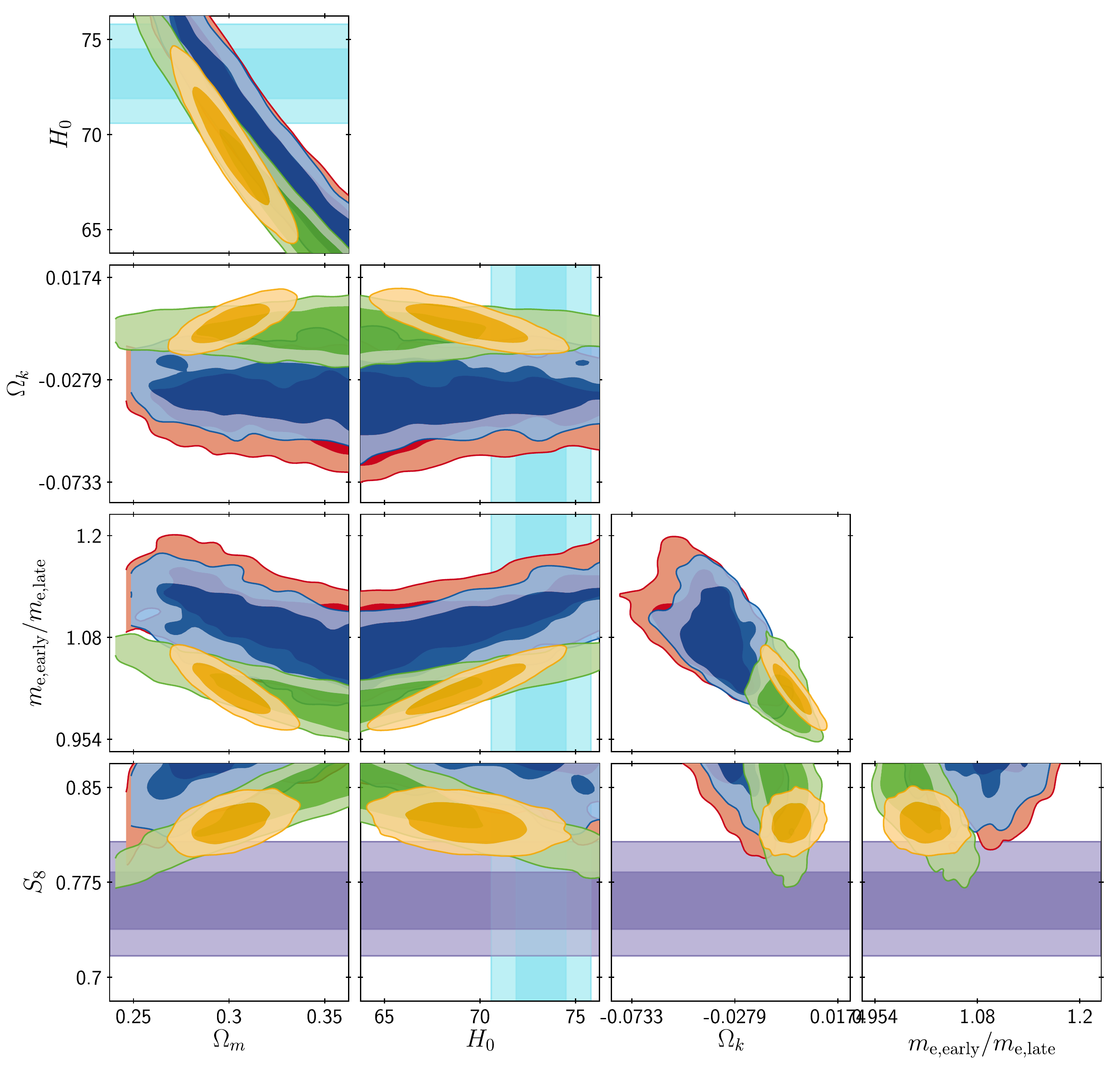}
    \includegraphics[height=6.5cm]{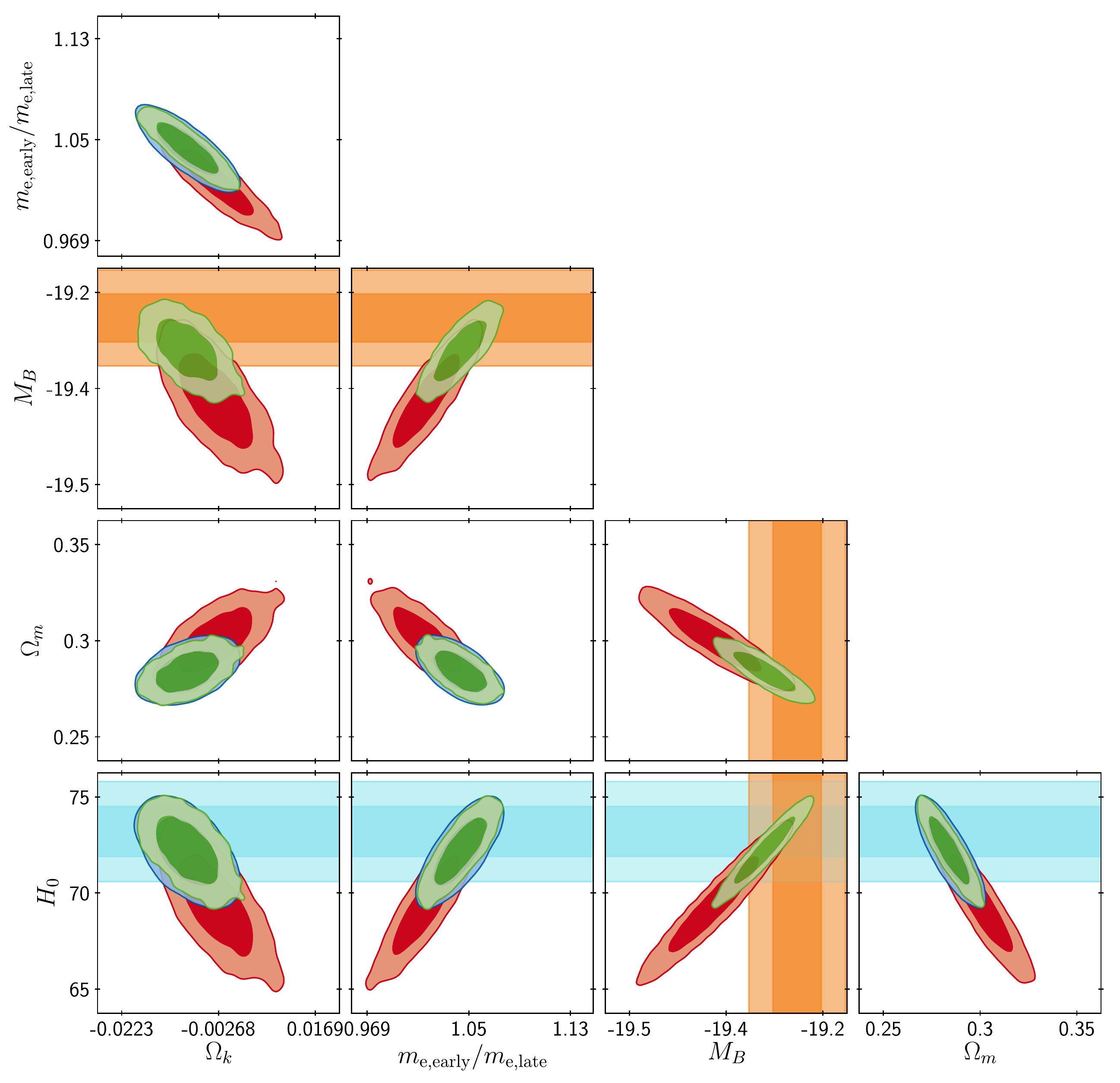}
    \caption{Varying $m_e$ + $\Omega_k$: {\it (Left)} 2D contours on ($\Omega_{\rm m}$, $H_0$, $S_8$, $m_e$, $\Omega_k$), for Planck + BAO data. {\it (Right)} 2D contours on ($\Omega_{\rm m}$, $M_{\rm B}$, $H_0$, $m_{\rm e, early}/m_{\rm e, late}$, $\Omega_k$), for Planck +BAO + Pantheon and/or SH0ES (treated as a measurement of either $H_0$ or $M_B$).}
    \label{fig:varyingmeok_lens_bao_sh0es}
\end{figure}

\newpage

\enlargethispage*{3cm}
\begin{figure}[H]
    \centering
    \includegraphics[height=1.5cm]{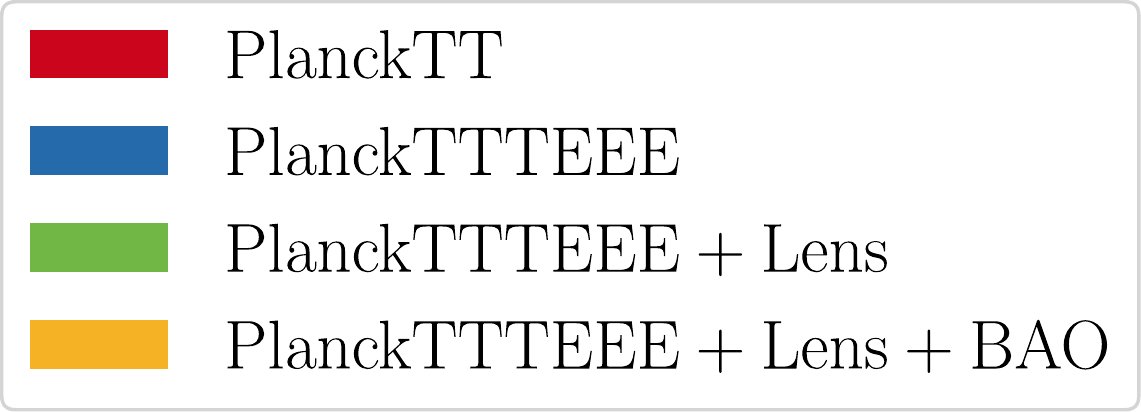}
    \hfil
    \includegraphics[height=1.125cm]{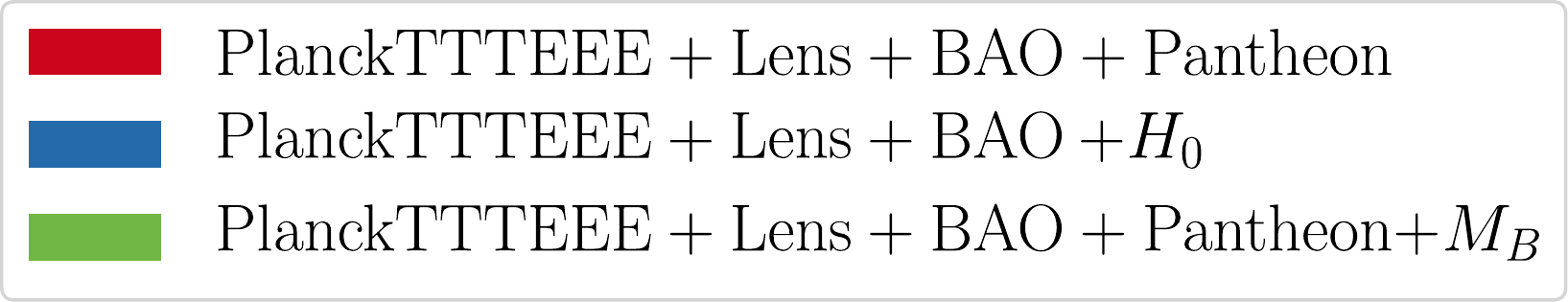}
\end{figure}
\vspace*{-1cm}

\subsection{\EDElong}

 \begin{figure*}[h]
     \centering
     \includegraphics[height=7cm]{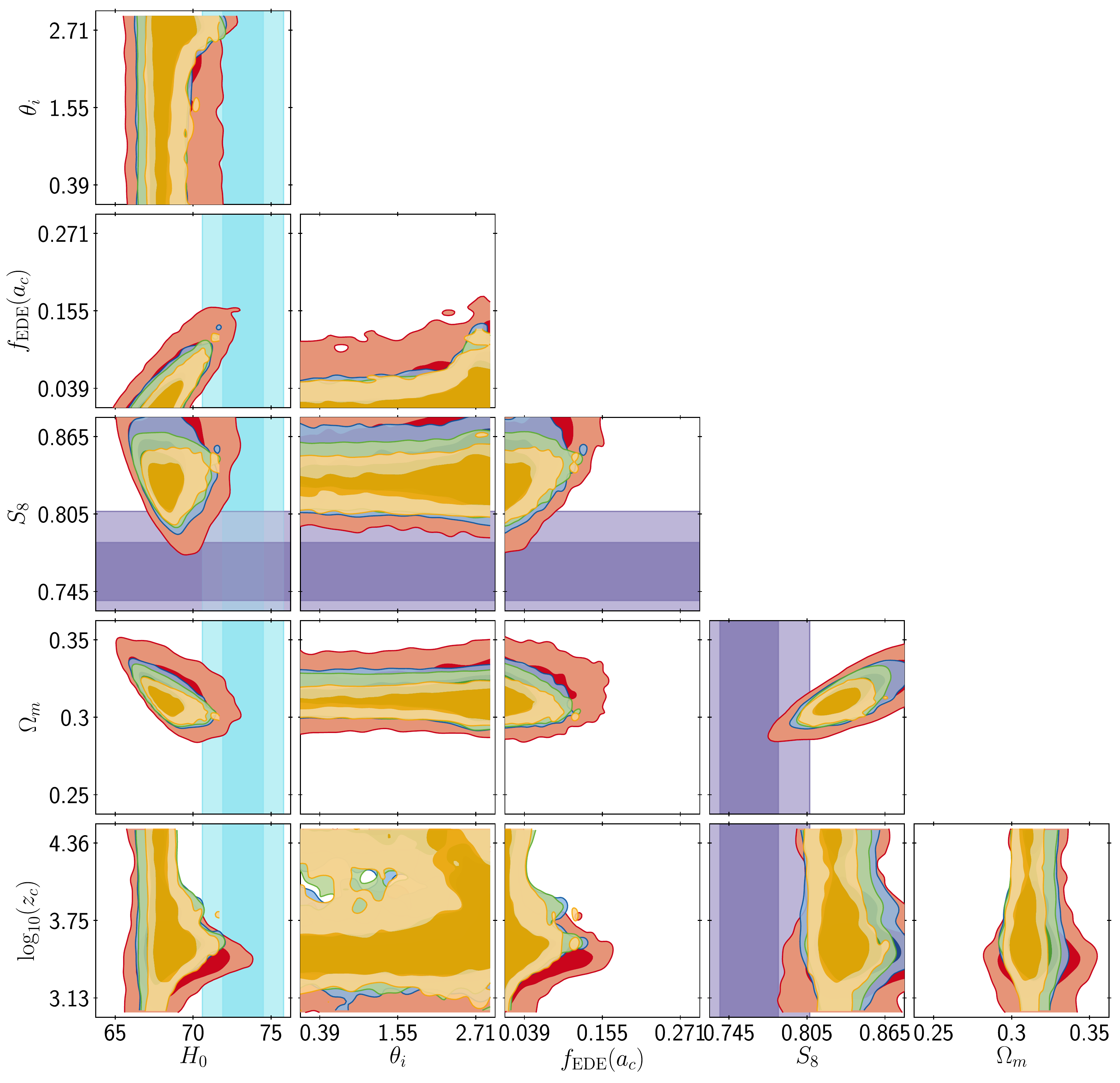}
     \includegraphics[height=7cm]{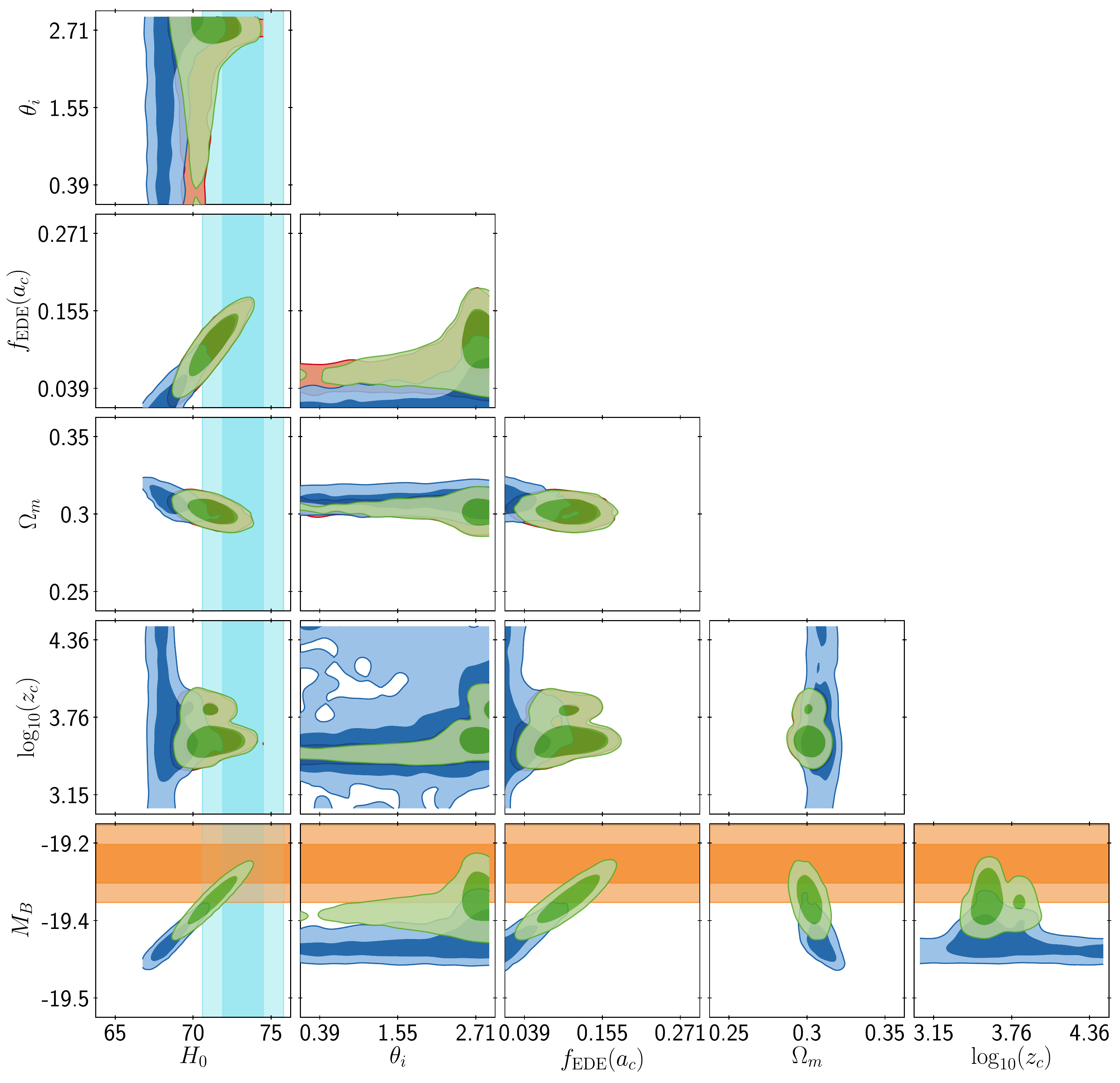}
     \caption{Early dark Energy: {\it (Left)} 2D contours on ($\Omega_{\rm m}$, $H_0$, $S_8$, $f_{\rm EDE}$, $\log_{10}(z_c)$, $\Theta_{i}$), for Planck + BAO data. {\it (Right)} 2D contours on ($\Omega_{\rm m}$, $M_{\rm B}$, $H_0$, $f_{\rm EDE}$, $\log_{10}(z_c)$, $\Theta_i$), for Planck +BAO + Pantheon and/or SH0ES (treated as a measurement of either $H_0$ or $M_B$).}
     \label{fig:ede}
 \end{figure*}

 \FloatBarrier
\vspace*{-0.5cm}

\subsection{\NEDElong}

 \begin{figure*}[h]
     \centering
     \includegraphics[height=7cm]{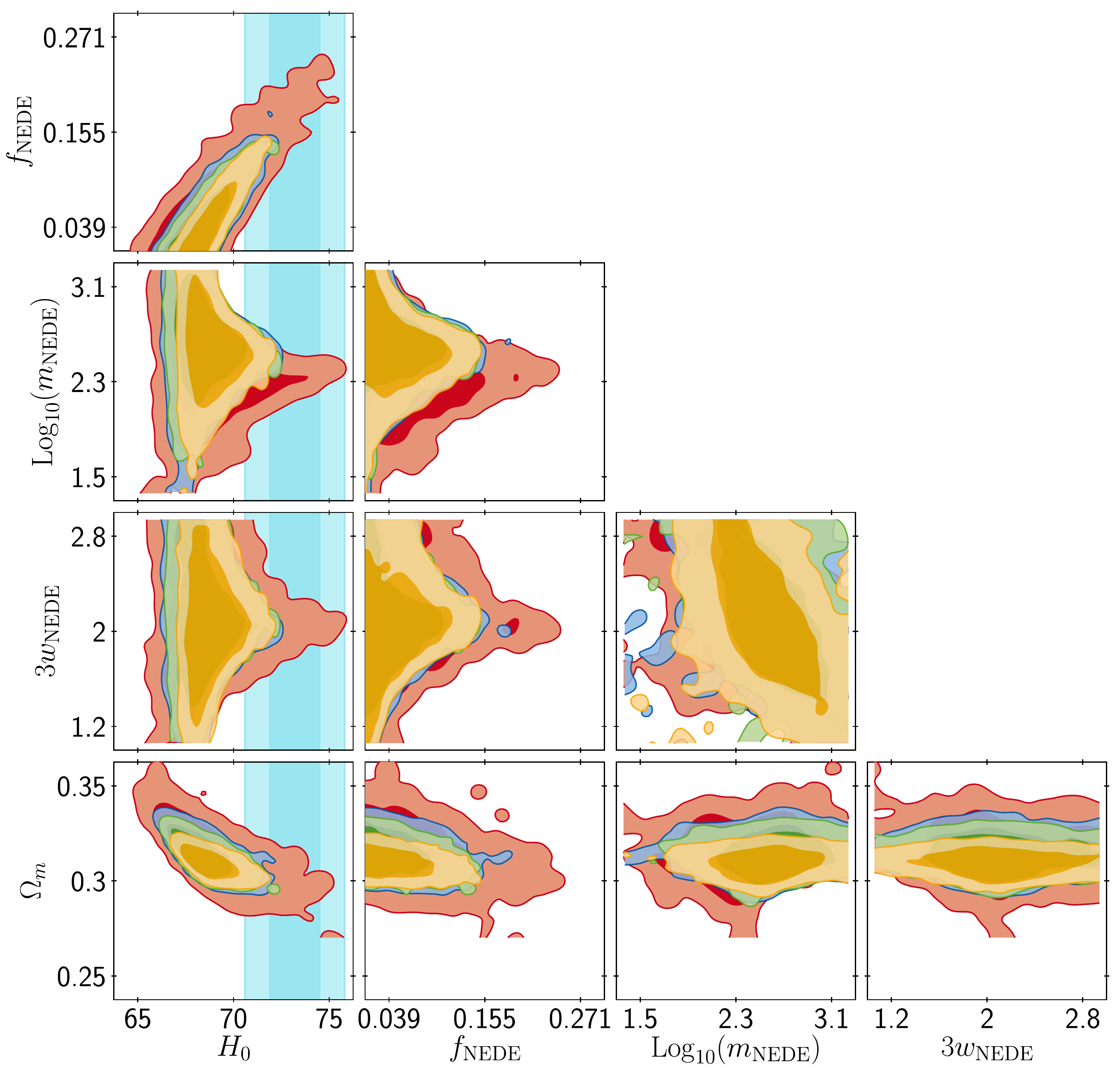}
     \includegraphics[height=7cm]{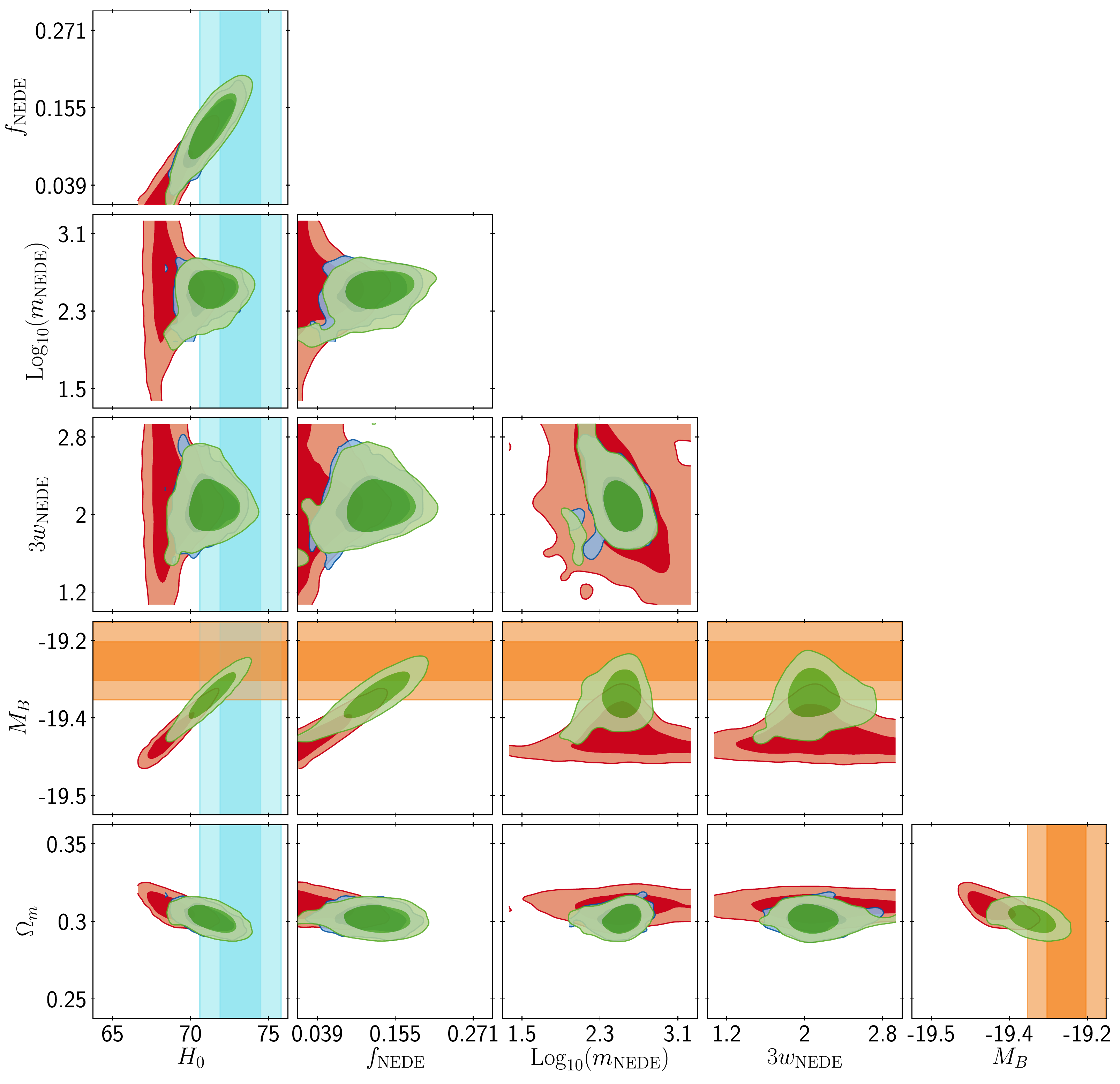}
     \caption{New Early dark Energy: {\it (Left)} 2D contours on ($\Omega_{\rm m}$, $H_0$, $S_8$, $f_{\rm NEDE}$, $\log_{10}(m_{\phi}/{\rm eV})$, $3w_{\rm NEDE}$), for Planck + BAO data. {\it (Right)} 2D contours on ($\Omega_{\rm m}$, $M_{\rm B}$, $H_0$, $f_{\rm NEDE}$, $\log_{10}(m_\mathrm{NEDE}/{\rm eV})$, $3w_{\rm NEDE}$), for Planck +BAO + Pantheon and/or SH0ES (treated as a measurement of either $H_0$ or $M_B$).\vspace*{-1cm}}
     \label{fig:ede}
 \end{figure*}
  \FloatBarrier

\pagebreak

\begin{figure}[H]

    \centering
    \includegraphics[height=1.5cm]{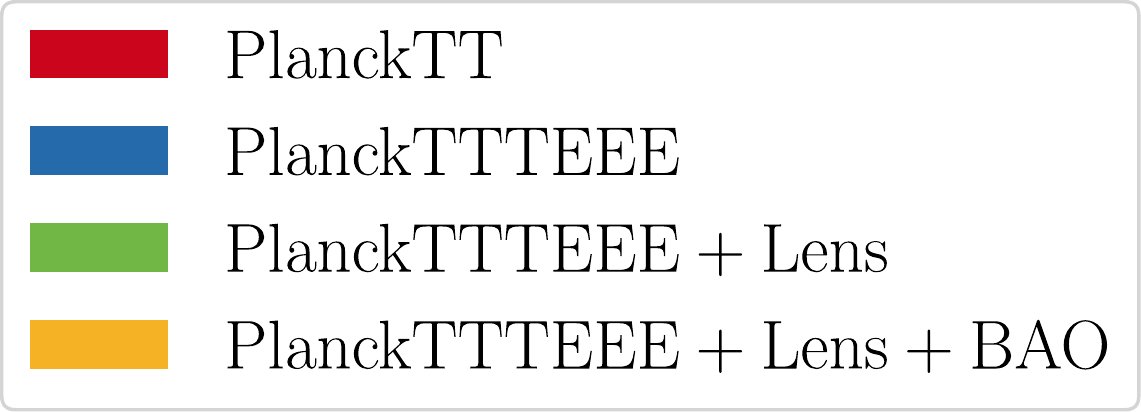}
    \hfil
    \includegraphics[height=1.125cm]{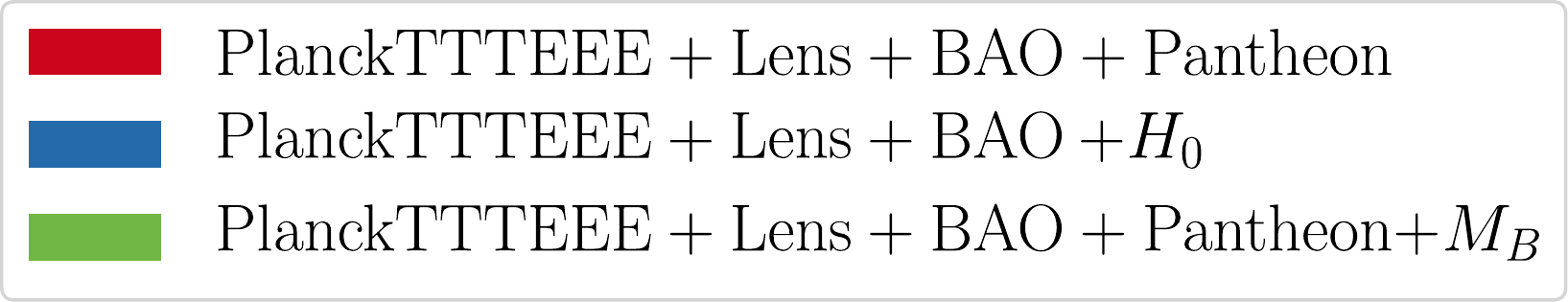}
\end{figure}
\vspace*{-1cm}

 \subsection{Early Modified Gravity}
 \begin{figure*}[h]
     \centering
     \includegraphics[height=7cm]{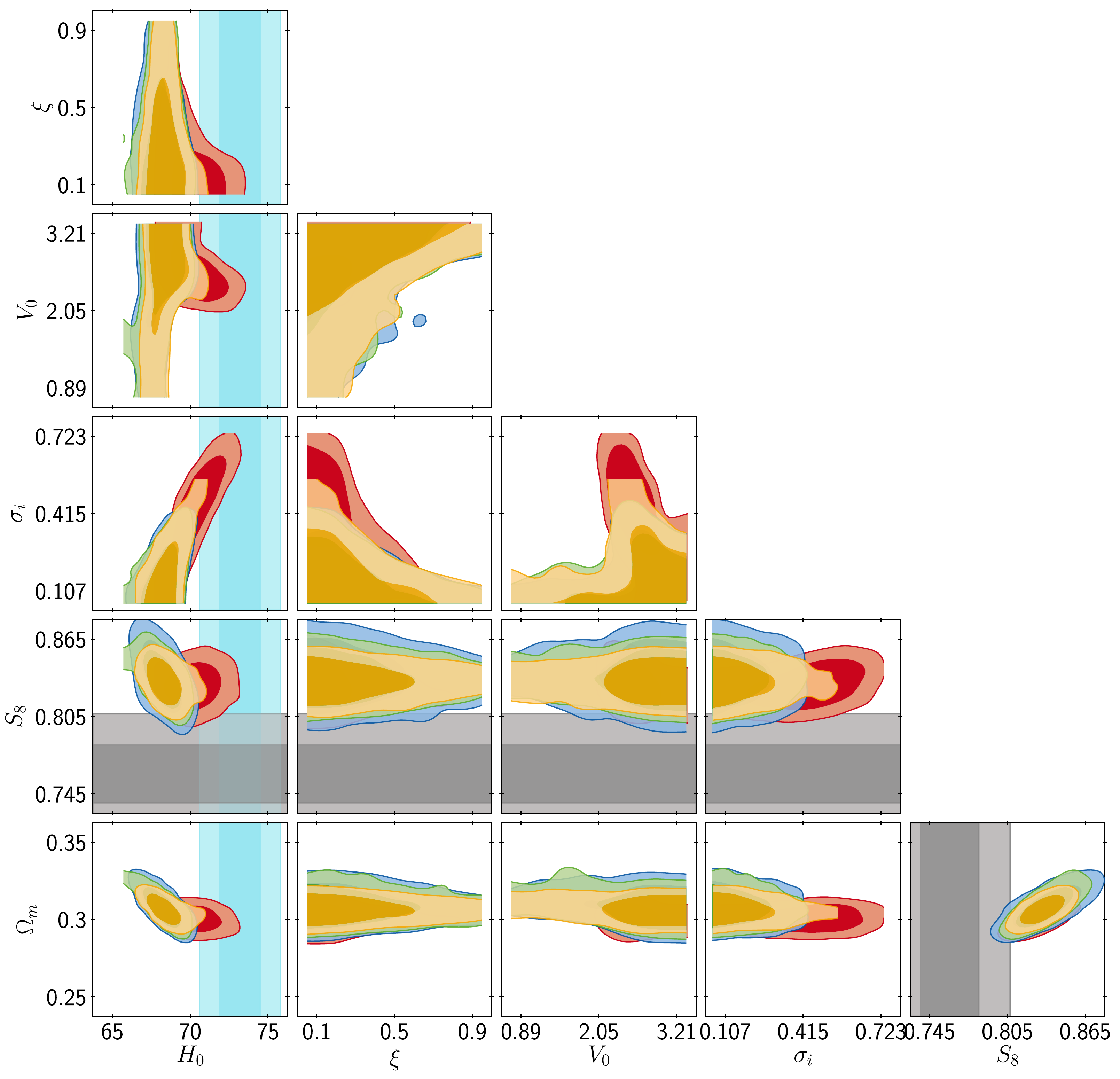}
     \includegraphics[height=7cm]{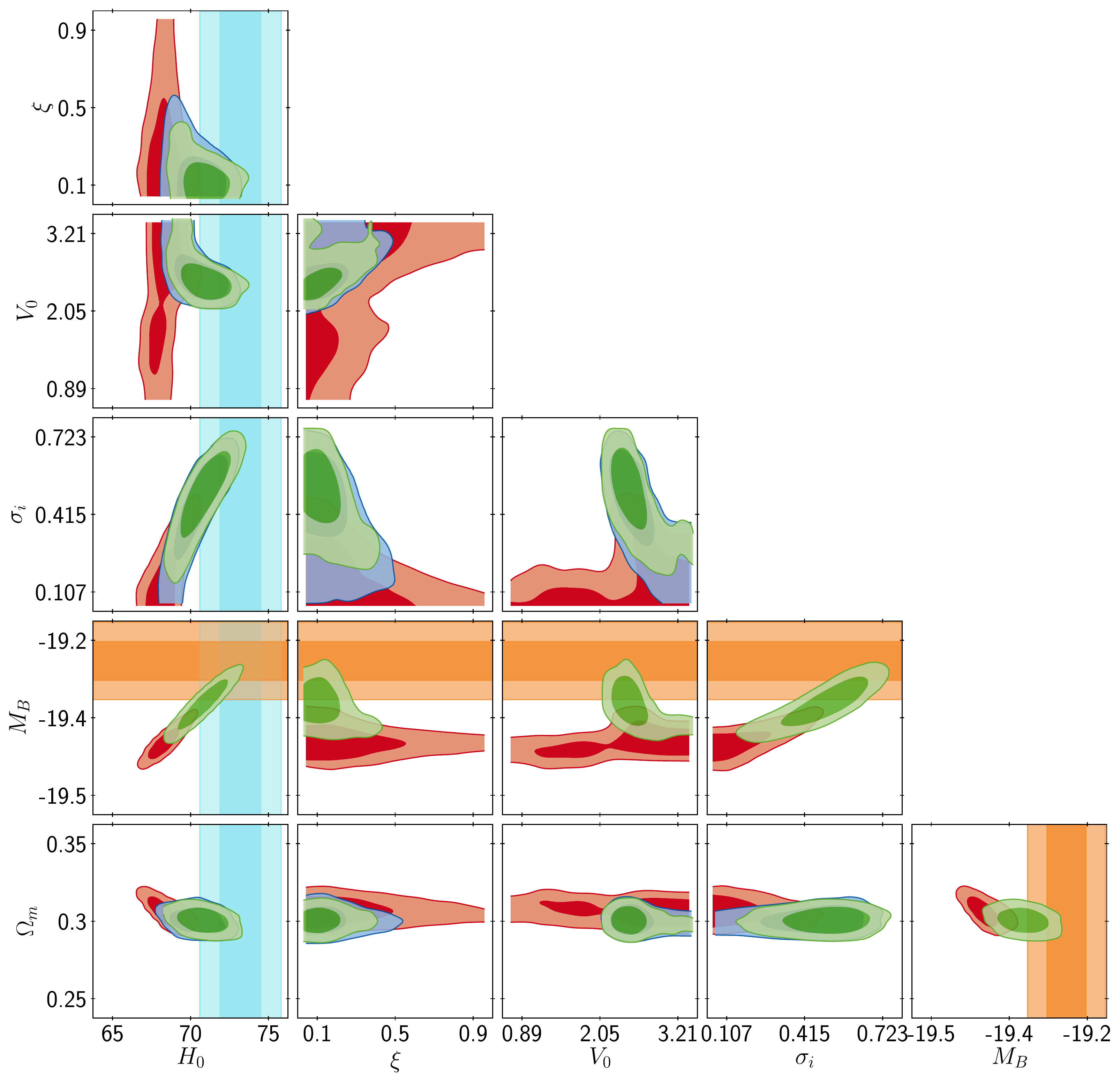}
     \caption{Early Modified Gravity: {\it (Left)} 2D contours on ($\Omega_{\rm m}$, $H_0$, $S_8$, $\xi$, $V_0$, $\sigma_i$), for Planck + BAO data. {\it (Right)} 2D contours on ($\Omega_{\rm m}$, $M_{\rm B}$, $H_0$,$\xi$, $V_0$, $\sigma_i$), for Planck +BAO + Pantheon and/or SH0ES (treated as a measurement of either $H_0$ or $M_B$).\vspace*{-0.5cm}}
     \label{fig:emg}
 \end{figure*}
 
\subsection{\CPLlong}

 \begin{figure*}[h]
     \centering
     \includegraphics[height=6.5cm]{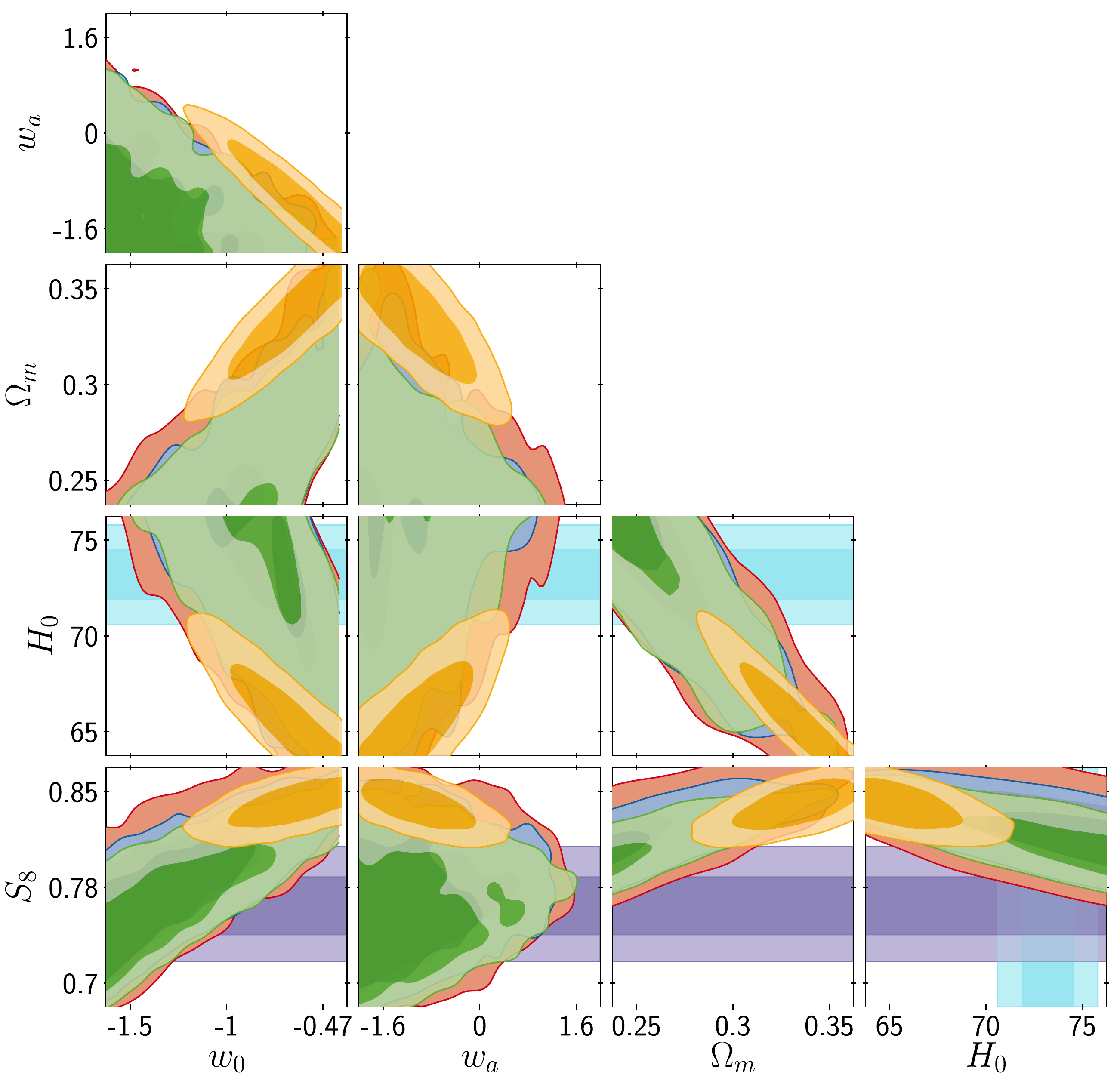}
     \includegraphics[height=6.5cm]{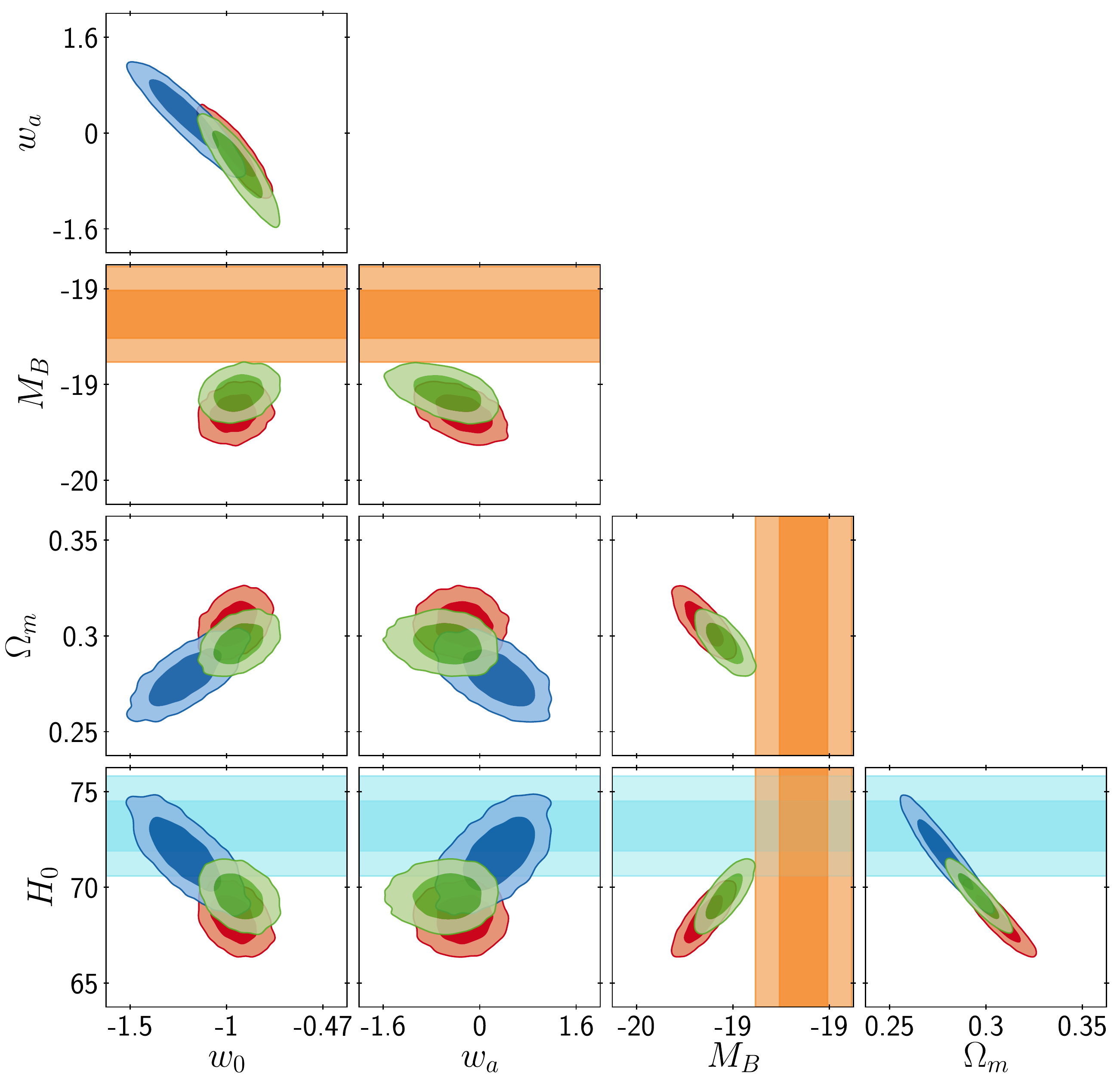}
     \caption{CPL Dark Energy: {\it (Left)} 2D contours on ($\Omega_{\rm m}$, $H_0$, $S_8$, $w_0$, $w_a$), for Planck + BAO data. {\it (Right)} 2D contours on ($\Omega_{\rm m}$, $M_{\rm B}$, $H_0$, $w_0$, $w_a$), for Planck +BAO + Pantheon and/or SH0ES (treated as a measurement of either $H_0$ or $M_B$).}
     \label{fig:CPL}
 \end{figure*}
 
 \FloatBarrier
 \pagebreak

\begin{figure}[H]

    \centering
    \includegraphics[height=1.5cm]{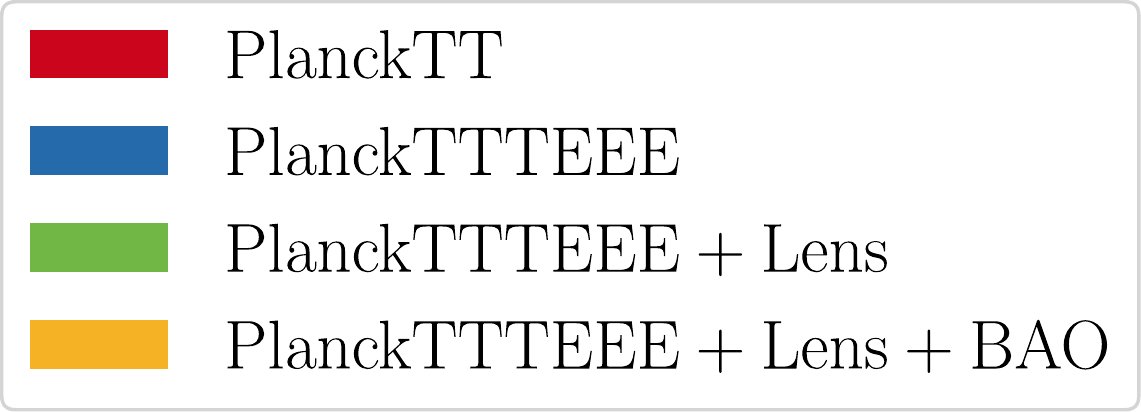}
    \hfil
    \includegraphics[height=1.125cm]{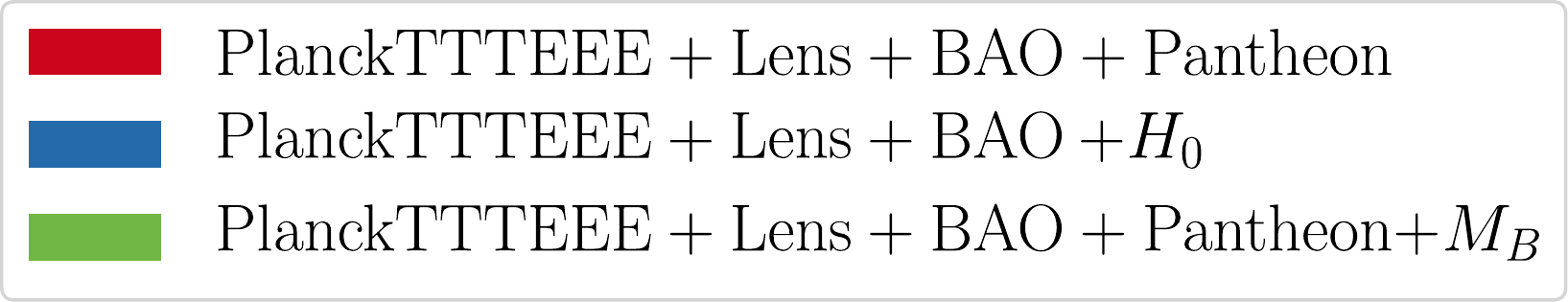}
\end{figure}

\enlargethispage*{2cm}


 \FloatBarrier
 
\vspace*{-0.5cm}
\subsection{\PEDElong}
\vspace*{-0.3cm}

\begin{figure}[h]
    \centering
    \includegraphics[height=3.6cm]{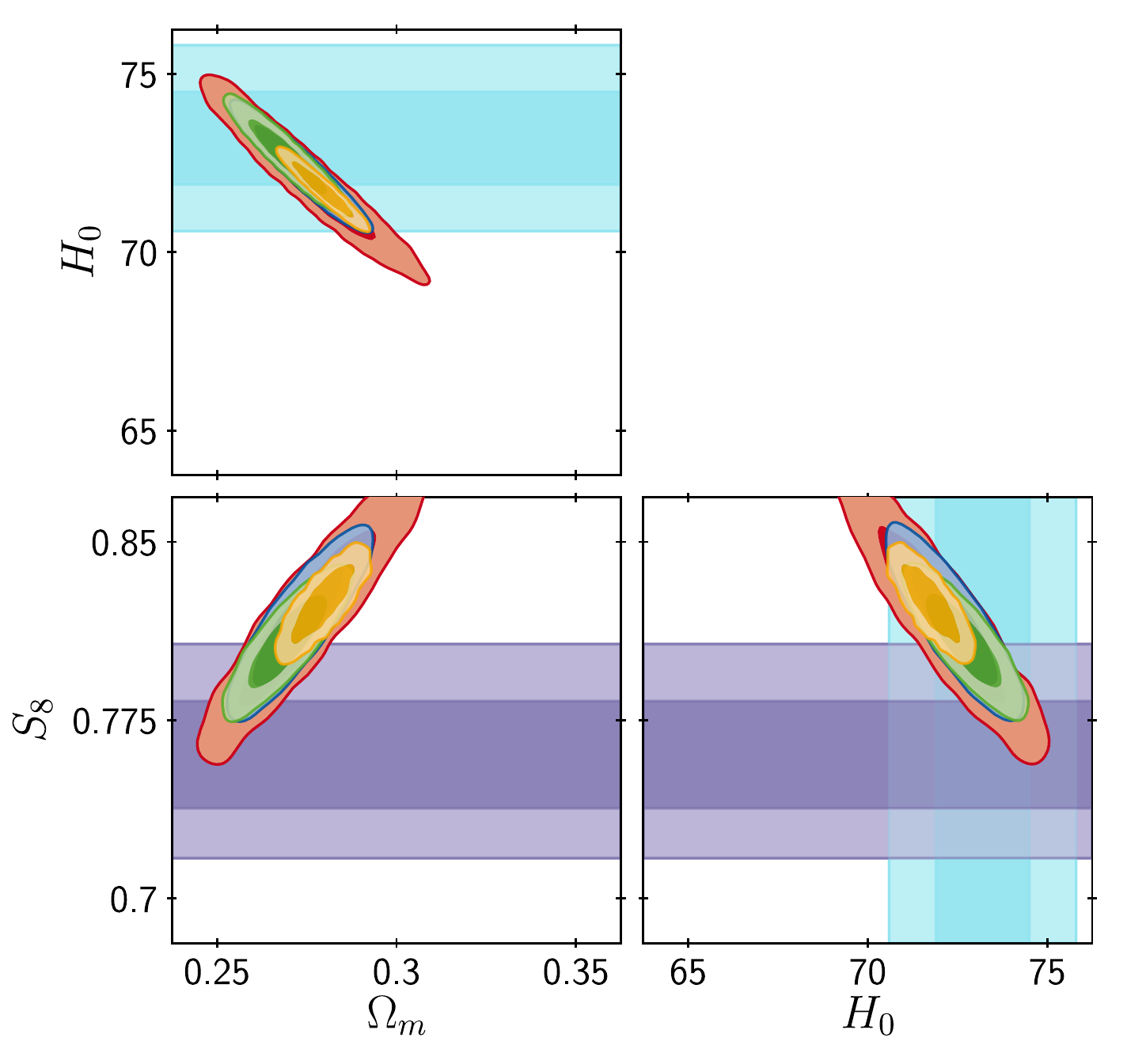}
    \includegraphics[height=3.6cm]{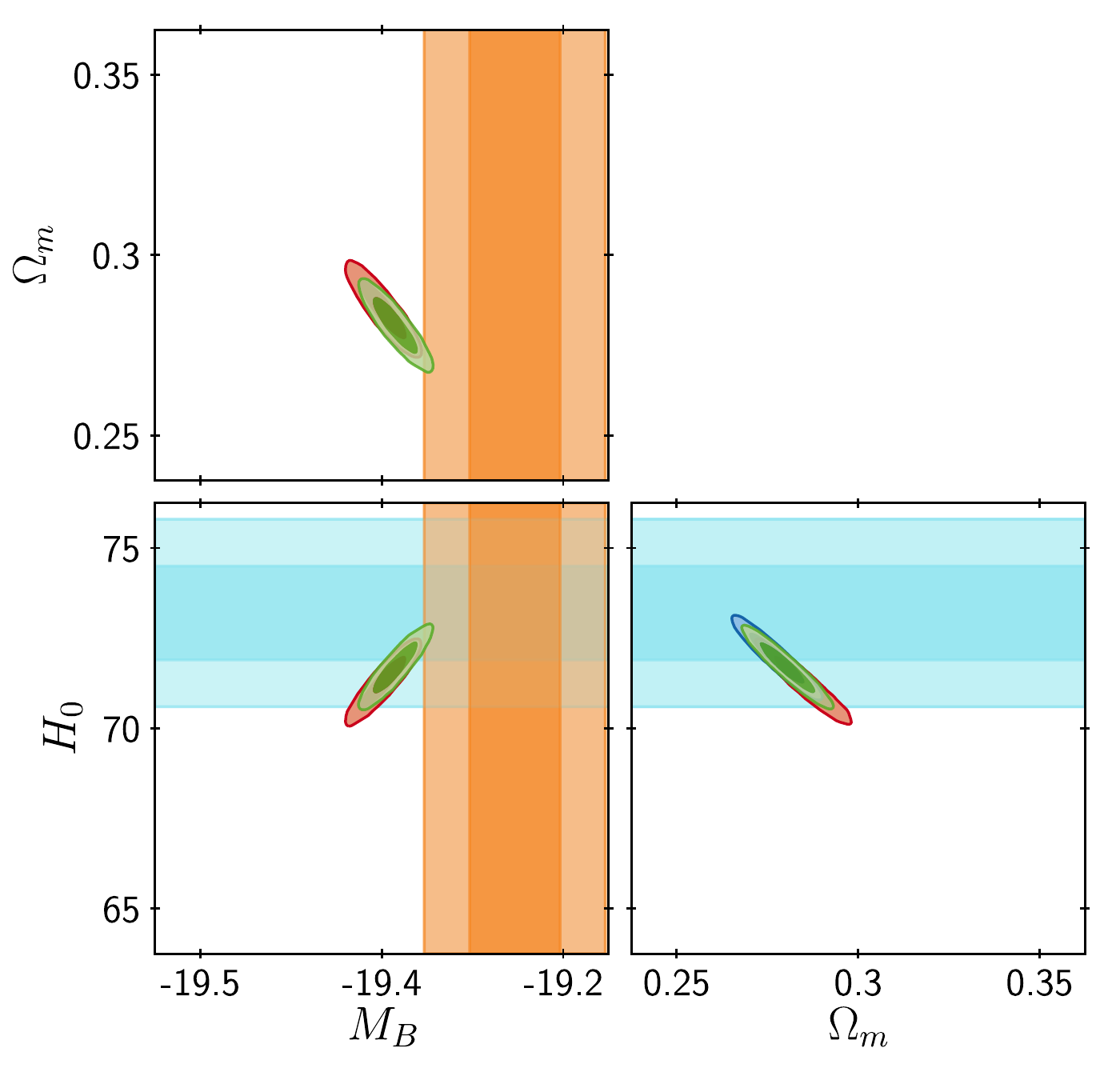}
    \caption{PEDE: {\it (Left)} 2D contours on ($\Omega_{\rm m}$, $H_0$, $S_8$), for Planck + BAO data. {\it (Right)} 2D contours on ($\Omega_{\rm m}$, $M_{\rm B}$, $H_0$), for Planck +BAO + Pantheon and/or SH0ES (treated as a measurement of either $H_0$ or $M_B$).}
    \label{fig:PEDE_AB}
\end{figure}

\vspace*{-0.5cm}
\subsection{\MEDElong}
\vspace*{-0.3cm}

\begin{figure}[h]
    \centering
    \includegraphics[height=3.6cm]{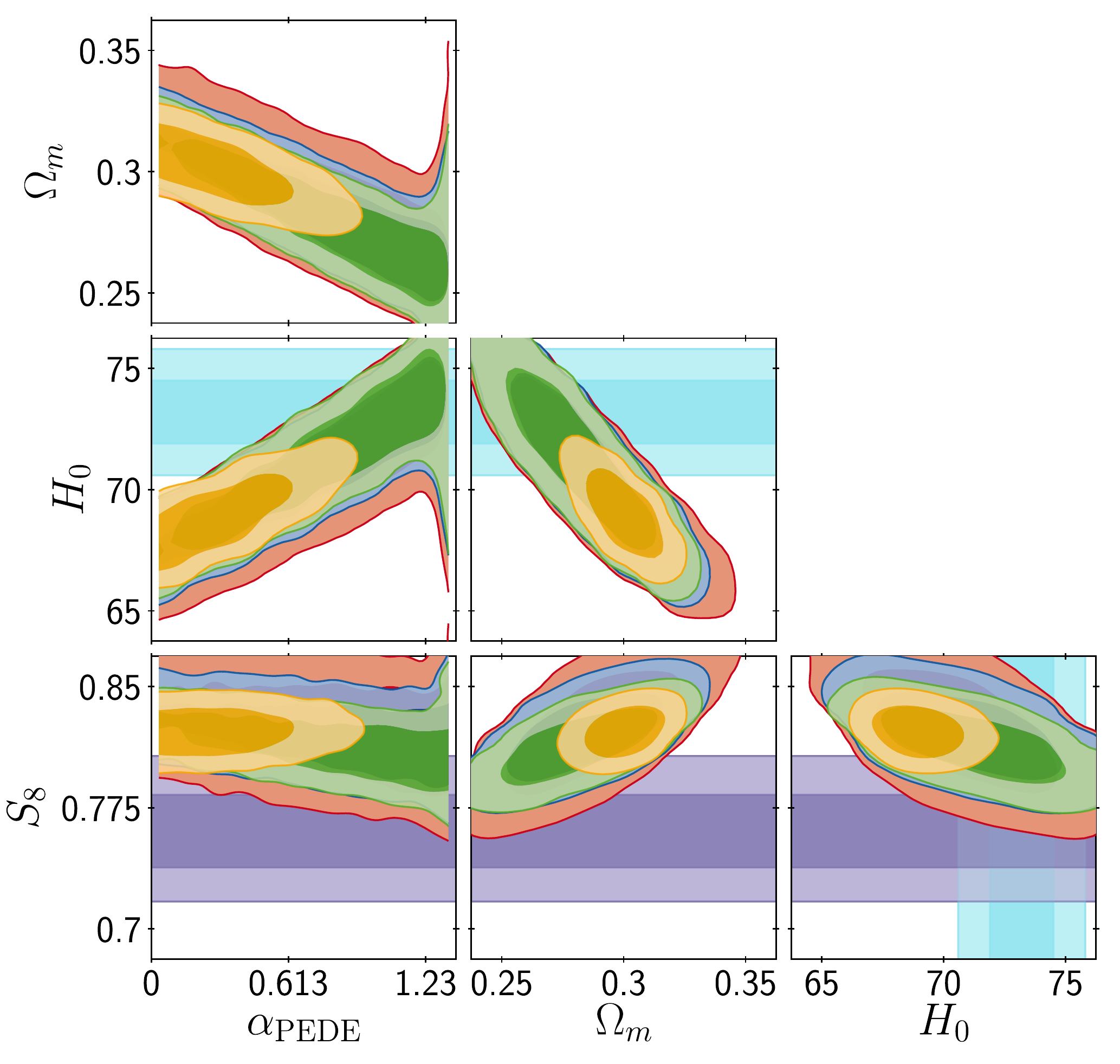}
    \includegraphics[height=3.6cm]{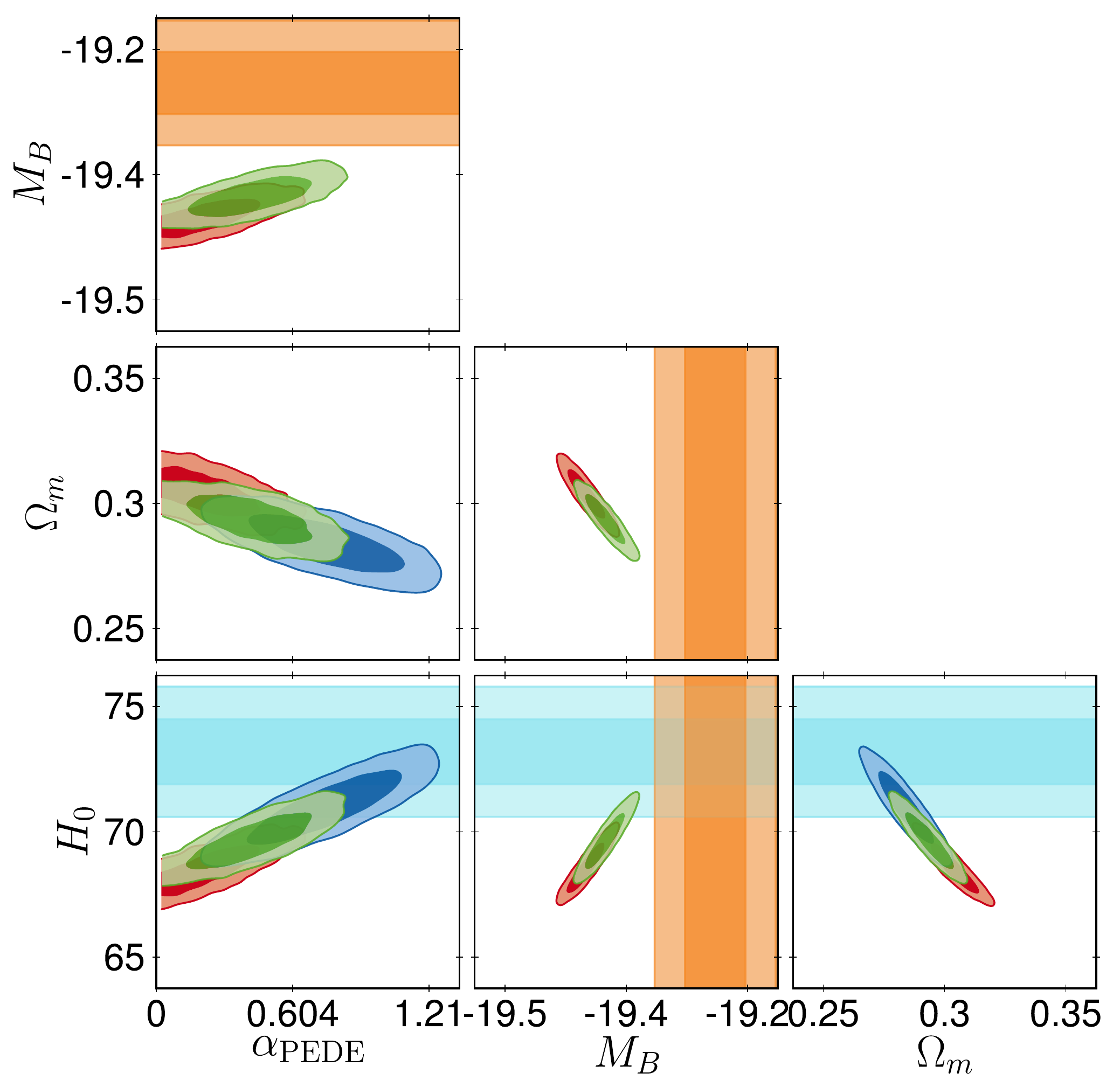}
    \setlength{\belowcaptionskip}{-10cm}
    \caption{PEDE Modification: {\it (Left)} 2D contours on ($\Omega_{\rm m}$, $H_0$, $S_8$, $\alpha_\mathrm{PEDE}$), for Planck + BAO data. {\it (Right)} 2D contours on ($\Omega_{\rm m}$, $M_{\rm B}$, $H_0$, $\alpha_\mathrm{PEDE}$), for Planck +BAO + Pantheon and/or SH0ES (treated as a measurement of either $H_0$ or $M_B$).}
    \label{fig:PEDE_AB}
\end{figure}

\FloatBarrier

\newpage

\enlargethispage*{2cm}
\vspace*{-1cm}

\begin{figure}[H]
    \centering
    \includegraphics[height=1.8cm]{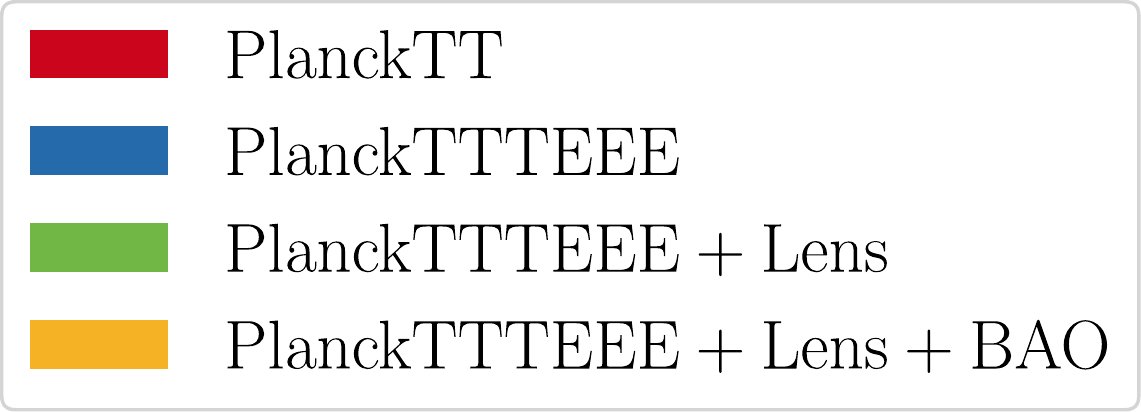}
    \hfil
    \includegraphics[height=1.35cm]{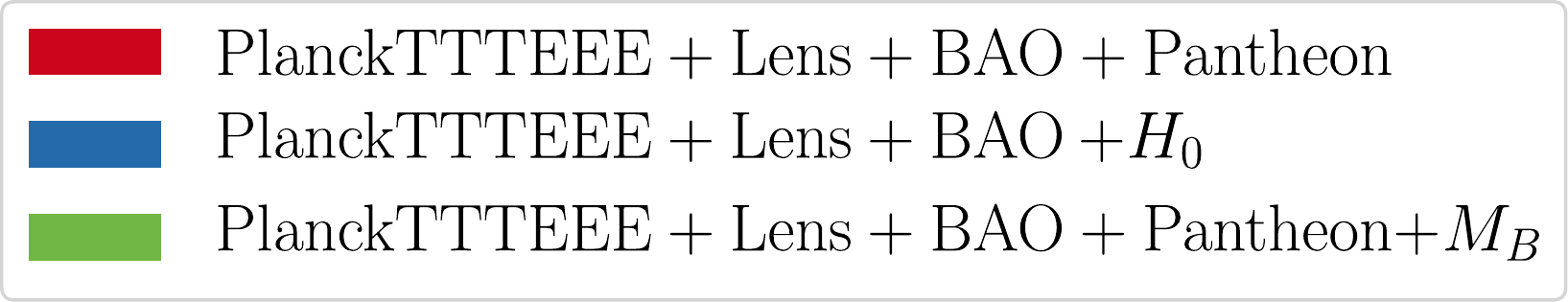}
\end{figure}

\subsection{\fracDMlong}

\begin{figure}[h]
    \centering
    \includegraphics[height=6.5cm]{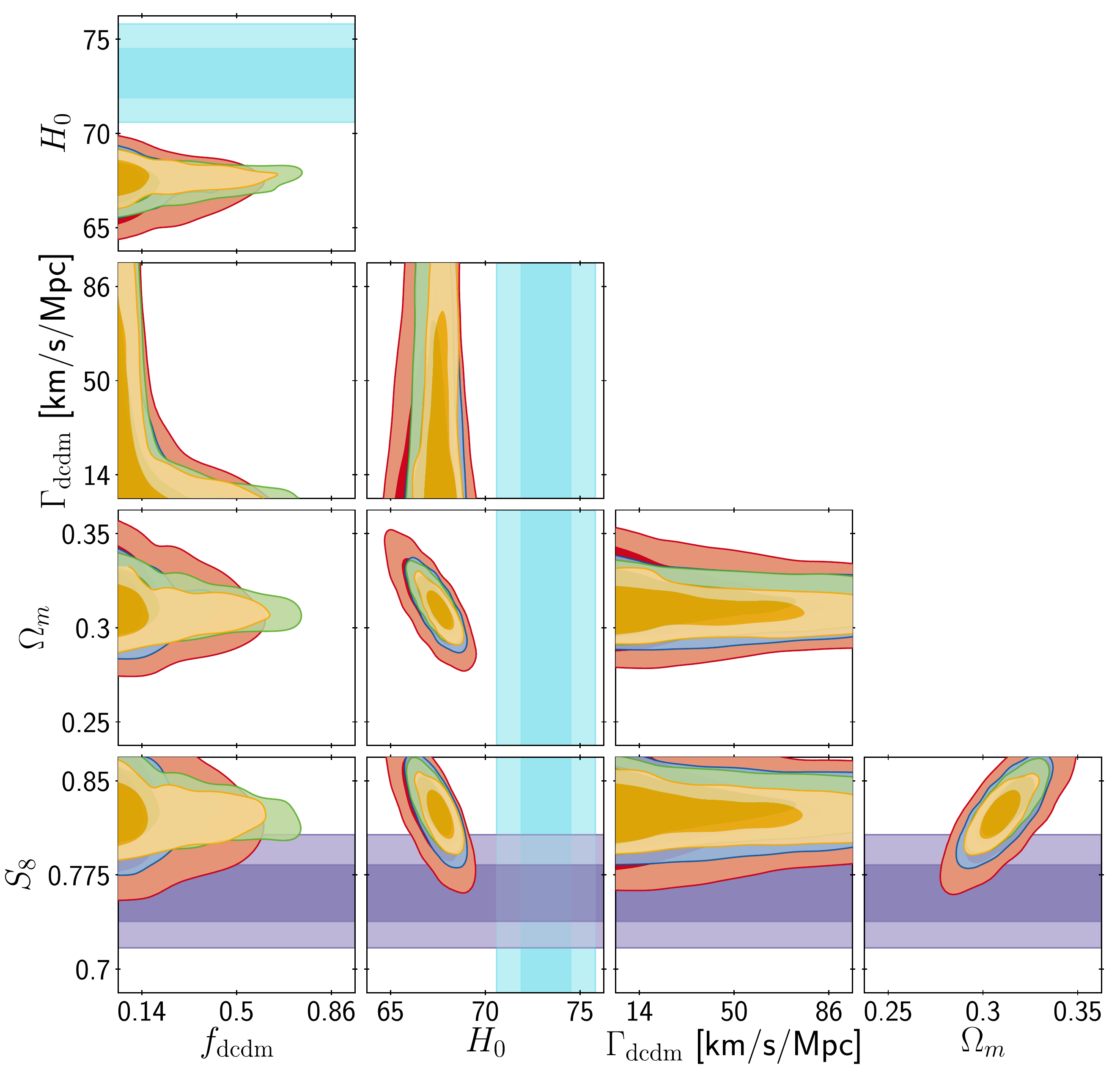}
    \includegraphics[height=6.5cm]{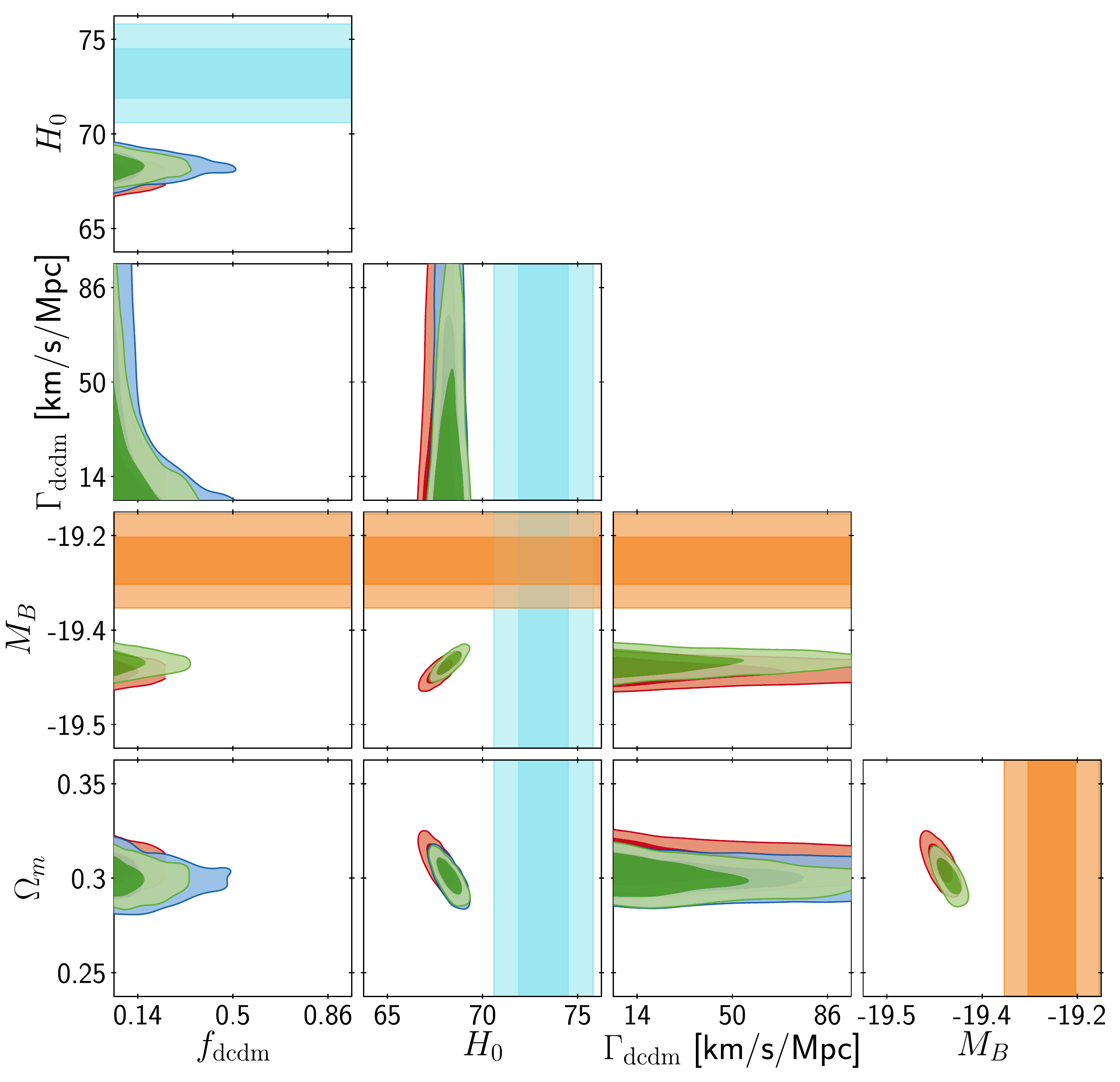}
    \caption{Fraction of Decaying DM to DR: {\it (Left)} 2D contours on ($\Omega_{\rm m}$, $H_0$, $S_8$, $f_{\rm dcdm}$, $\Gamma_{
    \rm dcdm}$), for Planck + BAO data. {\it (Right)} 2D contours on ($\Omega_{\rm m}$, $M_{\rm B}$, $H_0$, $f_{\rm dcdm}$, $\Gamma_{
    \rm dcdm}$), for Planck +BAO + Pantheon and/or SH0ES (treated as a measurement of either $H_0$ or $M_B$).}
    \label{fig:decay_std}
\end{figure} 

\FloatBarrier

\subsection{\DMDRWDMlong}
 \begin{figure*}[h]
     \centering
     \includegraphics[height=6.5cm]{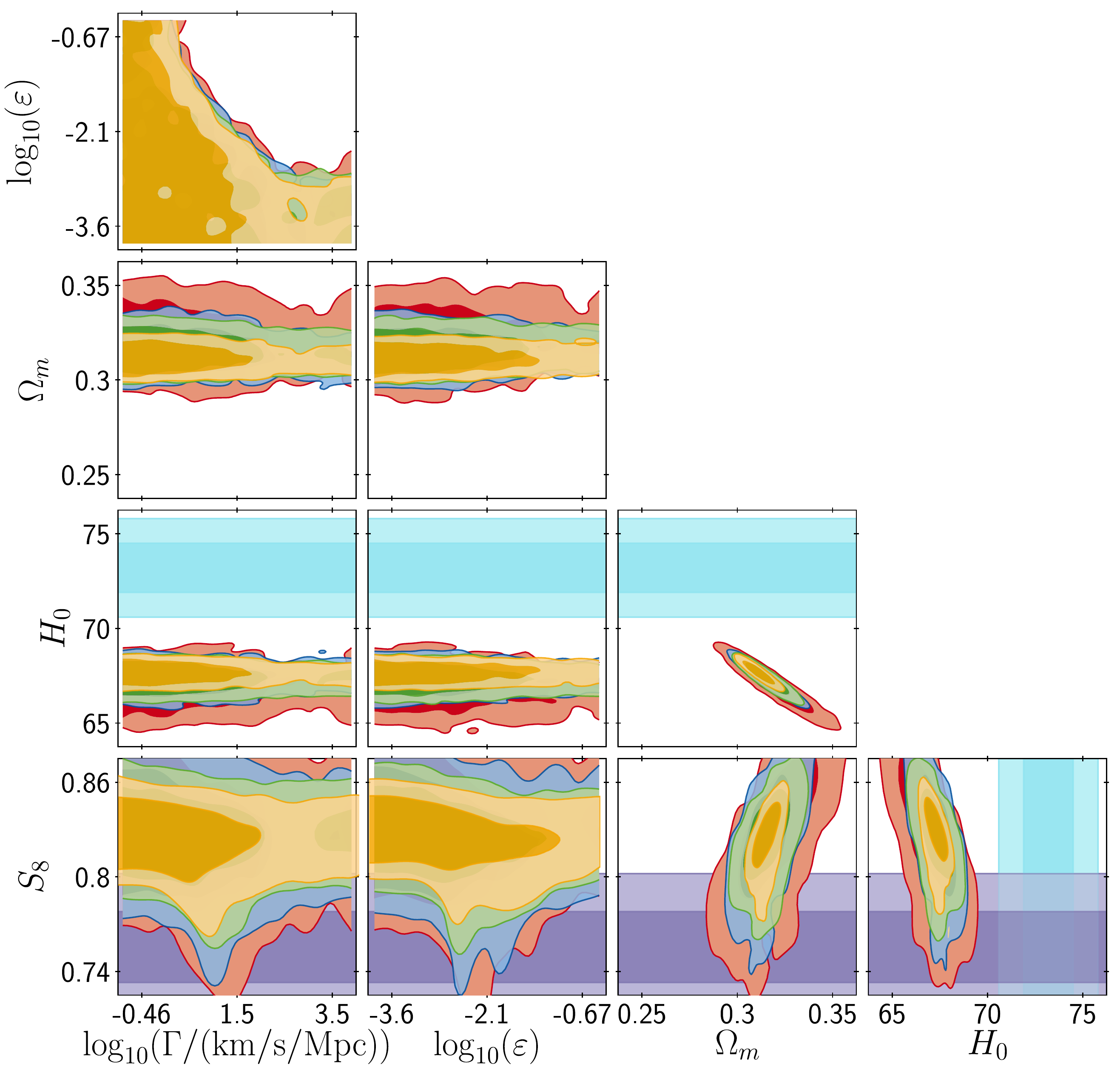}
     \includegraphics[height=6.5cm]{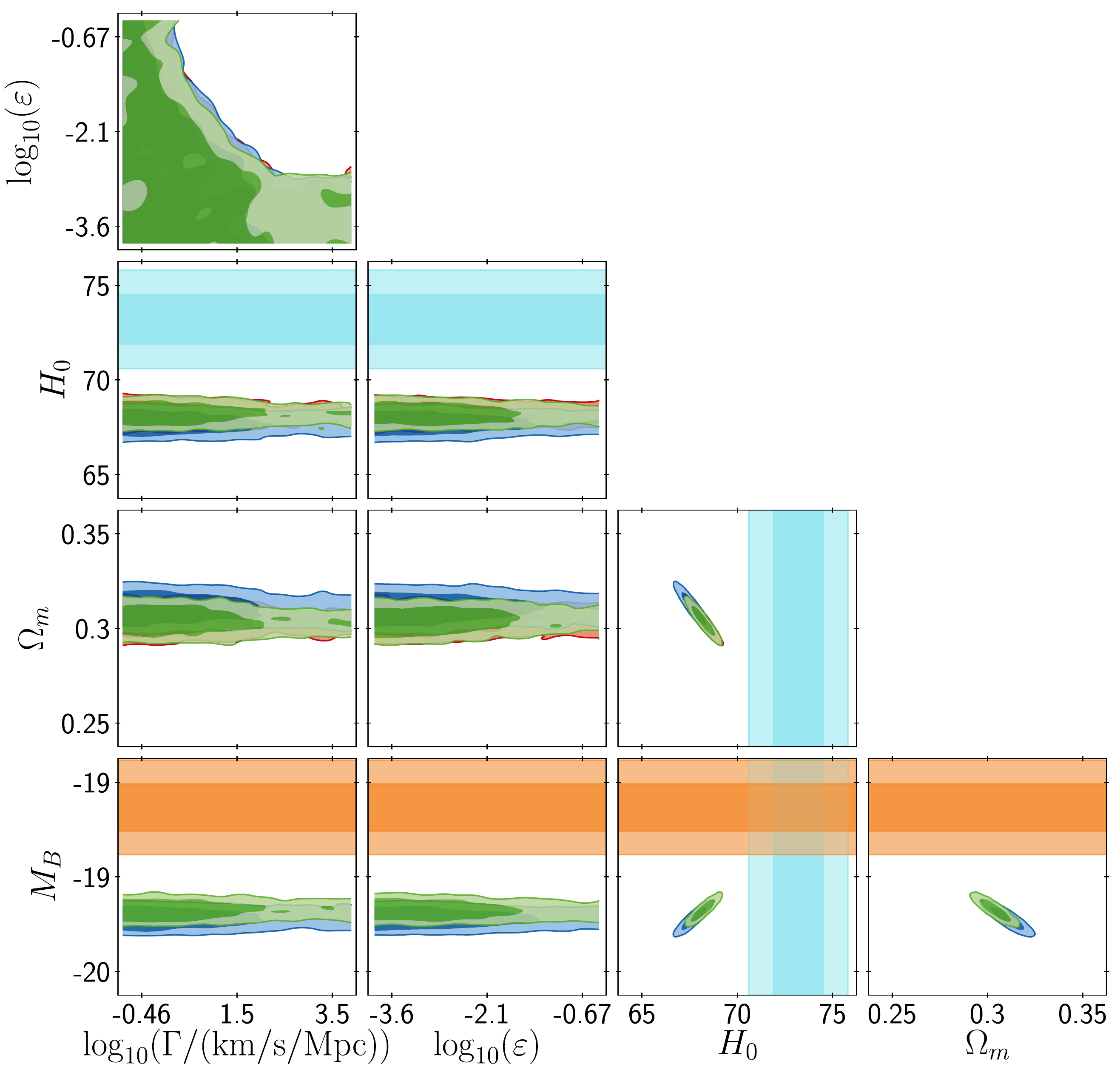}
     \caption{Decaying DM to DR and WDM: {\it (Left)} 2D contours on ($\Omega_{\rm m}$, $H_0$, $S_8$, $\log_{10}(\Gamma/[{\rm km/s/Mpc}])$, $\log_{10}(\epsilon)$), for Planck + BAO data. {\it (Right)} 2D contours on ($\Omega_{\rm m}$, $M_{\rm B}$, $H_0$, $\log_{10}(\Gamma/[{\rm km/s/Mpc}])$, $\log_{10}(\epsilon)$), for Planck +BAO + Pantheon and/or SH0ES (treated as a measurement of either $H_0$ or $M_B$).}
     \label{fig:dcdm}
 \end{figure*}
 
 \FloatBarrier
 \newpage

\newpage
\vspace*{-1cm}
\section{Test C: triangle plots \label{app:C}}  
 
\enlargethispage*{2cm}
\vspace*{-0.5cm}
\begin{figure}[H]
    \centering
    \includegraphics[height=1cm]{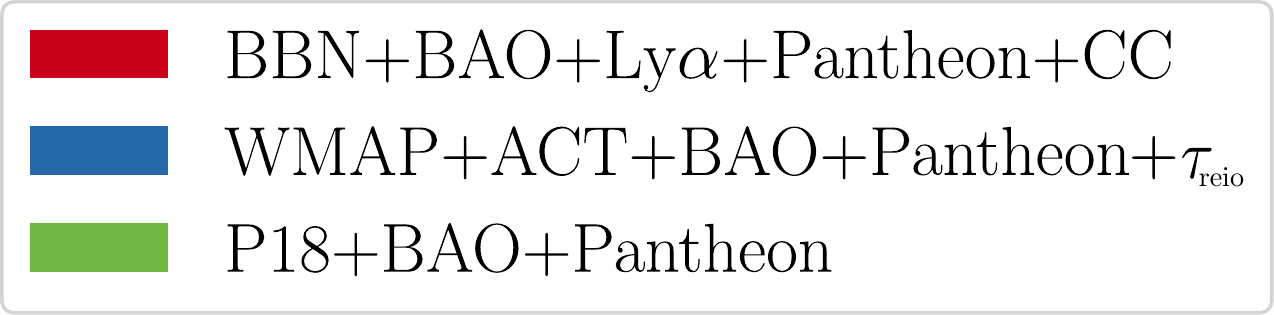}
\end{figure}
\vspace*{-0.5cm}

\subsection{Self-Interacting Dark Radiation}

\begin{figure}[h]
     \centering
     \includegraphics[height=6cm]{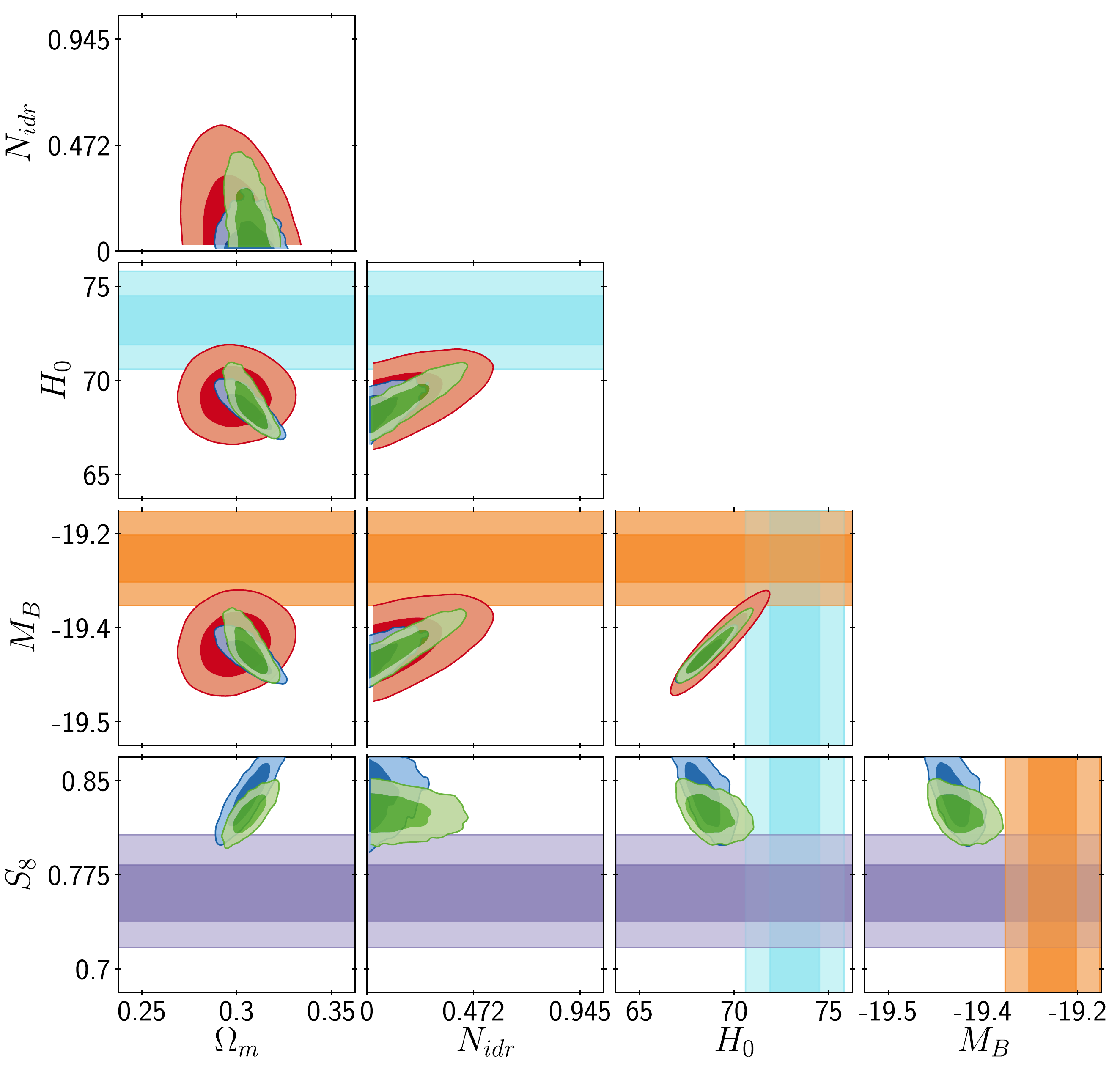}
     \caption{SIDR: 2D contours on ($\Omega_{\rm m}$, $H_0$, $M_{\rm B}$,  $\Delta N_{\rm idr}$) for a few data combinations not including Planck, compared with the results for our baseline data set (Planck + BAO + Pantheon). }
     \label{fig:SIDR_noPlanck}
 \end{figure}

\vspace*{-0.5cm}
\subsection{Majoron}

\begin{figure*}[h]
     \centering
     \includegraphics[height=9.7cm]{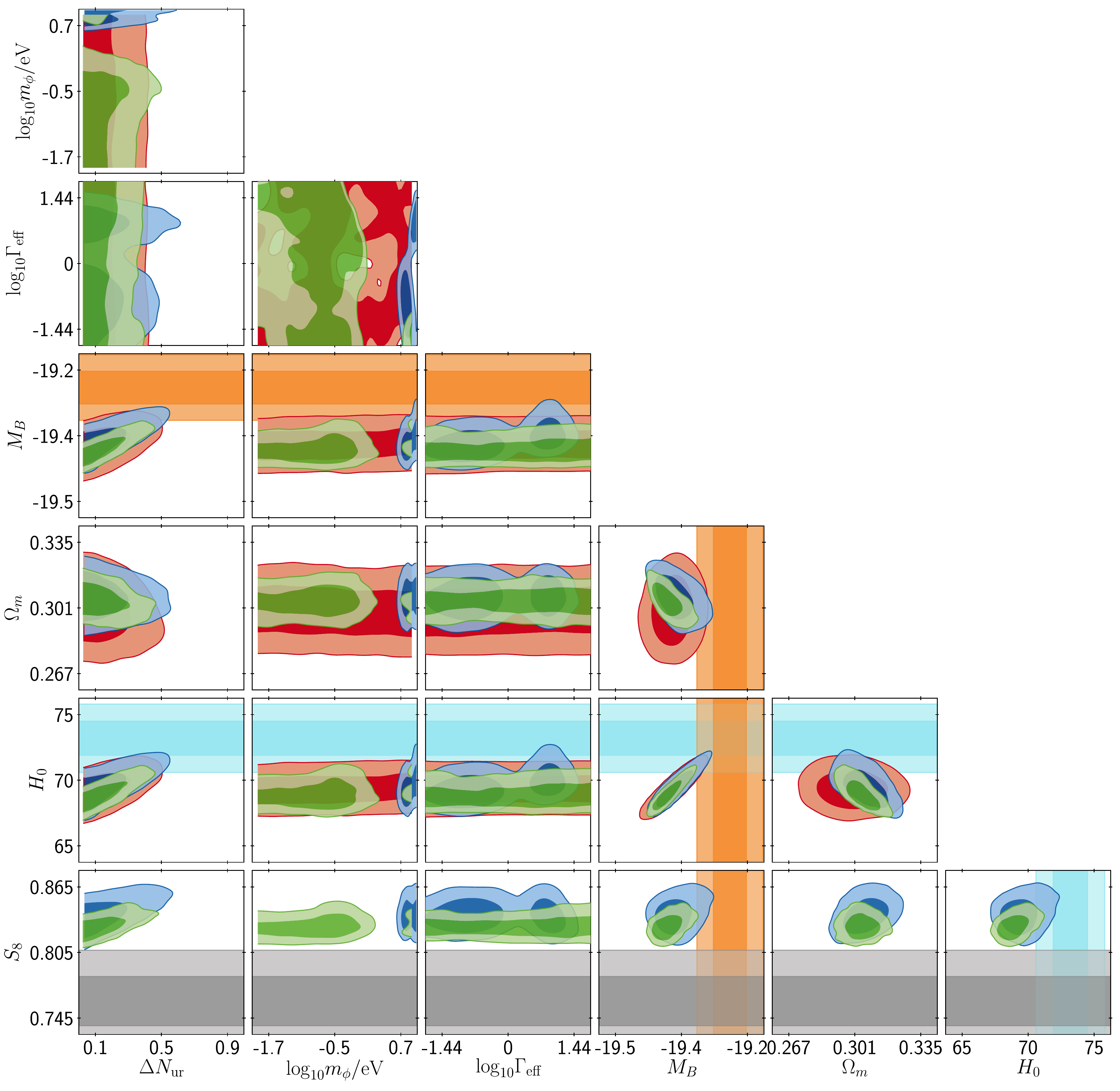}
    \setlength{\belowcaptionskip}{-2.5cm}
     \caption{Majoron: 2D contours on ($\Omega_{\rm m}$, $H_0$, $M_{\rm B}$,  $\Delta N_{\rm eff}$, $\log_{10} m_\phi$, $\log_{10} g_{\rm eff}$) for a few data combinations not including Planck, compared with the results for our baseline data set (Planck + BAO + Pantheon). }
    
     \label{fig:majoron_noP}
 \end{figure*}
 
\FloatBarrier
\pagebreak

 \vspace*{-0.5cm}
\enlargethispage*{1cm}
\begin{figure}[H]
    \centering
    \includegraphics[height=1.5cm]{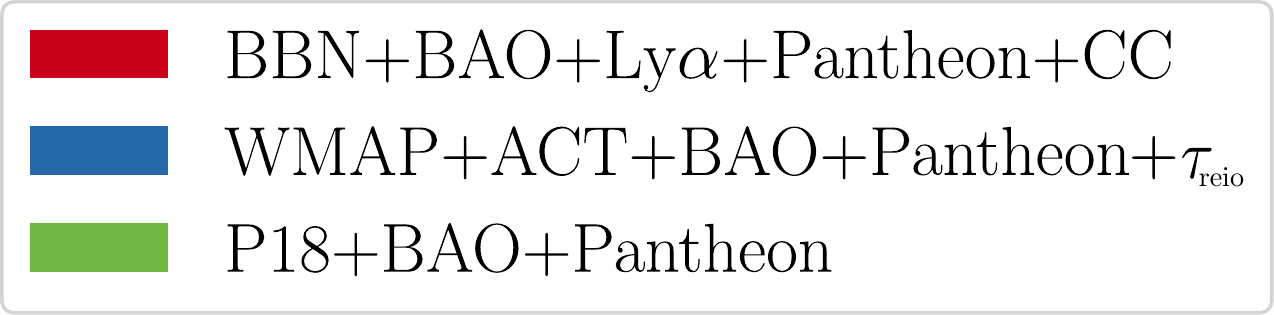}
\end{figure}
 \vspace*{-0.5cm}

\subsection{Primordial magnetic fields }

\begin{figure*}[h]
     \centering
     \includegraphics[height=6cm]{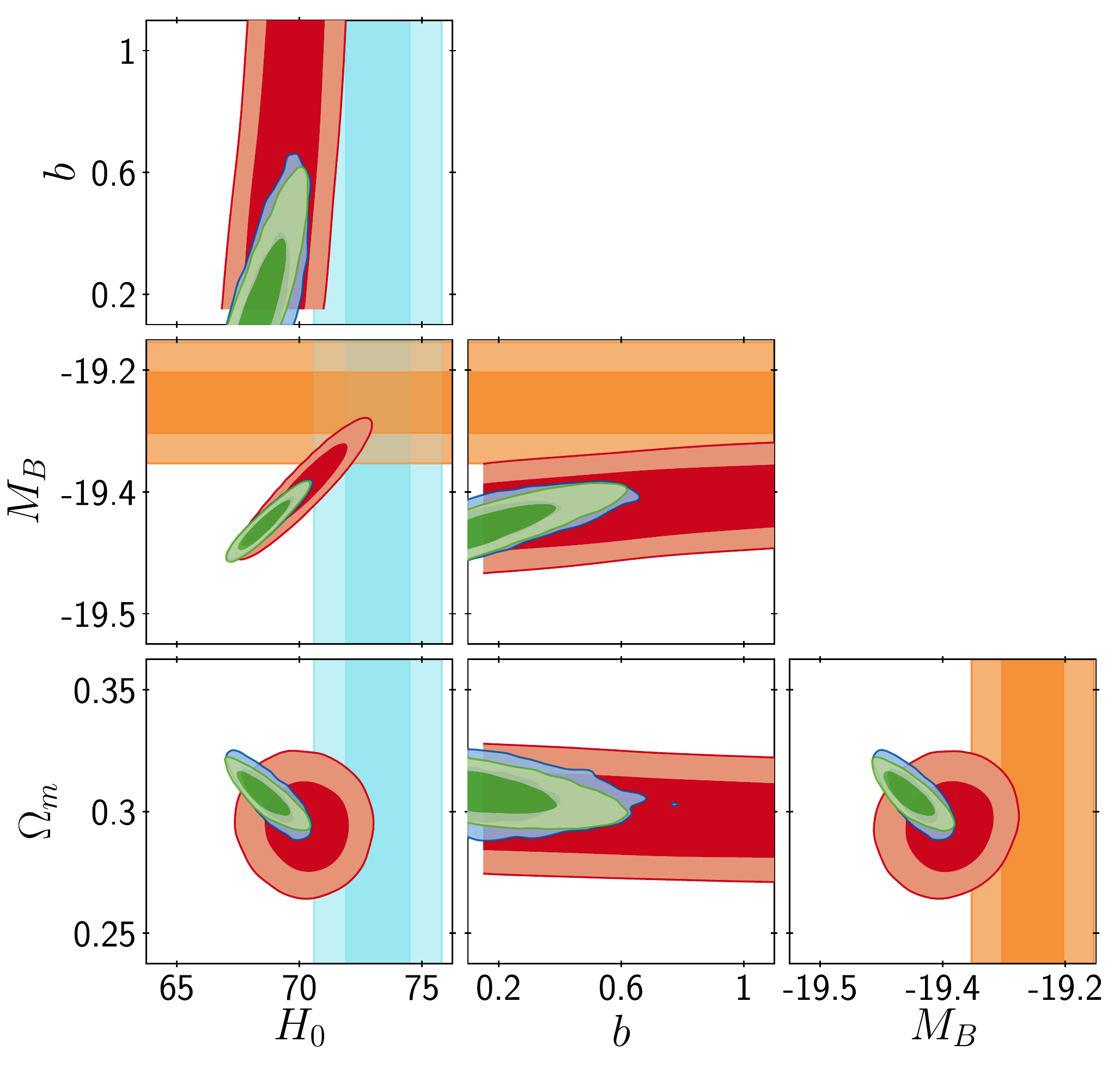}
     \caption{\Bshort: 2D contours on ($\Omega_{\rm m}$, $H_0$, $M_{\rm B}$, $S_8$, $b$) for a few data combinations not including Planck, compared with the results for our baseline data set (Planck + BAO + Pantheon).
     }
     \label{fig:primb_noP}
 \end{figure*}

\subsection{Varying effective electron mass}
\begin{figure}[h]
    \centering
    \includegraphics[height=6.5cm]{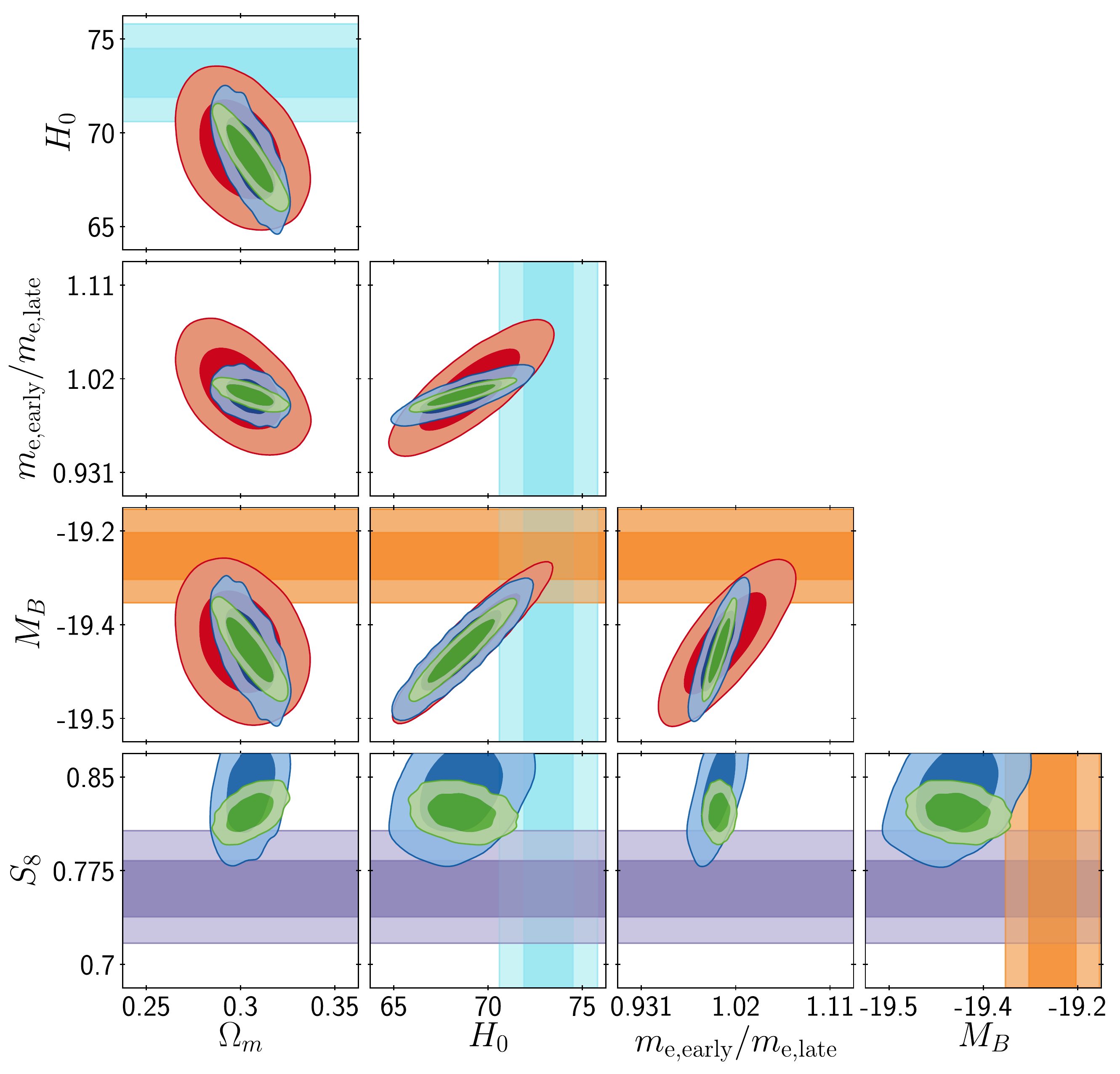}
    \caption{Varying $m_e$: 2D contours on ($\Omega_{\rm m}$, $H_0$, $M_{\rm B}$, $S_8$, $m_e$) for a few data combinations not including Planck, compared with the results for our baseline data set (Planck + BAO + Pantheon).}
    \label{fig:varme_noP}
\end{figure}

 \FloatBarrier
\pagebreak

\enlargethispage*{3cm}
\begin{figure}[H]
    \centering
    \includegraphics[height=1.5cm]{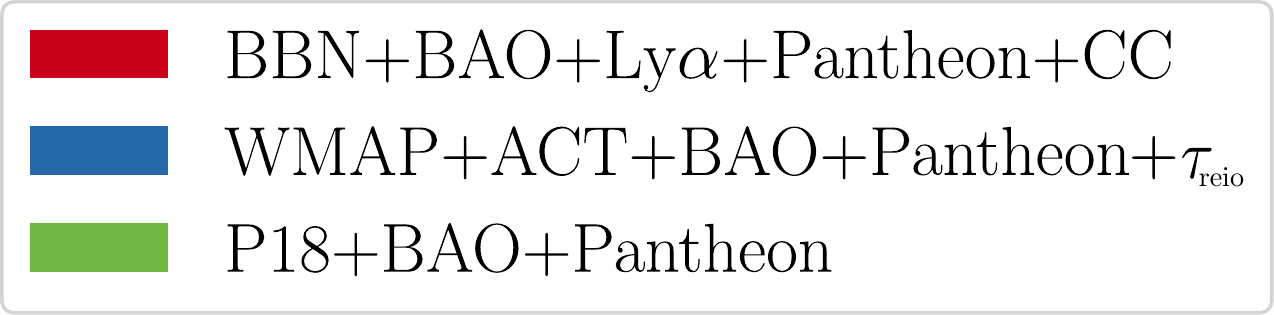}
\end{figure}
\vspace*{-1cm}
\subsection{Varying effective electron mass in curved universe}

\begin{figure}[h]
    \centering
    \includegraphics[height=7cm]{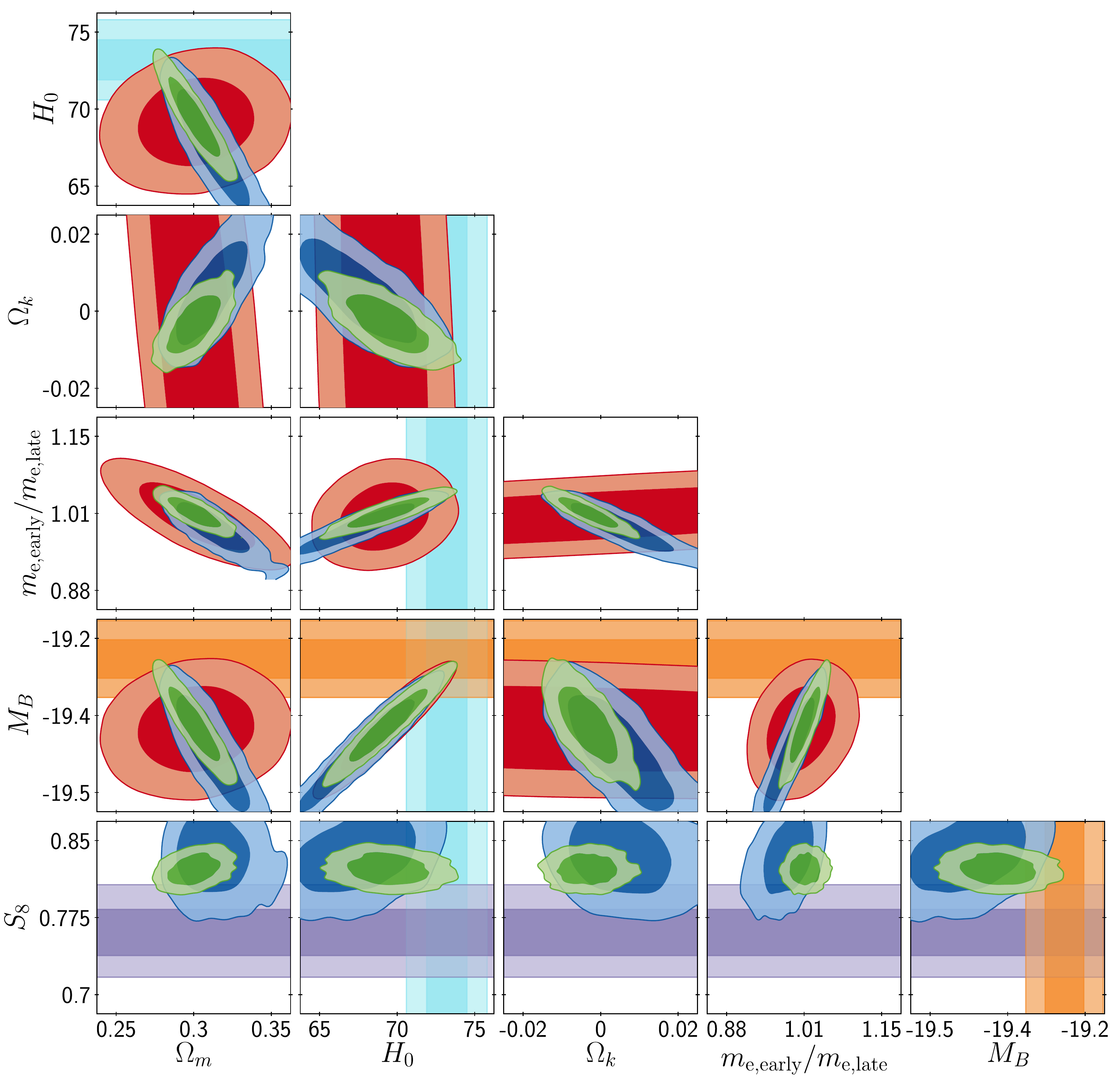}
    \caption{Varying $m_e$+$\Omega_k$: 2D contours on ($\Omega_{\rm m}$, $H_0$, $M_{\rm B}$, $S_8$, $m_e$, $\Omega_k$) for a few data combinations not including Planck, compared with the results for our baseline data set (Planck + BAO + Pantheon).}
    \label{fig:varmeok_noP}
\end{figure}

\subsection{Early Dark Energy}
\begin{figure}[h]
     \centering
     \hspace*{-1.5cm}\includegraphics[height=9cm]{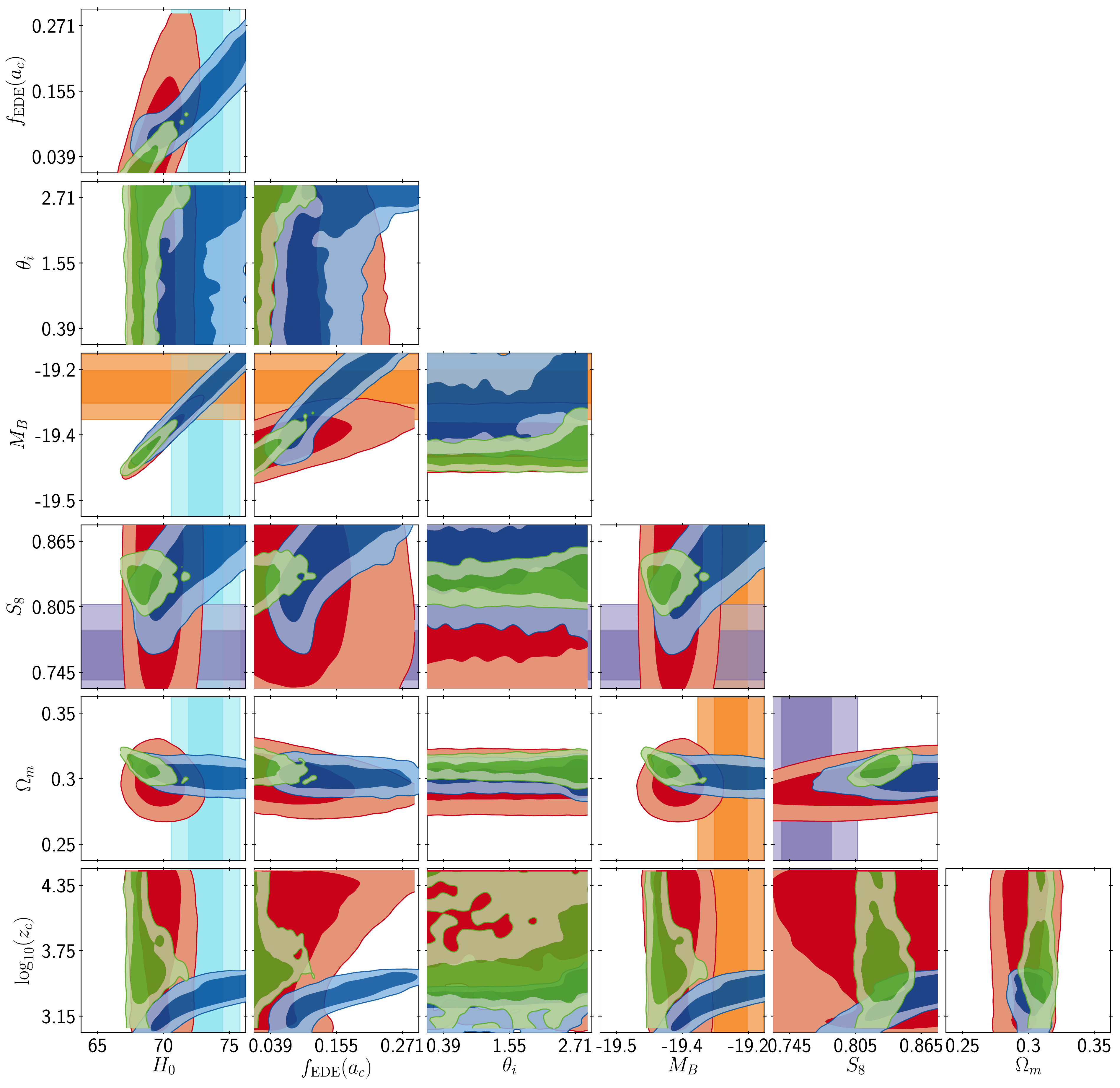}
    \setlength{\belowcaptionskip}{-2.5cm}
     \caption{\EDEmid{}: 2D contours on ($\Omega_{\rm m}$, $H_0$, $M_{\rm B}$, $S_8$, $f_{\rm EDE}$, $\log_{10}(z_c)$, $\theta_i$) for a few data combinations not including Planck, compared with the results for our baseline data set (Planck + BAO + Pantheon).
     }
     \label{fig:ede_noP}
 \end{figure}
 
 \FloatBarrier
\pagebreak

\enlargethispage*{3cm}
\begin{figure}[H]
    \centering
    \includegraphics[height=1.5cm]{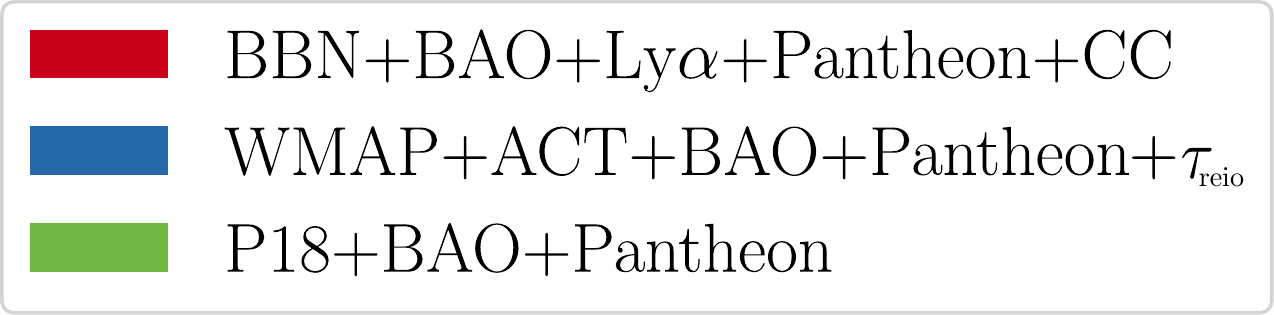}
\end{figure}
\vspace*{-1cm}

\subsection{New Early Dark Energy}

\begin{figure}[h]
     \centering
     \hspace*{-1.5cm}\includegraphics[height=8cm]{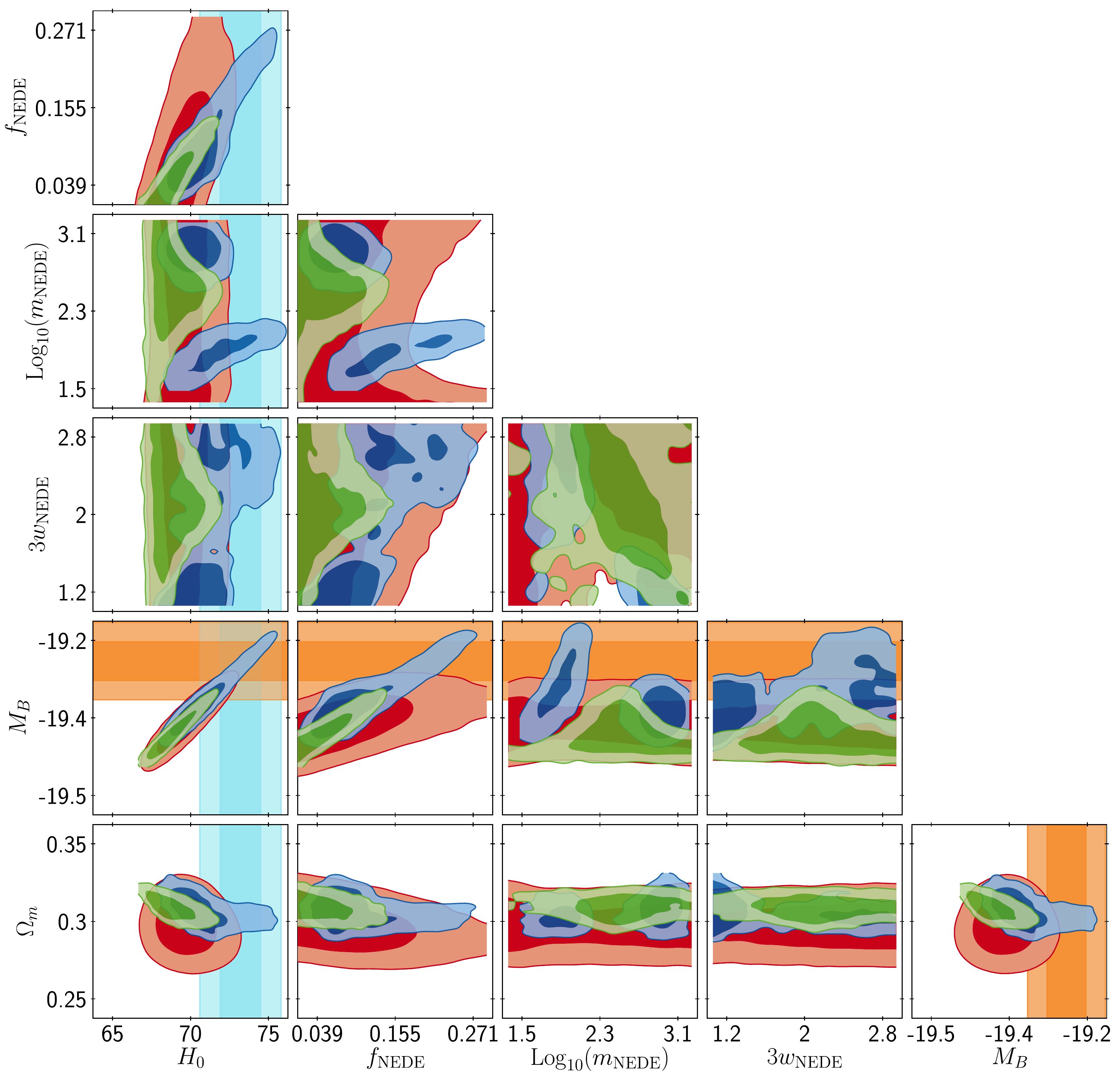}
     \caption{\NEDEmid{}: 2D contours on ($\Omega_{\rm m}$, $H_0$, $M_{\rm B}$, $S_8$, $f_{\rm NEDE}$, $\log_{10}(m_\mathrm{NEDE}/{\rm eV})$, $3w_{\rm NEDE}$) for a few data combinations not including Planck, compared with the results for our baseline data set (Planck + BAO + Pantheon).
     }
     \label{fig:ede_noP}
 \end{figure}
 
\vspace*{-0.5cm}
\subsection{Early Modified Gravity }

\begin{figure}[H]
     \centering
     \hspace*{-1.5cm}\includegraphics[height=8cm]{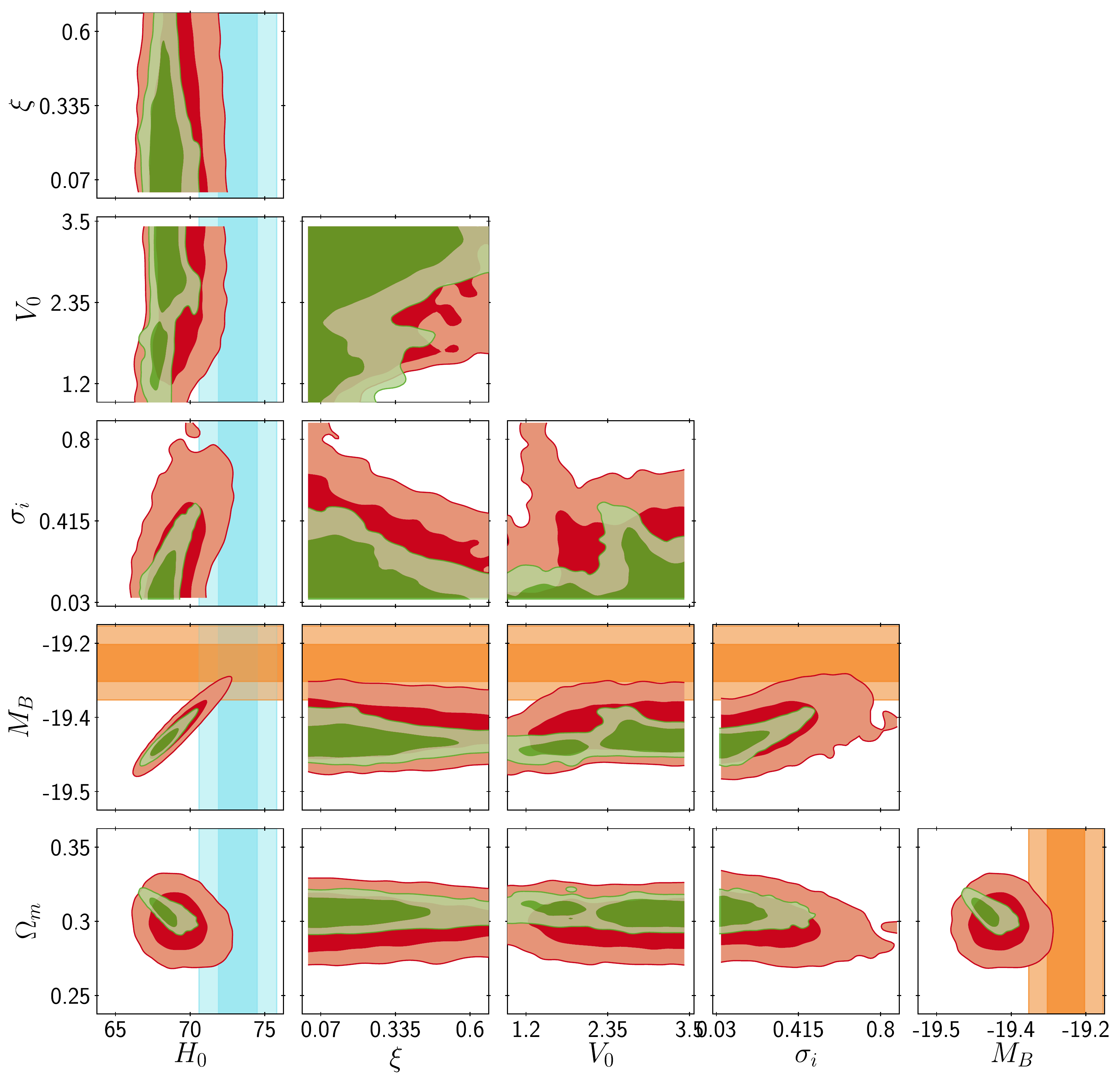}
     \caption{\EMGmid{}: 2D contours on ($\Omega_{\rm m}$, $H_0$, $M_{\rm B}$, $S_8$,  $\xi$, $V_0$, $\sigma_i$) for a few data combinations not including Planck, compared with the results for our baseline data set (Planck + BAO + Pantheon).
     \vspace*{-1cm}}
     \label{fig:emg_noP}
\end{figure}
 
\FloatBarrier
\pagebreak
%
%
\pagebreak
 
\newpage
\vspace*{-1cm}

\section{Test D: triangle plots \label{app:D}} 

\enlargethispage*{3cm}
\vspace*{-0.5cm}
\begin{figure}[H]
    \centering
    \includegraphics[height=2cm]{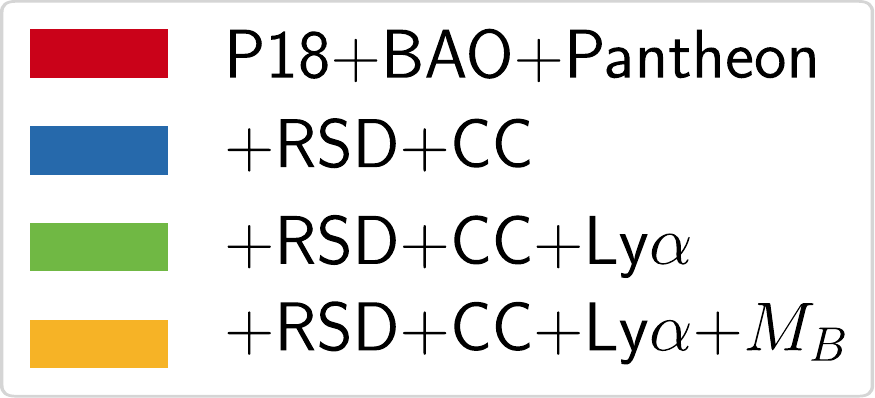}
\end{figure}
\vspace*{-0.5cm}

\subsection{Self-Interacting Dark Radiation}

\begin{figure}[h]
     \centering
     \includegraphics[height=6cm]{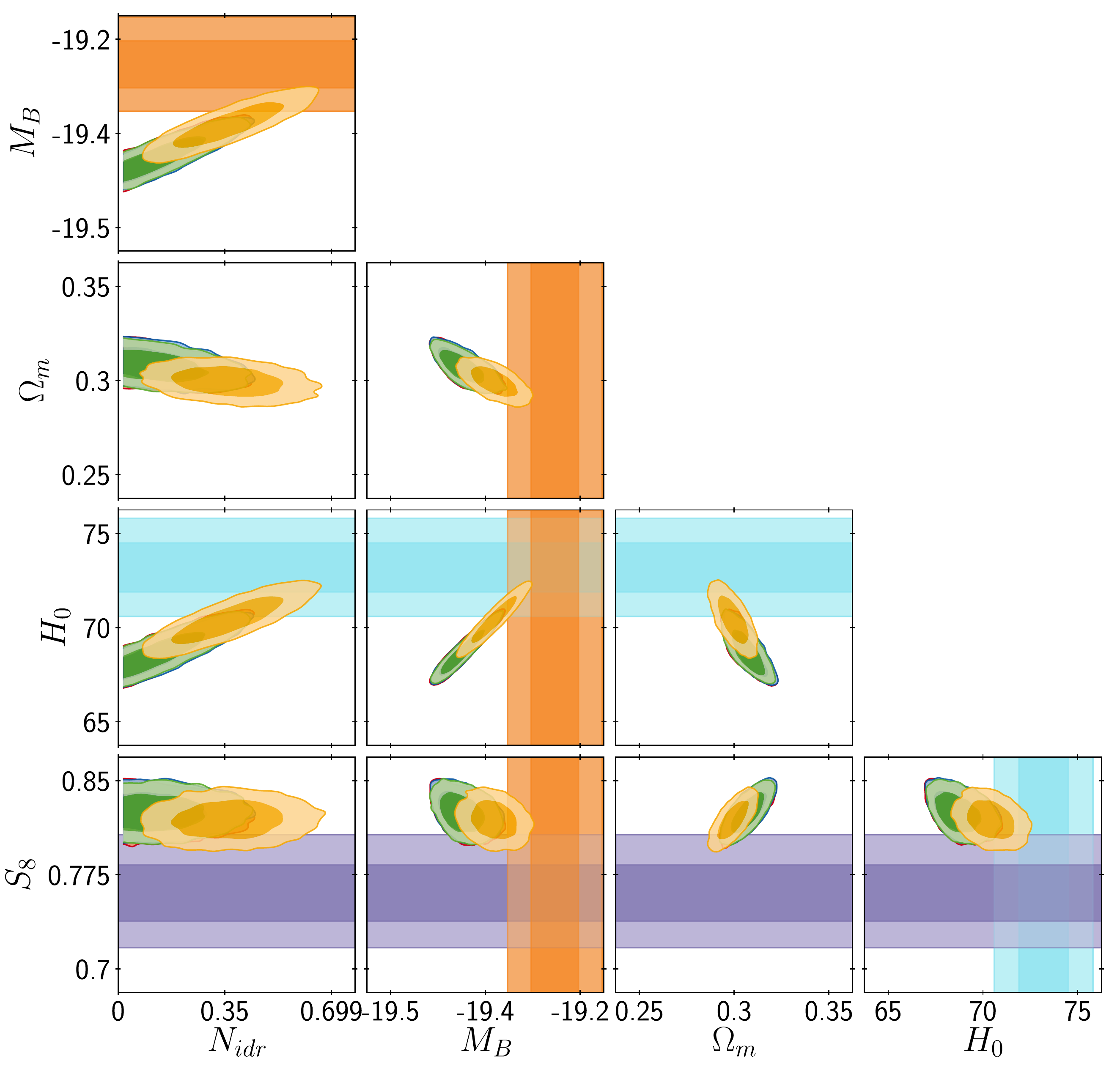}
     \caption{SIDR: 2D contours on ($\Omega_{\rm m}$, $H_0$, $M_{\rm B}$,  $\Delta N_{\rm idr}$) for the baseline data set Planck + BAO + Pantheon and with additional data: Redshift Space Distortions, Cosmic Chronometers, BAO-Ly$\alpha$, SH0ES (treated a  measurement of $M_{\rm B}$). }
    
     \label{fig:SIDR_extended}
 \end{figure}

\subsection{Majoron}
 
\begin{figure}[h]
    \centering
    \includegraphics[height=8cm]{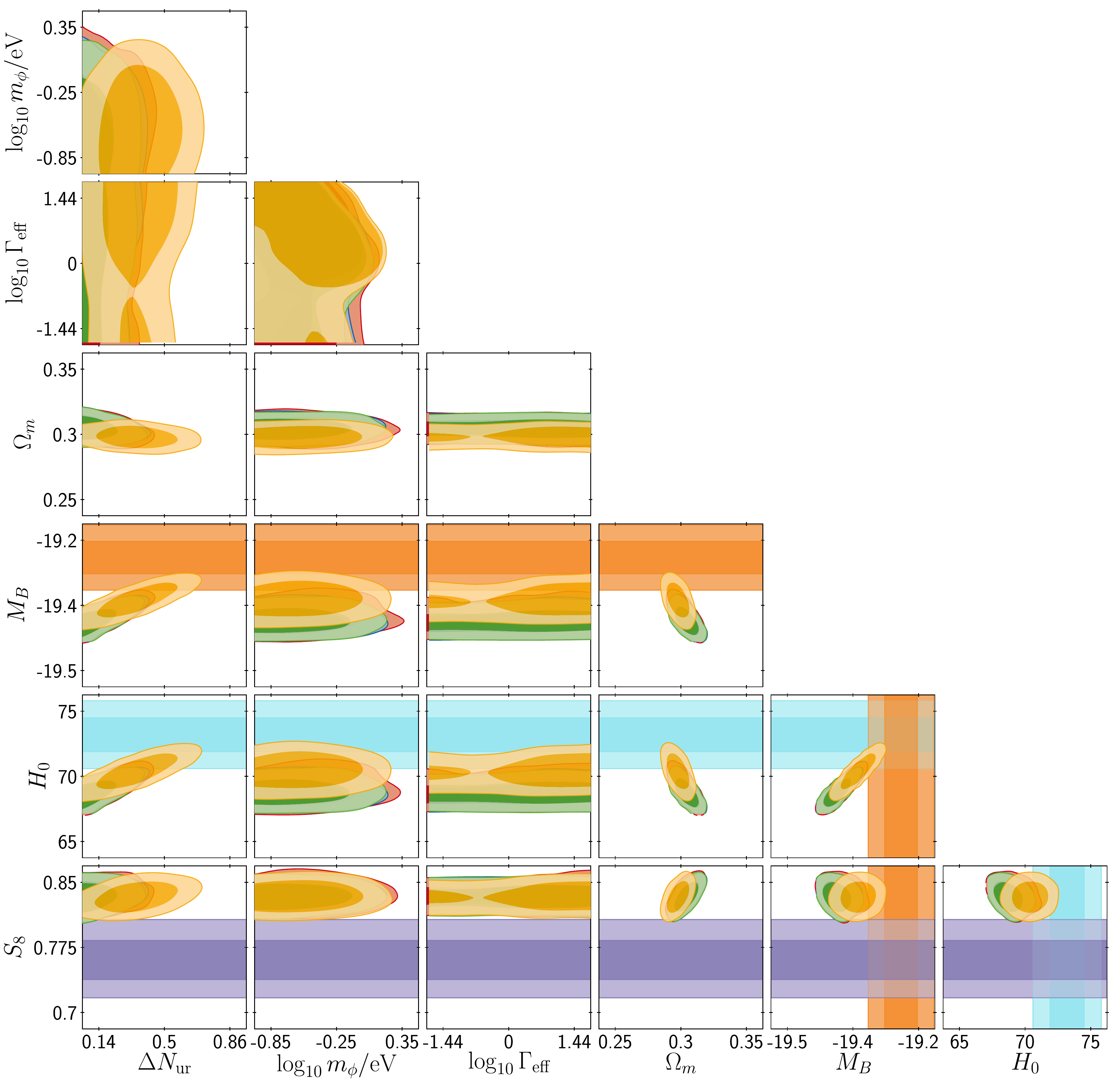}
    \setlength{\belowcaptionskip}{-2.5cm}
    \caption{Majoron: 2D contours on ($\Omega_{\rm m}$, $H_0$, $M_{\rm B}$, $S_8$, $\Delta N_{\rm eff}$, $\log_{10} m_\phi$, $\log_{10} g_{\rm eff}$), for the baseline data set Planck + BAO + Pantheon and with additional data: Redshift Space Distortions, Cosmic Chronometers, BAO-Ly$\alpha$, SH0ES (treated a  measurement of $M_{\rm B}$).}
    \label{fig:Maj_sn_gold}
\end{figure}

\FloatBarrier

\enlargethispage*{2cm}
\pagebreak

\begin{figure}[H]
    \centering
    \includegraphics[height=1.5cm]{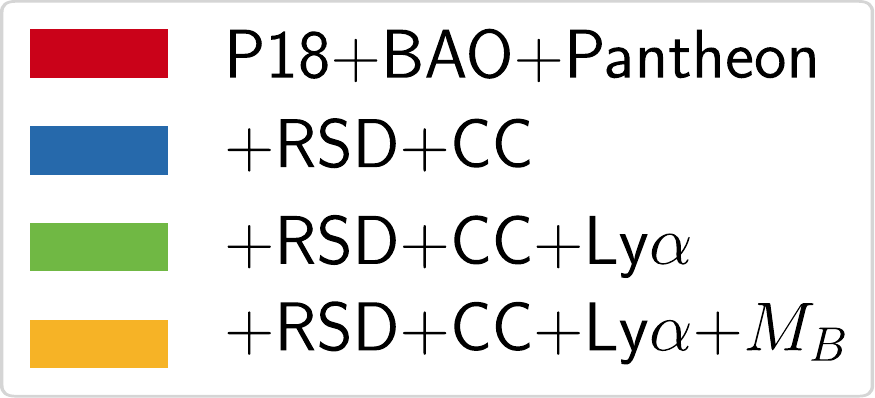}
\end{figure}
\vspace*{-0.5cm}
\subsection{Primordial magnetic fields }
\vspace*{-0.3cm}

\begin{figure}[h]
	\centering
	\includegraphics[height=6cm]{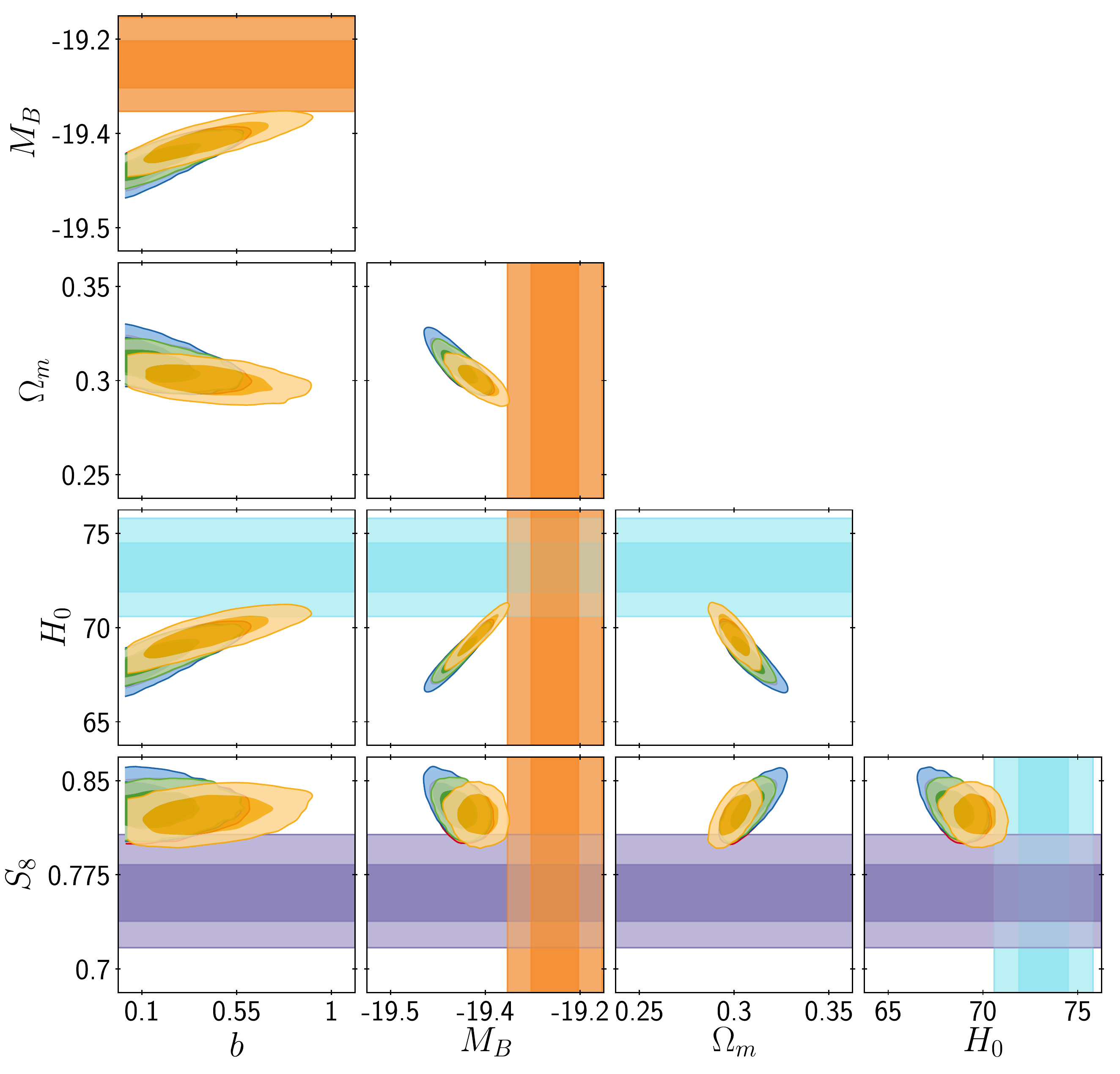}
	\caption{\Bshort: 2D contours on ($\Omega_{\rm m}$, $H_0$, $M_{\rm B}$, $S_8$, $b$), for the baseline data set Planck + BAO + Pantheon and with additional data: Redshift Space Distortions, Cosmic Chronometers, BAO-Ly$\alpha$, SH0ES (treated a  measurement of $M_{\rm B}$).}
	\label{3-zones_3}
\end{figure}

\FloatBarrier

\subsection{Varying effective electron mass}

\begin{figure}[h]
    \centering
    \includegraphics[height=6.5cm]{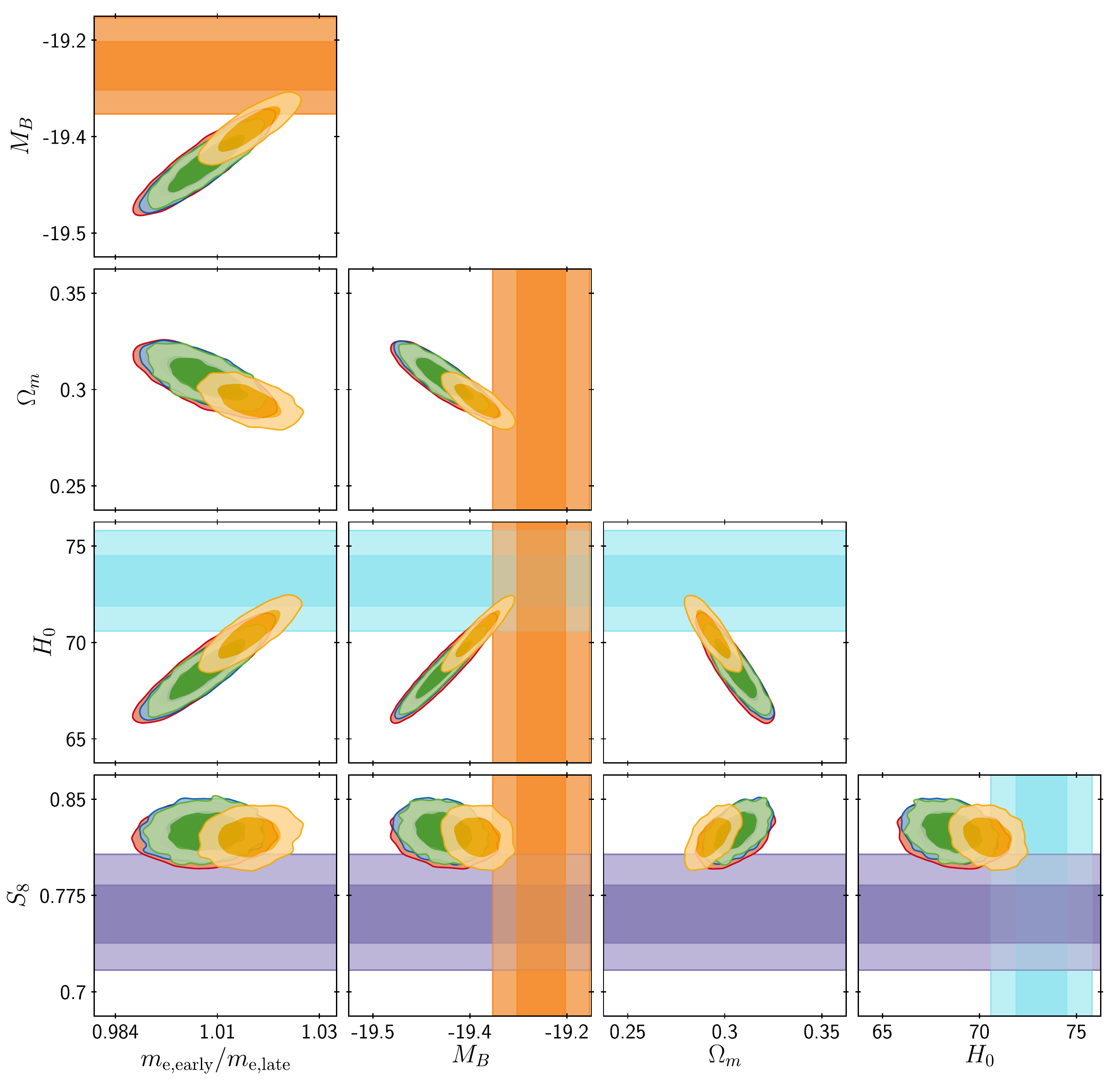}
    \caption{Varying $m_e$: 2D contours on ($\Omega_{\rm m}$, $H_0$, $M_{\rm B}$, $S_8$, $m_e$), for the baseline data set Planck + BAO + Pantheon and with additional data: Redshift Space Distortions, Cosmic Chronometers, BAO-Ly$\alpha$, SH0ES (treated a  measurement of $M_{\rm B}$).}
    \label{fig:varme_kids}
\end{figure}

\vspace*{-0.5cm}
\FloatBarrier
\pagebreak
\enlargethispage*{3cm}
\begin{figure}[H]
    \centering
    \includegraphics[height=1.5cm]{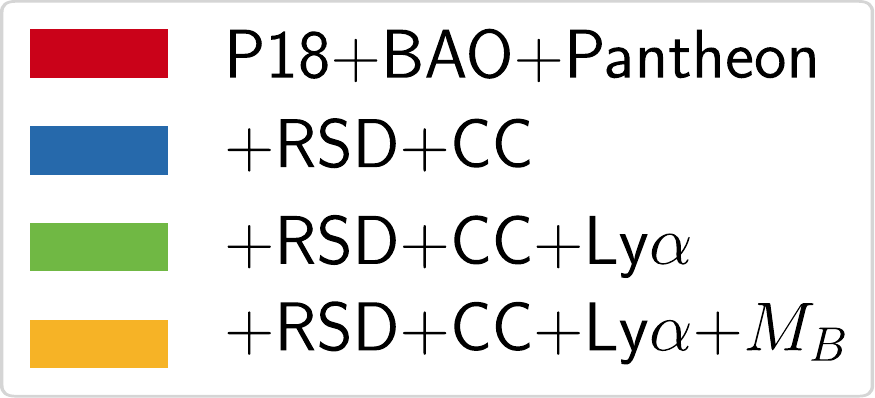}
\end{figure}
\vspace*{-1cm}

\subsection{Varying effective electron mass in curved universe}

\begin{figure}[h]
    \centering
    \includegraphics[height=8cm]{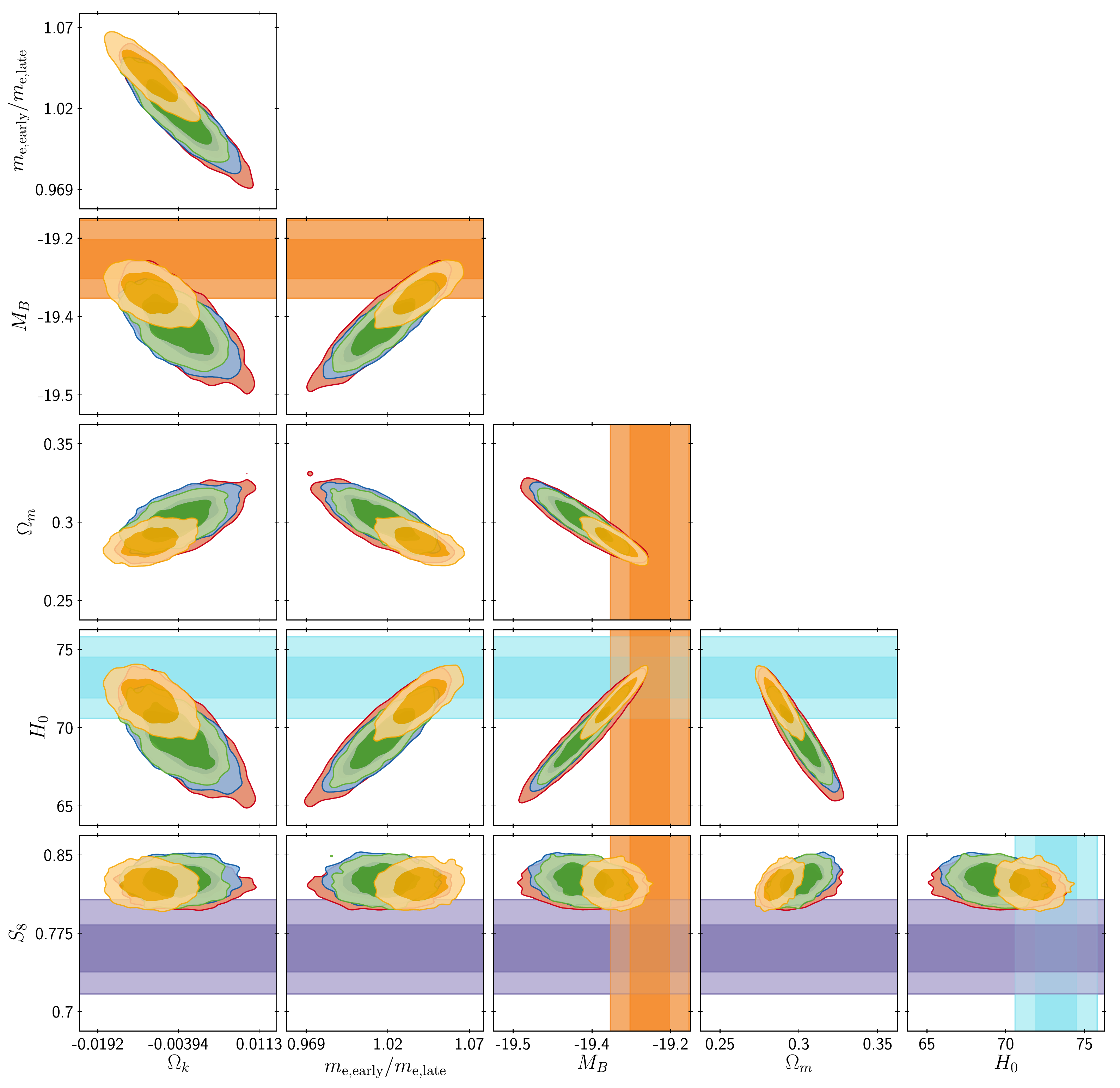}
    \caption{Varying $m_e$ + $\Omega_k$: 2D contours on ($\Omega_{\rm m}$, $H_0$, $M_{\rm B}$, $S_8$, $m_e$, $\Omega_k$), for the baseline data set Planck + BAO + Pantheon and with additional data: Redshift Space Distortions, Cosmic Chronometers, BAO-Ly$\alpha$, SH0ES (treated a  measurement of $M_{\rm B}$).
    }
    \label{fig:varmeok_kids}
\end{figure}

\vspace*{-0.5cm}
\subsection{Early Dark Energy}

\begin{figure}[h]
    \centering
    \includegraphics[height=8cm]{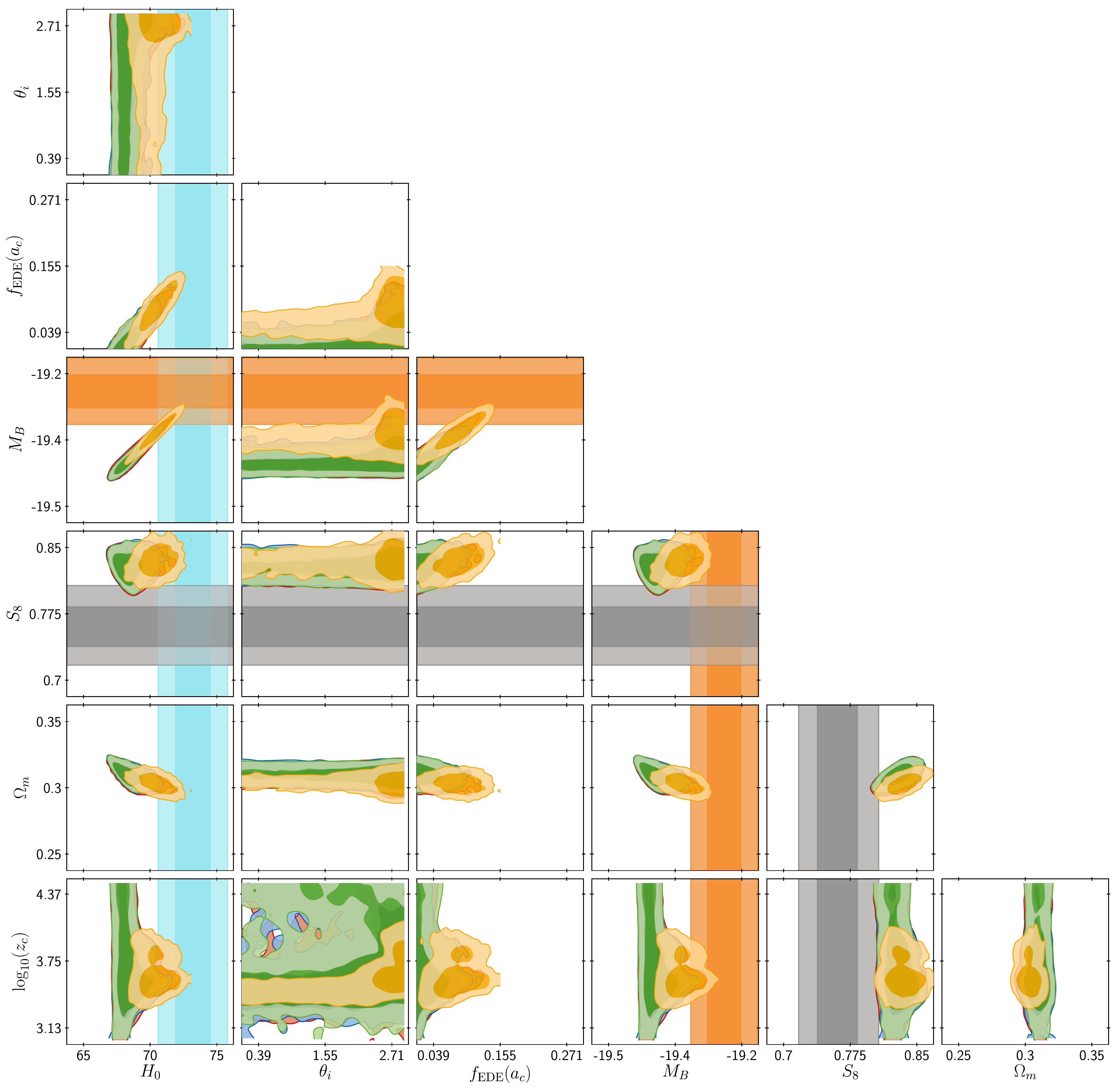}
    \setlength{\belowcaptionskip}{-4cm}
    \caption{Early Dark Energy: : 2D contours on ($\Omega_{\rm m}$, $H_0$, $M_{\rm B}$, $S_8$, $f_{\rm EDE}$, $\log_{10}(z_c)$, $\theta_{\rm ini}$), for the baseline data set Planck + BAO + Pantheon and with additional data: Redshift Space Distortions, Cosmic Chronometers, BAO-Ly$\alpha$, SH0ES (treated a  measurement of $M_{\rm B}$).}
    \label{fig:ede_kids}
\end{figure}

\FloatBarrier
\pagebreak
\begin{figure}[H]
    \centering
    \includegraphics[height=1.5cm]{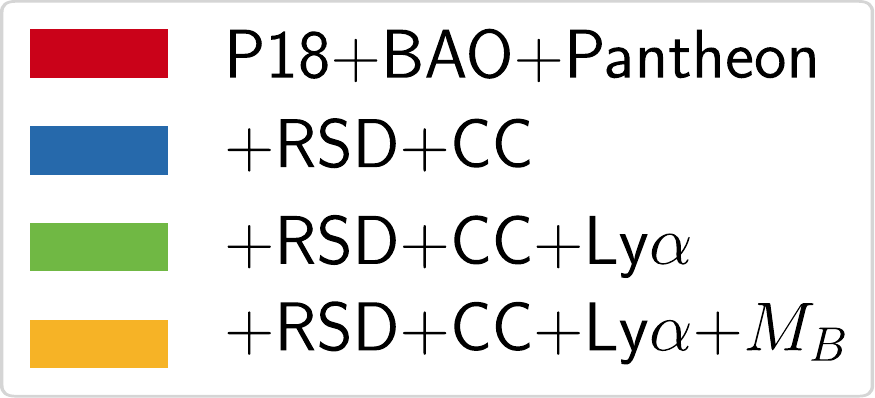}
\end{figure}
\vspace*{-1cm}

\enlargethispage*{2cm}
\subsection{New Early Dark Energy}

\begin{figure}[h]
    \centering
    \includegraphics[height=8cm]{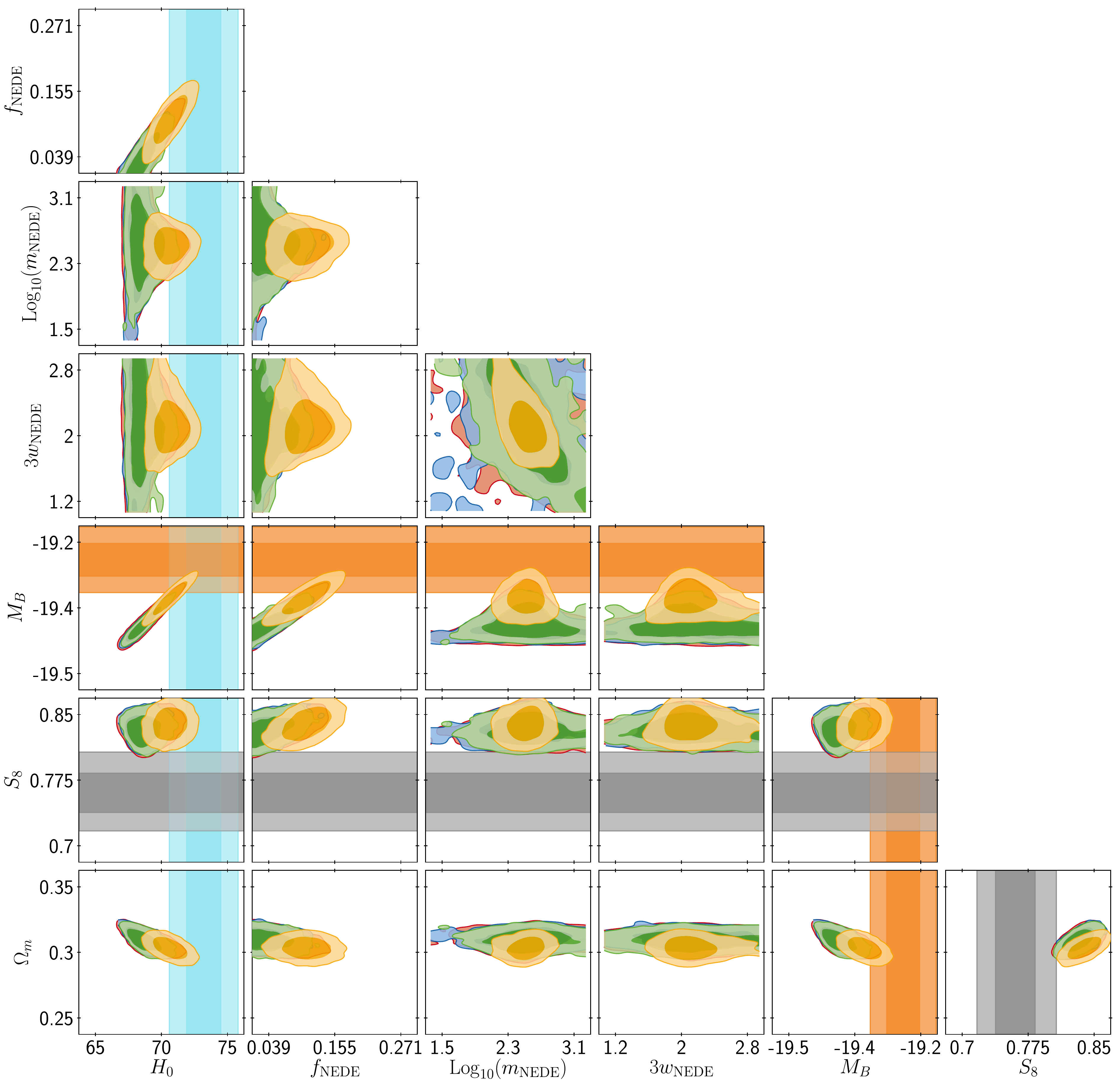}
    \caption{
    New Early dark Energy: 2D contours on ($\Omega_{\rm m}$, $H_0$, $M_{\rm B}$, $S_8$, $f_{\rm NEDE}$, $\log_{10}(m_\mathrm{NEDE}/{\rm eV})$, $3w_{\rm NEDE}$), for the baseline data set Planck + BAO + Pantheon and with additional data: Redshift Space Distortions, Cosmic Chronometers, BAO-Ly$\alpha$, SH0ES (treated a  measurement of $M_{\rm B}$).}
    \label{fig:nede_kids}
\end{figure}

\vspace*{-0.5cm}
\subsection{Early Modified Gravity }

\begin{figure}[H]
    \centering
    \includegraphics[height=8cm]{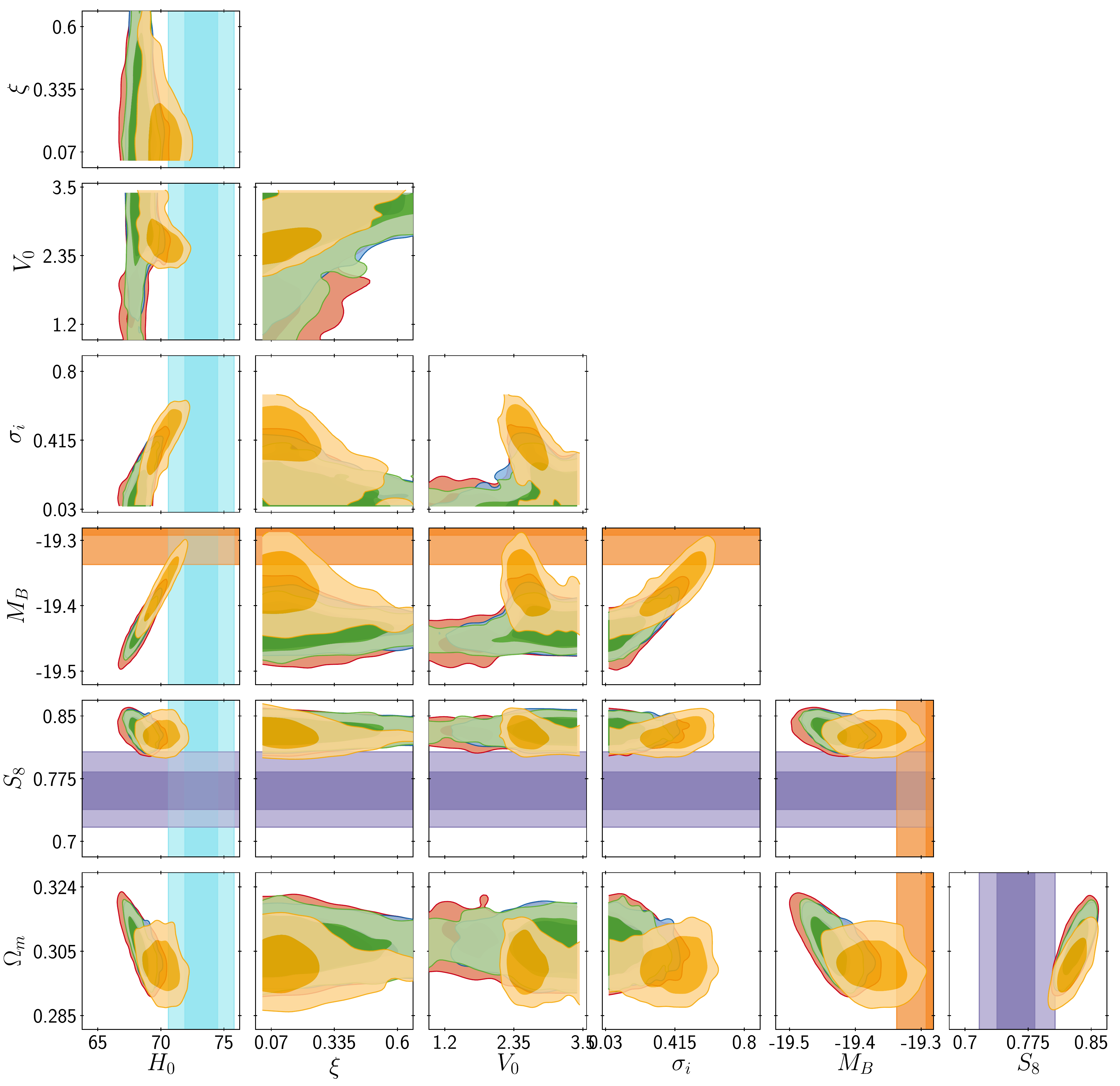}
    \caption{
    Early Modified Gravity : 2D contours on ($\Omega_{\rm m}$, $H_0$, $M_{\rm B}$, $S_8$, $\xi$, $V_0$, $\sigma_i$), for the baseline data set Planck + BAO + Pantheon and with additional data: Redshift Space Distortions, Cosmic Chronometers, BAO-Ly$\alpha$, SH0ES (treated a  measurement of $M_{\rm B}$).}
    \label{fig:emg_kids}
\end{figure}

\FloatBarrier
\pagebreak



\newpage

\section{Implementation details \label{app:E}}
In this last section we aim to give a few details on the used likelihoods and data in order to enhance the reproducibility. Additionally, we discuss the choice of using a non-linear correction code such as {\tt halofit} \cite{Takahashi:2012em,Ali-Haimoud:2012fzp}.
\subsection{Code and likelihood versions}
For the minimization of the $\chi^2$ values considered in this analysis, we attempted to use the minimization algorithm Migrad from the Minuit suite~\cite{James:1975dr,iminuit}. However, we found that such a minimization depending on gradient descent (computed from numerical derivatives) appears to quite easily get stuck in local minima, while methods like Nelder-Mead minimization have been found to be comparatively slow. Here instead, a modified Metropolis-Hastings method has been adopted for the minimization. In this method, one runs multiple small chains ($\sim 1000$points)  with successively decreasing step-size (specified with the parameter {\tt -f} within MontePython), and successively enhanced sensitivity to likelihood differences. The latter is accomplished by a simple modification of the Metropolis-Hastings algorithm. Instead of computing the acceptance probability as $p = \exp(-\Delta \ln \mathcal{L})$, we simply enhance the likelihood difference by a factor $p' = \exp(-F \Delta \ln \mathcal{L})$, where we adopt different factors $F$ in accordance with the proposal stepsize. This allows us to avoid getting trapped in local minima, while also being sensitive to small differences in the likelihood, thus reaching the minimum more effectively than in a normal MCMC approach. In many cases, the minimum found by this modified approach surpassed the minimum initially found by the Migrad algorithm (by $\Delta \chi^2 \sim -1$ or more).

For this work, modified versions of {\tt class} \cite{Lesgourgues:2011re,Blas:2011rf} have been used, with most modifications linked to in the text. For visibility, we summarized the use of models and corresponding {\tt class} versions in \cref{tab:classversions}. We have used the MontePython version {\tt v3.3} throughout \cite{Audren:2012wb,Brinckmann:2018cvx}. However, for the purpose of this work we have additionally coded several likelihood versions (update of the cosmic clocks, RSD likelihood, SH0ES prior likelihood, updated $\tau_\mathrm{reio}$ prior for ACT), which are available upon request.

\begin{table}[h]
    \centering
    \hspace*{-1em}
    \small
    \resizebox{0.85\textwidth}{3.25cm}{
    \begin{tabular}{c|c c}
         Name & {\tt class} version & public? \\ \hline
         \lcdmtable &\texttt{v2.9.4} & yes \\
         \nefftable &\texttt{v2.9.4} & yes \\
         \idrtable &\texttt{v2.9.4} & yes \\
         \schmaltztable &\texttt{v2.9.4} & yes \\
         \equitable &\texttt{v2.9.4} & yes \\
         \SINtable & \url{https://github.com/PoulinV/class_interacting_neutrinos}  & yes  \\
         \majorontable & modified \texttt{v2.9.4}  & to be released \\
         \primtable & \url{https://github.com/GuillermoFrancoAbellan/class_clumpy} & yes \\
         \varyingmetable & \texttt{v3.1} & yes \\
         \varyingmeomegaktable & \texttt{v3.1} & yes \\ 
         \edetable & \url{https://github.com/PoulinV/AxiCLASS} & yes \\
         \nedetable & \url{https://github.com/flo1984/TriggerCLASS} & yes \\
         \emgtable & modified \texttt{HiCLASS} \texttt{v2.7.2} & on request \\
         \cpltable &\texttt{v2.9.4} & yes \\
         \pedetable & modified \texttt{v2.9.4} & on request \\
         \medetable& modified \texttt{v2.9.4} & on request \\
         \fracdmdecaytable & modified \texttt{v2.9.4}  & on request \\
         \dmdecaytable & \url{https://github.com/PoulinV/class_decays} & yes \\
    \end{tabular}
    }
    \caption{Summary of the publication status of the different {\tt class} versions employed within this work.}
    \label{tab:classversions}
\end{table}

\subsection{Results with Halofit}

The use of non-linear fitting codes like {\tt halofit} \cite{Takahashi:2012em,Ali-Haimoud:2012fzp} is in general not recommended for extensions of the $\Lambda$CDM model. As such, all results presented within this work are generally shown \emph{without} including {\tt halofit}. However, Planck data are sensitive even to small changes of the lensing power induced by non-linear corrections of the matter power spectrum. As such, not using {\tt halofit} incurs small systematic differences in the likelihoods and best-fits. For this purpose, we have decided to recreate our main results with data-set \DB~using the {\tt halofit} routine to predict the non-linear corrections relevant for CMB lensing. The results are shown in \cref{tab:halofit_Mb}. 
\enlargethispage*{2cm}
\begin{table}[]
    \centering
    \renewcommand{\arraystretch}{1.2}
    \hspace*{-0.5cm}
    \scalebox{0.85}{
    \begin{filecontents*}{sheets/MbTable_halofit.csv}
Name,Npar,Mbstring,TensVal,Tension,Chi2,DChi2 wrt LCDM,AIK wrt LCDM,Passes DAIK Mb,String DAIK,Passes Sigma Mb,String SigmaH0,Passes both Mb,Chi2 wrt noSH0ES,Tension,Tensval,Passes Chi2,PassChi2,AIC or Chi2,PassBoth,Delta Crit 1 Crit 3,AIC AND Chi2,Passes all,Medal
\lcdmtable,0,-19.416 \pm 0.012,4.4092850522997,4.4 \sigma,3826.26,0.00,0.00,FALSE,\color{Red} X,FALSE,\color{Red} X,\color{Red} X ,19.18,4.37940293240793,4.4 \sigma,FALSE,\color{Red} X,FALSE,\color{Red} X,-0.006777089604626,FALSE,FALSE,\nomedal
\nefftable,1,-19.395 \pm 0.019,3.61131137333291,3.6 \sigma,3820.35,-5.92,-3.92,FALSE,\color{Red} X,FALSE,\color{Red} X,\color{Red} X ,13.09,3.61800968670896,3.6 \sigma,FALSE,\color{Red} X,FALSE,\color{Red} X,0.001854814687405,FALSE,FALSE,\nomedal
\idrtable,1,-19.385 \pm 0.024,3.19023185828377,3.2 \sigma,3817.77,-8.50,-6.50,FALSE,\color{Red} X,FALSE,\color{Red} X,\color{Red} X ,10.43,3.22927084274299,3.2 \sigma,FALSE,\color{Red} X,FALSE,\color{Red} X,0.012237036740087,FALSE,FALSE,\nomedal
\equitable,2,-19.413 \pm 0.036,3.27411706314673,3.3 \sigma,3819.86,-6.40,-2.40,FALSE,\color{Red} X,FALSE,\color{Red} X,\color{Red} X ,11.52,3.39400305669518,3.4 \sigma,FALSE,\color{Red} X,FALSE,\color{Red} X,0.036616281958233,FALSE,FALSE,\nomedal
\schmaltztable,2,-19.388 \pm 0.026,3.19807688448604,3.2 \sigma,3818.47,-7.80,-3.80,FALSE,\color{Red} X,FALSE,\color{Red} X,\color{Red} X ,11.30,3.36168450455487,3.4 \sigma,FALSE,\color{Red} X,FALSE,\color{Red} X,0.051158125954536,FALSE,FALSE,\nomedal
\SINtable,3,-19.440 \pm 0.038,3.69031817070912,3.7 \sigma,3822.69,-3.57,2.43,FALSE,\color{Red} X,FALSE,\color{Red} X,\color{Red} X ,16.67,4.08289113251873,4.1 \sigma,FALSE,\color{Red} X,FALSE,\color{Red} X,0.106379163977121,FALSE,FALSE,\nomedal
\majorontable,3,-19.380 \pm 0.027,2.96438781085131,3.0 \sigma,3812.57,-13.69,-7.69,TRUE,\color{Green} \checkmark,TRUE,\color{Green} \checkmark,\color{Green} \checkmark,7.96,2.82134719593318,2.8 \sigma,TRUE,\color{Green} \checkmark,TRUE,\color{Green} \checkmark,-0.048253003333274,TRUE,TRUE,\silvermedal
\primtable,1,-19.390^{+0.018}_{-0.024},3.53727518037797,3.5 \sigma,3817.01,-9.25,-7.25,TRUE,\color{Green} \checkmark,FALSE,\color{Red} X,\color{Green} \checkmark,12.79,3.57631094844959,3.6 \sigma,FALSE,\color{Red} X,TRUE,\color{Green} \checkmark,0.01103554744289,FALSE,FALSE,\onzemedal
\varyingmetable,1,-19.391 \pm 0.034,2.92248859480687,2.9 \sigma,3815.90,-10.37,-8.37,TRUE,\color{Green} \checkmark,TRUE,\color{Green} \checkmark,\color{Green} \checkmark,9.24,3.03929297757413,3.0 \sigma,TRUE,\color{Green} \checkmark,TRUE,\color{Green} \checkmark,0.039967438358807,TRUE,TRUE,\goldmedal
\varyingmeomegaktable,2,-19.368 \pm 0.048,2.04011204369655,2.0 \sigma,3810.58,-15.68,-11.68,TRUE,\color{Green} \checkmark,TRUE,\color{Green} \checkmark,\color{Green} \checkmark,4.08,2.01997046622197,2.0 \sigma,TRUE,\color{Green} \checkmark,TRUE,\color{Green} \checkmark,-0.009872780045008,TRUE,TRUE,\goldmedal
\edetable,3,-19.390^{+0.016}_{-0.033},3.60954610682971,3.6 \sigma,3806.24,-20.02,-14.02,TRUE,\color{Green} \checkmark,FALSE,\color{Red} X,\color{Green} \checkmark,3.04,1.74453432181772,1.7 \sigma,TRUE,\color{Green} \checkmark,TRUE,\color{Green} \checkmark,-0.516688727561383,TRUE,FALSE,\silvermedal
\nedetable,3,-19.380^{+0.021}_{-0.038},3.18886068128501,3.2 \sigma,3809.27,-16.99,-10.99,TRUE,\color{Green} \checkmark,FALSE,\color{Red} X,\color{Green} \checkmark,3.49,1.86884269000898,1.9 \sigma,TRUE,\color{Green} \checkmark,TRUE,\color{Green} \checkmark,-0.413946585695336,TRUE,FALSE,\silvermedal
\emgtable,3,-19.397^{+0.014}_{-0.026},3.85283348505724,3.9 \sigma,3809.92,-16.35,-10.35,TRUE,\color{Green} \checkmark,FALSE,\color{Red} X,\color{Green} \checkmark,5.66,2.37840913217215,2.4 \sigma,TRUE,\color{Green} \checkmark,TRUE,\color{Green} \checkmark,-0.382685719121646,TRUE,FALSE,\silvermedal
\cpltable,2,-19.396 \pm 0.020,3.61122032365849,3.6 \sigma,3821.69,-4.57,-0.57,FALSE,\color{Red} X,FALSE,\color{Red} X,\color{Red} X ,15.75,3.96884479414352,4.0 \sigma,FALSE,\color{Red} X,FALSE,\color{Red} X,0.099031473693835,FALSE,FALSE,\nomedal
\pedetable,0,-19.349 \pm 0.013,2.67085183996509,2.7 \sigma,3831.25,4.99,4.99,FALSE,\color{Red} X,TRUE,\color{Green} \checkmark,\color{Green} \checkmark,7.39,2.71791290194136,2.7 \sigma,TRUE,\color{Green} \checkmark,FALSE,\color{Red} X,0.017620244325081,FALSE,FALSE,\nomedal
\medetable,1,-19.400 \pm 0.022,3.62104967146468,3.6 \sigma,3823.25,-3.01,-1.01,FALSE,\color{Red} X,FALSE,\color{Red} X,\color{Red} X ,16.90,4.11153723276173,4.1 \sigma,FALSE,\color{Red} X,FALSE,\color{Red} X,0.135454524460764,FALSE,FALSE,\nomedal
\dmdecaytable,2,-19.410^{+0.012}_{-0.013},4.24931598718297,4.2 \sigma,3822.72,-3.54,0.46,FALSE,\color{Red} X,FALSE,\color{Red} X,\color{Red} X ,18.37,4.28553784399262,4.3 \sigma,FALSE,\color{Red} X,FALSE,\color{Red} X,0.008524161751893,FALSE,FALSE,\nomedal
\fracdmdecaytable,2,-19.410 \pm 0.011,4.28150738684098,4.3 \sigma,3826.96,0.70,4.70,FALSE,\color{Red} X,FALSE,\color{Red} X,\color{Red} X ,19.42,4.40681290730615,4.4 \sigma,FALSE,\color{Red} X,FALSE,\color{Red} X,0.029266683236445,FALSE,FALSE,\nomedal
\end{filecontents*}
\csvreader[/csv/head=true,tabular={l c|  c c c c | r r c | c c},/csv/respect dollar=false,/csv/respect backslash=false,
    table head={ Model& $\Delta N_{\rm param}$ & $M_B$ &  \begin{tabular}{@{}c@{}}Gaussian \\ Tension\end{tabular} & \begin{tabular}{@{}c@{}}$Q_{\rm DMAP}$ \\ Tension\end{tabular} & & $\Delta \chi^2$  & $\Delta$AIC & & \begin{tabular}{@{}c@{}}Finalist\end{tabular} & \\ \hline}]{sheets/MbTable_halofit.csv}{1=\name,2=\nparam,3=\mbsig,5=\mbtens,7=\Dchi,8=\DAIK,12=\Mbtest,10=\DAIKtest,13=\anytest,16=\newmbtens,18=\newMbtest,20=\newanytest,24=\mname}{ \name & \nparam & $\mbsig$ & $\mbtens$ & $\newmbtens$ & $\newMbtest$ & $\Dchi$  &  $\DAIK$ & $\DAIKtest $ & $\newanytest$ & \mname}}
    \caption{Same as \cref{tab:summary_MB}, but including non-linear corrections of the matter power spectrum from {\tt halofit}, and corresponding changes in the CMB lensing. We observe no significant difference in any of our conclusions compared to \cref{tab:summary_MB}. The only notable difference is the non-passing of the \idrshort{} model.\label{tab:halofit_Mb}}
    \renewcommand{\arraystretch}{1.0}
\end{table}
\end{appendix}

\newpage

\end{document}